\documentclass[12pt]{article}
\usepackage{amssymb, amsmath}
\usepackage{graphicx, graphics}
\usepackage{eucal, eufrak}
\usepackage{dsfont}

\renewcommand\baselinestretch{2}\small\normalsize

\renewcommand\baselinestretch{1}\small\normalsize

\begin{document}
\thispagestyle{empty}
\begin{center}
\large
University of Wroclaw\\
Faculty of Physics and Astronomy\\
Institute of Theoretical Physics\\ \vspace{50mm} \Huge
{\bf{Stability and properties of striped phases in systems of
interacting fermions
or hard-core bosons}}\\
{\LARGE{Volodymyr Derzhko}}\\
\end{center}
\vspace{20mm}
\begin{center}
{\Large{Thesis submitted for the degree of Doctor of Physical Sciences
at the University of Wroclaw}}
\end{center}
\vspace{35mm}
\begin{flushright}
\Large
Supervisor:\\
dr hab. Janusz J\c{e}drzejewski
\end{flushright}
\vspace{25mm}
\begin{center}
\large Wroclaw, 2006
\end{center}
\clearpage

\thispagestyle{empty} $$$$ \clearpage

\thispagestyle{empty}
\begin{center}
\large
Uniwersytet Wroc{\l}awski\\
Wydzia{\l} Fizyki i Astronomii\\
Instytut Fizyki Teoretycznej\\ \vspace{50mm} \Huge
{\bf{Stabilno{\'s}{\'c} i w{\l}asno{\'s}ci faz pasemkowych w
uk{\l}adach oddzia{\l}uj{\c{a}}cych fermion{\'o}w lub bozon{\'o}w z
twardym rdzeniem}}\\
{\LARGE{Volodymyr Derzhko}}\\
\end{center}
\vspace{20mm}
\begin{center}
{\Large{Praca doktorska}}
\end{center}
\vspace{30mm}
\begin{flushright}
\Large
praca wykonana pod kierunkiem:\\
dr. hab. Janusza J\c{e}drzejewskiego
\end{flushright}
\vspace{25mm}
\begin{center}
\large Wroc{\l}aw, 2006
\end{center}
\normalsize \clearpage

\thispagestyle{empty} $$$$ \clearpage

\thispagestyle{empty}
\begin{abstract}
In this thesis we deal with the specific collective phenomena in
condensed matter --- striped-structures formation. Such structures
are observed in different branches of condensed matter physics, like
surface physics or physics of high-temperature
superconductors. These quasi-one-dimensional objects appear in
theoretical analyses as well as in computer simulations of
different theoretical models.
Here, the main topic of interest is the stability of striped
structures in certain quantum models, where a tendency towards
crystallization competes with a tendency towards phase separation,
and some basic properties of these structures.

We consider two strongly correlated two-component quantum systems,
consisting of quantum mobile particles and immobile ones.
By immobile particles we mean those particles whose occupation numbers are
invariant with respect to the hamiltonian evolution. But this does not mean that
these occupation numbers serve as a kind of a fixed disorder field.
They are varied to reach the minimum of the energy of the total system
(an analog of annealing).
The both systems are described by Falicov-–Kimball-like Hamiltonians on
a square lattice, extended by direct short-range interactions
between the immobile particles, that favor phase separation.
In the first system the mobile
particles are spinless fermions while in the second one they are
hardcore bosons. We construct rigorously ground-state phase diagrams
of the both systems in the strong-coupling regime and at half
filling. Two main conclusions are drawn. Firstly, short-range
interactions in quantum gases are sufficient for the appearance of
charge stripe-ordered phases. When they occur, a first order phase transition
between a checkerboard-crystal phase and phase-separated state
(a segregated phase) is impossible:
by varying the intensity of a direct
nearest-neighbor interaction between the immobile particles, the
both systems can be driven from a segregated phase via striped phases
(for instance, via a diagonal-striped phase in the case of fermions,
and via vertical-
(horizontal-) striped phases in the case of hardcore bosons)
to the checkerboard phase.
Secondly, the phase diagrams of the two systems (mobile fermions or
mobile hardcore bosons) are definitely different. However, if the
strongest effective interaction in the fermionic case gets
frustrated gently, then the phase diagram becomes similar to that of
the bosonic case.

We show that any anisotropy of nearest-neighbor hopping eliminates
the $\pi/2$-rotation degeneracy of the so called dimeric and
axial-stripe phases and orients them in the direction of a weaker
hopping. Moreover, due to the same anisotropy the obtained phase
diagrams of fermions  show a tendency to become similar to those of
hardcore bosons.

Finally, introducing a next-nearest-neighbor hopping, small enough
not to destroy the striped structure, we examine rigorously how the
presence of the next-nearest-neighbor hopping anisotropy reduces the
$\pi/2$-rotation degeneracy of the diagonal-striped phase. The
effect appears to be similar to that in the case of anisotropy of
the nearest-neighbor hopping: the stripes are oriented in the
direction of the weaker next-nearest-neighbor hopping.
\end{abstract}

\clearpage

\thispagestyle{empty} $$$$ \clearpage

\pagenumbering{arabic} \tableofcontents

\clearpage
\section{Introduction}
In the passing decade, some specific phases, having
quasi-one-dimensional structure, the so called striped phases, have
been a highly debated subject in condensed-matter physics.
Apparently, a broad interest in such phases was initiated by reports
presenting experimental evidence for the existence of charge stripes
in doped layered perovskites, some of which constitute materials
exhibiting high-temperature superconductivity \cite{TBSL1,TSANU1}.

However, striped phases had been observed much earlier, for instance
in physisorbed monolayers on metallic surfaces (see \cite{KNSZGC1}
and references therein), or in ultrathin magnetic films \cite{SW1}.
Many other instances of experimental observations of stripe-ordered
phases are listed in \cite{MP1}, where Monte-Carlo studies of
formation of striped phases in a continuous gas with hardcore and
short-range repulsive interactions are reported. Theoretical
descriptions, including computer simulations, of these phenomena
involve various kinds of classical lattice-gas models (also termed
classical spin models) with competing interactions: for physisorbed
monolayers see Ref.\cite{Sasaki1}, for ultrathin magnetic films see
Ref.\cite{BMWB1}. Striped phases have been also studied in the
framework of quantum-spin models, like $XY$ model for instance
\cite{SDSS1}. Interesting theoretical considerations of stripe
phases in a two-dimensional electron gas, physically realizable in
MOSFET's, based on classical spin models with competing
interactions, can be found in Ref.\cite{Spivak1}.

Quite interestingly, theoretical studies of striped phases in
systems of strongly correlated electrons
\cite{PR1,ZG1,Machida1,KMNF1} have preceded experimental
observations of such phases. But only after those observations, the
discussion became much more vigorous, and the nature of
stripe-ordered phases started to attract attention of numerous
researchers. In the context of the Hubbard model, a comprehensive
review of the problem can be found in \cite{Oles1}. The existence of
the same kind of striped phases was investigated also in the
$t$--$J$ model \cite{WS1,WS2,TV1}. A bird's eye view on the problem
of stripe-ordered phases in high-temperature superconductors, but
emphasizing its general relevance for contemporary condensed-matter
physics can be found in \cite{EKT1}. The general relevance of
striped phases is underlined also in \cite{ZH1}, where they are
viewed as an emergent phenomenon resulting from collective motions
of microscopic particles, somewhat analogous to quasi-particles like
phonons.

The Hubbard or $t$--$J$-like models belong to the most realistic
models, in the framework of which the problem of striped-ordered
phases in doped layered perovskites can be investigated (see a
review paper by Ole{\'s} \cite{Oles1}). In the both models, the spin
and the charge degrees of freedom are taken into account, and it is
believed that it is the competition between these degrees of freedom
that is decisive for the formation of striped phases. The major
difference between Hubbard, $t$--$J$, and similar quantum-particle
models for stripe formation, on one side, and the classical or
quantum spin models (mentioned above) on the second side, is the
role of the kinetic energy. It is absent in the second group of
models while it plays a crucial role in the first group. The model
studied by us in this thesis belong to the first group.

The question of formation of striped phases is closely related to
the question of relative stability of these phases against mixtures
of an electron-reach phase and a hole-reach phase (segregated
phases). Due to the tiny energy differences between both phases, the
results obtained by means of approximate methods, which introduce
hardly-controllable errors, are disparate. Therefore, as pointed out
in \cite{ZH1,LFB1} further careful studies are necessary to settle
the problem of formation of stripe-ordered phases. One of possible
ways of attacking this problem is to formulate analog problems in
less realistic but simpler models, where some control over the
results obtained by means of various methods of statistical physics
and many-body physics can be gained. This leads hopefully to a
deeper insight into the, mentioned above, stability problem.

Such an approach has been adopted by many researchers, using models
of the both groups mentioned above. Here are two examples concerning
classical spin models. To investigate the phenomenon of frustrated
phase separation in high-temperature superconductors, L\"{o}w et al
\cite{LEFK1} consider a spin 1 two-dimensional Ising model with
short-range ferromagnetic coupling competing with long-range
antiferromagnetic Coulomb interactions, where spin variables are
coarsegrained representations of the local density of mobile holes.
By means of combined analytical and numerical techniques they show
that the transition between ferromagnetic and antiferromagnetic
states proceeds via numerous phases, among which there are striped
phases. In a recent paper, Valdez-Balderas and Stroud \cite{VS1}
investigate, by means of Monte Carlo techniques, a competition
between superconductivity and other types of order in two
dimensions. Each site of the underlying lattice, occupied by a
classical $XY$ spin (plane rotator), is interpreted by them as a
mobile, positively charged superconducting domain, and the
orientation of the spin as a phase of the superconducting order
parameter of this domain. Vacant sites represent negatively charged
nonsuperconducting domains. The nearest neighbor spins are coupled
ferromagnetically (a representation of Josephson tunneling) and the
occupied sites repel each other with a kind of a screened Coulomb
interaction. By varying the relative strength of the two competing
interactions, the  system is driven from a phase separated state to
a checkerboard-like state via a series of complex patterns of
self-organization, including some kinds of striped phases.

And here are examples concerning quantum-particle models. Buhler et
al \cite{BYM1} have found, by means of Monte Carlo simulations, that
upon hole doping antiferromagnetic spin domains and charge stripes,
whose properties are in very good agreement with experiments, appear
in a spin-fermion model for cuprates. Using the so called restricted
phase diagrams, the stability problem of charge-stripe phases has
been studied in the spinless Falicov--Kimball model by
Lema{\'{n}}ski et al \cite{LFB1,LFB2}. In their study the formation
of charge stripe-ordered phases can be looked upon as a way the
system interpolates between a periodic charge-density wave phase
(the chessboard phase) and the segregated phase (a mixture of
completely filled and completely empty phases), as the degree of
doping varies. A considerable reduction of the Hilbert space
dimension in a spinless fermion model with infinite nearest-neighbor
repulsion (as compared to a Hubbard model) has been exploited by
Zhang and Henley \cite{ZH1,ZH2} to study carefully, by means of an
exact diagonalization technique, the formation of charge
stripe-ordered phases upon doping. They addressed also an
interesting question of the role of quantum statistics in the
problem of striped phases, by replacing fermion particles with
hardcore-boson ones.

Our work has been inspired mainly by the recent studies of
Lema{\'{n}}ski et al \cite{LFB1,LFB2} and by Zhang and Henley
\cite{ZH1,ZH2}. To investigate the key question, whether charge
stripe-ordered phases are stable compared to a phase-separated
state, we study rigorously the strong-coupling limit of an extended
spinless Falicov--Kimball model on a square lattice, with mobile
particles being spinless fermions or hardcore bosons. The usual
spinless Falicov--Kimball  Hamiltonian (such as that studied in
\cite{FLU1}) has been augmented by a direct, Ising-like interaction
between the immobile particles.

In the framework of striped-structures formations, one of the
interesting questions, the influence of hopping anisotropy on
striped phases, was investigated by means of the Hartree--Fock
method in \cite{Oles1,RNO1}. They showed that under an anisotropy in
hopping intensities, stripes become oriented in the direction of
weaker hopping. This problem is investigated here rigorously, using
the introduced model.

The thesis is organized as follows. In the next section we give some
insight into the Falicov--Kimball model, providing its main
properties, and we introduce technics which are used for deriving
our results. Then, in Section 3 we present the considered models and
construct phase diagrams due to truncated effective interactions.
After that, in Sections 4 and 5, we analyze the influence of hopping
anisotropy on striped phases. Finally, we draw conclusions and
provide a summary. Various technical details a placed in Appendices
(from A to E). Our original results are contained in Sections 3,4,
and 5, and have been published in \cite{DJ2,DJ3,Derzhko},
respectively.

\clearpage
\section{The Falicov--Kimball models}
In this section we give the definition of the spinless
Falicov--Kimball model and it basic properties. First of all, we
discuss possible interpretations of the model in respect to history
of its development. We also shortly mention about today's topics of
interest. In the second paragraph we present some useful properties
of the model. The following paragraphs deal with technics we use to
investigate the model: we make a sketch of a perturbation scheme and
discuss the $m$-potential method in more details.

\subsection{Definition}
The spinless Falicov--Kimball model, on arbitrary lattice $\Lambda$
is described by the Hamiltonian:
\begin{eqnarray}
H_{FK}=-\sum\limits_{(x,y)  \subset \Lambda}
\left(t_{xy}c^{+}_{x}c_{y} +h.c.\right)+U\sum\limits_{x\in\Lambda}
\left( c^{+}_{x}c_{x} - \frac{1}{2} \right) \cdot s_{x}, \label{FK1}
\end{eqnarray}
where sites on lattice $\Lambda$ are denoted as $x,y,\ldots$, whose
number is $|\Lambda|$, and the first summation is over the pairs of
different sites $x$ and $y$, with each pair counted once. The system
described by Hamiltonian (\ref{FK1}) consists of two sorts of
spinless particles: quantum which can hop (usually between
nearest-neighbors sites) on the lattice with hopping amplitude
$t_{xy}=t^{*}_{yx}$ (we call them {\em{electrons}}) and classical
which are immobile ({\em{ions}}). The electrons, as quantum
particles, are described by creation and annihilation operators on
site $x$: $c^{+}_{x}$ and $c_{x}$, respectively, which satisfy the
appropriate commutation relations: canonical anticommutation
relations for spinless fermions or commutation relations of spin
$1/2$ operators $S^{+}_{x}$, $S^{-}_{x}$, $S^{z}_{x}$ for hard-core
bosons. The total electron-number operator is
$N_{e}=\sum_{x}n_{x}^{e}=\sum_{x}c^{+}_{x}c_{x}$ and corresponding
electron density is $\rho_{e}=N_{e}/|\Lambda|$. Classical ions, are
described by pseudo-spin (or simply, {\em{spin}}) $s_{x}$ on cite
$x$: if $s_{x}=1$, site $x$ is occupied by an ion and, if $s_{x}=-1$
the site $x$ is empty. Ions are described in terms of spins for
convenience: instead of spin variable $s_{x}$ we can use
occupation-number variable $n^{i}_{x}=(s_{x}+1)/2$.  The collection
of spins values $\{s_{x}\}_{x\in\Lambda}$ is called the {\em{ion
configuration}}. The total number of ions is
$N_{i}=\sum_{x}n_{x}^{i}=\sum_{x}(s_{x}+1)/2$ and the ion density is
$\rho_{i}=N_{i}/|\Lambda|$. There is no direct interaction either
between the electrons or between the ions, although there is the
simple on-site interaction between the two kinds of particles with
coupling constant $U$. It is clear that the particle-number
operators $N_{e}$, $N_{i}$, and spins $s_{x}$ are conserved.
Therefore, the description of the classical subsystem in terms of
the ion configurations $S =\left\{ s_{x} \right\}_{x \in \Lambda}$
remains valid. Whenever periodic configurations of pseudo-spins are
considered, it is assumed that $\Lambda$ is sufficiently large, so
that it accommodates an integer number of elementary cells.

The Falicov--Kimball model was first considered by Hubbard
\cite{Hubbard1} and Gutzwiller \cite{Gutzwiller1} as a
simplification of the one-band Hubbard model,
\begin{eqnarray}
H_{Hubbard}=-\sum\limits_{\sigma=\uparrow,\downarrow}\sum\limits_{(x,y)
\subset \Lambda} t^{(\sigma)}_{xy}
c^{+}_{x\sigma}c_{y\sigma}+2U\sum\limits_{x\in\Lambda}n_{x\uparrow}n_{x\downarrow},
\label{Hubbardh}
\end{eqnarray}
where $t^{(\uparrow)}_{xy}=0$, and $t^{(\downarrow)}_{xy}=t_{xy}$
for arbitrary sign of coupling constant $U$. This is an
approximation of the Hubbard model in the case of infinitely heavy
spin-up electrons.

In 1969 this model was introduced by Falicov and Kimball in order to
explain metal--insulator transitions in mixed valence compounds of
rare-earth materials \cite{FK1}. Experiments suggested that these
transitions occur only due to the interactions between electrons:
localized immobile $f$-electrons (ions) and $d$-electrons in
Bloch-like states. The model described by (\ref{FK1}) is the
simplification of the original model, suggested in \cite{FK1}.

In 1986, Kennedy and Lieb reinvestigated Falicov--Kimball model as a
``static Hubbard model''. It was used in \cite{KL1} and \cite{Lieb1}
to study crystallization. Some rigorous results concerning the model
were presented in these works for the first time. Mainly, they
proved the existence of the phase transition in the model for any
value of $U$ and for sufficiently low temperature.

In the last years the Falicov--Kimball model is intensively studied
in the frame of the dynamical mean field theory, where properties of
the system could be calculated exactly in the limit of infinite
dimension (see review \cite{FZ1}).

We would like to mention about the numerical studies of the
Falicov--Kimball-like models, which are provided on finite clusters
and are performed by Farka{\v{s}}ovsk{\'{y}} and co-workers (see for
e.g. \cite{CF1}, \cite{FCT1}, \cite{Farkasovsky1}).

Due to the relative simplicity of the model, a variety of
modifications have been applied to it. First of all it can be
considered on different types of lattices. In \cite{GMMU}, the model
was considered on a triangular lattice. Due to increasing interest
of such systems on lattices such as Kagom{\'{e}} or pyrochlore,
studies of the Falicov--Kimball model on them are going to be
instructive.

An interesting problem arises on changing the statistic of quantum
particles. In \cite{GMMU} the electron subsystem, whose particles
obey Fermi statistic where replaced by a system of hard-core bosons.
Since two bosons cannot occupy the same site, because of hard-core
interaction, the subsystem consists of interacting particles. A
comparison of this two statistics might be useful for computer
simulations, where fermions are often replaced by hard-core bosons,
in order to make computations easier.

One of the alterations, which was studied recently, is the so-called
{\em{correlated hopping}}, i.e. the hopping intensity $t_{xy}$
depends on occupation numbers of ions on sites $x$ and $y$.
Influence of the correlated hopping on the phase diagram of the
simple Falicov--Kimball model was investigated in \cite{GL1} in the
case of 1D model, and in \cite{WL1} and \cite{WL2} in 2D.

Another simple modification, which perhaps makes the model more
realistic, is introduction of a spin degree of freedom. It was
studied by Brandt et al. in \cite{BFH1} and \cite{BF1}. Recently,
using perturbation technique, the Falicov--Kimball model in case of
infinite Coulomb repulsion between classical ions with different
spin orientation at the same site was investigated in \cite{LW1}.
The model with an additional spin-depended Ising-type interaction
(Ising--Falicov--Kimball model) was studied recently in
\cite{Lemanski1}.

Historical review and list of new trends, as well as extended
comments, concerning  the field of the Falicov--Kimball model are
given in a short guide \cite{JL1}.

\subsection{Symmetries and other basic properties}
We are interested in ground-state properties of the Falicov--Kimball
model. We describe its ground state from the point of ion subsystem,
so the main question is the ground-state configuration of ions. Here
the model on a square lattice is considered, though many properties
of the model can be extended to a wider class of latices. The most
convenient ensemble for our purposes is the grand-canonical
ensemble. Let,
\begin{eqnarray}
H\left( \mu_{e}, \mu_{i} \right)=H_{FK}-\mu_{e}N_{e}-\mu_{i}N_{i},
\label{FK-GC}
\end{eqnarray}
where $\mu_{e}$, $\mu_{i}$ are chemical potentials of electrons and
ions, respectively.

First, we notice, that for a fixed configuration
$\{s_{x}\}_{x\in\Lambda}$ Hamiltonian (\ref{FK1}), in the case when
electrons obey Fermi statistic, describes a system of free Fermi
particles in an external field $U\cdot s_{x}$. The Hamiltonian
(\ref{FK1}) is the second-quantized version of the single particle
Hamiltonian
\begin{eqnarray}
h=-T+US, \label{1partham}
\end{eqnarray}
where $h$ is $|\Lambda|\times|\Lambda|$-matrix with the elements
$h_{xy}=t_{xy}+Us_{x}\delta_{xy}$. However, the configuration of
ions $\{s_{x}\}_{x\in\Lambda}$ is not fixed in the system: the
configuration of ions should minimize the whole Hamiltonian $H\left(
\mu_{e}, \mu_{i} \right)$, and the main question of describing the
ion subsystem is to find such a configuration for specific values of
$\left( \mu_{e}, \mu_{i} \right)$. This follows from the form of a
partition function,
\begin{eqnarray}
Z_{\Lambda}=\sum_{S} \, {\rm{Tr}}\, \exp{\left[-\beta H\left(
\mu_{e}, \mu_{i} \right) \right]},
\end{eqnarray}
where the sum runs over all the configurations and the trace is
taken over the fermion Fock space ($\beta$ is the inverse
temperature). In other words, if $E_{S}\left( \mu_{e}, \mu_{i}
\right)$ is the ground-state energy of $H\left( \mu_{e}, \mu_{i}
\right)$ for a given configuration $S$ of the ions, then we are
looking for the ground-state energy of $H\left( \mu_{e}, \mu_{i}
\right)$, which is given by,
\begin{eqnarray}
E_{G}\left( \mu_{e}, \mu_{i} \right)=\min_{S} E_{S} \left( \mu_{e},
\mu_{i} \right).
\end{eqnarray}
The minimum is attained at the set $G$ of the ground-state
configurations of ions. The task is to determine the set $G$ in the
$\left( \mu_{e}, \mu_{i} \right)$-plane.

For this purpose, it is useful to examine the spectrum of the
Hamiltonian (\ref{1partham}) (in the case of fermions). When
$\max{|t_{xy}|}=t$ and $|U|>zt$, where $z$ is the coordination
number of the underlaying lattice ($z=4$, for square lattice), there
is a gap in the spectrum $(-|U|+zt, |U|-zt)$ (the so-called
{\em{universal gap\/}}). This gap is of great importance, because if
we suppose the chemical potential of electrons $\mu_{e}$ to be in
the universal gap,
\begin{eqnarray}
\mu_{e}\in (-|U|+zt, |U|-zt), \hspace{20mm} (|U|>zt),
\label{unigap1}
\end{eqnarray}
then the ground-state energy expansion, we use below, is absolutely
convergent, and uniformly in $\Lambda$. Quite interestingly, a
similar condition holds in the case when quantum particles are
hard-core bosons,
\begin{eqnarray}
\mu_{e}\in (-|U|+4zt, |U|-4zt), \hspace{20mm} (|U|>4zt).
\label{unigap2}
\end{eqnarray}
This condition was obtained by Messager and Miracle-Sol{\'{e}} in
\cite{MM1}, using a cluster expansion. Under conditions
(\ref{unigap1}) and (\ref{unigap2}), for $U$ positive (negative) and
in the zero-temperature limit, the {\em{half-filling condition\/}},
$N_{e}+N_{i}=|\Lambda|$ ({\em{neutrality condition\/}},
$N_{e}=N_{i}$), is implied, either for fermions or hard-core bosons.

In studies of grand-canonical phase diagrams an important role is
played by unitary transformations ({\em hole--particle
transformations}) that exchange particles and holes: $c^{+}_{x}c_{x}
\rightarrow 1 - c^{+}_{x}c_{x}$, for electrons, and $s_{x}
\rightarrow -s_{x}$, for ions.

The peculiarity of the model is that the case of attraction ($U<0$)
and the case of repulsion ($U>0$) are related by a unitary
transformation (the hole--particle transformation for ions): if $S
=\left\{ s_{x} \right\}_{x \in \Lambda}$ is a ground-state
configuration at $\left( \mu_{e}, \mu_{i} \right)$ for $U>0$, then
$-S =\left\{- s_{x} \right\}_{x \in \Lambda}$ is the ground-state
configuration at $\left( \mu_{e}, -\mu_{i} \right)$ for $U<0$. This
property holds for both statistics, and does not depend on the form
of hopping term. Consequently, without any loss of generality one
can fix the sign of the coupling constant $U$. In the following
paragraphs we choose $U>0$ and express all the other parameters of
the Hamiltonian (\ref{FK-GC}) in the units of $U$, i.e. formally we
set $U=1$, preserving previous notations.

In the case, when only nearest-neighbor (n.n.) hopping terms are
present, i.e. $t_{xy}=0$ for all $\langle x,y \rangle_{i}$,
$i\geqslant 2$, this transformation for some $\left( \mu^{0}_{e},
\mu^{0}_{i} \right)$ leave the Hamiltonian $H \left( \mu_{e},
\mu_{i} \right)$ invariant. An example of such transformation for
the electrons is given by $c_{x}^{+} \rightarrow \epsilon_{x}
c_{x}$, where $\epsilon_{x} = 1$ for bosons; for fermions
$\epsilon_{x} = 1$ at the even sublattice of $\Lambda$ and
$\epsilon_{x} = -1$ at the odd one. Since $H_{FK}$ is invariant
under the joint hole--particle transformation of mobile and
localized particles (when only n.n. hopping is non-zero), $H \left(
\mu_{e}, \mu_{i} \right)$ is hole--particle invariant at the point
$(0,0)$. At the hole--particle symmetry point, the system under
consideration has very special properties, which simplify studies of
its phase diagram \cite{KL1}. Moreover, by means of the defined
above hole--particle transformations one can determine a number of
symmetries of the grand-canonical phase diagram \cite{GJL}. With the
sign of $U$ fixed, applying joint hole--particle transformation for
electrons and ions, one concludes that there is an {\em inversion
symmetry\/} of the grand-canonical phase diagram. If $S$ is a
ground-state configuration at $\left( \mu_{e}, \mu_{i} \right)$,
then $-S$ is the ground-state configuration at $\left( -\mu_{e},
-\mu_{i} \right)$. Therefore, it is enough to determine the phase
diagram in a half-plane specified by fixing the sign of one of the
chemical potentials.

Let us note, that in the case when the next-nearest-neighbor
(n.n.n.) hopping amplitude is nonzero, $t_{xy}\ne 0$ for $\langle
x,y \rangle_{2}$, Hamiltonian (\ref{FK-GC}) is not invariant with
respect to the joint hole--particle transformation for any values of
$\mu_{e}$ and $\mu_{i}$.

Let  us consider symmetry properties of the model with the following
restrictions: n.n. hopping is supposed to be independent of
direction, i.e. $t_{xy}=t$, for $\langle x,y \rangle_{1}$, and
n.n.n. hopping amplitude differs $t_{xy}=\{t_{+}, t_{-}\}$ in two
different directions. Moreover, we consider the system of fermions
only. Then applying the hole--particle transformation for electrons,
i.e. $c_{x}\rightarrow \varepsilon_{x}c_{x}^{+}$, and the
hole--particle transformation for ions, one finds that if $S$ is the
ground-state configuration at $(t,t_{+},t_{-},\mu_{e},\mu_{i})$,
then $-S$ is the ground-state configuration at
$(t,-t_{+},-t_{-},-\mu_{e},-\mu_{i})$. Thus, it is enough to
consider only one sign of n.n.n. hopping. We shall consider the case
of positive n.n.n. hopping intensities: $t_{+},t_{-}>0$. Using
another hole--particle transformation for electrons,
$c_{x}\rightarrow c_{x}^{+}$, and the hole-particle transformation
for ions, we obtain that if $S$ is the ground-state configuration at
$(t,t_{+},t_{-},\mu_{e},\mu_{i})$, then $-S$ is the ground-state
configuration at $(-t,-t_{+},-t_{-},-\mu_{e},-\mu_{i})$. Applying
consecutively the two joint (with respect to electrons and ions)
hole--particle transformations, we obtain that if $S$ is the
ground-state configuration at $(t,t_{+},t_{-},\mu_{e},\mu_{i})$, it
is also the ground-state configuration at
$(-t,t_{+},t_{-},\mu_{e},\mu_{i})$. So the relative sign of n.n. and
n.n.n. hopping amplitudes does not play any role in the model.

Above, we have briefly discussed those properties of our model,
which shall be useful in the sequel. For more extended picture of
the Falicov--Kimball model, its modifications, results and
applications, readers are referred to a series of reviews
\cite{JL1}, \cite{GM1}, \cite{Gruber1} and \cite{GU1}.

\subsection{Strong-coupling expansion of the ground-state energy at
half-filling --- the method} There are different ways of obtaining
ground-state energy expansion. In the case of fermions, when $zt<1$
and $|\mu_{e}|<1-zt$, the ground-state energy of the model can be
written in the form \cite{GJL,GM1}:
\begin{eqnarray}
E_{S} \left( \mu_{e}, \mu_{i}
\right)=-\frac{1}{2}{\rm{Tr}}|h|-(\mu_{i}-\mu_{e})N_{i}-\mu_{e}|\Lambda|,
\label{GSEinit}
\end{eqnarray}
where the one-particle Hamiltonian $h=-T+US$ is given by
(\ref{1partham}), and $|h|=\sqrt{h^{2}}$. The trace of $|h|$, can be
rewritten in the form
\begin{eqnarray}
{\rm{Tr}}|h|={\rm{Tr}}\sqrt{I+\Delta},
\end{eqnarray}
where, $I$ is the unity operator, $\Delta=T^{2}-(TS+ST)$ stands for
small perturbation, $S^{2}=I$. Expanding the square root into a
power series of $\Delta$, one obtains the ground-state energy
expansion. Such a ``square root expansion'' was applied in
\cite{GJL}.

An alternative way for fermions, the ``resolvent expansion'', was
used in \cite{GMMU}. Ground-state energy (\ref{GSEinit}) can be
rewritten in the form
\begin{eqnarray}
E_{S} \left( \mu_{e}, \mu_{i}
\right)=-\frac{1}{2}\sum\limits_{x\in\Lambda}s_{x}+\int_{\mathcal{C}}\frac{dz}{2\pi
i}{\rm{Tr}}\left[\frac{z}{z-h} \right]
-(\mu_{i}-\mu_{e})N_{i}-\mu_{e}|\Lambda|, \label{GSEres}
\end{eqnarray}
where $\mathcal{C}$ is a contour in the complex plain enclosing all
the negative eigenvalues of $h$. In order to calculate the above
integral trace of resolvent is expanded into a power series of
$t_{xy}$.

Unfortunately, the above methods are not applicable in the case when
quantum hopping particles are hard-core bosons. To obtain the
ground-state energy expansion in this case, a closed-loop expansion
is used, which was first applied in \cite{MM1} for fermions, and
then used for hard-core bosons in \cite{GMMU}.

We use another, more powerful method, the method of
unitary-equivalent interactions, which is suitable not only for both
systems, fermions and hard-core bosons, but can also be applied to a
wide class of Hamiltonians. It was proposed in a series of papers
\cite{DFF1}, \cite{DFFR1}, \cite{FR1} and summarized in \cite{DFF2}.
The main advantage of this method is that the unitary
transformations are applied locally, and that is why the method is
well-defined: the expansion is convergent, uniformly in $|\Lambda|$.

The idea of the method is the following. Considering the
Hamiltonians which are sums of local terms, $H=\sum_{X}H_{X}$ ($X$
is a finite set of sites in $\Lambda$), they are supposed to consist
of two parts: classical, which is diagonal in some basis, and small
quantum perturbation, which is off-diagonal. With $\lambda$ being
the small parameter considered in the model, we are looking for some
unitary operator $U_{\Lambda}^{(n)}(\lambda)$, which makes the
original Hamiltonian $H_{\Lambda}(\lambda)$ block-diagonal to order
$\lambda^{n}$ in some basis, i.e.
\[
H^{(n)}_{\Lambda}(\lambda)=U_{\Lambda}^{(n)}(\lambda)H_{\Lambda}(\lambda)
\left.U_{\Lambda}^{(n)}\right.^{*}(\lambda)=H^{(n)}_{0\Lambda}(\lambda)+
V^{(n)}_{\Lambda}(\lambda),
\]
where the leading block-diagonal part $H^{(n)}_{0\Lambda}(\lambda)$
contains terms up to the $n$-th order in $\lambda$, and off-diagonal
part $V^{(n)}_{\Lambda}(\lambda)$ consists of ($n+1$)-th and higher
order terms. Following this procedure, the convergence of the
expansion can be rigorously proven. Moreover, by applying quantum
Pirogov--Sinai theory, it was established that some parts of the
grand-canonical zero-temperature phase diagram of the leading part
is modified ``slightly'' when the whole expansion is taken into
account at zero or low-temperature regimes \cite{DFF2}.

Here we present a scheme of obtaining the ground-state energy
expansion for the Falicov--Kimball model up to the fourth order. We
consider the case where nearest-neighbor and next-nearest-neighbor
hopping intensities are anisotropic.

\subsubsection{Local projections}
\begin{figure}[th]
\centering \includegraphics[width=0.5\textwidth]{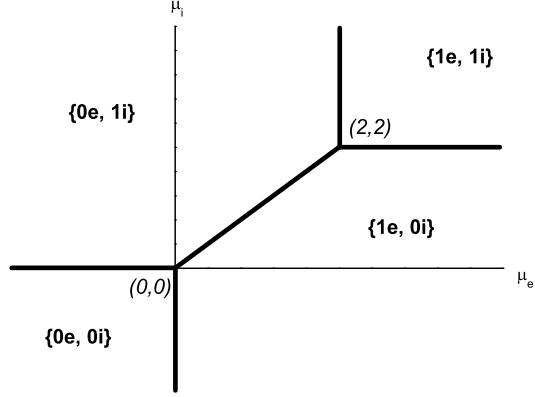}
\caption{Ground-state phase diagram for
$H_{0}=\sum\limits_{x\in\Lambda} \left(
2n^{e}_{x}n^{i}_{x}-\mu_{e}n^{e}_{x}-\mu_{i}n^{i}_{x} \right)$.
Numbers of electrons and ions at each site of lattice $\Lambda$,
which determine the state of the system, are given in curly
brackets.} \label{classinteract}
\end{figure}

First we consider the system, where quantum hopping particles obey
Fermi statistic. Following \cite{DFF2}, we rewrite FK Hamiltonian
(\ref{FK-GC}) in the form convenient for perturbation (up to some
constant), i.e. ``classical part'' plus ``small quantum
perturbation'':
\begin{eqnarray}
H_{FK}=H_{0}+V_{1}, \label{PertrFK1}
\end{eqnarray}
where the classical part,
\begin{eqnarray}
H_{0}=\sum\limits_{x\in\Lambda}\Phi_{0x}, \label{PertrClPrt1}
\end{eqnarray}
with the classical on-site potential,
\begin{eqnarray}
\Phi_{0x}=2n^{e}_{x}n^{i}_{x}-\mu_{e}n^{e}_{x}-\mu_{i}n^{i}_{x};
\label{PertrClPrt2}
\end{eqnarray}
and the quantum perturbation,
\begin{eqnarray}
V_{1}&=&\sum\limits_{\langle x,y \rangle_{1,h}\subset
\Lambda}Q_{\langle x,y \rangle_{1,h}} + \sum\limits_{\langle x,y
\rangle_{1,v}\subset\Lambda}Q_{\langle x,y \rangle_{1,v}} +
\sum\limits_{\langle x,y \rangle_{2,+}\subset\Lambda}Q_{\langle x,y
\rangle_{2,+}} + \sum\limits_{\langle x,y
\rangle_{2,-}\subset\Lambda}Q_{\langle x,y
\rangle_{2,-}}\nonumber\\
&=&\sum\limits_{i}\sum\limits_{\langle x,y
\rangle_{i}\subset\Lambda}Q_{\langle x,y \rangle_{i}}.
\label{PertrQuPrt1}
\end{eqnarray}
Operators $Q_{\langle x,y \rangle_{i}}$ are defined in the following
way:
\begin{eqnarray}
Q_{\langle x,y \rangle_{1,h}}&=& t_{h} \left(
c^{+}_{x}c_{y}+c^{+}_{y}c_{x} \right), \nonumber \\
Q_{\langle x,y \rangle_{1,v}}&=& t_{v} \left(
c^{+}_{x}c_{y}+c^{+}_{y}c_{x} \right),
\nonumber \\
Q_{\langle x,y \rangle_{2,+}}&=& t_{+} \left(
c^{+}_{x}c_{y}+c^{+}_{y}c_{x} \right), \nonumber \\
Q_{\langle x,y \rangle_{2,-}}&=& t_{-} \left(
c^{+}_{x}c_{y}+c^{+}_{y}c_{x} \right), \label{PertrQuPrt2}
\end{eqnarray}
and the role of the small parameter $\lambda$ is played by
$t=\max{\{t_{h}, t_{v}, t_{+}, t_{-}\}}$, (the so-called
{\em{strong-coupling regime}}). The method is applied in the
situation where for certain value of energy of the classical part
there are infinitely many states. Then, it is possible that the
quantum perturbation reduces this degeneracy by splitting the energy
levels. Let us look at the classical part, or the ``zeroth order''.
The phase diagram of the classical interaction
(Fig.~\ref{classinteract}) is partitioned into four parts.
On each boundary line between any two phases the degeneracy is
infinite. We are interested in the case of half-filling which
corresponds to the boundary line $\mu_{e}=\mu_{i}$, ($0<\mu_{e}<2$),
the configurations on this line with the lowest energy are
characterized by the condition: each site of lattice $\Lambda$ is
occupied by one particle only, either ion or electron.

On the line $\mu_{e}=\mu_{i}$ we introduce the projection onto the
ground state at site $x$:
\begin{eqnarray}
P^{0}_{x}=n^{e}_{x}\left( 1-n^{i}_{x} \right) +
n^{i}_{x}\left(1-n^{e}_{x} \right) = \left( n^{e}_{x}-n^{i}_{x}
\right)^{2}. \label{proj0}
\end{eqnarray}
The projection onto excited states $P_{x}^{1}=1-P_{x}^{0}$. It is
easy to extend these projections to the ground-state projection on
$Y\subset\Lambda$:
\begin{eqnarray}
P^{0}_{Y}=\prod\limits_{x\in Y}P^{0}_{x}. \label{proj00}
\end{eqnarray}
In this case unity is given by $\mathds{1}_{Y}=P^{0}_{Y}+P^{1}_{Y}$,
and hence the definition of the projection onto exited states at the
set $Y$:
\begin{eqnarray}
P^{1}_{Y}=\mathds{1}_{Y}-P^{0}_{Y} \hspace{20mm}\mbox{(note, that }
P^{1}_{Y}\ne\prod\limits_{x\in Y}P^{1}_{x}\mbox{)}. \label{proj1}
\end{eqnarray}

For calculation we need the following simple relations:
\begin{eqnarray}
c^{+}_{x}P^{0}_{x}&=&P^{1}_{x}c^{+}_{x},\nonumber\\
c^{+}_{x}P^{1}_{x}&=&P^{0}_{x}c^{+}_{x},\label{rel1}
\end{eqnarray}
If $y\ne x$, then $c^{+}_{x}$ or $c_{x}$ commute with any projection
at site $y$. These relations give immediately very useful properties
for any operator $Q_{\langle x,y\rangle_{i}}$, with pair $x\ne y$ on
lattice $\Lambda$:
\begin{eqnarray}
P_{\langle x,y\rangle_{i}}^{0} Q_{\langle x,y\rangle_{i}}&
=& Q_{\langle x,y\rangle_{i}} P_{x}^{1}P_{y}^{1}, \nonumber \\
P_{\langle x,y\rangle_{i}}^{1} Q_{\langle x,y\rangle_{i}}& =&
Q_{\langle x,y\rangle_{i}} \left( 1- P_{x}^{1}P_{y}^{1} \right).
\label{rel2}
\end{eqnarray}
Another pairs of relations which correspond to (\ref{rel1}) and
(\ref{rel2}) are obtained by Hermitian conjugation.

Using these properties and the partition of unity, for $Y=\langle
x,y \rangle_{i}$, we can rewrite $Q_{\langle x,y\rangle_{i}}$:
\begin{eqnarray}
Q_{\langle x,y\rangle_{i}}=Q^{00}_{\langle
x,y\rangle_{i}}+Q^{01}_{\langle x,y\rangle_{i}} +Q^{11}_{\langle
x,y\rangle_{i}}, \label{partpert}
\end{eqnarray}
for $i=\left\{ 1h, 1v, 2+, 2- \right\}$; where
\begin{eqnarray}
Q^{00}_{\langle x,y\rangle_{i}}&=&P^{0}_{\langle
x,y\rangle_{i}}Q_{\langle x,y\rangle_{i}}P^{0}_{\langle
x,y\rangle_{i}}=0
\nonumber \\
Q^{01}_{\langle x,y\rangle_{i}}&=&P^{0}_{\langle
x,y\rangle_{i}}Q_{\langle x,y\rangle_{i}}P^{1}_{\langle
x,y\rangle_{i}}+P^{1}_{\langle x,y\rangle_{i}}Q_{\langle
x,y\rangle_{i}}P^{0}_{\langle x,y\rangle_{i}}
\nonumber \\
&=&P^{0}_{x}P^{0}_{y}Q_{\langle
x,y\rangle_{i}}P^{1}_{x}P^{1}_{y}+P^{1}_{x}P^{1}_{y}Q_{\langle
x,y\rangle_{i}}P^{0}_{x}P^{0}_{y},
\nonumber \\
Q^{11}_{\langle x,y\rangle_{i}}&=&P^{1}_{\langle
x,y\rangle_{i}}Q_{\langle x,y\rangle_{i}}P^{1}_{\langle
x,y\rangle_{i}}.
\end{eqnarray}
So the whole Hamiltonian (\ref{PertrFK1}) is rewritten in the form:
\begin{eqnarray}
H_{FK}=H_{0}+V^{01}+V^{11},
\end{eqnarray}
where,
\begin{eqnarray}
V^{01}&=& \sum\limits_{\langle x,y \rangle_{1,h}} Q^{01}_{\langle
x,y\rangle_{1,h}} + \sum\limits_{\langle x,y
\rangle_{1,v}}Q^{01}_{\langle x,y\rangle_{1,v}} +
\sum\limits_{\langle x,y \rangle_{2,+}}Q^{01}_{\langle
x,y\rangle_{2,+}} + \sum\limits_{\langle x,y
\rangle_{2,-}}Q^{01}_{\langle x,y\rangle_{2,-}}=\nonumber
\\
&=&\sum\limits_{i} \sum\limits_{\langle x,y \rangle_{i}}
Q^{01}_{\langle x,y\rangle_{i}},
\nonumber \\
V^{11}&=& \sum\limits_{\langle x,y\rangle_{1,h}}Q^{11}_{\langle
x,y\rangle_{1,h}} + \sum\limits_{\langle x,y
\rangle_{1,v}}Q^{11}_{\langle x,y\rangle_{1,v}} +
\sum\limits_{\langle x,y \rangle_{2,+}}Q^{11}_{\langle
x,y\rangle_{2,+}} + \sum\limits_{\langle x,y
\rangle_{2,-}}Q^{11}_{\langle x,y\rangle_{2,-}}=\nonumber
\\&=&\sum\limits_{i} \sum\limits_{\langle x,y \rangle_{i}}
Q^{11}_{\langle x,y\rangle_{i}}.
\end{eqnarray}

\subsubsection{First-order unitary transformation for fermions}
We are looking for a unitary transformation, that eliminates the
first-order off-diagonal term $V^{01}$. The transformation can be
written in the form:
\begin{eqnarray}
U^{(1)}=\exp{\left( S^{(1)} \right)}, \label{1ut}
\end{eqnarray}
where the operator $S^{(1)}=S^{(1)}\left( t_{h}, t_{v}, t_{+}, t_{-}
\right)$ is of order one, i.e. it depends on
$t_{h}^{a}t_{v}^{b}t_{+}^{c}t_{-}^{d}$ with $a+b+c+d=1$. Applying
the first unitary transformation (\ref{1ut}) and using the
Lie--Schwinger series, the transformed Hamiltonian can be written as
\begin{eqnarray}
H^{(1)} & = & e^{S^{(1)}} H_{FK} e^{-S^{(1)}}=\nonumber\\
& = & H_{FK} + \left[ S^{(1)}, H_{FK} \right] +\frac{1}{2!}\left[
S^{(1)},
\left[ S^{(1)}, H_{FK} \right] \right] + \ldots =\nonumber\\
& = & \sum\limits^{\infty}_{n=0}\frac{1}{n!} \, {\rm{ad}}^{n}
S^{(1)}(H_{FK}), \label{1utexp}
\end{eqnarray}
where, the operation ${\rm{ad}} A(B)$ is defined as follows
\begin{eqnarray*}
{\rm{ad}}^{0} A(B)&=&B, \\
{\rm{ad}}^{1} A(B)&=&\left[ A,B \right], \\
{\rm{ad}}^{n} A(B)&=&\left[ A, {\rm{ad}}^{n-1} A(B) \right].
\end{eqnarray*}
Rewriting (\ref{1utexp}),
\begin{eqnarray}
H^{(1)}&=&H_{0}+V^{01}+V^{11}+ {\rm{ad}}^{1} S^{(1)}\left( H_{0}
\right) + {\rm{ad}}^{1} S^{(1)}\left( V^{01}+V^{11} \right)+\nonumber \\
&&+\sum\limits_{n=2}^{\infty} \frac{1}{n!} \, {\rm{ad}}^{n}
S^{(1)}\left( H_{0}+V^{01}+V^{11} \right),
\end{eqnarray}
and choosing the operator $S^{(1)}$ in the way to eliminate
$V^{01}$,
\[
{\rm{ad}}^{1} H_{0}\left(S^{(1)}\right)=V^{01},
\]
we obtain:
\[
H^{(1)}=H_{0}+V^{11}+ V_{2} +\sum\limits_{n=2}^{\infty} \frac{1}{n!}
\, {\rm{ad}}^{n} S^{(1)} \left( \frac{n}{n+1} V^{01}+V^{11} \right).
\]
In the above formula,
\[
V_{2}= {\rm{ad}} S^{(1)} \left( \frac{1}{2} V^{01}+V^{11} \right) .
\]

Now our problem is to find the operator $S^{(1)}$. If we suppose it
to be a sum of local operators,
\begin{eqnarray}
S^{(1)}=\sum\limits_{i}\sum\limits_{\langle x,y
\rangle_{i}}S^{(1)}_{\langle x,y \rangle_{i}},
\end{eqnarray}
in order to satisfy the condition $[H_{0},S^{(1)}]=V^{01}$,
$S^{(1)}_{\langle x,y \rangle_{i}}$ can be chosen in the form:
\begin{eqnarray}
S^{(1)}_{\langle x,y \rangle_{i}} = \sum\limits_{i}{\rm{ad}}^{-1}
{\overline{H}}_{0,\langle x,y \rangle_{i}} \left( Q_{\langle x,y
\rangle_{i}}^{01} \right),
\end{eqnarray}
where ${\overline{H}}_{0,\langle x,y
\rangle_{i}}=\Phi_{0x}+\Phi_{0y}$. Introducing the spectral
decomposition of ${\overline{H}}_{0,\langle x,y \rangle_{i}}$,
\[
{\overline{H}}_{0,\langle x,y \rangle_{i}}= \sum\limits_{j}
E^{(j)}_{\langle x,y \rangle_{i}}P^{(j)}_{\langle x,y \rangle_{i}},
\]
$S^{(1)}$ can be rewritten as
\begin{eqnarray}
S^{(1)}_{\langle x,y \rangle_{i}} =\sum\limits_{i}
\sum\limits_{\substack{j,k\\(j\ne k)}} \frac{P^{(j)}_{\langle x,y
\rangle_{i}} Q_{\langle x,y \rangle_{i}}^{01} P^{(k)}_{\langle x,y
\rangle_{i}}}{E^{(j)}_{\langle x,y \rangle_{i}}-E^{(k)}_{\langle x,y
\rangle_{i}}}.
\end{eqnarray}

In the spectral decomposition of $\Phi_{0x}$ there are three
eigensubspaces with eigenvalues $E^{(l)}_{x}$ and with the
corresponding projections $P^{(l)}_{x}$, $l=0,1,2$:
\begin{alignat*}{2}
E^{(0)}_{x} & =-\bar{\mu}, & P_{x}^{(0)} & =
n^{e}_{x}+n^{i}_{x}-2n^{e}_{x}n^{i}_{x},\\
E^{(1)}_{x} & =0, &  P_{x}^{(1)} & =
1 - n^{e}_{x} - n^{i}_{x} + n^{e}_{x}n^{i}_{x},\\
E^{(2)}_{x} & =2(1-\bar{\mu}), & \qquad P_{x}^{(2)} & =
n^{e}_{x}n^{i}_{x},
\end{alignat*}
where $\bar{\mu}=\mu_{i}=\mu_{e}$. Note, that
$P^{(0)}_{x}=P^{0}_{x}$, and $P_{x}^{(1)}+P_{x}^{(2)}=P_{x}^{1}$.
Then, the spectral decomposition of ${\overline{H}}_{0,\langle x,y
\rangle_{i}}$ consists of six eigenvalues $E^{(l)}_{\langle x,y
\rangle_{i}}$ and eigenprojections $P^{(l)}_{\langle x,y
\rangle_{i}}$, $l=0,\ldots,5$:
\begin{alignat*}{2}
E^{(0)}_{\langle x,y \rangle_{i}} & =-2\bar{\mu}, & P^{(0)}_{\langle
x,y \rangle_{i}} & =
P_{x}^{(0)}P_{y}^{(0)}, \displaybreak[0]\\
E^{(1)}_{\langle x,y \rangle_{i}} & =-\bar{\mu}, & P^{(1)}_{\langle
x,y \rangle_{i}} & = P_{x}^{(0)}P_{y}^{(1)} +
P_{x}^{(1)}P_{y}^{(0)}, \displaybreak[0]\\
E^{(2)}_{\langle x,y \rangle_{i}} & =2-3\bar{\mu}, &
P^{(2)}_{\langle x,y \rangle_{i}} & = P_{x}^{(0)}P_{y}^{(2)} +
P_{x}^{(2)}P_{y}^{(0)}, \displaybreak[0]\\
E^{(3)}_{\langle x,y \rangle_{i}} & =0, & P^{(3)}_{\langle x,y
\rangle_{i}} & =
P_{x}^{(1)}P_{y}^{(1)}, \displaybreak[0]\\
E^{(4)}_{\langle x,y \rangle_{i}} & =2(1-\bar{\mu}), &
P^{(4)}_{\langle x,y \rangle_{i}} & = P_{x}^{(1)}P_{y}^{(2)} +
P_{x}^{(2)}P_{y}^{(1)}, \displaybreak[0]\\
E^{(5)}_{\langle x,y \rangle_{i}} & =4(1-\bar{\mu}), & \qquad
P^{(5)}_{\langle x,y \rangle_{i}} & = P_{x}^{(2)}P_{y}^{(2)}.
\end{alignat*}
Considering the products $P^{(j)}_{\langle x,y
\rangle_{i}}Q^{01}_{\langle x,y \rangle_{i}}P^{(k)}_{\langle x,y
\rangle_{i}}$, one finds that only two of them are nonzero:
$P^{(0)}_{\langle x,y \rangle_{i}}Q^{01}_{\langle x,y
\rangle_{i}}P^{(4)}_{\langle x,y \rangle_{i}}$ and $P^{(4)}_{\langle
x,y \rangle_{i}}Q^{01}_{\langle x,y \rangle_{i}}P^{(0)}_{\langle x,y
\rangle_{i}}$, for any $i$. Using the projections defined above, we
can write the expression for operator $S^{(1)}$:
\begin{eqnarray}
S^{(1)}= - \frac{1}{2} \sum\limits_{i}\sum\limits_{\langle x,y
\rangle_{i}} \left( P^{0}_{x}P^{0}_{y} Q^{01}_{\langle x,y
\rangle_{i}} P^{1}_{x}P^{1}_{y} - P^{1}_{x}P^{1}_{y} Q^{01}_{\langle
x,y \rangle_{i}} P^{0}_{x}P^{0}_{y} \right). \label{1ops}
\end{eqnarray}

\begin{figure}[th]
\includegraphics[width=0.175\textwidth]{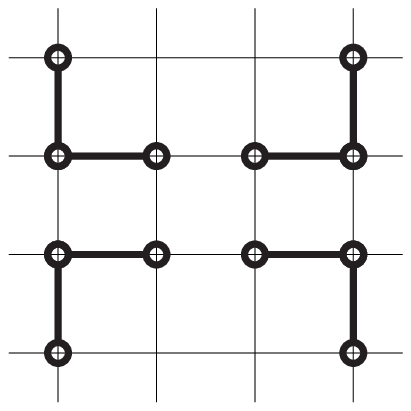}
\hspace{20mm}
\includegraphics[width=0.133\textwidth]{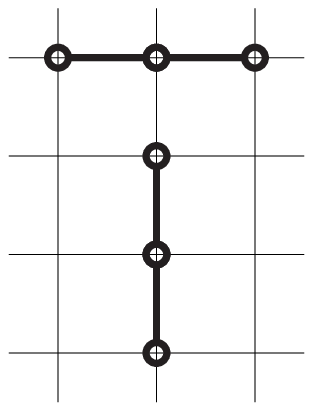}
\centering \caption{The set of paths over which the sum $\sum_{x:
\langle x,y \rangle_{1}}\sum_{z: \langle y,z\rangle_{1}}$ runs in
(\ref{prodways}). Sums run over the set of paths shown here, and
their translations. Sites $x$ and $z$ are at the ends of each path
($x$ can be chosen arbitrary), and site $y$ is always between them.}
\label{fig-ways21}
\end{figure}

\begin{figure}[th]
\centering \includegraphics[width=0.717\textwidth]{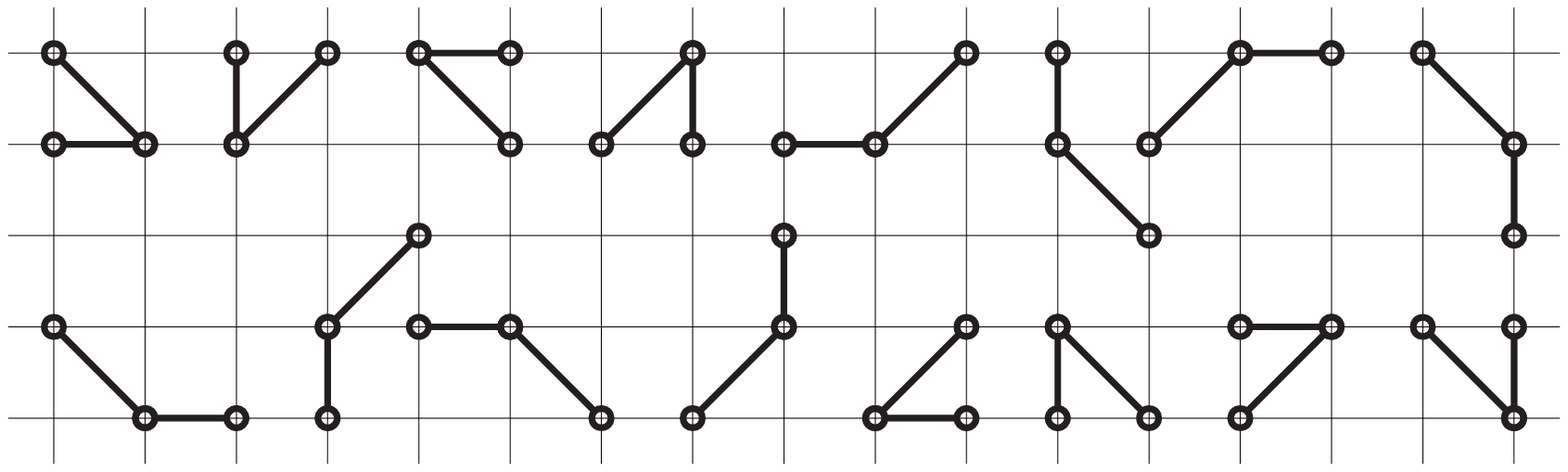}
\caption{The set of paths over which the sum $\sum_{x: \langle x,y
\rangle_{1}}\sum_{z: \langle y,z\rangle_{2}}$ runs in
(\ref{prodways}). Sums run over the set of paths shown here, and
their translations. Sites $x$ and $z$ are at the ends of each path
($x$ can be chosen arbitrary), and site $y$ is always between them.}
\label{fig-ways22}
\end{figure}

\begin{figure}[th]
\includegraphics[width=0.383\textwidth]{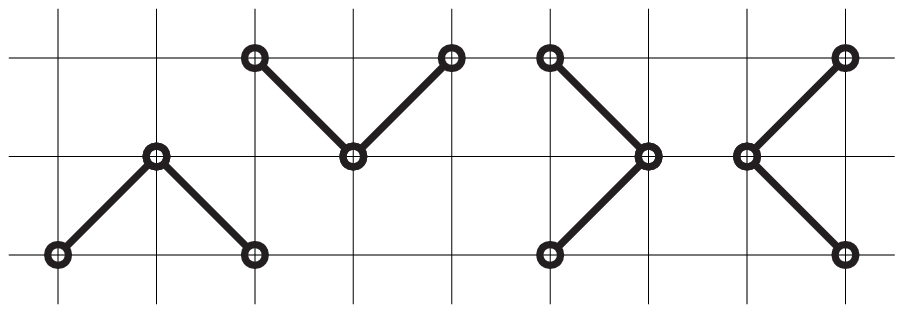}
\hspace{20mm}
\includegraphics[width=0.258\textwidth]{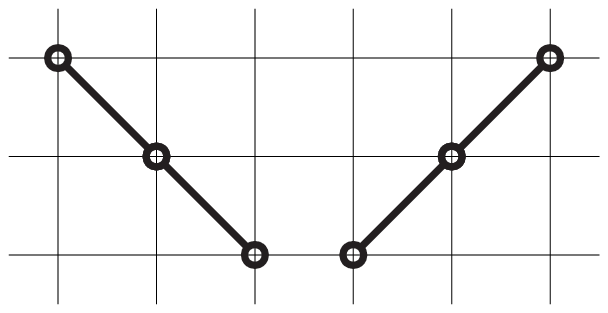}
\centering \caption{The set of paths over which the sum $\sum_{x:
\langle x,y \rangle_{2}}\sum_{z: \langle y,z\rangle_{2}}$ runs in
(\ref{prodways}). Sums run over the set of paths shown here, and
their translations. Sites $x$ and $z$ are at the ends of each path
($x$ can be chosen arbitrary), and site $y$ is always between them.}
\label{fig-ways23}
\end{figure}

To proceed, we have to consider commutators
${\rm{ad}}^{1}A(B)=AB-BA$, where $A=\sum_{i}\sum_{\langle x,y
\rangle_{i}}A_{\langle x,y \rangle_{i}}$ and
$B=\sum_{j}\sum_{\langle m,n \rangle_{j}}B_{\langle m,n
\rangle_{j}}$. First, let us note that the commutator for $\left[
A_{\langle x,y \rangle_{i}},B_{\langle m,n \rangle_{j}}  \right]$
vanishes unless $\langle x,y \rangle_{i}\cap\langle m,n
\rangle_{j}\ne\emptyset$. In this case, the product,
\begin{eqnarray}
AB  = \sum\limits _{\substack{i,j\\(i\leqslant j)}}\left(
\delta_{ij}  \sum\limits_{x: \langle x,y \rangle_{i}} A_{\langle x,y
\rangle_{i}}B_{\langle x,y \rangle_{i}}+ \sum\limits_{x: \langle x,y
\rangle_{i}} \sum\limits_{z: \langle y,z \rangle_{j}} \left(
A_{\langle x,y \rangle_{i}}B_{\langle y,z \rangle_{j}} + A_{\langle
y,z \rangle_{j}}B_{\langle x,y \rangle_{i}} \right) \right).
\label{prodways}
\end{eqnarray}
The first term in the brackets runs over all the possible pairs of
sites, i.e. $i=\{1h,1v,2+,2-\}$. In the second term we introduce the
extended indices of summation, $x: \langle x,y \rangle_{i}$ and $z:
\langle y,z \rangle_{j}$, in order to stress that first sum runs
over all $x\in\Lambda$ and pair $(x,y)$ is the n.n. pair of the
$i$-type, and the second sum runs over $z$ (shown on
Figs.~\ref{fig-ways21},\ref{fig-ways22},\ref{fig-ways23}),  such
that pair $(y,z)$ is the n.n. pair of $j$-th type. Whenever we skip
the index of direction (like `$h$' or `$-$') this means that the
summation is over all the possible directions. So, the notation
$\sum_{\langle x,y \rangle_{2}}$ stands for sum over all n.n.n.
pairs, each counted once.

The second-order term is $V_{2}$, which after somewhat lengthy
calculations can be written in the form:
\begin{align*}
V^{00}_{2}= & -\frac{1}{2} \sum\limits_{i} \sum\limits_{x: \langle
x,y \rangle_{i}} P^{0}_{x}P^{0}_{y}Q^{2}_{\langle x,y
\rangle_{i}}P^{0}_{x}P^{0}_{y}  , \displaybreak[0]
\\
V^{01}_{2}= & -\frac{1}{2} \sum\limits_{i \leqslant j}
\sum\limits_{x: \langle x,y \rangle_{i}} \sum\limits_{z: \langle y,z
\rangle_{j}} \left( P^{0}_{x}P^{0}_{y}P^{0}_{z} \left[ Q_{\langle
x,y \rangle_{i}}Q_{\langle y,z \rangle_{j}} + Q_{\langle y,z
\rangle_{j}}Q_{\langle x,y \rangle_{i}} \right]
P^{1}_{x}P^{0}_{y}P^{1}_{z}\right.
\\
&\left.\qquad \qquad \qquad \qquad +P^{1}_{x}P^{0}_{y}P^{1}_{z}
\left[ Q_{\langle x,y \rangle_{i}}Q_{\langle y,z
\rangle_{j}}+Q_{\langle y,z \rangle_{j}}Q_{\langle x,y \rangle_{i}}
\right]
P^{0}_{x}P^{0}_{y}P^{0}_{z} \right) , \displaybreak[0] \\
V^{11}_{2}= & \sum\limits_{i \leqslant j} \Biggl(
\frac{\delta_{ij}}{2} \sum\limits_{x: \langle x,y \rangle_{i}}
P^{1}_{x}P^{1}_{y}Q^{2}_{\langle
x,y \rangle_{i}}P^{1}_{x}P^{1}_{y} - \\
&-\frac{1}{2} \sum\limits_{x: \langle x,y \rangle_{i}}
\sum\limits_{z: \langle y,z \rangle_{j}} \Bigl[
P^{0}_{x}P^{0}_{y}P^{1}_{z}Q_{\langle x,y \rangle_{i}}Q_{\langle y,z
\rangle_{j}}P^{1}_{x}P^{0}_{y}P^{0}_{z} -
P^{1}_{x}P^{1}_{y}P^{0}_{z}Q_{\langle x,y \rangle_{i}}Q_{\langle y,z
\rangle_{j}}P^{0}_{x}P^{1}_{y}P^{1}_{z}
-  \\
& \phantom{-\frac{1}{2} \sum\limits_{x: \langle x,y \rangle_{i}}
\sum\limits_{z: \langle y,z \rangle_{j}} \Bigl[}-
P^{0}_{x}P^{1}_{y}P^{1}_{z}Q_{\langle y,z \rangle_{j}}Q_{\langle x,y
\rangle_{i}}P^{1}_{x}P^{1}_{y}P^{0}_{z} +
P^{1}_{x}P^{0}_{y}P^{0}_{z}Q_{\langle y,z \rangle_{j}}Q_{\langle x,y
\rangle_{i}}P^{0}_{x}P^{0}_{y}P^{1}_{z}
+ \\
&\phantom{-\frac{1}{2} \sum\limits_{x: \langle x,y \rangle_{i}}
\sum\limits_{z: \langle y,z \rangle_{j}} \Bigl[}-
P^{1}_{x}P^{1}_{y}P^{1}_{z} \left[ Q_{\langle x,y
\rangle_{i}}Q_{\langle y,z \rangle_{j}}+Q_{\langle y,z
\rangle_{j}}Q_{\langle x,y \rangle_{i}} \right]
P^{0}_{x}P^{1}_{y}P^{0}_{z} - \\
&\phantom{-\frac{1}{2} \sum\limits_{x: \langle x,y \rangle_{i}}
\sum\limits_{z: \langle y,z \rangle_{j}}
\Bigl[}-P^{0}_{x}P^{1}_{y}P^{0}_{z} \left[ Q_{\langle x,y
\rangle_{i}}Q_{\langle y,z \rangle_{j}}+Q_{\langle y,z
\rangle_{j}}Q_{\langle x,y \rangle_{i}} \right]
P^{1}_{x}P^{1}_{y}P^{1}_{z} \Bigr] \Biggr).
\end{align*}

We have got all the elements which are necessary for obtaining the
effective Hamiltonian up to the third order (the second-order
ground-state energy expansion):
\[
E^{(2)}_{S} = P^{0}_{\Lambda}H^{(1)}P^{0}_{\Lambda} =
P^{0}_{\Lambda} \left( H_{0}+V^{00}_{2} \right) P^{0}_{\Lambda}.
\]
Hence,
\[
E^{(2)}_{S} = \sum\limits_{x} \Phi_{0x} -\frac{1}{2} \sum\limits_{i}
\sum\limits_{x: \langle x,y \rangle_{i}}
P^{0}_{x}P^{0}_{y}Q^{2}_{\langle x,y\rangle_{i}}P^{0}_{x}P^{0}_{y}.
\]
In the ground state $n^{e}_{x}+n^{i}_{x}=1$, so
\[
Q^{2}_{\langle x,y\rangle_{i}} = t_{i}^{2} \left[ n^{e}_{x} \left(
1-n^{e}_{y} \right) + n^{e}_{y} \left( 1-n^{e}_{x} \right) \right] =
-\frac{t_{i}^{2}}{2} \left( s_{x}s_{y} -1 \right) ,
\]
\begin{eqnarray}
V_{2}^{00}=\sum\limits_{i}\frac{t_{i}^{2}}{4} \sum\limits_{\langle
x,y \rangle_{i}} \left( s_{x}s_{y}-1 \right) \label{corr2}
\end{eqnarray}
and finally the second-order ground-state energy expansion is of the
form:
\begin{eqnarray}
E^{(2)}_{S} = -\frac{1}{2} \sum\limits_{x} \left( \mu s_{x} + \nu
\right) + \sum\limits_{i}\frac{t_{i}^{2}}{4} \sum\limits_{\langle
x,y \rangle_{i}} \left( s_{x}s_{y}-1 \right),
\end{eqnarray}
where $\mu=\mu_{i}-\mu_{e}$ and $\nu=\mu_{i}+\mu_{e}$.

\subsubsection{Second-order unitary transformation for fermions}
Applying the second unitary transformation, we obtain:
\begin{eqnarray}
H^{(2)} & = & e^{S^{(2)}} H^{(1)} e^{-S^{(2)}} \nonumber\\
&=& H^{(1)} + {\rm{ad}}^{1} S^{(2)} \left( H^{(1)} \right) +
\frac{1}{2!} \, {\rm{ad}}^{2} S^{(2)} \left( H^{(1)} \right) +
\ldots
\nonumber\\
&=& H_{0} + V^{11} + V^{00}_{2} + V^{01}_{2} + V^{11}_{2} +
{\rm{ad}}^{1} S^{(2)} \left( H_{0} \right)
+ \frac{1}{2!} \, {\rm{ad}}^{2} S^{(1)} \left( \frac{2}{3}V^{01}V^{11} \right) + \nonumber\\
&&  + {\rm{ad}}^{1} S^{(2)} \left( V^{11} \right) + \frac{1}{3!} \,
{\rm{ad}}^{3} S^{(1)} \left( \frac{3}{4}V^{01} + V^{11} \right) +
{\rm{ad}}^{1} S^{(2)} \left( V^{00}_{2} + V^{01}_{2} +
V^{11}_{2} \right) +\nonumber\\
&& +  \frac{1}{2!} \, {\rm{ad}}^{2} S^{(2)} \left( H_{0} \right) +
\tilde{V}^{(5)}, \label{exp2}
\end{eqnarray}
where $\tilde{V}^{(5)}$ is the remainder of the fifth order:
\begin{eqnarray*}
\tilde{V}^{(5)} &=& \sum\limits_{n \geqslant 4} \frac{1}{n!} \,
{\rm{ad}}^{n} S^{(1)} \left( \frac{n}{n+1} V^{01} +V^{11} \right)+
\\&+& {\rm{ad}}^{1} S^{(2)} \left( \sum\limits_{n \geqslant 2}
\frac{1}{n!} \, {\rm{ad}}^{n} S^{(1)} \left( \frac{n}{n+1} V^{01}
+V^{11} \right)
\right) + \\
& +&\frac{1}{2!} \, {\rm{ad}}^{2} S^{(2)} \left( V^{11} + V^{00}_{2}
+ V^{01}_{2} + V^{11}_{2} + \sum\limits_{n \geqslant 2} \frac{1}{n!}
\, {\rm{ad}}^{n} S^{(1)} \left( \frac{n}{n+1} V^{01} +V^{11} \right)
\right) + \\
& +&\sum\limits_{m \geqslant 3} \frac{1}{m!} \, {\rm{ad}}^{m}
S^{(2)} \left( H_{0} + V^{11} + V^{00}_{2} + V^{01}_{2} + V^{11}_{2}
+ \sum\limits_{n \geqslant 2} \frac{1}{n!} \, {\rm{ad}}^{n} S^{(1)}
\left( \frac{n}{n+1} V^{01} +V^{11} \right) \right).
\end{eqnarray*}
In order to eliminate the second-order off-diagonal term
$V_{2}^{01}$, we put:
\[
{\rm{ad}}^{1} H_{0} \left( S^{(2)} \right) = V^{01}_{2},
\]
then
\[
H^{(2)} = H_{0} + V^{11} + V^{00}_{2} + V^{11}_{2} + V_{3} + V_{4} +
\tilde{V}^{(5)},
\]
and
\begin{eqnarray}
V_{3} & = &  {\rm{ad}}^{2} S^{(1)} \left( \frac{2}{3!} V^{01} +
\frac{1}{2!} V^{11} \right) + {\rm{ad}}^{1} S^{(2)} \left( V^{11}
\right) , \label{v3} \\
V_{4} & = &  {\rm{ad}}^{3} S^{(1)} \left( \frac{3}{4!} V^{01} +
\frac{1}{3!} V^{11} \right) + {\rm{ad}}^{1} S^{(2)} \left(
V^{00}_{2} + \frac{1}{2} V^{01}_{2} + V^{11}_{2} \right) .
\label{v4}
\end{eqnarray}

The operator $S^{(2)}$ can be chosen in the form:
\[
S^{(2)}=\sum\limits_{i \leqslant j} \sum\limits_{x: \langle x,y
\rangle_{i}} \sum\limits_{z: \langle y,z \rangle_{j}} S^{(2)}_{
\langle x,y \rangle_{i} \cup \langle y,z \rangle_{j}},
\]
where, analogously as for $S^{(1)}$,
\[
S^{(2)}_{ \langle x,y \rangle_{i} \cup \langle y,z \rangle_{j}} =
\sum\limits_{\alpha,\beta} P^{(\alpha)}_{\langle x,y \rangle_{i}
\cup \langle y,z \rangle_{j}} \frac{V^{01}_{2,\langle x,y
\rangle_{i} \cup \langle y,z \rangle_{j}}}{E^{(\alpha)}_{\langle x,y
\rangle_{i} \cup \langle y,z \rangle_{j}}-E^{(\beta)}_{\langle x,y
\rangle_{i} \cup \langle y,z \rangle_{j}}}P^{(\beta)}_{\langle x,y
\rangle_{i} \cup \langle y,z \rangle_{j}}.
\]
The spectral decomposition used in the above formula, is given by
\[
\Phi_{0x} + \Phi_{0y} + \Phi_{0z} = \sum\limits_{\alpha}
E^{(\alpha)}_{\langle x,y \rangle_{i} \cup \langle y,z \rangle_{j}}
P^{(\alpha)}_{\langle x,y \rangle_{i} \cup \langle y,z \rangle_{j}}.
\]
Operator $S^{(2)}$ is constructed in the similar way as $S^{(1)}$
was. It reads:
\begin{eqnarray}
S^{(2)}&=&\frac{1}{4}\sum\limits_{i \leqslant j} \sum\limits_{x:
\langle x,y \rangle_{i}} \sum\limits_{z: \langle y,z
\rangle_{j}}\left( P_{\langle x,y \rangle_{i} \cup \langle y,z
\rangle_{j}}^{0} \left[ Q_{\langle x,y \rangle_{i}} Q_{\langle y,z
\rangle_{j}} + Q_{\langle y,z \rangle_{j}} Q_{\langle x,y
\rangle_{i}} \right] P_{\langle x,y \rangle_{i} \cup \langle y,z
\rangle_{j}}^{1} - \right.
\nonumber \\
&& \left.  \qquad\qquad\qquad -P_{\langle x,y \rangle_{i} \cup
\langle y,z \rangle_{j}}^{1} \left[ Q_{\langle x,y \rangle_{i}}
Q_{\langle y,z \rangle_{j}} + Q_{\langle y,z \rangle_{j}} Q_{\langle
x,y \rangle_{i}} \right] P_{\langle x,y \rangle_{i} \cup \langle y,z
\rangle_{j}}^{0} \right).
\end{eqnarray}

The next step, consists in calculating double commutators,
$[A,[B,C]]$ that appear in in (\ref{v3}) and (\ref{v4}). Each
commutant $X$, of these commutators, has the form $X=\sum_{x:\langle
x,y \rangle_{i}} X_{\langle x,y \rangle_{i}}$. Applying the same
argument as before (commutators vanish unless commutants are defined
on subsets intersection of which is not empty), we restrict
ourselves to the products where the sets of sites, on which each
multiplier is defined, forms a connected path, i.e. each site on a
path has nearest neighbor, or next-nearest neighbor, or both.
Another thing, worth mentioning, is that, even considering the
fourth order only, we do not need to calculate all the expressions
in (\ref{v3}) and (\ref{v4}). The result, we are interested in, is
projected onto ground state, hence, some of the terms vanish. The
following property of operators $Q_{\langle x,y \rangle_{i}}$ is
useful. Because of relations (\ref{rel1}) it is clear, that if we
consider the path, $path=\{i_{1},i_{2},\ldots,i_{q}\}$, the
expression
\begin{eqnarray}
P^{0}_{path}Q_{\langle i_{1},i_{2} \rangle_{a}}Q_{\langle
i_{3},i_{4} \rangle_{b}} \cdots Q_{\langle i_{q-1},i_{q}
\rangle_{z}} P^{0}_{path} \ne 0, \label{closedpathdef}
\end{eqnarray}
only if each of indices $i_{j}$ in the subscript of $Q$ repeated
even number of times. With this condition satisfied, let us note
that any operator $A^{01}=P^{0}AP^{1}+P^{1}AP^{0}$ or
$A^{11}=P^{1}AP^{1}$ is equal to zero after projecting onto ground
states. We will use these properties during calculations third- and
fourth order terms.

\begin{figure}[th]
\centering
\includegraphics[width=0.342\textwidth]{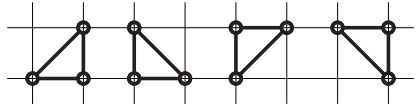} \caption{The set of
paths of $P_{1}$-type which appear in the third-order calculations.}
\label{path1}
\end{figure}

\paragraph{Third order.} The calculation of the third order terms is
the calculation of $V_{3}$. First note, that in our case there is
only one type of path that consists of three pairs of n.n. or n.n.n
sites and product of three operators $Q$ on it satisfies the
condition (\ref{closedpathdef}) (see Fig.~\ref{path1}).
In addition, after ground-state projection only $V^{00}_{3}$ is
nonzero. Using the multiplication rule in a way described above, we
obtain:
\begin{eqnarray*}
V^{00}_{3}&=& \frac{1}{4}\sum\limits_{P_{1}}
P^{0}_{x}P^{0}_{y}P^{0}_{z} \left( Q_{1xy}Q_{1yz}Q_{2xz} +
Q_{1yz}Q_{1xy}Q_{2xz} +
Q_{1xy}Q_{2xz}Q_{1yz} + \right. \\
&&\left. \hspace{20mm} + Q_{1yz}Q_{2xz}Q_{1xy} +
Q_{2xz}Q_{1xy}Q_{1yz} + Q_{2xz}Q_{1yz}Q_{1xy} \right)
P^{0}_{x}P^{0}_{y}P^{0}_{z},
\end{eqnarray*}
or in terms of spin variables:
\begin{eqnarray}
V^{00}_{3}&=& \frac{3}{4}\left[ -t_{h}t_{v}\left( t_{+}+t_{-}
\right)\sum\limits_{x}s_{x}
+\frac{t_{h}t_{v}t_{+}}{2}\sum\limits_{P_{1,+}}s_{P_{1,+}}
+\frac{t_{h}t_{v}t_{-}}{2}\sum\limits_{P_{1,-}}s_{P_{1,-}} \right],
\label{corr3}
\end{eqnarray}
where the set of paths $P_{1}$ splits into two parts, $P_{1,+}$ and
$P_{1,-}$, where n.n.n. part of the path has the slope $+1$ and
$-1$, respectively. The $s_{P_{1,i}}$ stands for the product of
spins which belong to $P_{1,i}$. This is the third-order correction
to the effective Hamiltonian.

\begin{figure}[th]
\centering
\includegraphics[width=0.758\textwidth]{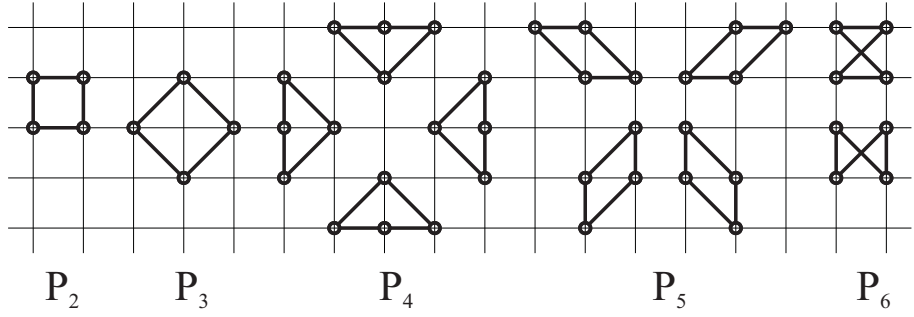} \caption{The sets of paths
$P_{2}$, ... , $P_{6}$ which appear in the fourth order
calculations.} \label{path2}
\end{figure}

\paragraph{Fourth order.} To consider fourth-order products, we
start with describing connected paths that satisfy condition
(\ref{closedpathdef}) for a product of four operators $Q$. There are
three types of such paths: $X$-type path, i.e the path consisting of
the pair (n.n. or n.n.n.) of two sites $X$; $XY$-type path, i.e. the
path consisting of three sites that formed by two pairs $X$ and $Y$;
$XYZW$-type path, i.e. the path  consisting of four sites which are
in the form of closed loop and are formed by four different pairs of
sites, --- $X$, $Y$, $Z$ and $W$. The corresponding product of $Q$'s
to the $X$-type path is $Q^{4}_{X}$. The product proportional to
$Q^{2}_{X}Q^{2}_{Y}$ corresponds to $XY$-type path. The product of
the type $Q_{X}Q_{Y}Q_{Z}Q_{W}$ corresponds to the $XYZW$-type path.
Only these types of products contribute to the ground-state
projection of $V_{4}$. Another thing connected with calculations, is
that some terms, like ${\rm{ad}}^{1} S^{(2)}(V_{2}^{11})$, vanish on
projecting onto the ground-state. Performing all necessary
calculations, we finally obtain:
\begin{align*}
8 \cdot V^{00}_{4} &= \sum\limits_{\alpha} \sum\limits_{x: \langle
x,y \rangle_{\alpha}} Q^{4}_{\langle x,y \rangle_{\alpha}}
+ \displaybreak[0]\\
&+\sum\limits_{\substack{\alpha,\beta\\(\alpha\leqslant\beta)}}
\sum\limits_{x: \langle x,y \rangle_{\alpha}} \sum\limits_{z:
\langle y,z \rangle_{\beta}} \left( 2Q^{2}_{\langle x,y
\rangle_{\alpha}}Q^{2}_{\langle y,z \rangle_{\beta}} - Q_{\langle
x,y \rangle_{\alpha}}Q^{2}_{\langle y,z \rangle_{\beta}}Q_{\langle
x,y \rangle_{\alpha}} - Q_{\langle y,z
\rangle_{\beta}}Q^{2}_{\langle x,y \rangle_{\alpha}}Q_{\langle y,z
\rangle_{\beta}} \right)
- \displaybreak[0]\\
&-\sum\limits^{6}_{i=2} \sum\limits_{x: x \in P_{i}} \Biggl(
\frac{1}{2}\cdot \left[ Q_{\langle x,y \rangle_{\alpha}}Q_{\langle
z,j \rangle_{\gamma}}+Q_{\langle z,j \rangle_{\gamma}}Q_{\langle x,y
\rangle_{\alpha}} \right] \cdot \left[ Q_{\langle y,z
\rangle_{\beta}}Q_{\langle x,j \rangle_{\omega}}+Q_{\langle x,j
\rangle_{\omega}}Q_{\langle y,z \rangle_{\beta}} \right]
+ \\
& \phantom{-\sum\limits^{6}_{i=2} \sum\limits_{x: x \in
P_{i}}\Biggl(}+ \frac{1}{2}\cdot \left[ Q_{\langle y,z
\rangle_{\beta}}Q_{\langle x,j \rangle_{\omega}}+Q_{\langle x,j
\rangle_{\omega}}Q_{\langle y,z \rangle_{\beta}} \right] \cdot
\left[ Q_{\langle x,y \rangle_{\alpha}}Q_{\langle z,j
\rangle_{\gamma}}+Q_{\langle z,j \rangle_{\gamma}}Q_{\langle x,y
\rangle_{\alpha}} \right]
+\displaybreak[0]\\
& \phantom{-\sum\limits^{6}_{i=2} \sum\limits_{x: x \in
P_{i}}\Biggl(}+ \left[ Q_{\langle x,y \rangle_{\alpha}}Q_{\langle
y,z \rangle_{\beta}}+Q_{\langle y,z \rangle_{\beta}}Q_{\langle x,y
\rangle_{\alpha}} \right] \cdot \left[ Q_{\langle z,j
\rangle_{\gamma}}Q_{\langle x,j \rangle_{\omega}}+Q_{\langle x,j
\rangle_{\omega}}Q_{\langle z,j \rangle_{\gamma}} \right]
+\\
& \phantom{-\sum\limits^{6}_{i=2} \sum\limits_{x: x \in
P_{i}}\Biggl(}+ \left[ Q_{\langle y,z \rangle_{\beta}}Q_{\langle z,j
\rangle_{\gamma}}+Q_{\langle z,j \rangle_{\gamma}}Q_{\langle y,z
\rangle_{\beta}} \right] \cdot \left[ Q_{\langle x,j
\rangle_{\omega}}Q_{\langle x,y \rangle_{\alpha}}+Q_{\langle x,y
\rangle_{\alpha}}Q_{\langle x,j \rangle_{\omega}} \right]
+\\
& \phantom{-\sum\limits^{6}_{i=2} \sum\limits_{x: x \in
P_{i}}\Biggl(}+ \left[ Q_{\langle z,j \rangle_{\gamma}}Q_{\langle
x,j \rangle_{\omega}}+Q_{\langle x,j \rangle_{\omega}}Q_{\langle z,j
\rangle_{\gamma}} \right] \cdot \left[ Q_{\langle x,y
\rangle_{\alpha}}Q_{\langle y,z \rangle_{\beta}}+Q_{\langle y,z
\rangle_{\beta}}Q_{\langle x,y \rangle_{\alpha}} \right]
+\\
& \phantom{-\sum\limits^{6}_{i=2} \sum\limits_{x: x \in
P_{i}}\Biggl(}+ \left[ Q_{\langle x,j \rangle_{\omega}}Q_{\langle
x,y \rangle_{\alpha}}+Q_{\langle x,y \rangle_{\alpha}}Q_{\langle x,j
\rangle_{\omega}} \right] \cdot \left[ Q_{\langle y,z
\rangle_{\beta}}Q_{\langle z,j \rangle_{\gamma}}+Q_{\langle z,j
\rangle_{\gamma}}Q_{\langle y,z \rangle_{\beta}} \right] \Biggr) .
\end{align*}
The paths $P_{i}$ are shown in Fig.~\ref{path2}.
We group the terms and rewrite it using spins variables. Note, that
during this procedure the summation over $P_{6}$ is replaced.
Summing up, the fourth-order correction to the ground-state energy
expansion, reads
\begin{align}
V^{00}_{4}&= \left( t_{h}^{4}+t_{v}^{4}+t_{+}^{4}+t_{-}^{4} \right)
\frac{|\Lambda |}{32}
-\displaybreak[0]\nonumber\\
&- \left( \frac{3}{16}t_{h}^{4}+\frac{3}{8}t^{2}_{h}t^{2}_{v}+
\frac{3}{16}\left(2t^{2}_{h}-t_{v}^{2} \right)\left(
t_{+}^{2}+t_{-}^{2} \right) +\frac{1}{8} \left(3t_{h}^{2}+2t_{v}^{2}
\right)t_{+}t_{-} \right) \sum\limits_{\langle x,y
\rangle_{1,h}}s_{x}s_{y}
+\displaybreak[0]\nonumber\\
&- \left( \frac{3}{16}t_{v}^{4}+\frac{3}{8}t^{2}_{h}t^{2}_{v}+
\frac{3}{16}\left(2t^{2}_{v}-t_{h}^{2} \right)\left(
t_{+}^{2}+t_{-}^{2} \right) +\frac{1}{8} \left(3t_{v}^{2}+2t_{h}^{2}
\right)t_{+}t_{-} \right) \sum\limits_{\langle x,y
\rangle_{1,v}}s_{x}s_{y}
+\displaybreak[0]\nonumber\\
&+ \left( \frac{3}{16}t_{h}^{2}t_{v}^{2} -\frac{3}{16}\left(
t_{h}^{2}+t_{v}^{2} \right)\left( 2t_{+}^{2}+t_{+}t_{-} \right)
-\frac{3}{16}t^{4}_{+}- \frac{3}{8}t_{+}^{2}t_{-}^{2} \right)
\sum\limits_{\langle x,y \rangle_{2,+}}s_{x}s_{y}
+\displaybreak[0]\nonumber\\
&+ \left( \frac{3}{16}t_{h}^{2}t_{v}^{2} -\frac{3}{16}\left(
t_{h}^{2}+t_{v}^{2} \right)\left( 2t_{-}^{2}+t_{+}t_{-} \right)
-\frac{3}{16}t^{4}_{-}- \frac{3}{8}t_{+}^{2}t_{-}^{2} \right)
\sum\limits_{\langle x,y \rangle_{2,-}}s_{x}s_{y}
+\displaybreak[0]\nonumber\\
&+ \left( \frac{1}{8}t_{h}^{4}-\frac{1}{8}t_{h}^{2}t_{+}t_{-}
+\frac{3}{16}t_{+}^{2}t_{-}^{2} \right) \sum\limits_{\langle x,y
\rangle_{3,h}}s_{x}s_{y} + \left(
\frac{1}{8}t_{v}^{4}-\frac{1}{8}t_{v}^{2}t_{+}t_{-}
+\frac{3}{16}t_{+}^{2}t_{-}^{2} \right) \sum\limits_{\langle x,y
\rangle_{3,v}}s_{x}s_{y}
+\displaybreak[0]\nonumber\\
&+\frac{3}{16}t_{h}^{2}t_{+}^{2}\sum\limits_{\langle x,y
\rangle_{4,h+}}s_{x}s_{y}
+\frac{3}{16}t_{h}^{2}t_{-}^{2}\sum\limits_{\langle x,y
\rangle_{4,h-}}s_{x}s_{y}
+\frac{3}{16}t_{v}^{2}t_{+}^{2}\sum\limits_{\langle x,y
\rangle_{4,v+}}s_{x}s_{y}
+\frac{3}{16}t_{v}^{2}t_{-}^{2}\sum\limits_{\langle x,y
\rangle_{4,v-}}s_{x}s_{y}+ \displaybreak[0]\nonumber\\
&+ \frac{1}{8}t_{+}^{4} \sum\limits_{\langle x,y
\rangle_{5,+}}s_{x}s_{y} + \frac{1}{8}t_{-}^{4} \sum\limits_{\langle
x,y \rangle_{5,-}}s_{x}s_{y}+ \frac{1}{16}\left[ t_{h}^{2}t_{v}^{2}
+ \left( t_{h}^{2}+t_{v}^{2} \right) t_{+}t_{-} \right]
\sum\limits_{P_{2}} \left( 5s_{P_{2}}+1 \right)
+\displaybreak[0]\nonumber\\
&+ \frac{1}{16}t_{+}^{2}t_{-}^{2} \sum\limits_{P_{3}} \left(
5s_{P_{3}}+1 \right) +\frac{1}{16}t_{h}^{2}t_{+}t_{-}
\sum\limits_{P_{4,h}} \left( 5s_{P_{4,h}}+1 \right)
+\frac{1}{16}t_{v}^{2}t_{+}t_{-} \sum\limits_{P_{4,v}} \left(
5s_{P_{4,v}}+1 \right)
+\displaybreak[0]\nonumber\\
&+ \frac{1}{16}t_{h}^{2}t_{+}^{2} \sum\limits_{P_{5,h+}} \left(
5s_{P_{5,h+}}+1 \right) + \frac{1}{16}t_{h}^{2}t_{-}^{2}
\sum\limits_{P_{5,h-}} \left( 5s_{P_{5,h-}}+1 \right)
+\displaybreak[0]\nonumber\\
&+ \frac{1}{16}t_{v}^{2}t_{+}^{2} \sum\limits_{P_{5,v+}} \left(
5s_{P_{5,v+}}+1 \right) + \frac{1}{16}t_{v}^{2}t_{-}^{2}
\sum\limits_{P_{5,v-}} \left( 5s_{P_{5,v-}}+1 \right). \label{corr4}
\end{align}

The whole ground-state energy expansion up to the fourth order,
\begin{eqnarray}
E^{(4)}_{S}=H_{0}+V^{00}_{2}+V^{00}_{3}+V^{00}_{4} + R^{(4)},
\label{effh}
\end{eqnarray}
where $V^{00}_{2}$, $V^{00}_{3}$ and $V^{00}_{4}$ are given by
(\ref{corr2}), (\ref{corr3}) and (\ref{corr4}), respectively. The
remainder $R^{(4)}$ is independent of the chemical potentials, and
collects all the terms proportional to
$t_{h}^{a}t_{v}^{b}t_{+}^{c}t_{-}^{d}$ , with $a+b+c+d = 5, 6,
\ldots$.

\subsubsection{The case of hard-core bosons}
On replacing the hopping fermions with hard-core bosons, the
commutation rules change: operators of creation and annihilation at
different sites commute. Using the same technic as in the case of
fermions we obtain the ground-state energy expansion in the case of
hard-core bosons. For the system of hard-core boson we consider only
the case $t_{+}=t_{-}=0$. The second order correction to the
``classical'' part $H_{0}$ remains the same as for fermions. The
third order disappears (due to the absence of n.n.n. hopping, on the
square lattice the path that satisfy condition (\ref{closedpathdef})
for a product of three $Q$ operators does not exist), and the
fourth-order correction, reads:
\begin{align}
\left( V^{00}_{4} \right)_{hcb} =& \frac{t^{4}}{16} | \Lambda | -
\left( \frac{3t_{h}^{4}}{16} + \frac{t_{h}^{2}t_{v}^{2}}{8} \right)
\sum\limits_{\langle x,y \rangle_{1,h}}s_{x}s_{y} - \left(
\frac{3t_{v}^{4}}{16} + \frac{t_{h}^{2}t_{v}^{2}}{8} \right)
\sum\limits_{\langle x,y \rangle_{1,v}}s_{x}s_{y} +
\frac{5t_{h}^{2}t_{v}^{2}}{16} \sum\limits_{\langle x,y
\rangle_{2}}s_{x}s_{y} + \nonumber
\\
+& \frac{t_{h}^{4}}{8} \sum\limits_{\langle x,y
\rangle_{3,h}}s_{x}s_{y} + \frac{t_{v}^{4}}{8} \sum\limits_{\langle
x,y \rangle_{3,v}}s_{x}s_{y} - \frac{t_{h}^{2}t_{v}^{2}}{16}
\sum\limits_{P_{2}}\left( s_{P_{2}} + 5 \right). \label{corr4b}
\end{align}

\subsubsection{Building the ground-state phase diagram}
We make some remarks on the expression of the ground-state energy
expansion (\ref{effh}) in both cases: hopping fermions and hard-core
bosons (in the case of hard-core bosons $V^{00}_{3}=0$ and
$V^{00}_{4}$ in (\ref{effh}) is replaced with $(V^{00}_{4})_{hcb}$,
given by (\ref{corr4b})). In the above expressions we keep even the
terms, that are independent of spin variables, hence have no
influence on the phase diagram and so can be omitted. The effective
Hamiltonian (\ref{effh}) depends only on $\mu=\mu_{i}-\mu_{e}$,
which we call {\em{the unique chemical potential parameter}}.

The procedure of building the phase diagram is recursive. The phase
diagram of the effective interaction truncated at the order
$k^{\prime}$ is constructed on the basis of the phase diagram
obtained at the preceding order $k$: the conditions imposed on the
ground-state configurations by the $k$-th order effective
Hamiltonian have to be obeyed by the ground-state configurations of
the $k^{\prime}$-th order effective Hamiltonian. In other words, the
$k^{\prime}$-th order terms of the expansion cannot change the
hierarchy of configuration's energies established by the $k$-th
order effective Hamiltonian; they can only split the energies of
configurations in cases of degeneracy.

\subsection{$m$-potential method}
To proceed, we need to introduce some definitions. First, as was
already mentioned, we denote the configuration of spins on the
lattice $\Lambda$ as $S=\{ s_{x} \}_{x\in\Lambda}$, and $s_{x}$
takes two values, either $+1$ or $-1$, at any site $x$. We call
$S_{A}$ the restriction of the configuration $S$ to the subset of
sites $A\subset\Lambda$ with $|A|$ sites, i.e. $S_{A}=\{ s_{x}
\}_{x\in A}$. Instead of talking about functions of spin variables
$s$, it is convenient to use the function of configuration $S$ (spin
$s_{x}$ depends on the configuration at site $x$, $S_{x}$). For
$A\subset\Lambda$, let $f_{A}$ be a function of configurations such
that,
\[
f_{A}(S)=f_{A}(S_{A}), \hspace{20mm} \mbox{for any }S.
\]

\subsubsection{Definition of $m$-potentials}
The concept of the $m$-potential was introduced in \cite{Slawny},
and its idea is the following. Suppose that for some
$A\subset\Lambda$, it is possible to write a Hamiltonian, which is a
function of configurations, in the form,
\begin{eqnarray}
H_{\Lambda}(S)=\sum\limits_{A}H_{A}(S),
\end{eqnarray}
where by the sum $\sum_{A}H_{A}$ we understand the summation over
all translations of $A$ in $\Lambda$. Assume, that there exists a
certain set of periodic configurations $\{\bar{S}\}$ (global
configurations) which minimize the potential $H_{A}(S)$ for
$\{\bar{S}_{A}\}$ (local configurations). If it holds for all
translations of $A$ in $\Lambda$, then we say that $H_{A}$ is an
$m$-potential and the set $\{\bar{S}\}$ is the set of the
ground-state configurations.

Let us present two examples. Consider the antiferromagnetic Ising
model on a square lattice ($J>0$),
\begin{eqnarray}
H_{Ising}=J\sum\limits_{\langle x,y \rangle_{1}} s_{x}s_{y}.
\label{Ising}
\end{eqnarray}
The potential $H_{\langle x,y \rangle_{1}}(S)=Js_{x}s_{y}$ is an
$m$-potential. Indeed, the local minimum for this potential is for
configurations of nearest-neighbor pairs of spins where one is up
and the second is down: $\{-1,1\}$ or $\{1,-1\}$. There exist two
global configurations, called the chessboard configurations, which
minimize the potentials $H_{\langle x,y \rangle_{1}}$ for all
$\langle x,y \rangle_{1}\subset\Lambda$: if we choose any
nearest-neighbor pair of sites, then the local configuration of the
pair will be one of those which minimize the $H_{\langle x,y
\rangle_{1}}$. In other words, if we can extend the local
configuration (which is the ground-state configuration locally) to
the global one, in a way that other local configurations with higher
energy do not appear, we call the local potential $H_{A}$ the
$m$-potential, and the global configuration is the ground-state
configuration because every potential under the sum has minimum on
it.

Now, we consider the same Ising model (\ref{Ising}) on a triangular
lattice. As in the case of square lattice, local nearest-neighbor
configurations remain the same. But the problem appears while trying
to extend the local configurations to the whole lattice $\Lambda$:
there is no such global configuration, which is ground-state
configuration locally. This means that $H_{\langle x,y \rangle_{1}}$
is not an $m$-potential in this case (rewriting (\ref{Ising}) as a
sum over elementary triangles, it appears that the corresponding
potential is $m$-potential, however the number of ground states is
infinite).

Practically, potentials depend on some parameters and often they are
$m$-potentials only for some range of these parameters. In order to
find the complete phase diagram, i.e. all ground-state
configurations for the whole range of the parameters, we are forced
to build new $m$-potential. There is no recipe for constructing
$m$-potentials. The only way to deal with the problem, as far as we
know, is to introduce the so-called zero-potentials.

\subsubsection{Definition of zero-potentials}
The idea is simple: if some local potential does not satisfy the
$m$-potential condition, then let's change it by adding zero, which
means that we change it locally, but in a way that the total
Hamiltonian remains the same. So the most general definition of
zero-potential reads: a potential $K_{A}$, is a zero potential if it
satisfies the zero-potential condition:
\begin{eqnarray}
\sum\limits_{A}K_{A}(S)=0, \hspace{20mm} \mbox{for any } S.
\label{0potcond}
\end{eqnarray}
Now, it is clear that adding such a potential to the $H_{\Lambda}$,
\begin{eqnarray}
H_{\Lambda}(S)=\sum\limits_{A}H_{A}(S)=\sum\limits_{A}\left(
H_{A}(S)+K_{A}(S)\right)=\sum\limits_{A}\widetilde{H}_{A}(S) ,
\hspace{20mm} \mbox{for any }S, \label{newpot}
\end{eqnarray}
modifies only the local potential $H_{A}$. Under the zero-potential
condition (\ref{0potcond}), $K_{A}$ has no definite structure. In
the sequel we shall consider two ways of constructing $K_{A}$.

\subsubsection{First kind zero-potentials}
In physics literature one can find two different ways of
constructing zero-potentials. The first method was proposed in
\cite{GJL}, where a part of the ground-state phase diagram of the
Falicov--Kimball model was obtained for the first time in the fourth
order. To illustrate the idea of this method we provide an example.
Let us consider Hamiltonian $H_{\Lambda}$ with a local potential
$H_{A}$ which is not an $m$-potential. Now we take some function (or
several functions) of configurations on a set of sites $A$. For
example, one of the simplest choices is $\alpha\sum\limits_{x\in
A}s_{x}$, where $\alpha$ is a number. We add and subtract this
function, so the Hamiltonian does not change,
\begin{eqnarray}
H_{\Lambda}=\sum\limits_{A}\left( H_{A} + \alpha\sum\limits_{x\in
A}s_{x} - \alpha\sum\limits_{x\in A}s_{x} \right).
\end{eqnarray}
Next we chose a new set of sites $A^{\prime}$ (or several sets), and
rewrite the Hamiltonian in the form:
\begin{eqnarray}
H_{\Lambda}&=&\sum\limits_{A}\left( H_{A} + \alpha\sum\limits_{x\in
A}s_{x} \right) - \alpha\cdot|A|\sum\limits_{x\in \Lambda}s_{x} =
\sum\limits_{A}\left( H_{A} + \alpha\sum\limits_{x\in A}s_{x}
\right) + \nonumber \\
&&+ \sum\limits_{A^{\prime}} \left( -\alpha\frac{|A|}{|A^{\prime}|}
\right) \sum\limits_{x\in A^{\prime}}
s_{x}=\sum\limits_{A}\widetilde{H}_{A}(\alpha) +
\sum\limits_{A^{\prime}}\widehat{H}_{A^{\prime}}(\alpha).
\label{fmetzp}
\end{eqnarray}
The problem is to find such a value of the coefficient $\alpha$ that
converts potentials $\widetilde{H}_{A}$ and
$\widehat{H}_{A^{\prime}}$ into $m$-potentials. Here it means that
the $m$-potential condition must be satisfied for the both types of
sets $A$ and $A^{\prime}$: the restriction of a global ground-state
configuration to any $A$ and $A^{\prime}$ should minimize the
corresponding potentials $\widetilde{H}_{A}$ and
$\widehat{H}_{A^{\prime}}$.

\subsubsection{Second kind zero-potentials}
\begin{figure}[ht]
\centering \includegraphics[width=0.17\textwidth]{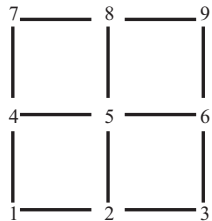}
\caption{The sites of $T$-plaquette are labeled from the left to the
right, starting at the bottom left corner and ending in the upper
right one.} \label{tblock}
\end{figure}

Another method, was proposed in \cite{Kennedy1} and in \cite{GMMU}.
In contrast to the first method, where zero-potentials are of the
form $f_{A}(S)-f_{A}(S)=0$ (i.e. the zero-potential condition is
satisfied locally), no such restrictions are applied to its form.
However, it is convenient to choose a zero-potential in a way that
leaves the Hamiltonian's symmetry unchanged. We show here how to
obtain zero potentials for the case when $A$ is a
$3\times3$-plaquette, later on called the $T$-plaquette
(Fig.~\ref{tblock}).
We assume that $H_{T}$ does not change under rotations by $\pi/2$
and reflections.

Any function $f(S)$ can be rewritten in a form,
\begin{eqnarray}
f(S)=\sum_{S^{\prime}}\lambda(S^{\prime}) \chi (S^{\prime},S),
\label{characterfunct1}
\end{eqnarray}
where $S^{\prime}$ runs over all possible configurations,
$\lambda(S^{\prime})$ is a value of function $f(S)$ for specific
configuration $S^{\prime}$ and $\chi(S^{\prime},S)$ is an indicator
function of configuration $S^{\prime}$, i.e. it is equal to $1$ when
$S=S^{\prime}$ and is zero otherwise.

The construction of an indicator function $\chi(S^{\prime},S)$ for
specific configuration $S^{\prime}$ is simple. At a site $x$ the
indicator function for spin up is $\chi_{x}(+1, S_{x})=(1+s_{x})/2$,
for spin down is $\chi_{x}(-1, S_{x})=(1-s_{x})/2$. The indicator
function at $A$, is defined as
\begin{eqnarray}
\chi_{A}(S^{\prime}_{A},S_{A})=\prod_{x\in A}
\chi_{x}(S^{\prime}_{x},S_{x}). \label{chfunc}
\end{eqnarray}
Performing the summation in (\ref{characterfunct1}) and taking into
account the form of the indicator functions (\ref{chfunc}), any
function $f_{A}(S_{A})$, can be written as,
\begin{eqnarray}
f_{A}(S_{A})&=&\lambda_{0}+\sum\limits_{x\in A}\lambda_{x}s_{x}+
\sum\limits_{\{x,y\}\subset A} \lambda_{xy} s_{x}s_{y}+
\sum\limits_{\{x,y,z\}\subset A} \lambda_{xyz}
s_{x}s_{y}s_{z}+\ldots +
\lambda_{A}\prod\limits_{x \in A}s_{x}=\nonumber\\
&=&\sum\limits_{X\subseteq A}\lambda_{X}s_{X}(S),
\label{characterfunct2}
\end{eqnarray}
where $\lambda_{X}$ are the values which determine the function
$f_{A}$. Notation $s_{X}(S)$ stands for the product of all spins
which belong to $X$, $s_{X}=\prod_{x\in X}s_{x}$. For $X=\emptyset$,
the coefficient $\lambda_{X}=\lambda_{0}$ and $s_{X}=1$.

Another way to show the identity (\ref{characterfunct2}) and give a
formula for finding the coefficients in (\ref{characterfunct2}) is
to consider the functions $s_{X}(S)$ instead of indicator functions
$\chi(S^{\prime},S)$. If $N_{-}$ is the number of spins in $S_{X}$
oriented down, then $s_{X}(S)=(-1)^{N_{-}}$. Let us define the
scalar product in the set of functions of configurations:
\begin{eqnarray}
\langle f_{A} , g_{B} \rangle =\left( \frac{1}{2}
\right)^{|\Lambda|} \sum\limits_{S^{\prime}} f_{A}(S^{\prime})
g_{B}(S^{\prime}),
\end{eqnarray}
where $S^{\prime}$ runs over all possible configurations on
$\Lambda$. Now we show, that
\begin{eqnarray}
\langle s_{X} , s_{Y}\rangle = \left\{
\begin{array}{lr}
1, & \mbox{if    } X=Y, \\
0, & \mbox{if    }  X \ne Y.
\end{array}
\right. \label{ortgons}
\end{eqnarray}
Let us consider the product,
\[
s_{X}(S)s_{Y}(S)=s_{X\cdot Y}(S),
\]
which follows simply from the form of $s_{X}(S)$; here $X\cdot
Y=(X\cup Y)\setminus(X\cap Y)$. Using the fact, that for any set X,
\begin{eqnarray}
\sum\limits_{S}s_{X}(S)= \left\{
\begin{array}{rl}
2^{|\Lambda|}, & \mbox{if   } X=\emptyset\\
0, & \mbox{if   } X\ne\emptyset
\end{array}
\right.
\end{eqnarray}
we obtain expression (\ref{ortgons}). Since the functions $s_{X}$
are orthonormal and there are as many $s_{X}$ as $\chi_{\Lambda}$,
they constitute an orthonormal basis, and any function $f_{A}$, can
be expanded as follows,
\begin{eqnarray*}
f_{A}(S)=\sum\limits_{X \subseteq A} \lambda_{X}s_{X}(S)
\end{eqnarray*}
which is equivalent to (\ref{characterfunct2}), where
\begin{eqnarray}
\lambda_{X}= \langle s_{X}, f_{A} \rangle .
\end{eqnarray}

We are looking for a zero-potential $K_{T}(S)$ in a form
(\ref{characterfunct2}). We can always set $\lambda_{0}=0$ although
it can change the Hamiltonian only by a constant. So for a
$T$-plaquette, which we will use later, we choose the zero-potential
in a form:
\begin{eqnarray}
K_{T}(S)&=&\sum\limits_{\emptyset\ne A \subseteq T}
\lambda_{A} s_{A}(S)= \nonumber \\
&=& \sum_{x \in T}\lambda_{x}s_{x} + \sum_{\{x,y\}\subset
T}\lambda_{xy}s_{x}s_{y}+ \sum_{\{x,y,z\}\subset
T}\lambda_{xyz}s_{x}s_{y}s_{z} + \ldots +\lambda_{T}s_{T} .
\end{eqnarray}
Now applying to these objects the symmetry operations of the
Hamiltonian and the zero-potential condition (\ref{0potcond}) we can
eliminate some of the arbitrary coefficients $\lambda_{A}$. For
instance, we consider the first term:
\begin{eqnarray}
\sum_{x \in T}\lambda_{x}s_{x}=\lambda_{1}s_{1}+ \lambda_{2}s_{2}+
\lambda_{3}s_{3}+ \lambda_{4}s_{4}+ \lambda_{5}s_{5}+
\lambda_{6}s_{6}+ \lambda_{7}s_{7}+ \lambda_{8}s_{8}+
\lambda_{9}s_{9}. \label{zeroone}
\end{eqnarray}
Demanding $\sum_{x \in T}\lambda_{x}s_{x}$ to be invariant with
respect to symmetry operations (rotations or reflections), we obtain
\begin{eqnarray}
\lambda_{1}=\lambda_{3}=\lambda_{7}=\lambda_{9} , \hspace{20mm}
\lambda_{2}=\lambda_{4}=\lambda_{6}=\lambda_{8} .
\end{eqnarray}
The zero-potential condition (\ref{0potcond}) now reads:
\begin{eqnarray*}
\sum_{T}\sum_{x\in
T}\lambda_{x}s_{x}=\sum\limits_{i}\lambda_{i}\sum_{x\in\Lambda}s_{x}=0,
\end{eqnarray*}
and holds for any configuration $S$. That directly implies the
condition
\begin{eqnarray}
\sum\limits_{i}\lambda_{i}=0.
\end{eqnarray}
Expressing $\lambda_{5}$ in terms of $\lambda_{1}$ and
$\lambda_{2}$,
\begin{eqnarray}
\lambda_{5}=-4\left(\lambda_{1}+\lambda_{2}\right),
\end{eqnarray}
only two coefficients remain independent, and the expression is of
the form:
\begin{eqnarray}
\sum_{x \in T}\lambda_{x}s_{x}=\lambda_{1} \left(s_{1}+ s_{3}+
s_{7}+ s_{9} - 4s_{5} \right) + \lambda_{2} \left( s_{2}+ s_{4}+
s_{6}+ s_{8} - 4s_{5} \right).
\end{eqnarray}
Similarly, we can obtain other zero-potentials applying the
procedure sketched above to products of nearest-neighbors spins,
next-nearest-neighbors, etc. Finally, we can rewrite $K_{T}(S)$ in
the form
\begin{eqnarray}
K_{T}(S)=\sum_{i}\alpha_{i}k_{T}^{(i)}(S), \label{0potform}
\end{eqnarray}
where
\begin{eqnarray}
k^{(1)}_{T}(S)&=&s_{1}+ s_{3}+ s_{7}+ s_{9} - 4s_{5}, \nonumber \\
k^{(2)}_{T}(S)&=&s_{2}+ s_{4}+ s_{6}+ s_{8} - 4s_{5},
\end{eqnarray}
are the only two symmetric zero-potentials of the form
(\ref{zeroone}). For a $T$-plaquette, there are $16$ independent
symmetrical zero potentials. The complete list of them is presented
in Appendix A.

Now, having zero potentials, we are trying to find such a vector
$\vec{\alpha}=(\alpha_{1}, \alpha_{2}, \ldots)$ that makes the total
potential (the original potential $H_{A}$ plus the zero potential
$K_{A}$) an $m$-potential. Components $\alpha_{i}$ are called the
zero-potential coefficients. Both, in the first and the second
methods, there are no restrictions for $\vec{\alpha}$ but we
suppose, that they are functions of the Hamiltonian's parameters
$p_{i}$. For convenience, we write the set of parameters $p_{i}$ as
a vector $\vec{p}=(p_{1}, p_{2}, \ldots)$.

The both methods have their own advantages and disadvantages. For
our purposes the second one is more suitable for two reasons. First
of all, adding a zero-potential conserves the symmetry of the
system. As a result, the number of block configurations whose
energies we have to compare is reduced significantly. The second
reason is that extending some local configurations to the whole
lattice is much simpler in the case when we have one type of blocks,
say $T$-plaquettes only. If other blocks are present, like
$A^{\prime}$ in the first method, then we have to check if potential
$\widehat{H}_{A^{\prime}}$ in (\ref{fmetzp}), is an $m$-potential
for a configuration we have built using ground-state configurations
of the potential $\widetilde{H}_{A}$, or vice versa.

\subsubsection{Some remarks on constructing zero-potentials}
Let the Hamiltonian, a function of configurations, be the linear
function of parameters $p_{i}$ (by linear function we understand
function which is convex and concave simultaneously), and it is
possible to rewrite it in the following form
\begin{eqnarray}
H_{\Lambda}(\vec{p};S)=\sum\limits_{A}H_{A}(\vec{p};S),
\label{Hamiltabc}
\end{eqnarray}
where $H_{A}$ is linear in $\vec{p}$ and is not an $m$-potential.
Adding the zero-potential $K_{A}$ in the form (\ref{0potform}), we
suppose that a zero-potential vector $\vec{\alpha}$ is the linear
function of a vector of parameters $\vec{p}$,
\begin{eqnarray}
H_{\Lambda}(\vec{p} ;S)=\sum\limits_{A}\left( H_{A}(\vec{p} ;S) +
K_{A}(\vec{\alpha}(\vec{p}) ;S) \right) =
\sum\limits_{A}\widetilde{H}_{A}(\vec{p} ; \vec{\alpha}(\vec{p}) ;S)
,
\end{eqnarray}
and we are looking for a vector $\vec{\alpha}$ for which the above
potential is an $m$-potential for any vector of parameters
$\vec{p}$. Unfortunately, we hardly ever succeed in finding a
zero-potential vector which converts $\widetilde{H}_{A}$ into an
$m$-potential for the whole range of the parameters; usually
coefficients $\vec{\alpha}(\vec{p})$ can be found in a certain range
of the parameters only. So sometimes we need several sets of
zero-potential coefficients to cover the whole range of the
considered parameters. Such an approach is used in Section 3.

In studies of the minimum of the potentials, the following
observation is useful. Consider two points in the parameters space,
$\vec{x}$ and $\vec{y}$, and such $\vec{\alpha}^{\prime}$,
$\vec{\alpha}^{\prime\prime}$ in each point, respectively, that the
potentials $\widetilde{H}_{A}(\vec{x} ; \vec{\alpha}^{\prime};S)$
and $\widetilde{H}_{A}(\vec{y} ; \vec{\alpha}^{\prime\prime};S)$ are
minimized at some sets of local configurations
$G_{A}(\vec{x},\vec{\alpha}^{\prime})$ and
$G_{A}(\vec{y},\vec{\alpha}^{\prime\prime})$, respectively, and
$G_{A}(\vec{x},\vec{\alpha}^{\prime})\cap
G_{A}(\vec{y},\vec{\alpha}^{\prime\prime}) \ne \emptyset$. There
exist only one linear function $\vec{\alpha}(\vec{p})$ on the line
segment between points $\vec{x}$ and $\vec{y}$ whose values are
equal to $\vec{\alpha}^{\prime}$ at $\vec{x}$ and
$\vec{\alpha}^{\prime\prime}$ at $\vec{y}$:
\begin{eqnarray}
\vec{\alpha}\Bigl( t\vec{x}+(1-t)\vec{y} \Bigr)= t \cdot
\vec{\alpha}^{\prime} +(1-t) \cdot \vec{\alpha}^{\prime\prime},
\hspace{10mm} \mbox{for any } t\in [0,1]. \label{alphadef}
\end{eqnarray}
This definition ensures that potential $\widetilde{H}_{A}$ is linear
in $\vec{p}$:
\begin{eqnarray}
\widetilde{H}_{A}\Bigl( t\vec{x}+(1-t)\vec{y}, \vec{\alpha}\left(
t\vec{x}+(1-t)\vec{y} \right); S \Bigr)=\nonumber\\
= t \cdot \widetilde{H}_{A}(\vec{x}, \vec{\alpha}(\vec{x});S) +(1-t)
\cdot \widetilde{H}_{A}(\vec{y}, \vec{\alpha}(\vec{y});S),
\hspace{10mm} \mbox{for any  } t\in [0,1]. \label{linearh}
\end{eqnarray}

We claim that a set of block configurations that realizes the
minimum of energy $\widetilde{H}_{A}(\vec{p};
\vec{\alpha}(\vec{p});S)$ on the line segment between $\vec{x}$ and
$\vec{y}$, is
\[
G_{A}\bigl(t\vec{x}+(1-t)\vec{y}, \vec{\alpha}\left(
t\vec{x}+(1-t)\vec{y} \right)\bigr)=G_{A}(\vec{x},
\vec{\alpha}(\vec{x})) \cap G_{A}(\vec{y},\vec{\alpha}(\vec{y})),
\]
for $0<t<1$.

Indeed, let $\widetilde{S}\in G_{A}(\vec{x}, \vec{\alpha}(\vec{x}))
\cap G_{A}(\vec{y},\vec{\alpha}(\vec{y}))$, i.e.:
\begin{eqnarray}
\widetilde{H}_{A}(\vec{x},
\vec{\alpha}(\vec{x});\widetilde{S})=\min_{S}\widetilde{H}_{A}(\vec{x},
\vec{\alpha}(\vec{x});S) \hspace{3mm}\mbox{ and }\hspace{3mm}
\widetilde{H}_{A}(\vec{y},
\vec{\alpha}(\vec{y});\widetilde{S})=\min_{S}\widetilde{H}_{A}(\vec{y},
\vec{\alpha}(\vec{y});S). \label{linearhh}
\end{eqnarray}
Then by (\ref{linearh}) and (\ref{linearhh}),
\begin{multline*}
\widetilde{H}_{A}\Bigl( t\vec{x}+(1-t)\vec{y}, \vec{\alpha}\left(
t\vec{x}+(1-t)\vec{y} \right); \widetilde{S} \Bigr) = \\
\shoveright{= t \cdot \widetilde{H}_{A}(\vec{x},
\vec{\alpha}(\vec{x});\widetilde{S}) +(1-t) \cdot
\widetilde{H}_{A}(\vec{y}, \vec{\alpha}(\vec{y});\widetilde{S})}
\\
\shoveright{= t \cdot \min_{S}\widetilde{H}_{A}(\vec{x},
\vec{\alpha}(\vec{x});S) + (1-t) \cdot
\min_{S}\widetilde{H}_{A}(\vec{y},
\vec{\alpha}(\vec{y});S)} \\
\shoveright{\leqslant t \cdot \widetilde{H}_{A}(\vec{x},
\vec{\alpha}(\vec{x});S) +(1-t) \cdot \widetilde{H}_{A}(\vec{y},
\vec{\alpha}(\vec{y});S)}
\\
= \widetilde{H}_{A}\Bigl( t\vec{x}+(1-t)\vec{y}, \vec{\alpha}\left(
t\vec{x}+(1-t)\vec{y} \right); S \Bigr),
\end{multline*}
for any configuration $S$ and any $t\in[0,1]$, that is,
\[
G_{A}(\vec{x}, \vec{\alpha}(\vec{x})) \cap
G_{A}(\vec{y},\vec{\alpha}(\vec{y}))\subseteq
G_{A}\bigl(t\vec{x}+(1-t)\vec{y}, \vec{\alpha}\left(
t\vec{x}+(1-t)\vec{y} \right)\bigr),
\]
for any $t\in[0,1]$.

Now we shall demonstrate that for $t\in(0,1)$ the opposite inclusion
holds. Let
\[
\widehat{S}\in G_{A}\bigl(t\vec{x}+(1-t)\vec{y},
\vec{\alpha}\left( t\vec{x}+(1-t)\vec{y} \right)\bigr)
\]
that is,
\begin{multline*}
\widetilde{H}_{A}\Bigl( t\vec{x}+(1-t)\vec{y}, \vec{\alpha}\left(
t\vec{x}+(1-t)\vec{y} \right); \widehat{S} \Bigr) \leqslant \\
\leqslant \tilde{H}_{A}\Bigl( t\vec{x}+(1-t)\vec{y},
\vec{\alpha}\left( t\vec{x}+(1-t)\vec{y} \right); S \Bigr),
\end{multline*}
for any $S$. In particular, the above inequality holds for
$\widetilde{S}\in G_{A}(\vec{x}, \vec{\alpha}(\vec{x})) \cap
G_{A}(\vec{y},\vec{\alpha}(\vec{y}))$, and can be rewritten as
\begin{multline}
t \left[ \widetilde{H}_{A}(\vec{x},
\vec{\alpha}(\vec{x});\widehat{S})-\widetilde{H}_{A}(\vec{x},
\vec{\alpha}(\vec{x});\widetilde{S}) \right] +\\
+ (1-t)\left[ \widetilde{H}_{A}(\vec{y},
\vec{\alpha}(\vec{y});\widehat{S})-\widetilde{H}_{A}(\vec{y},
\vec{\alpha}(\vec{y});\widetilde{S}) \right] \leqslant0,
\label{proenone}
\end{multline}
for any $t\in(0,1)$. Simultaneously, since each of the energy
differences (in square brackets) in (\ref{proenone}) is nonnegative,
\begin{multline}
t \left[ \widetilde{H}_{A}(\vec{x},
\vec{\alpha}(\vec{x});\widehat{S})-\widetilde{H}_{A}(\vec{x},
\vec{\alpha}(\vec{x});\widetilde{S}) \right]+\\
+ (1-t)\left[ \widetilde{H}_{A}(\vec{y},
\vec{\alpha}(\vec{y});\widehat{S})-\widetilde{H}_{A}(\vec{y},
\vec{\alpha}(\vec{y});\widetilde{S}) \right] \geqslant0.
\label{proentwo}
\end{multline}
By (\ref{proenone}) and (\ref{proentwo}),
\begin{multline*}
t \left[ \widetilde{H}_{A}(\vec{x},
\vec{\alpha}(\vec{x});\widehat{S})-\widetilde{H}_{A}(\vec{x},
\vec{\alpha}(\vec{x});\widetilde{S}) \right] +\\
+ (1-t)\left[ \widetilde{H}_{A}(\vec{y},
\vec{\alpha}(\vec{y});\widehat{S})-\widetilde{H}_{A}(\vec{y},
\vec{\alpha}(\vec{y});\widetilde{S}) \right] =0,
\end{multline*}
with $t>0$ and $1-t>0$, hence
\begin{eqnarray*}
\widetilde{H}_{A}(\vec{x},
\vec{\alpha}(\vec{x});\widehat{S})-\widetilde{H}_{A}(\vec{x},
\vec{\alpha}(\vec{x});\widetilde{S})=0
\hspace{9mm}\mbox{and}\hspace{9mm} \widetilde{H}_{A}(\vec{y},
\vec{\alpha}(\vec{y});\widehat{S})-\widetilde{H}_{A}(\vec{y},
\vec{\alpha}(\vec{y});\widetilde{S}) =0,
\end{eqnarray*}
which implies in turn that $\widehat{S}\in G_{A}(\vec{x},
\vec{\alpha}(\vec{x})) \cap G_{A}(\vec{y},\vec{\alpha}(\vec{y}))$.
Therefore,
\[
G_{A}\bigl(t\vec{x}+(1-t)\vec{y}, \vec{\alpha}\left(
t\vec{x}+(1-t)\vec{y} \right)\bigr)\subset G_{A}(\vec{x},
\vec{\alpha}(\vec{x})) \cap G_{A}(\vec{y},\vec{\alpha}(\vec{y})).
\]

\begin{figure}[t]
\centering \includegraphics[width=0.5\textwidth]{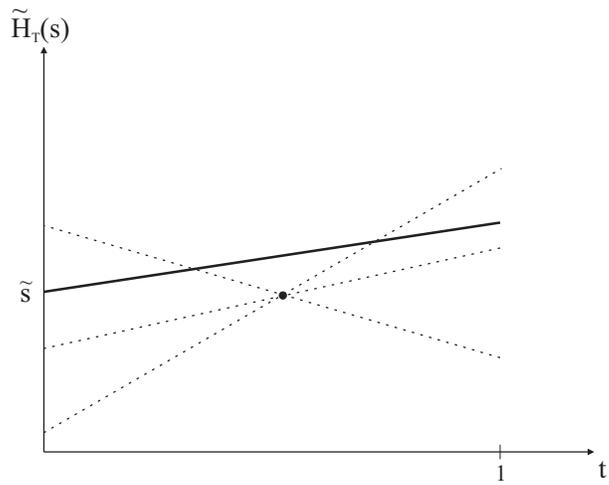}
\caption{The illustration of the proof concerning $\alpha$.}
\label{Fig-proof}
\end{figure}

This proof has a simple geometrical interpretation
(Fig.~\ref{Fig-proof}).
All energies are linear functions of $t$, so if there exist a point
at $t\in (0,1)$ where some block configuration $\widehat{S} \ne
\widetilde{S}$ has a lower energy than that of $\widetilde{S}$, then
at least at one of the points, $\vec{x}$ or $\vec{y}$, the energy of
$\widehat{S}$ is lower than the energy of local ground state
configuration.

So, there is no need to find zero-potential coefficients everywhere:
in order to find them for some interval $(\vec{x},\vec{y})$, it is
enough to find them at $\vec{x}$ and at $\vec{y}$, where the sets of
local ground state configurations are $G_{\vec{x}}$ and
$G_{\vec{y}}$, respectively. The construction described above
guarantees that with $\vec{\alpha}\left( t\vec{x} + (1-t)\vec{y}
\right)$ for $t \in (0,1)$, the set of local ground state
configurations $G_{[\vec{x},\vec{y}]}$ on the line segment
$(\vec{x},\vec{y})$ is given by $G_{[\vec{x},\vec{y}]}=G_{\vec{x}}
\cap G_{\vec{y}}$. Such an approach is used in Section 4 and 5,
where zero-potential coefficients are given only on some
semi-infinite lines and at certain points.

\clearpage
\section{Stability of striped phases}
In this Section\footnote{This section is based on \cite{DJ2}.} we
consider two strongly correlated quantum systems, described by
Falicov--Kimball-like Hamiltonians on a square lattice, extended by
direct short-range interactions between the immobile particles. In
the first system hopping particles are spinless fermions while in
the second one they are hardcore bosons. Ground-state phase diagrams
of the both systems in the strong-coupling regime and at
half-filling are constructed rigorously. Two main conclusions are
drawn. Firstly, short-range interactions in quantum gases are
sufficient for the appearance of charge stripe-ordered phases. By
varying the intensity of a direct nearest-neighbor interaction
between the immobile particles, the both systems can be driven from
a phase-separated state (the segregated phase) to a crystalline
state (the chessboard phase) and these transitions occur necessarily
via charge-stripe phases: via a diagonal striped phase in the case
of fermions and via vertical (horizontal) striped phases in the case
of hardcore bosons. Secondly, the phase diagrams of the two systems
(mobile fermions or mobile hardcore bosons) are definitely
different. However, if the strongest effective interaction in the
fermionic case gets frustrated gently, then the phase diagram
becomes similar to that of the bosonic case.

\subsection{Introducing the modifications to the Falicov--Kimball model}
The task is to generate the physical situation, where the striped
phases appear, in the framework of Falicov--Kimball model. For
technical reasons, our analysis is restricted to the case of
half-filling, where the well-defined expansion can be written.
However, it is now well established (proven) \cite{FLU1,FLU2} that,
in the fermion spinless Falicov--Kimball model the phase-separated
state, the segregated phase, is stable only off the half-filling.
Thus, an additional interaction is needed to stabilize the
segregated phase at half-filling.

In the regime of singly occupied sites and for strong-coupling, the
strongest effective interaction in the Hubbard model is the
Heisenberg antiferromagnetic interaction. Doping of holes can be
seen as means to weaken the strong tendency toward antiferromagnetic
ordering and to make possible a phase separated state with
hole-reach and electron-reach regions. Under similar conditions, in
the spinless Falicov--Kimball model an analogous role is played by
the Ising-like n.n. repulsive interaction between the immobile
particles that favors a chessboard-like ordering. The effect of
weakening of the strong tendency toward chessboard ordering can be
achieved by an extra Ising-like n.n. interaction that compensates
the strongest repulsive interaction, and consequently permits the
system to reach a phase-separated state, the segregated phase. On
varying, in a suitable interval of values, the corresponding
interaction constant, which is our control parameter, we can study
how the both systems ``evolve'' from the crystalline chessboard
phase to the segregated one.

Here, we would like to make a remark concerning physical
interpretations of our extended Falicov--Kimball model. The
classical spin models, like those studied in \cite{LEFK1,VS1} can be
considered as classical effective models of crystallization that
contain the necessary ``ingredients'' of such models. Namely, in
these models there are two competing interactions, one favoring
phase separation into macroscopic regions, which are either empty or
completely filled with particles, the other favoring periodic
arrangements of particles. Our extended Falicov--Kimball model, with
hopping fermions, can be thought of as a microscopic quantum model
of crystallization. In this case, the necessary ``ingredients'' are:
the tendency towards crystalline orders, coming from the kinetic
energy of itinerant electrons and from the screened (in our case on
site) electron-ion Coulomb interaction, a kind of Van der Walls
attractive forces, favoring phase separated state, and finally the
hard-core repulsion provided by the  underlying lattice and the
Pauli principle.

In what follows, the modified Hamiltonian reads:
\begin{eqnarray}
H=H_{FK}+V, \label{ourhamilt}
\end{eqnarray}
where $H_{FK}$ is given by (\ref{FK1}), and a direct Ising-like
interaction $V$ between the immobile particles is of the form
\begin{eqnarray}
V=\frac{W}{8} \sum\limits_{\langle x,y \rangle_{1}} s_{x}s_{y}
-\frac{\tilde{\varepsilon}}{16} \sum\limits_{\langle x,y
\rangle_{2}} s_{x}s_{y}. \label{ourinter}
\end{eqnarray}
Here $W$ stands for an intensity of direct n.n. interaction between
ions. We also include a subsidiary direct interaction between the
immobile particles, an Ising-like n.n.n. interaction with intensity
$\tilde{\varepsilon}$, much weaker than the n.n. interaction. This
interaction can reinforce or frustrate the n.n. interaction,
depending on the sign of its interaction constant. It appears that
on varying $W$, the systems (fermion or boson) characterized by
different values of n.n.n. interactions, may undergo different
sequences of transitions between the chessboard phase and the
segregated one. We find, in particular, that if we set appropriate
values of the n.n.n. interaction in the fermion system and in the
boson one, then the both systems settle in the same phases, for
typical values of the control parameter. On the other hand, the
n.n.n. interaction enables us to demonstrate that the ground-state
phase diagrams in the cases of mobile fermions and mobile hardcore
bosons are definitely different.

To simplify the model, we consider the case where only n.n. hopping
is nonzero and does not depend on direction. So we put
$t_{h}=t_{v}=t$, and $t_{+}=t_{-}=0$. We are interesting in
investigation of the ground-state phase diagrams of our systems, so
the grand-canonical formalism is used. We use the expansion
(\ref{effh}) which was obtained in the strong-coupling regime and at
half-filling. In this case, with the expansion terms up to order
four shown explicitly, it reads:
\begin{eqnarray}
E^{fermion}_{S}\left( \mu \right) &=& -\frac{\mu}{2} \sum\limits_{x}
s_{x} +\left[ \frac{t^{2}}{4}- \frac{9t^{4}}{16} + \frac{W}{8}
\right] \sum\limits_{\langle x,y \rangle_{1}} s_{x}s_{y} + \left[
\frac{3t^{4}}{16} -\frac{\tilde{\varepsilon}}{16} \right]
\sum\limits_{\langle x,y \rangle_{2}} s_{x}s_{y} +
\nonumber \\
&&+\frac{t^{4}}{8} \sum\limits_{\langle x,y \rangle_{3}} s_{x}s_{y}
+ \frac{t^{4}}{16} \sum\limits_{P_{2}} \left(5s_{P}+1\right) +
R^{(4)}, \label{Expfmn}
\end{eqnarray}
for fermions, and
\begin{eqnarray}
E^{boson}_{S}\left( \mu \right) &=& -\frac{\mu}{2} \sum\limits_{x}
s_{x} +\left[ \frac{t^{2}}{4}- \frac{5t^{4}}{16} + \frac{W}{8}
\right] \sum\limits_{\langle x,y \rangle_{1}} s_{x}s_{y} + \left[
\frac{5t^{4}}{16} -\frac{\tilde{\varepsilon}}{16} \right]
\sum\limits_{\langle x,y \rangle_{2}} s_{x}s_{y} +
\nonumber \\
&&+\frac{t^{4}}{8} \sum\limits_{\langle x,y \rangle_{3}} s_{x}s_{y}
- \frac{t^{4}}{16} \sum\limits_{P_{2}} \left(5+s_{P}\right) +
{\tilde{R}}^{(4)}, \label{Exphcb}
\end{eqnarray}
for hardcore bosons, up to a term independent of the ion
configuration and the chemical potentials. The remainders $R^{(4)}$,
${\tilde{R}}^{(4)}$, are independent of the chemical potential $\mu$
and the parameters $W$ and $\tilde{\varepsilon}$, and collect those
terms of the expansion that are proportional to $t^{2m}$, with
$m=3,4,\ldots$.

Due to the convergence of the expansions (\ref{Expfmn}) and
(\ref{Exphcb}), it is possible to establish rigorously a part of the
phase diagram (that is the ground-state configurations of ions are
determined everywhere in the $\left( \mu_{e}, \mu_{i}
\right)$-plane, except some small regions), by determining the phase
diagram of the expansion truncated at the order $k$, that is
according to the $k$-th order effective Hamiltonians
$(E^{fermion}_{S})^{(k)} \left( \mu \right)$ and
$(E^{boson}_{S})^{(k)} \left( \mu \right)$. In order to construct a
phase diagram according to a  $k$-th order effective Hamiltonian, we
use the $m$-potential method.

\subsection{Zeroth- and second-order effective interactions --- phase diagram}
As was discussed above, up to the second order the effective
Hamiltonians for fermions and hardcore bosons are the same. Hence,
the discussion that follows applies to the both cases, and the
common effective Hamiltonians are denoted as $E_{S}^{(0)} \left( \mu
\right)$, $E_{S}^{(2)} \left( \mu \right)$.

\begin{figure}[th]
\includegraphics[width=0.47\textwidth]{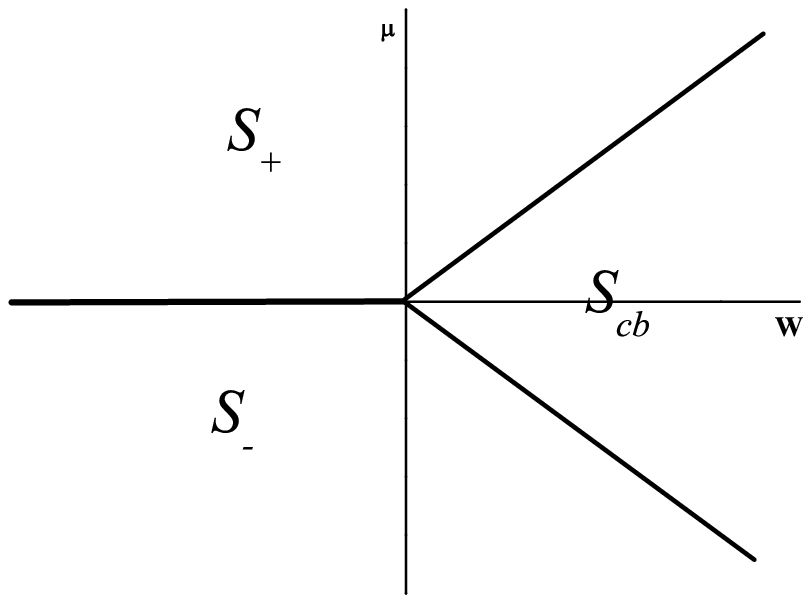}
\hfill
\includegraphics[width=0.47\textwidth]{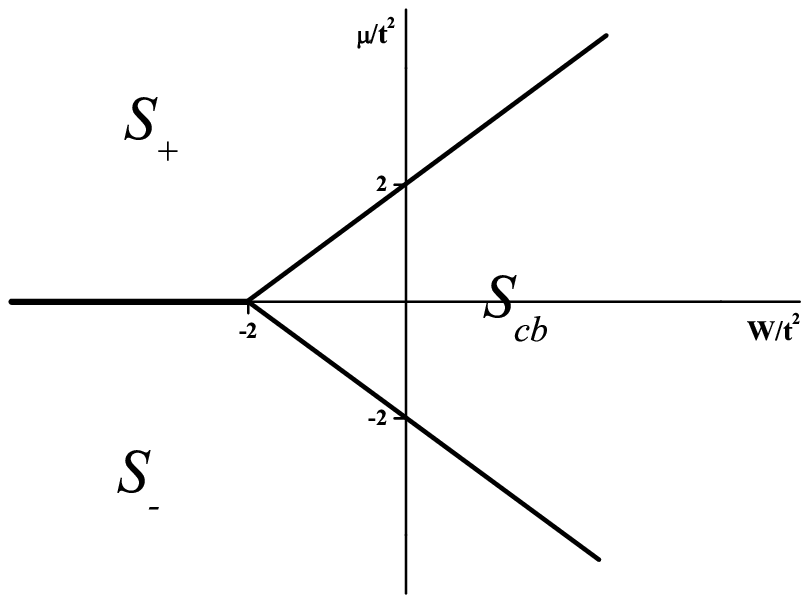}
\\
\parbox[t]{0.47\textwidth}{\caption{{\small{Ground-state
phase diagram of $E^{(0)}_{S} \left( \mu \right)$ for fermion and
hard-core boson systems.}}} \label{zerord}} \hfill
\parbox[t]{0.47\textwidth}{\caption{{\small{Ground-state
phase diagram of  $E^{(2)}_{S} \left( \mu \right)$ for fermion and
hard-core boson systems.}}} \label{secord}}
\end{figure}

In the zeroth order,
\begin{eqnarray}
\label{E0} E_{S}^{(0)} \left( \mu \right) & = & -\frac{\mu}{2}
\sum\limits_{x} \left( s_{x} +1 \right) + \frac{W}{8}
\sum\limits_{\langle x,y \rangle_{1}} s_{x}s_{y}
-\frac{\tilde{\varepsilon}}{16}
\sum\limits_{\langle x,y \rangle_{2}} s_{x}s_{y} \nonumber \\
&=& \sum\limits_{P_{2}} H_{P_{2}}^{(0)},
\end{eqnarray}
where
\begin{eqnarray}
\label{HP20} H_{P_{2}}^{(0)} =  -\frac{\mu}{8}
\sideset{}{'}\sum\limits_{x} \left( s_{x} +1 \right) + \frac{W}{16}
\sideset{}{'} \sum\limits_{\langle x,y \rangle_{1}} s_{x}s_{y}
-\frac{\tilde{\varepsilon}}{16} \sideset{}{'} \sum\limits_{\langle
x,y \rangle_{2}} s_{x}s_{y},
\end{eqnarray}
and the primed sums in (\ref{HP20}) are restricted to a plaquette
$P_{2}$.  For any $\tilde{\varepsilon} > 0$, the plaquette
potentials $H_{P_{2}}^{(0)}$ are minimized by the restrictions to a
plaquette $P_{2}$ of a few periodic configurations of ions on
$\Lambda$. Namely, $S_{-}$ --- the empty configuration ($S_{+}$ ---
the completely filled configuration), where $s_{x}=-1$ at every site
($s_{x}=+1$ at every site), and the chessboard configurations
$S_{cb}^{e}$, where $s_{x}=\epsilon_{x}$, and $S_{cb}^{o}$ (where
$s_{x}=-\epsilon_{x}$), with $\epsilon_{x}=1$ if $x$ belongs to the
even sublattice of $\Lambda$ and $\epsilon_{x}=-1$ otherwise.
Moreover, out of the restrictions of those configurations to a
plaquette $P_{2}$, $S_{-|P_{2}}$, $S_{+|P_{2}}$, $S^{e}_{cb|P_{2}}$,
and $S_{cb|P_{2}}^{o}$, only four ground-state configurations on
$\Lambda$ can be built, which coincide with the four configurations
named above. Clearly, this is due to the direct n.n.n. attractive
interaction between the ions. The section of the phase diagram in
the $(W,\mu,\tilde{\varepsilon})$ space by a plane with
$\tilde{\varepsilon} > 0$, according to the effective Hamiltonian
$E^{(0)}_{S} \left( \mu \right)$, is shown in Fig.~\ref{zerord}.

In the second order,
\begin{eqnarray}
\label{E2} E_{S}^{(2)} \left( \mu \right) & = & -\frac{\mu}{2}
\sum\limits_{x} \left( s_{x} +1 \right) + \left[ \frac{t^{2}}{4}+
\frac{W}{8} \right]\sum\limits_{\langle x,y \rangle_{1}} s_{x}s_{y}
-\frac{\tilde{\varepsilon}}{16} \sum\limits_{\langle x,y
\rangle_{2}} s_{x}s_{y}
\nonumber \\
&=& \sum\limits_{P_{2}} H_{P_{2}}^{(2)},
\end{eqnarray}
where $H_{P_{2}}^{(2)}$,
\begin{eqnarray}
\label{HP22} H_{P_{2}}^{(2)} =  -\frac{\mu}{8} \sideset{}{'}
\sum\limits_{x} \left( s_{x} +1 \right) + \left[\frac{t^{2}}{8}+
\frac{W}{16} \right] \sideset{}{'} \sum\limits_{\langle x,y
\rangle_{1}} s_{x}s_{y} -\frac{\tilde{\varepsilon}}{16}
\sideset{}{'} \sum\limits_{\langle x,y \rangle_{2}} s_{x}s_{y},
\end{eqnarray}
and the primed sums are restricted to a plaquette $P_{2}$. Since
$E_{S}^{(2)}$ differs from $E_{S}^{(0)}$ by the strength of n.n.
interactions, the  only effect on the phase diagram is that the
coexistence lines are translated by the vector $(-2t^{2},0,0)$. The
section of the phase diagram in the $(W,\mu,\tilde{\varepsilon})$
space by a plane with $\tilde{\varepsilon} > 0$, according to the
effective Hamiltonian $E^{(2)}_{S} \left( \mu \right)$, is shown in
Fig.~\ref{secord}.

\subsection{Fourth-order effective interactions --- phase diagram}
It follows from the investigation of  the phase diagram up to the
second order that the degeneracy is finite, independently of the
size of $\Lambda$, everywhere except the coexistence lines in the
($\tilde{\varepsilon}=0$)-plane, where it grows exponentially with
$|\Lambda|$. Only there the effect of the fourth order interactions
can be most significant. The meeting point of the coexistence lines
in the ($\tilde{\varepsilon}=0$)-plane, where the coexistence line
of $S_{+}$ and $S_{-}$ sticks to the stability domain of chessboard
configurations, appears to be particularly interesting. In that
point, the energies of all the configurations are the same. In what
follows, we shall study the phase diagrams of spinless fermions and
spinless hardcore bosons up to the fourth order, in a neighborhood
of radius $O(t^{4})$ of the point $(-2t^{2},0,0)$. In this
neighborhood, it is convenient to introduce new coordinates,
$(W,\mu,\tilde{\varepsilon}) \to (\omega ,\delta,\varepsilon)$,
\begin{eqnarray}
\label{newcoor} W = -2t^{2}+ t^{4}\omega , \hspace{5mm} \mu = t^{4}
\delta , \hspace{5mm} \tilde{\varepsilon} = t^{4} \varepsilon ,
\end{eqnarray}
and a new (equivalent to $(E^{fermion}_{S})^{(4)}$) $t$-independent
effective Hamiltonian, $(H^{fermion}_{{\mathrm{eff}}})^{(4)}$,
\begin{eqnarray}
(E^{fermion}_{S})^{(4)} \left( 0, \delta \right) = \frac{t^{4}}{2}
(H^{fermion}_{{\mathrm{eff}}})^{(4)}. \label{Heff4}
\end{eqnarray}

Then, in the spirit of the $m$-potential method, we express
$(H^{fermion}_{{\mathrm{eff}}})^{(4)}$ by the potentials
$(H^{fermion}_{T})^{(4)}$,
\begin{equation}
\label{HT} (H^{fermion}_{{\mathrm{eff}}})^{(4)} = \sum\limits_{T}
(H^{fermion}_{T})^{(4)},
\end{equation}
where, in  terms of the new variables,
\begin{eqnarray}
\label{H4fmn-1} (H^{fermion}_{T})^{(4)} &=& -\delta \left( s_{5} +1
\right) + \frac{1}{24} \left( \omega-\frac{9}{2} \right)
\sideset{}{''}\sum\limits_{\langle x,y \rangle_{1}} s_{x}s_{y} +
\frac{1}{32} \left(3 - \varepsilon \right)
\sideset{}{''} \sum\limits_{\langle x,y \rangle_{2}} s_{x}s_{y} + \nonumber \\
& & \frac{1}{12} \sideset{}{''} \sum\limits_{\langle x,y
\rangle_{3}} s_{x}s_{y} + \frac{1}{32} \sideset{}{''}
\sum\limits_{P_{2}} \left( 5s_{P_{2}}+1 \right) .
\end{eqnarray}
In (\ref{HT}) and (\ref{H4fmn-1}), $T$ stands for a $T$-plaquette of
a square lattice (see Fig.~\ref{tblock}). The double-primed sums are
restricted to a $T$-plaquette. For bosons, we introduce
$(H^{boson}_{{\mathrm{eff}}})^{(4)}$ and $(H^{boson}_{T})^{(4)}$ in
the same manner as for fermions, with
\begin{eqnarray}
\label{H4hcb-1} (H^{boson}_{T})^{(4)} &=& -\delta \left( s_{5} +1
\right) + \frac{1}{24} \left( \omega -\frac{5}{2} \right)
\sideset{}{''} \sum\limits_{\langle x,y \rangle_{1}} s_{x}s_{y} +
\frac{1}{32} \left( 5 - \varepsilon \right)
\sideset{}{''} \sum\limits_{\langle x,y \rangle_{2}} s_{x}s_{y} + \nonumber \\
& & \frac{1}{12} \sideset{}{''} \sum\limits_{\langle x,y
\rangle_{3}} s_{x}s_{y} - \frac{1}{32} \sideset{}{''}
\sum\limits_{P_{2}} \left( s_{P_{2}}+5 \right).
\end{eqnarray}

We have to search for the lowest-energy configurations of
$(E^{fermion}_{S})^{(4)}$ and $(E^{boson}_{S})^{(4)}$ among all the
configurations. Consequently, the potentials
$(H^{fermion}_{T})^{(4)}$ and $(H^{boson}_{T})^{(4)}$ have to be
minimized over all the $T$-plaquette configurations. There are (up
to the symmetries of $H_{0}$) 102 different $T$-plaquette
configurations, shown in Fig.~\ref{blconf102} in Appendix B.

Unfortunately, in contrast to the lower-order cases, the potentials
$(H^{fermion}_{T})^{(4)}$ and $(H^{boson}_{T})^{(4)}$ turn out to be
the $m$-potentials only in a small part of
$(\omega,\delta,\varepsilon)$-space. This difficulty can be overcome
by introducing the zero-potentials $K^{(4)}_{T}$, that are invariant
with respect to the symmetries of $H$ and are given by
(\ref{0potform}). Coefficients $\alpha_{i}$, depending on
$(\omega,\delta,\varepsilon)$ in general, have to be determined in
the process of constructing the phase diagram, and the potentials
$k_{T}^{(i)}$ are listed in Appendix A. We set all $\alpha_{i}=0$
for $i\geqslant6$.

Following (\ref{newpot}), new candidates for $m$-potentials can be
introduced,
\begin{eqnarray}
(H^{fermion}_{{\mathrm{eff}}})^{(4)} = \sum\limits_{T} \left(
(H^{fermion}_{T})^{(4)} + K^{(4)}_{T} \right),
\end{eqnarray}
with an analogous representation of
$(H^{boson}_{{\mathrm{eff}}})^{(4)}$, and now the task is to
minimize the potentials $(H^{fermion}_{T})^{(4)} + K^{(4)}_{T}$ and
$(H^{boson}_{T})^{(4)} + K^{(4)}_{T}$ over all the $T$-plaquette
configurations.

In our study of the ground-state phase diagrams we limit ourselves
to the ($\delta=0$)-plane, where the both considered systems are
hole-particle invariant, and to the ($\varepsilon=0$)-plane. Our
analysis of the minima of the $T$-plaquette potentials,
$(H^{fermion}_{T})^{(4)}$ and $(H^{boson}_{T})^{(4)}$ augmented by
the zero-potentials $K^{(4)}_{T}$, shows that these planes are
partitioned into a finite number of open domains
${\mathcal{S}}_{D}$, each with its unique set of periodic
ground-state configurations on $\Lambda$, denoted also by
${\mathcal{S}}_{D}$. There is a finite number, independent of
$|\Lambda|$, of configurations in ${\mathcal{S}}_{D}$ and they are
related by the symmetries of $H$. The last two statements do not
apply to only one of the domains, denoted ${\mathcal{S}}_{d2}$,
which will be described in the sequel.

The domains ${\mathcal{S}}_{D}$ are characterized as follows: at
each point $p$ ($p=(\omega,\varepsilon)$ or $p=(\omega,\delta)$) of
a domain ${\mathcal{S}}_{D}$, there exist a set of coefficients
$\left\{ \alpha_{i}(p) \right\}$ such that the corresponding
potentials are minimized by a set ${\mathcal{S}}_{TD}(p)$ of
$T$-plaquette configurations. Moreover, from the configurations in
${\mathcal{S}}_{TD}(p)$ one can construct only the configurations in
${\mathcal{S}}_{D}$. The set of the restrictions to $T$-plaquettes
of the configurations from ${\mathcal{S}}_{D}$,
${\mathcal{S}}_{D|T}$, is contained in each set
${\mathcal{S}}_{TD}(p)$ with $p \in {\mathcal{S}}_{D}$. In
Table~\ref{tb1-4} of Appendix E we mark by asterisk the cases, where
the set ${\mathcal{S}}_{TD}(p)$ contains, besides
${\mathcal{S}}_{D|T}$, some additional $T$-plaquette configurations.

\begin{figure}
\centering\includegraphics[height=0.26\textheight]{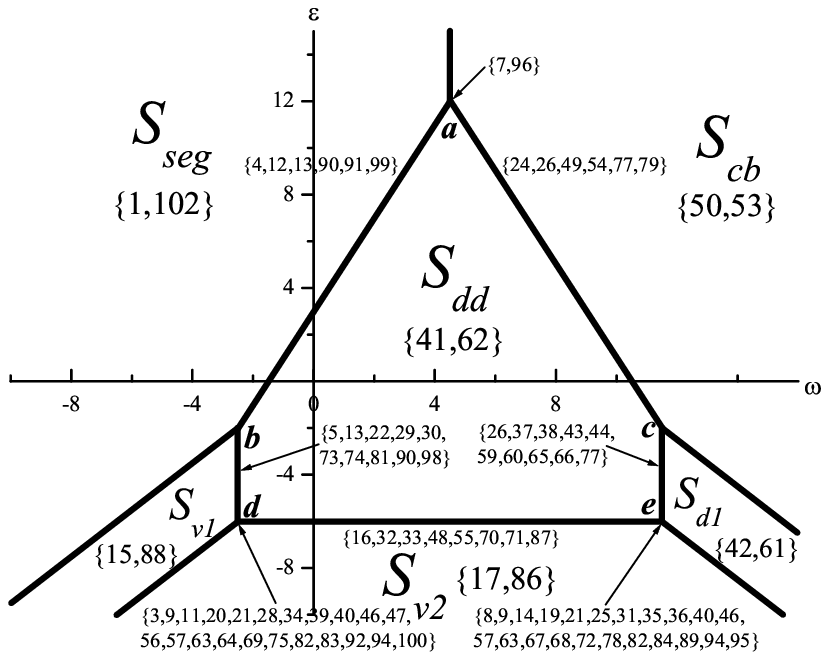}
\parbox[t]{\textwidth}{\caption{{\small{The ($\mu=0$)-phase diagram
of the effective Hamiltonian $(H^{fermion}_{{\mathrm{eff}}})^{(4)}$.
The numbers in curly brackets, displayed by the symbols of open
domains ${\mathcal{S}}_{D}$, denote the $T$-plaquette configurations
that minimize the $m$-potential in ${\mathcal{S}}_{D}$. The numbers
in curly brackets, displayed by the boundary-line segments or by the
arrows pointing towards boundary segments (or their crossing points)
identify the additional minimizing $T$-plaquette configurations. For
more comments see the text. The boundary-line segments can be
determined by means of their intersection points: ${\bf a} =
(9/2,12)$, ${\bf b} = (-5/2,-2)$, ${\bf c} = (23/2,-2)$, ${\bf d} =
(-5/2,-6)$, ${\bf e} = (23/2,-6)$, and by the slope $1$ of the
boundary of ${\mathcal{S}}_{v1}$, and the slope $-1$ of the boundary
of ${\mathcal{S}}_{d1}$.}}} \label{t1-40f}} \vfill
\centering\includegraphics[height=0.26\textheight]{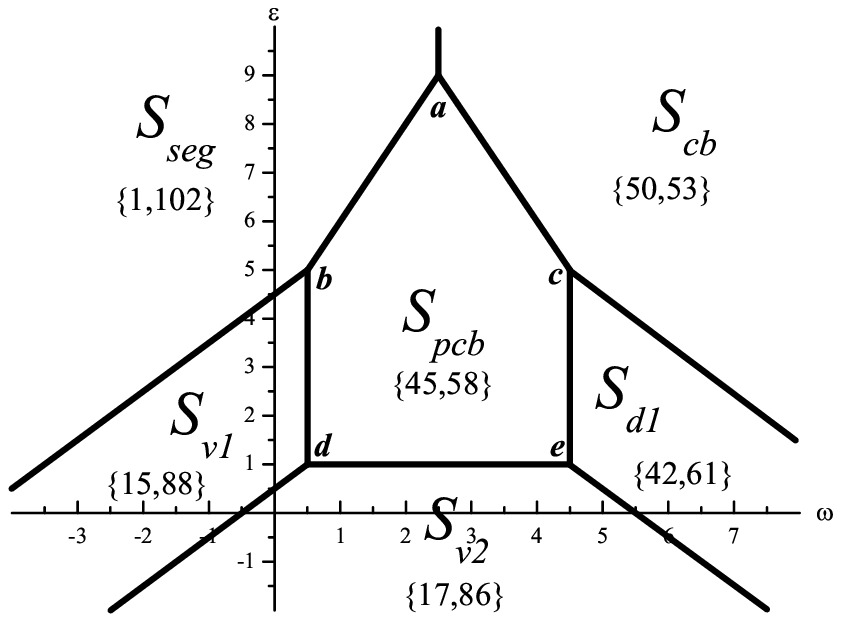}
\parbox[t]{\textwidth}{\caption{{\small{The ($\mu=0$)-phase diagram
of the effective Hamiltonian $(H^{boson}_{{\mathrm{eff}}})^{(4)}$.
The boundary-line segments can be determined by means of their
intersection points: ${\bf a} = (5/2,9)$, ${\bf b} = (1/2,5)$, ${\bf
c} = (9/2,5)$, ${\bf d} = (1/2,1)$, ${\bf e} = (9/2,1)$, and by the
slope $1$ of the boundary of ${\mathcal{S}}_{v1}$, and the slope
$-1$ of the boundary of ${\mathcal{S}}_{d1}$. For more explanations
see the description of Fig.~\ref{t1-40f}. }}} \label{t1-40b}}
\end{figure}

\begin{figure}
\centering\includegraphics[height=0.34\textheight]{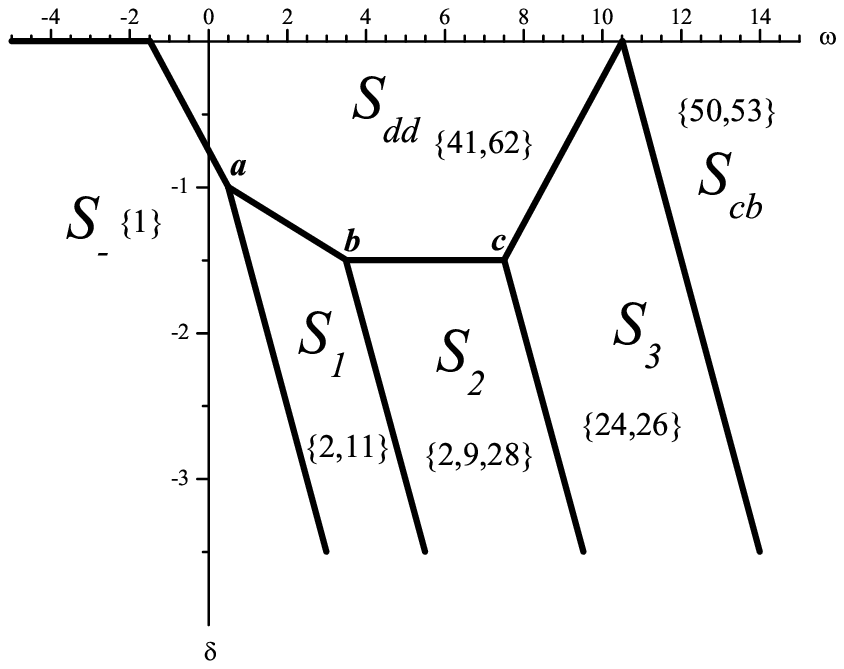}
\parbox[t]{\textwidth}{\caption{{\small{The ($\varepsilon=0$)-phase diagram
of the effective Hamiltonian $(H^{fermion}_{{\mathrm{eff}}})^{(4)}$.
The boundary-line segments can be determined by means of their
intersection points: ${\bf a} = (1/2,-1)$, ${\bf b} = (7/2,-3/2)$,
${\bf c} = (15/2,-3/2)$, and by the slope $-1$ of the boundaries of
${\mathcal{S}}_{1}$, ${\mathcal{S}}_{2}$, ${\mathcal{S}}_{3}$. For
more comments see the description of Fig.~\ref{t1-40f}.}}}
\label{t1-41f}} \vfill
\centering\includegraphics[height=0.34\textheight]{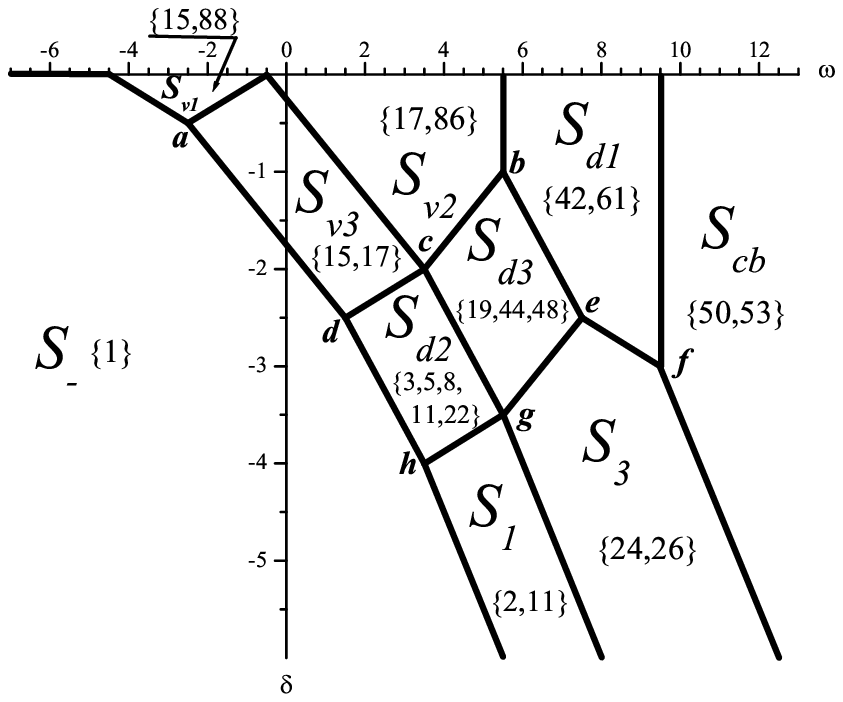}
\parbox[t]{\textwidth}{\caption{{\small{The ($\varepsilon=0$)-phase diagram
of the effective Hamiltonian $(H^{boson}_{{\mathrm{eff}}})^{(4)}$.
The boundary-line segments can be determined by means of their
intersection points: ${\bf a} = (-5/2,-1/2)$, ${\bf b} = (11/2,-1)$,
${\bf c} = (7/2,-2)$, ${\bf d} = (3/2,-5/2)$, ${\bf e} =
(15/2,-5/2)$, ${\bf f} = (19/2,-3)$, ${\bf g} = (11/2,-7/2)$, ${\bf
h} = (7/2,-4)$, and by the slope $-1$ of the boundaries of
${\mathcal{S}}_{1}$ and ${\mathcal{S}}_{3}$. For more comments see
the description of Fig.~\ref{t1-40f}}}} \label{t1-41b}}
\end{figure}

\begin{figure}
\centering \includegraphics[width=0.9\textwidth]{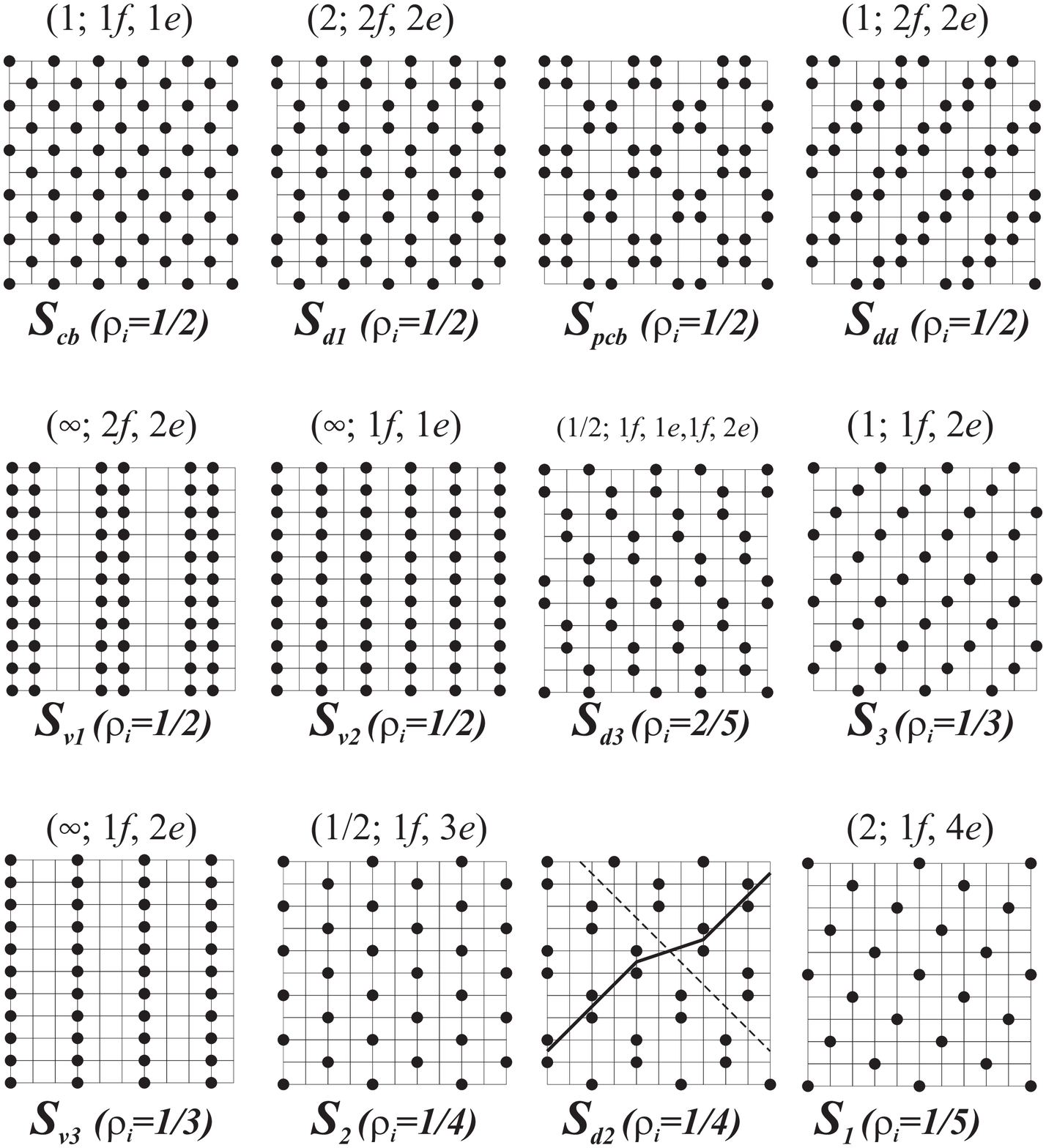}
\caption{{\small{Representative configurations of the sets
${\mathcal{S}}_{D}$  of ground-state configurations. The remaining
configurations of ${\mathcal{S}}_{D}$ can be obtained by applying
the symmetries of $H$ to the displayed configurations. As a
representative configuration of ${\mathcal{S}}_{d2}$, we show a
configuration with one defect line (the dashed line). The continuous
line is a guide for the eye. For more comments see the text.}}}
\label{conf1}
\end{figure}

Specifically, for $\delta=0$  the ground-state phase diagrams due to
the effective Hamiltonians $(H^{fermion}_{{\mathrm{eff}}})^{(4)}$
and $(H^{boson}_{{\mathrm{eff}}})^{(4)}$ are shown in
Fig.~\ref{t1-40f} and Fig.~\ref{t1-40b},
respectively, while the corresponding ground-state phase diagrams
for $\varepsilon =0$, in Fig.~\ref{t1-41f} and Fig.~\ref{t1-41b}.
In Fig.~\ref{conf1} we display the representatives of the sets
${\mathcal{S}}_{D}$ of ground-state configurations.
That is, the remaining configurations of ${\mathcal{S}}_{D}$ can be
obtained easily by applying the symmetries of $H_{0}$ to the
displayed configurations. The domain ${\mathcal{S}}_{seg}$, having
no representative in Fig.~\ref{conf1}, consists of the two
translation-invariant configurations $S_{+}$ and $S_{-}$, related by
the hole-particle transformation.

Only in the domain ${\mathcal{S}}_{d2}$, which appears in the phase
diagrams shown in Fig.~\ref{t1-41b}, that is off the hole-particle
symmetry plain, the situation is not that simple. In
${\mathcal{S}}_{d2}$ one can distinguish two classes,
${\mathcal{S}}_{d2a}$ and ${\mathcal{S}}_{d2b}$, of periodic
configurations with parallelogram elementary cells. A configuration
in ${\mathcal{S}}_{d2a}$ consists of vertical (horizontal) dimers of
filled sites that form a square lattice, where the sides of the
elementary squares have the length $2\sqrt{2}$ and the slope $\pm
1$. In a configuration of ${\mathcal{S}}_{d2b}$, the elementary
parallelograms formed by dimers have the sides of the length
$2\sqrt{2}$ and the slope $\pm 1$, and the sides of the length
$\sqrt{10}$ and the slope $\pm 1/3$. Two configurations, one from
${\mathcal{S}}_{d2a}$ and one from ${\mathcal{S}}_{d2b}$, having the
same kind of dimers (vertical or horizontal), can be merged together
along a ``defect line'' of the slope $\pm 1$ (dashed line in
Fig.~\ref{conf1}), as shown in Fig.~\ref{conf1}, without increasing
the energy. By introducing more defect lines one can construct many
ground-state configurations whose number scales with the size of the
lattice as $\exp{(const \sqrt{\Lambda})}$. While for a finite
lattice all the configurations are periodic, many of them become
aperiodic in the infinite volume limit.

As mentioned above, except ${\mathcal{S}}_{d2}$ all the other sets
${\mathcal{S}}_{D}$ contain exclusively periodic configurations. The
set  ${\mathcal{S}}_{pcb}$ (of plaquette-chessboard configurations)
contains configurations built out of elementary plaquettes with
filled sites, forming a square lattice according to the same rules
as filled sites form a square lattice in the chessboard
configurations from  ${\mathcal{S}}_{cb}$. The remaining sets
${\mathcal{S}}_{D}$ consist of configurations that have a
quasi-one-dimensional structure. That is, they are built out of
completely filled and completely empty lattice lines of given slope.
Such a configuration can be specified by the slope of the filled
lattice lines of the representative configuration and the succession
of filled (f) and empty (e) consecutive lattice lines in a period.
For instance, the representative configuration of
${\mathcal{S}}_{d1}$ (see Fig.~\ref{conf1}) is built out of filled
lines with the slope $2$ and, in the period, two consecutive filled
lines are followed by two consecutive empty lines, which is denoted
$(2;2f,2e)$. Similar description of the remaining
quasi-one-dimensional configurations is given in Fig.~\ref{conf1}.

The coefficients $\{\alpha_{i}\}$ for which the fourth order
potentials, $(H^{fermion}_{T})^{(4)} + K^{(4)}_{T}$ and
$(H^{boson}_{T})^{(4)} + K^{(4)}_{T}$, become $m$-potentials are
given in the tables collected in the Appendix E
(Tables~\ref{tb1-1}--\ref{tb1-4}).

Finally, a remark concerning ground-state configurations at the
boundaries between the open domains ${\mathcal{S}}_{D}$ is in order.
Let ${\mathcal{S}}_{D}$ and ${\mathcal{S}}_{D^{\prime}}$ be two
domains of the considered phase diagram, sharing a boundary. At this
boundary, the set of the minimizing $T$-plaquette configurations
contains always the subset ${\mathcal{S}}_{TD} \cup
{\mathcal{S}}_{TD^{\prime}}$, but it may contain also some
additional $T$-plaquette configurations of minimal energy.
Consequently, the set of the ground-state configurations at the
boundary contains always the subset ${\mathcal{S}}_{D} \cup
{\mathcal{S}}_{D^{\prime}}$ and typically a great many of other
ground-state configurations, whose number grows indefinitely with
the size of the lattice. In the considered diagrams, the only
exception is the boundary between ${\mathcal{S}}_{seg}$ and
${\mathcal{S}}_{cb}$, where the set of the ground-state
configurations amounts exactly to ${\mathcal{S}}_{seg} \cup
{\mathcal{S}}_{cb}$.

\subsection{Discussion of phase diagrams}
We start with a few comments concerning the phase diagrams due to
the fourth order effective interactions. In the fourth order
effective interaction, the parameters $\omega$ and $\varepsilon$
control the strength of n.n. and n.n.n. interactions, respectively.
The n.n. interaction is repulsive and favors the chessboard
configurations if $\omega > 9/2$, for fermions ($\omega
> 5/2$, for bosons). In the opposite case (attraction) it favors
the $S_{+}$, $S_{-}$ configurations. In turn, the n.n.n. interaction
is attractive if $\varepsilon > 3$, for fermions ($\varepsilon > 5$,
for bosons), and then it reinforces the tendency towards chessboard
and uniform, $S_{+}$, $S_{-}$, configurations. When it becomes
repulsive, it frustrates the n.n. interaction: the more negative it
is the larger $\omega$ is needed to stabilize the chessboard
configurations and the smaller $\omega$ is needed to stabilize the
$S_{+}$, $S_{-}$ configurations. Therefore, whatever the value of
$\varepsilon$ is, there is a sufficiently large $\omega$ such that
$(\varepsilon,\omega)\in {\mathcal{S}}_{cb}$ and a sufficiently
small $\omega$ such that $(\varepsilon,\omega)\in
{\mathcal{S}}_{seg}$.

Apparently, the fourth order phase diagrams in the cases of hopping
fermions and hopping bosons are quite similar if we compare the
geometry of the phase boundaries and the sets of ground states. The
main difference is in the domain occupying the central position: in
the case of fermions the ground-state configurations are
diagonal-stripe configurations, ${\mathcal{S}}_{dd}$, while in the
case of bosons this is the set ${\mathcal{S}}_{pcb}$ of
plaquette-chessboard configurations. These two sets of
configurations are ground-state configurations of the corresponding
fourth-order effective interactions with the site, n.n., and n.n.n.
terms dropped, which corresponds to the points $(9/2,3) \in
{\mathcal{S}}_{dd}$ and $(5/2,5) \in {\mathcal{S}}_{pcb}$. This
means that the fourth order phase diagrams can be looked upon as the
result of perturbing an Ising-like Hamiltonian, that consists of
pair interactions at distance of two lattice constants and plaquette
interactions, by n.n. and n.n.n. interactions. More generally, it is
easy to verify that the set of ground-state configurations of the
Ising-like Hamiltonian,
\begin{eqnarray}
\sum\limits_{\langle x,y \rangle_{3}} s_{x}s_{y} + \chi
\sum\limits_{P_{2}}  s_{P_{2}} ,
\end{eqnarray}
amount to ${\mathcal{S}}_{dd}$ if $\chi > 0$, and to
${\mathcal{S}}_{pcb}$ if $\chi < 0$. These ground states are stable
with respect to perturbations by n.n. and n.n.n. interactions, if
the corresponding interaction constants are in a certain vicinity of
zero. Otherwise, new sets of ground states, shown in
Fig.~\ref{t1-40f} and Fig.~\ref{t1-40b} emerge.

The basic question to be answered, before discussing the phase
diagrams due to the complete interaction, refers to the relation
between these phase diagrams and the diagrams due to the truncated
effective interactions, obtained in the previous section.

Firstly, we note, by inspection of the phase boundaries in
Fig.~\ref{t1-40f} and Fig.~\ref{t1-40b}, that the $T$-plaquette
configurations $\{45,58\}$ are missing in the phase diagram
corresponding to fermion mobile particles. In turn, the
$T$-plaquette configurations $\{41,62\}$ are missing in the phase
diagram corresponding to boson mobile particles. Consequently, the
ground-state configurations of ${\mathcal{S}}_{pcb}$ cannot appear
not only in the fourth order fermion phase diagram but also in the
fermion phase diagram of the complete interaction. Similarly, the
ground-state configurations of ${\mathcal{S}}_{dd}$ do not appear in
boson phase diagrams.

Secondly, by adapting the arguments presented in
\cite{Kennedy1,GMMU}, we can demonstrate, see for instance
\cite{DJ1}, that if the remainder $R^{(4)}$ is taken into account,
then there is a (sufficiently small) constant $t_{0}$ such that for
$t<t_{0}$ the phase diagram looks the same as the phase diagram due
to the effective interaction truncated at the fourth order, except
some regions of width $O(t^{6})$, located along the boundaries
between the domains, and except the domain ${\mathcal{S}}_{d2}$. For
$t<t_{0}$ and each domain ${\mathcal{S}}_{D}$, ${\mathcal{S}}_{D}
\neq {\mathcal{S}}_{d2}$, there is a nonempty two-dimensional open
domain ${\mathcal{S}}_{D}^{\infty}$ that is contained in the domain
${\mathcal{S}}_{D}$ and such that in ${\mathcal{S}}_{D}^{\infty}$
the set of ground-state configurations coincides with
${\mathcal{S}}_{D}$.

Now, consider our systems for specified particle densities,
$\rho_{e} = \rho_{i} = 1/2$. According to the phase diagrams in the
hole-particle symmetry plane ($\mu =\delta =0$), shown in
Fig.~\ref{t1-40f} and Fig.~\ref{t1-40b}, if $(\varepsilon,\omega)$
is in ${\mathcal{S}}_{seg}^{\infty}$, then the ground state is a
phase-separated state, where ion configurations are  mixtures of
$S_{+}$ and $S_{-}$ configurations, called the segregated phase.
Another phase-separated state, where the ion configurations are
mixtures of $S_{+}$, $S_{-}$, $S_{cb}^{e}$, and $S_{cb}^{o}$
configurations, exists at a line that is a small ($O(t^{2})$)
distortion of the boundary between the domains ${\mathcal{S}}_{seg}$
and ${\mathcal{S}}_{cb}$ \cite{Kennedy2}. In all the other domains
${\mathcal{S}}_{D}^{\infty}$ of these diagrams the ground-state
phase exhibits a crystalline long-range order of ions.

By the properties of the fourth order phase diagrams, whatever the
value of $\varepsilon$ is, there is a sufficiently large $\omega$
such that the systems are in the chessboard phase
($(\varepsilon,\omega)\in {\mathcal{S}}_{cb}^{\infty}$) and a
sufficiently small $\omega$ such that the systems are in the
segregated phase ($(\varepsilon,\omega)\in
{\mathcal{S}}_{seg}^{\infty}$). In between, the systems undergo a
series of phase transitions and visit various striped phases.
Consider first unbiased systems ($\varepsilon =0$). Then, it follows
from the phase diagram in Fig.~\ref{t1-40f}, that the fermion system
visits necessarily the diagonal striped phase, described by the
configurations in ${\mathcal{S}}_{dd}$. In turn, the phase diagram
in Fig.~\ref{t1-40b} implies that the boson system has to visit the
dimerized-chessboard phase, described by the configurations in
${\mathcal{S}}_{d1}$, then the vertical/horizontal striped phase,
described by the configurations in ${\mathcal{S}}_{v2}$, and after
that a dimerized vertical/horizontal striped phase, described by the
configurations in ${\mathcal{S}}_{v1}$. These scenarios are
preserved, if $\varepsilon \in (-2 + O(t^{2}),12 - O(t^{2}))$ in the
fermion case, and $\varepsilon \in (- \infty,1 - O(t^{2}))$ in the
boson case.

Note that the same sequence of transitions, as found in the unbiased
boson system, can be realized by the fermion system if $\varepsilon$
is sufficiently small ($\varepsilon < -6 - O(t^{2})$).

For somewhat larger values of $\varepsilon$, $\varepsilon \in (-6 +
O(t^{2}),-2 - O(t^{2}))$, on its way from the chessboard phase to
the segregated phase the fermion systems visits the dimerized
chessboard phase, then the diagonal striped phase, and after that
the dimerized vertical/horizontal striped phase. The
vertical/horizontal striped phase is closer to the segregated phase
than the diagonal striped phase. In case the both kinds of stripe
phases (vertical/horizontal and diagonal) appear in a phase diagram,
such a succession of striped phases is perhaps generic, see also
\cite{LFB1,SDHC1}.

For sufficiently large $\varepsilon$, the tendency toward the
chessboard and the uniform configurations is so strong that the both
systems jump directly from the chessboard phase to the segregated
phase.

Above, we have described the states of the considered systems for
typical values of the control parameter $\omega$. Only in small
intervals (whose width is of the order of $O(t^{2})$) of values of
$\omega$, about the transition points of the diagrams in
Fig.~\ref{t1-40f} and Fig.~\ref{t1-40b}, the states of the systems
remain undetermined.

For completeness we present also ground-state phase diagrams of the
both systems off the hole-particle symmetry plane, for
$\varepsilon=0$, see Fig.~\ref{t1-41f},~\ref{t1-41b}. Away from the
hole-particle symmetry plane new phases appear. In particular we
find the phases described by the quasi-one-dimensional
configurations ${\mathcal{S}}_{1}$, ${\mathcal{S}}_{2}$, and
${\mathcal{S}}_{3}$, well known from the studies of the phase
diagram of the spinless Falicov--Kimball model \cite{GMMU}. In the
boson phase diagram, Fig.~\ref{t1-41b}, there appears the domain
${\mathcal{S}}_{d2}$ that is not amenable to the kind of arguments
used in this thesis.

\clearpage
\section{Influence of nearest-neighbor anisotropy on axial striped phases}
In this Section\footnote{This section is based on \cite{DJ3}.} we
show that any anisotropy of nearest-neighbor hopping eliminates the
$\pi/2$-rotation degeneracy of the dimeric and axial-stripe phases
and orients them in the direction of a weaker hopping. Moreover, due
to the same anisotropy the obtained phase diagrams of fermions  show
a tendency to become similar to those of hardcore bosons.

\subsection{The effective interaction up to the fourth order}

We consider the model similar to (\ref{ourhamilt}): only a n.n.
hopping is taken into account and we allow for its anisotropy. We
set $t_{h} \equiv t$, and $t_{v} = \sqrt{\gamma}t$, with  $0 \leq
\gamma \leq 1$. In this case, up to a term independent of the ion
configurations and the chemical potentials, the ground-state energy
expansion (\ref{effh}) reads:
\begin{eqnarray}
\label{Expfmn1} E^{f}_{S}\left(\mu\right) &=&
\left(E^{f}_{S}\right)^{(4)}\left(\mu\right) +
\left(R_{S}^{f}\right)^{(4)},
\nonumber \\
\left(E^{f}_{S}\right)^{(4)}\left(\mu\right) &=&
-\frac{\mu}{2}\sum\limits_{x} \left( s_{x} + 1 \right)+ \left[
\frac{t^{2}}{4}- \frac{3t^{4}}{16}-\frac{3}{8}\gamma t^{4} +
\frac{W}{8} \right] \sum\limits_{\langle x,y \rangle_{1,h}}
s_{x}s_{y}+ \nonumber \\
&+& \left[ \gamma \frac{t^{2}}{4} -\frac{3}{8}\gamma t^{4} -
\gamma^{2}\frac{3t^{4}}{16} + \frac{W}{8} \right]
\sum\limits_{\langle x,y \rangle_{1,v}} s_{x}s_{y}+ \left[ \gamma
\frac{3t^{4}}{16}-\frac{\tilde{\varepsilon}}{16} \right]
\sum\limits_{\langle x,y \rangle_{2}} s_{x}s_{y} + \nonumber \\
&+& \frac{t^{4}}{8} \sum\limits_{\langle x,y \rangle_{3,h}}
s_{x}s_{y}+\gamma^{2} \frac{t^{4}}{8} \sum\limits_{\langle x,y
\rangle_{3,v}} s_{x}s_{y} + \gamma \frac{t^{4}}{16}
\sum\limits_{P_{2}} \left(1+5s_{P_{2}}\right),
\end{eqnarray}
in the case of hopping fermions, and
\begin{eqnarray}
\label{Exphcb1} E^{b}_{S}\left(\mu\right) &=&
\left(E^{b}_{S}\right)^{(4)}\left(\mu\right) +
\left(R_{S}^{b}\right)^{(4)},
\nonumber \\
\left(E^{b}_{S}\right)^{(4)}\left(\mu\right) &=&
-\frac{\mu}{2}\sum\limits_{x} \left( s_{x} + 1 \right)+ \left[
\frac{t^{2}}{4}- \frac{3t^{4}}{16} -\frac{1}{8}\gamma t^{4} +
\frac{W}{8} \right] \sum\limits_{\langle x,y \rangle_{1,h}}
s_{x}s_{y} + \nonumber \\
&+&  \left[ \gamma \frac{t^{2}}{4} -\frac{1}{8}\gamma t^{4} -
\gamma^{2}\frac{3t^{4}}{16} + \frac{W}{8} \right]
\sum\limits_{\langle x,y \rangle_{1,v}} s_{x}s_{y}+ \left[ \gamma
\frac{5t^{4}}{16}-\frac{\tilde{\varepsilon}}{16} \right]
\sum\limits_{\langle x,y \rangle_{2}} s_{x}s_{y} +\nonumber \\
&+& \frac{t^{4}}{8} \sum\limits_{\langle x,y \rangle_{3,h}}
s_{x}s_{y} + \gamma^{2} \frac{t^{4}}{8} \sum\limits_{\langle x,y
\rangle_{3,v}} s_{x}s_{y} - \gamma \frac{t^{4}}{16}
\sum\limits_{P_{2}} \left(5+s_{P_{2}}\right),
\end{eqnarray}
in the case of hopping hardcore bosons. As previously, the
remainders, $\left(R_{S}^{f}\right)^{(4)}$ and
$\left(R_{S}^{b}\right)^{(4)}$, are independent of the chemical
potentials and the parameters $W$ and $\tilde{\varepsilon}$, but in
this case, they depend on $\gamma$, and collect those terms of the
expansions that are proportional to $t^{2m}$, with $m=3,4,\ldots$.
The expressions (\ref{Expfmn1}) and (\ref{Exphcb1}) refer to
(\ref{Expfmn}) and (\ref{Exphcb}) for $\gamma=1$, respectively.

In the previous Section we have obtained the ground-state phase
diagrams, according to the fourth-order isotropic effective
Hamiltonians $(\gamma=1)$, for hole-particle symmetric systems ($\mu
=0$) with a weak (of fourth order) n.n.n. subsidiary interaction,
and for unsymmetrical systems without the subsidiary interaction
($\tilde{\varepsilon}=0$), see the top phase diagrams in
Figs.~\ref{t1-40f}-\ref{t1-41b}. Second-order phase diagrams of the
isotropic case consist exclusively of phases whose configurations
are invariant with respect to $\pi/2$-rotations, with {\em
macroscopic degeneracies} (i.e. the number of configurations grows
exponentially with the number of sites) at the boundaries of phase
domains. Stripe phases, whose configurations are not invariant with
respect to $\pi/2$-rotations, appear on perturbing the second-order
phase diagrams by the fourth-order isotropic interactions. Here, we
would like to observe the influence of a weak anisotropy of n.n.
hopping on these stripe phases. Since we are working with truncated
effective Hamiltonians, we have to assign an order to the deviation
of the anisotropy parameter $\gamma$ from the value $1$
(corresponding to the isotropic case). Therefore, we introduce an
anisotropy order, $a$, and a new anisotropy parameter, $\beta_{a}$:
\begin{eqnarray}
\gamma=1-\beta_{a}t^{a}. \label{parambeta}
\end{eqnarray}
The orders of the deviations can be neither too small, not to modify
the second-order effective Hamiltonians, nor too high, to effect the
considered effective Hamiltonian of the highest order. Since here,
the highest order of the effective Hamiltonians is $k=4$, the
weakest admissible deviation from the isotropic case corresponds to
the highest anisotropy order $a=2$. Then, we can consider an
intermediate deviation, i.e. $0<a<2$. The strongest deviation, i.e.
$a=0$, is not admissible, since it modifies the second-order
effective Hamiltonians.

\subsection{The smallest deviation from the isotropic case}
In the sequel, we drop the arguments of ground-state energies, that
is we set $\left( E^{f}_{S} \right)^{(4)}\left(\mu \right) \equiv
\left( E^{f}_{S} \right)^{(4)}$, etc. In the case of the smallest
deviation from the isotropic case the fourth-order effective
Hamiltonian for fermions reads:
\begin{eqnarray}
\label{4ordsmallani} \left( E^{f}_{S} \right)^{(4)} &=& \left.\left(
E^{f}_{S} \right)^{(4)}\right|_{\gamma=1} - \beta_{2}
\frac{t^{4}}{4} \sum\limits_{\langle x,y \rangle_{1,v}} s_{x}s_{y},
\end{eqnarray}
while for hardcore bosons only the first term, representing the
isotropic fourth-order effective Hamiltonian, has to be changed
properly.

Apparently, the effective Hamiltonians of order zero and two are
isotropic, and there is no difference between the cases of hopping
fermions and hopping bosons (see Figs.~\ref{zerord},~\ref{secord}).
To observe the effect of the fourth-order anisotropy term of
(\ref{4ordsmallani}) the construction of the fourth-order diagrams
has to be carried out again. As in the isotropic case, this is
facilitated by introducing new variables, $\omega$, $\delta$, and
$\varepsilon$ given by (\ref{newcoor}), and rewriting the fourth
order effective Hamiltonian in the form,
\begin{eqnarray}
\left(E^{f}_{S} \right)^{(4)} = \frac{t^{4}}{2} \sum\limits_{T}
\left( H^{f}_{T} \right)^{(4)},
\end{eqnarray}
where
\begin{eqnarray}
\left( H^{f}_{T} \right)^{(4)} = \left.\left( H^{f}_{T}
\right)^{(4)} \right|_{\gamma=1} - \frac{\beta_{2}}{12}
\sideset{}{''}\sum\limits_{\langle x,y \rangle_{1,v}} s_{x}s_{y},
\label{HTf4-ani1}
\end{eqnarray}
with analogous expressions in the bosonic case, and with the
isotropic potentials $\left.\left( H^{f}_{T} \right)^{(4)}
\right|_{\gamma=1}$, $\left.\left( H^{b}_{T} \right)^{(4)}
\right|_{\gamma=1}$ given by (\ref{H4fmn-1}) and (\ref{H4hcb-1}),
respectively.

We would like to get an idea of the phase diagram in the space of
the four energy parameters $(\omega , \varepsilon , \beta_{2},
\delta )$, that appear in the Hamiltonian. Due to the fact that the
domains occupied by the phases are polyhedral sets, this goal can be
achieved by studying phase diagrams in two-dimensional hyperplanes.
Of particular interest are those hyperplanes that result from
intersecting the four-dimensional space by hyperplanes $\delta=0$
and $\beta_2 = const$, and by hyperplanes $\varepsilon=0$ and
$\beta_2 = const$. A collection of such sections for suitable values
of  $\beta_2$, forming a finite increasing from zero sequence,
enable us to observe how the anisotropy effects our system if it is
hole-particle symmetric and if it is not, respectively.

\begin{figure}[p]
\begin{center}
\includegraphics[totalheight=0.28\textwidth,origin=c]{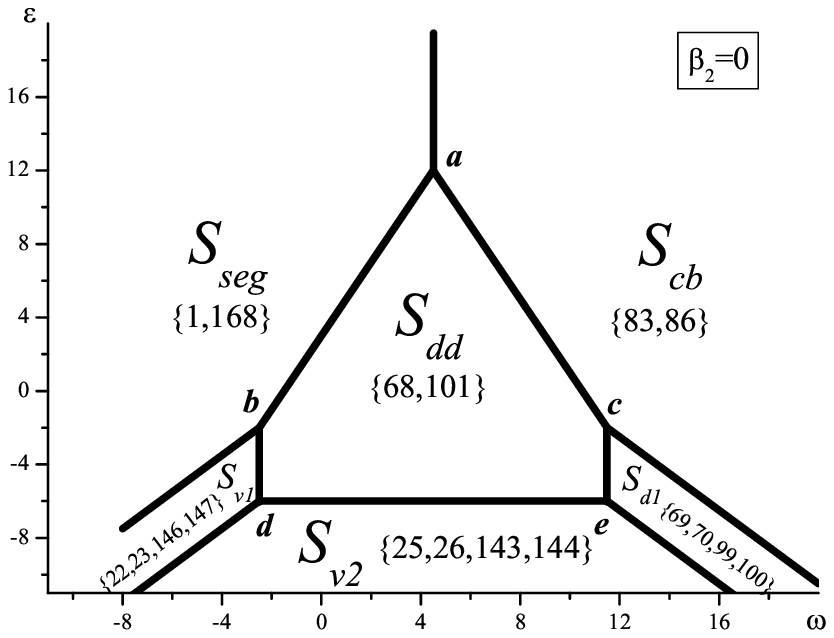}
\includegraphics[totalheight=0.28\textwidth,origin=c]{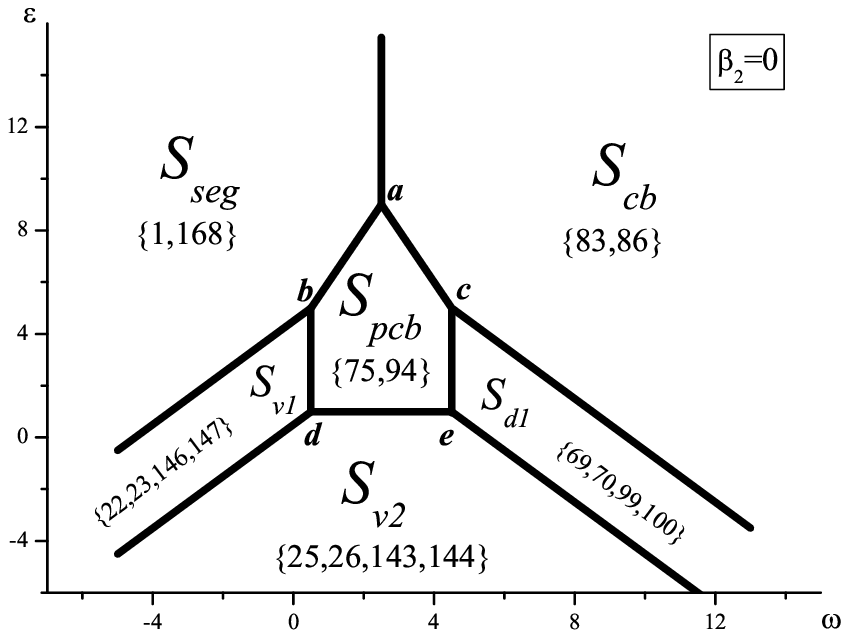}\\
\includegraphics[totalheight=0.28\textwidth,origin=c]{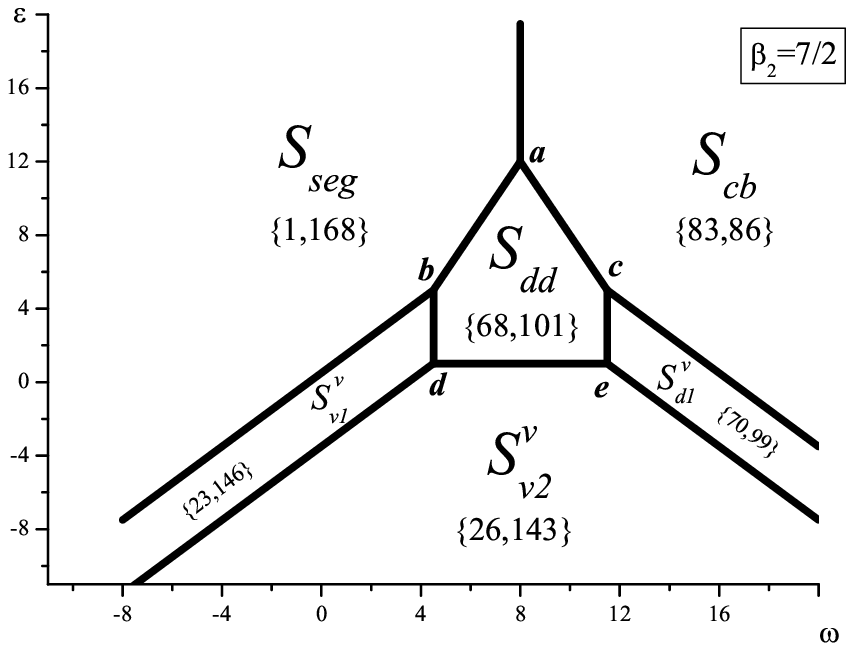}
\includegraphics[totalheight=0.28\textwidth,origin=c]{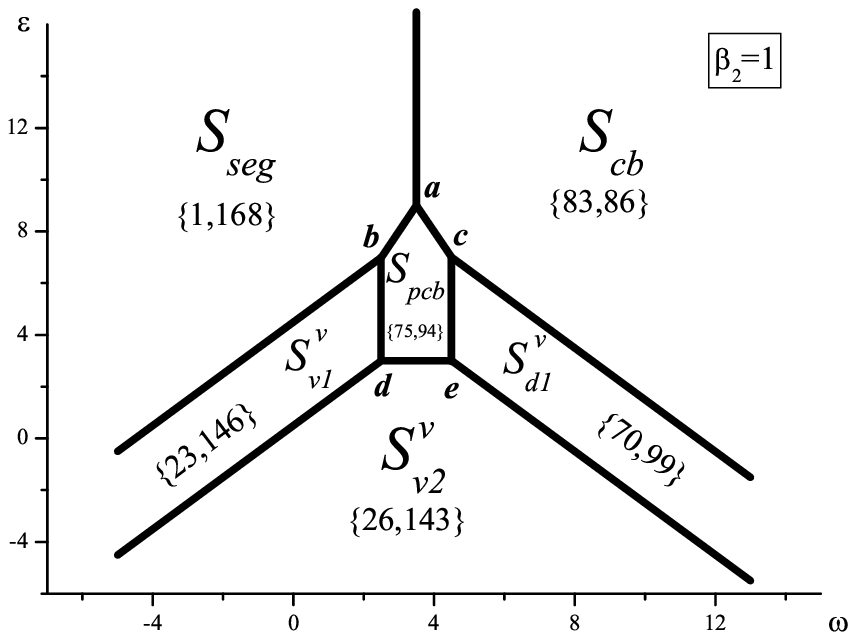}\\
\includegraphics[totalheight=0.28\textwidth,origin=c]{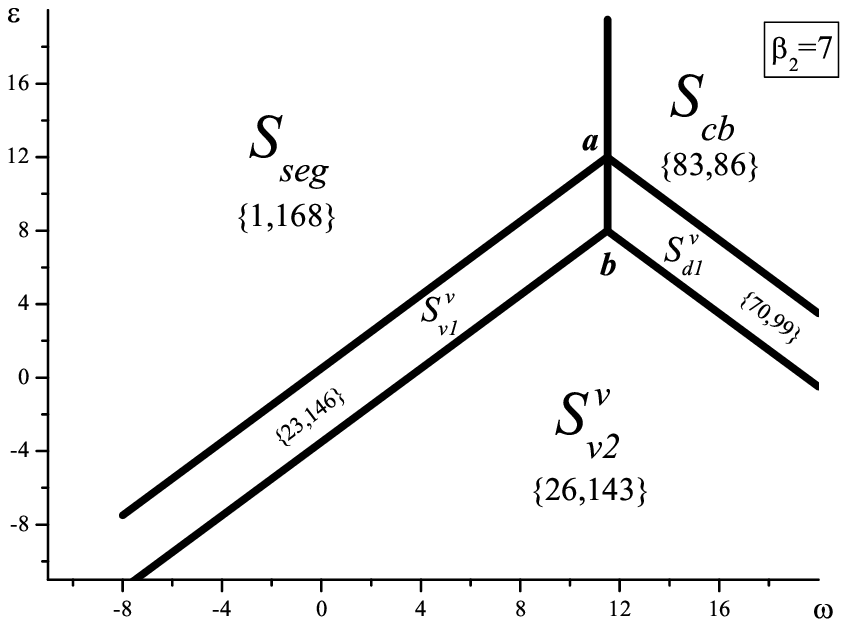}
\includegraphics[totalheight=0.28\textwidth,origin=c]{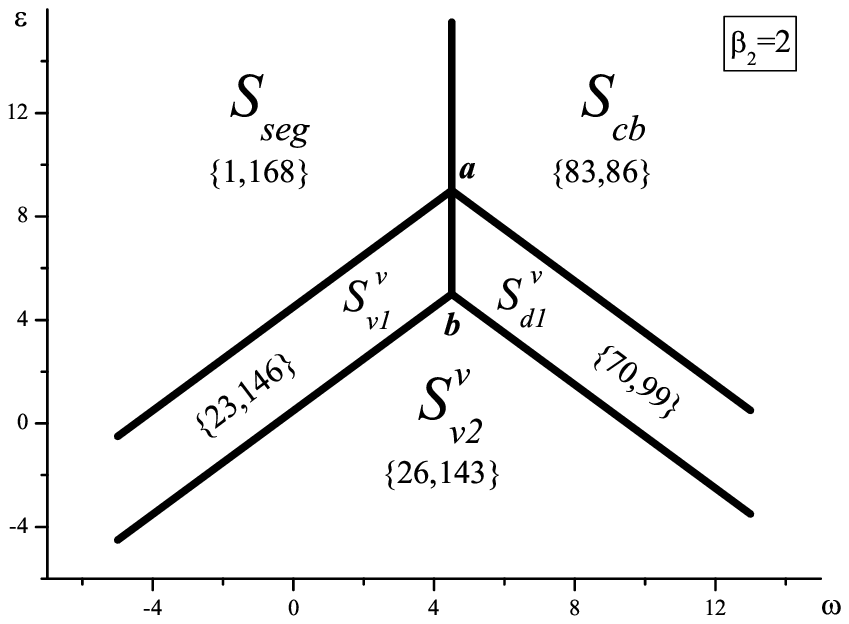}\\
\includegraphics[totalheight=0.28\textwidth,origin=c]{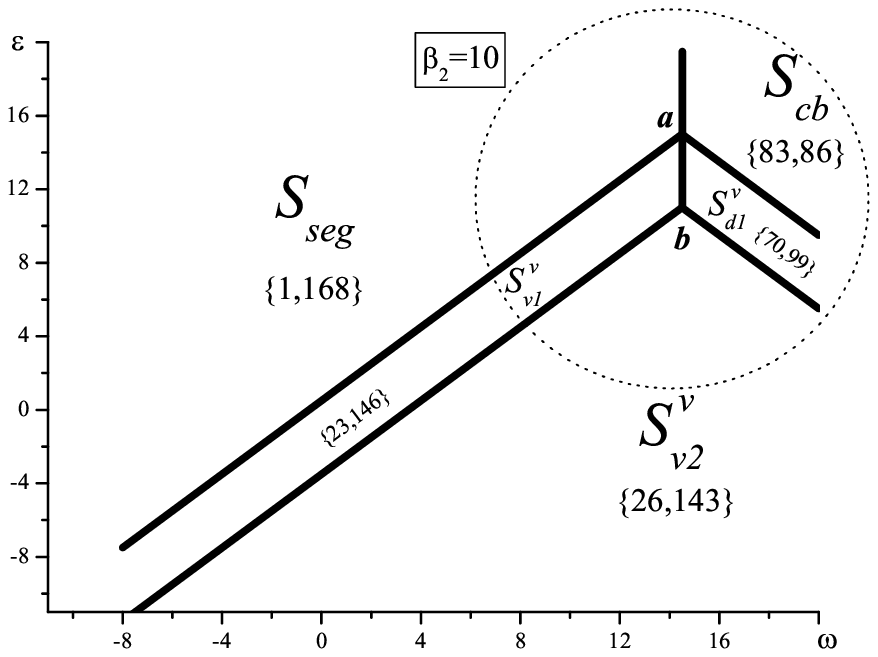}
\includegraphics[totalheight=0.28\textwidth,origin=c]{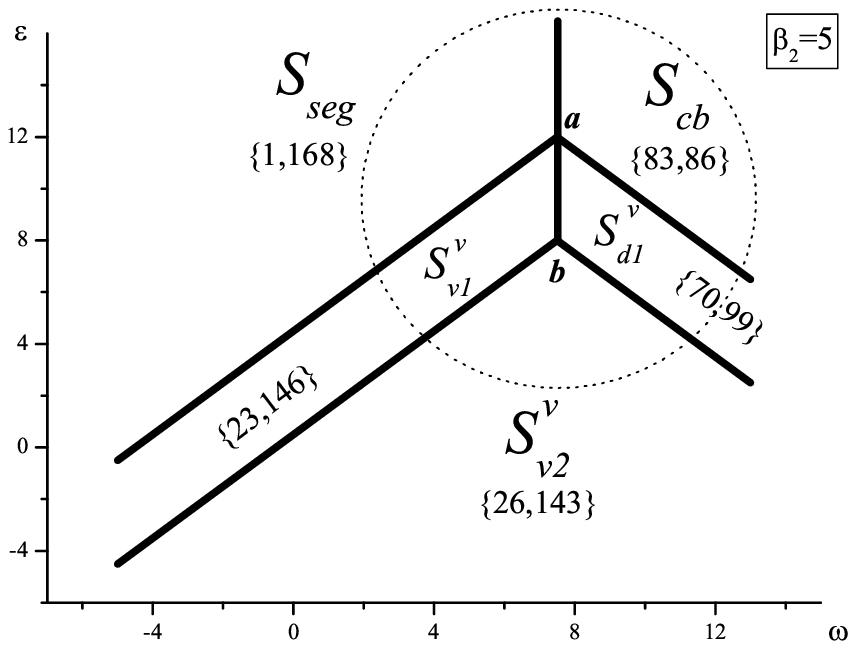}
\end{center}
\caption{\scriptsize{The case of the weakest anisotropy and of
hole-particle symmetric systems ($\mu =0$). Phase diagrams of
$\left(H_{T}^{f}\right)^{(4)}$ (given by (\ref{HTf4-ani1})) --- left
column, and $\left(H_{T}^{b}\right)^{(4)}$ --- right column, for an
increasing sequence of values of $\beta_2$. The representative ion
configurations of the displayed phases are shown in Fig.~\ref{conf1}
(for more comments see text). For fermions, the critical value is
$\beta_2=7$, while for bosons it is $\beta_2 =2$. The equations
defining the boundary lines of the phase domains are given in
Tab.~\ref{D-tb1} and Tab.~\ref{D-tb2} of Appendix D, while the
corresponding zero-potential coefficients $\{\alpha_{i} \}$ --- in
Tab.~\ref{E-tb8} -- Tab.~\ref{E-tb15} of Appendix E. In the bottom
diagrams, the regions surrounded by dotted circles, are reconsidered
in the case of an intermediate anisotropy. }} \label{ani-sppd}
\end{figure}

\begin{figure}[p]
\begin{center}
\includegraphics[totalheight=0.27\textwidth,origin=c]{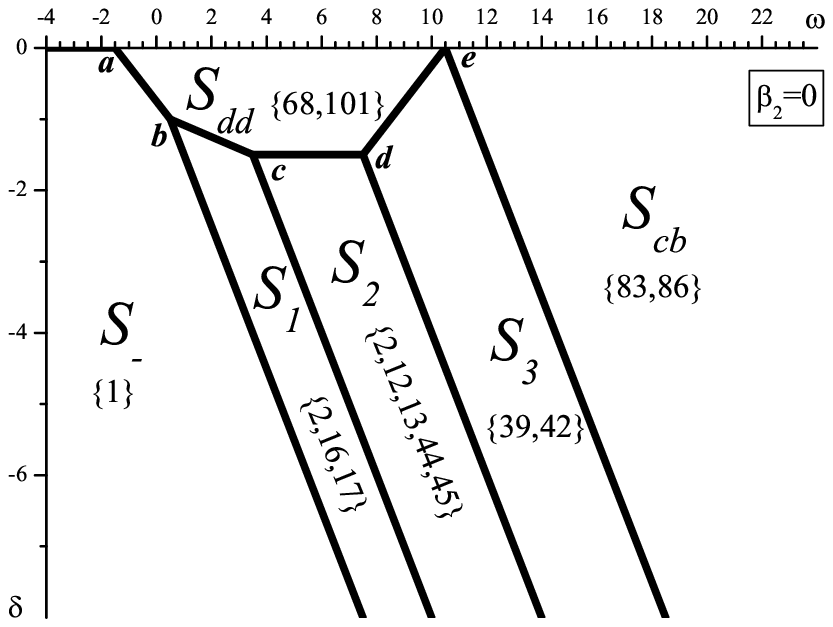}
\includegraphics[totalheight=0.27\textwidth,origin=c]{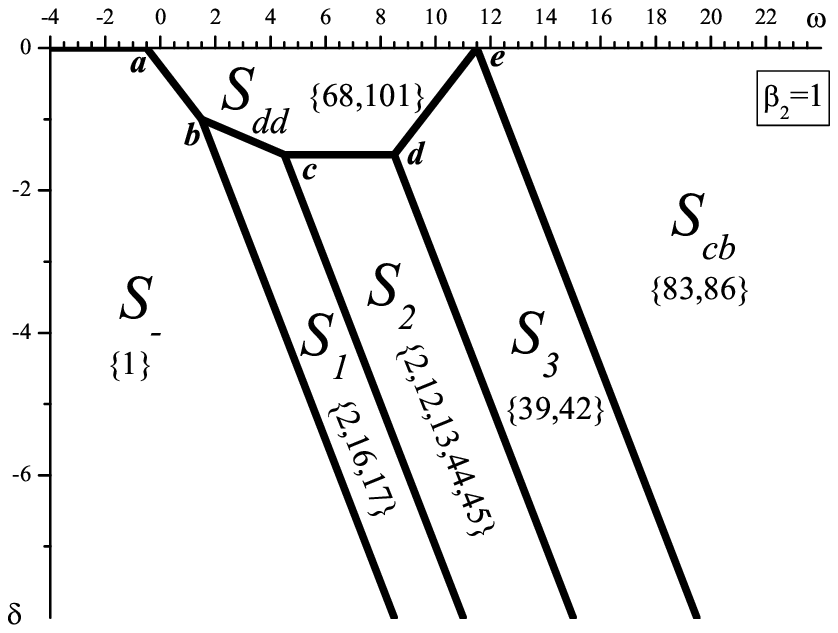}\\
\includegraphics[totalheight=0.27\textwidth,origin=c]{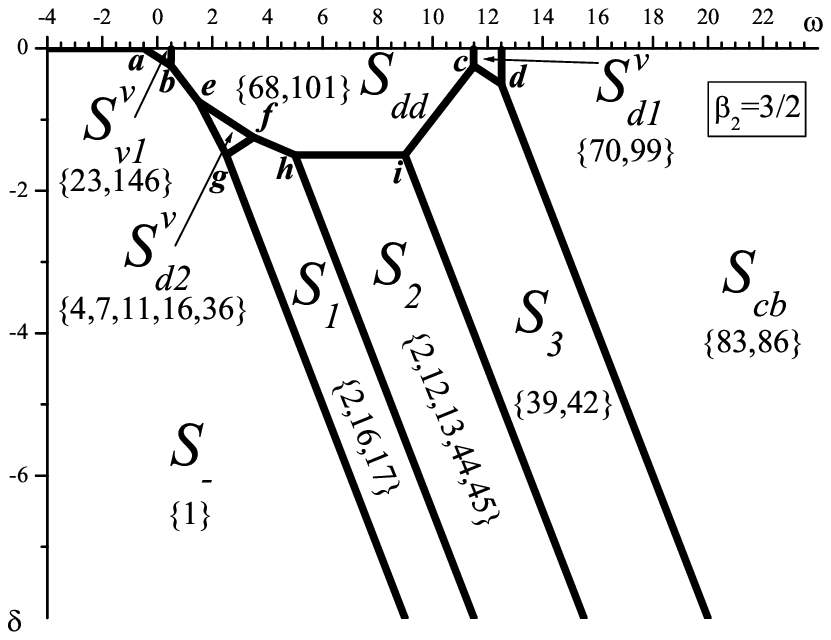}
\includegraphics[totalheight=0.27\textwidth,origin=c]{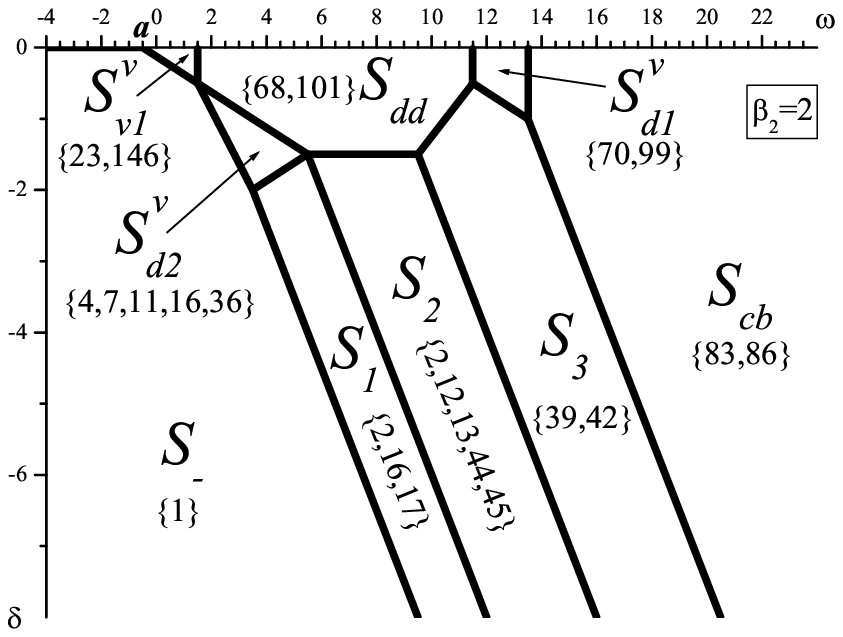}\\
\includegraphics[totalheight=0.27\textwidth,origin=c]{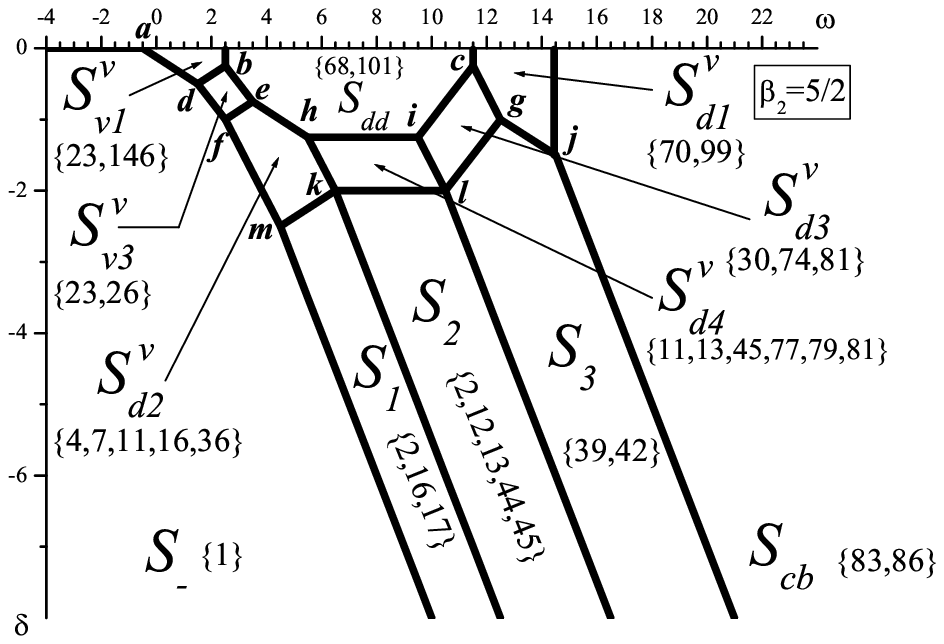}
\includegraphics[totalheight=0.27\textwidth,origin=c]{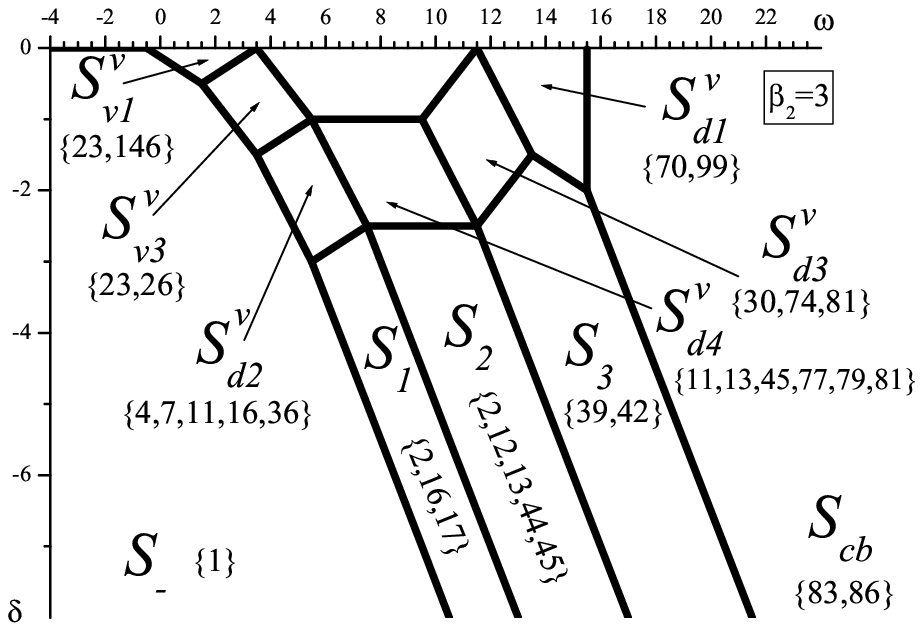}\\
\includegraphics[totalheight=0.27\textwidth,origin=c]{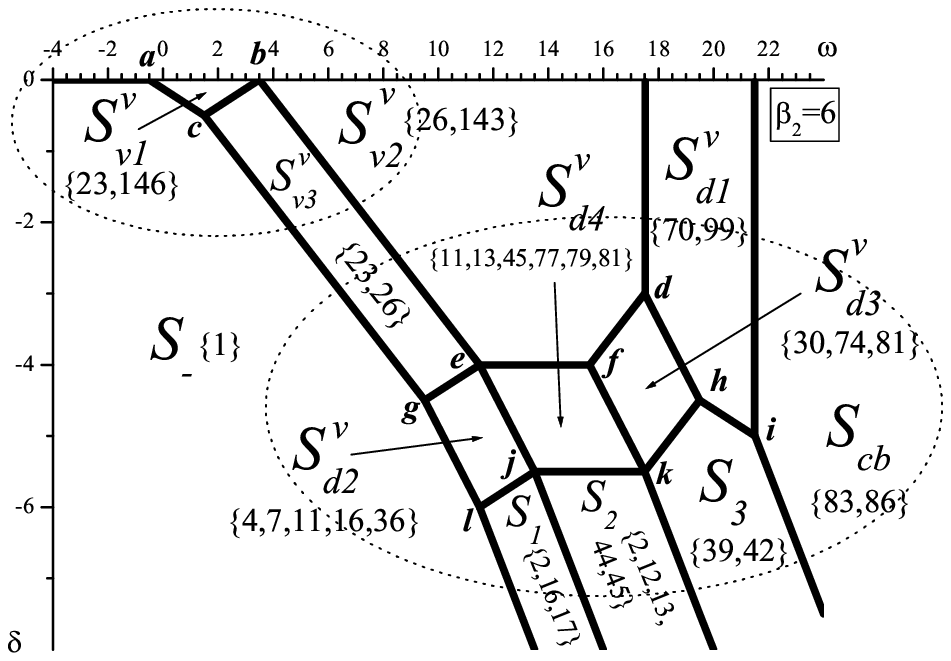}
\end{center}
\caption{\scriptsize{The case of the weakest anisotropy, off the
hole-particle symmetry, with $\varepsilon=0$. The phase diagram of
$\left(H_{T}^{f}\right)^{(4)}$ (given by (\ref{HTf4-ani1})) for an
increasing sequence of values of $\beta_2$. The representative ion
configurations of the displayed phases are shown in Fig.~\ref{conf1}
(for more comments see text). The critical values are: $\beta_2 =1$,
$\beta_2 =2$, and $\beta_2 =3$. In the blank region of the phase
diagram for the critical $\beta_2 =3$, the following $T$-plaquette
configurations have the minimal energy:
26,52,53,68,81,88,101,116,117,143 (see Fig.~\ref{bc168}). By means
of these $T$-plaquette configurations one can construct
${\mathcal{S}}^{v}_{v2}$, ${\mathcal{S}}_{dd}$, and many other
configurations. The equations defining the boundary lines of the
phase domains are given in Tab.~\ref{D-tb3} of Appendix D, while the
corresponding zero-potential coefficients $\{ \alpha_{i} \}$ --- in
Tab.~\ref{E-tb16} -- Tab.~\ref{E-tb22} of Appendix E. In the bottom
diagram, the regions surrounded by dotted ellipses are reconsidered
in the case of an intermediate anisotropy. }} \label{ani-epspdf}
\end{figure}

\begin{figure}[ht]
\includegraphics[width=0.43\textwidth]{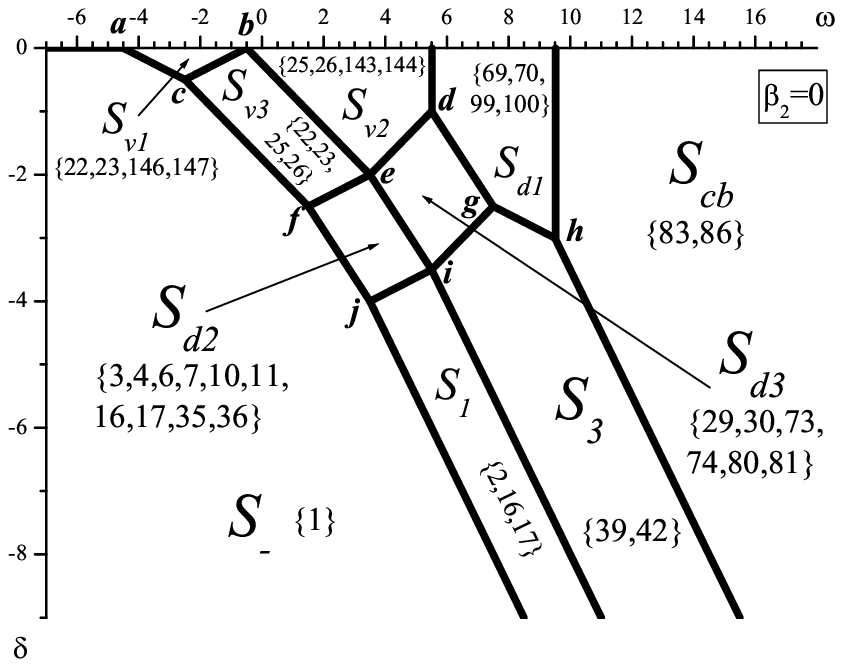}
\hfill
\includegraphics[width=0.47\textwidth]{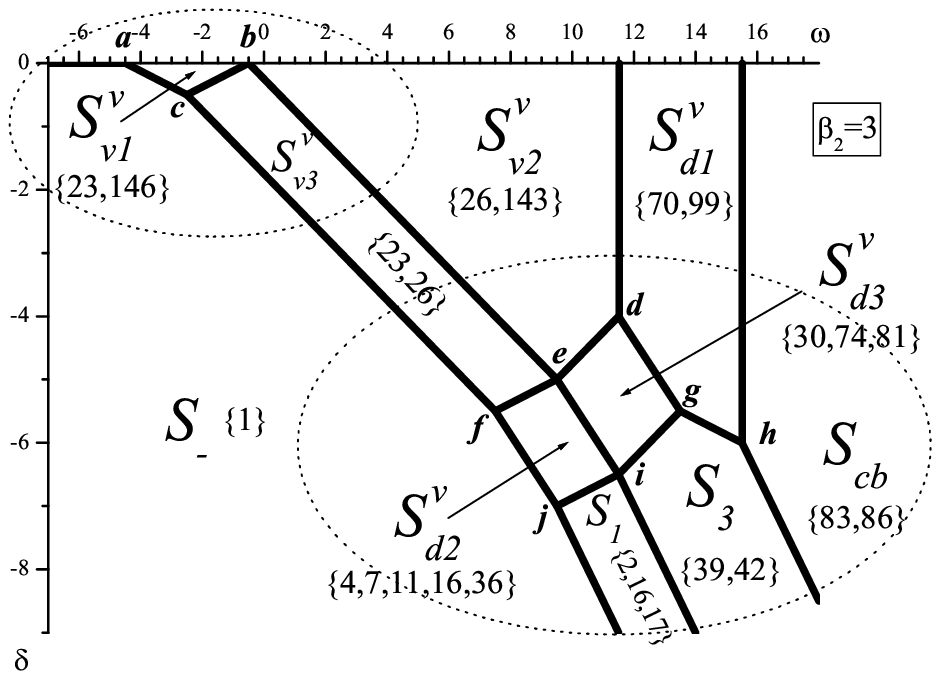}
\caption{\scriptsize{The case of the weakest anisotropy, off the
hole-particle symmetry, with $\varepsilon=0$. The phase diagram of
$\left(H_{T}^{b}\right)^{(4)}$ (given by (\ref{HTf4-ani1})); the
isotropic diagram ($\beta_2=0$) and the anisotropic diagram
($\beta_2=3$). The representative ion configurations of the
displayed phases are shown in Fig.~\ref{conf1} (for more comments
see text). Here, no critical values have been detected. The
equations defining the boundary lines of the phase domains are given
in Tab.~\ref{D-tb4} of Appendix D, while the corresponding
zero-potential coefficients $\{ \alpha_{i} \}$ --- in
Tab.~\ref{E-tb23} and Tab.~\ref{E-tb24} of Appendix E. In the right
diagram, the regions surrounded by dotted ellipses are reconsidered
in the case of an intermediate anisotropy. }} \label{ani-epspdb}
\end{figure}

Specifically, in Fig.~\ref{ani-sppd} we show phase diagrams in the
hole-particle-symmetric case, while in the absence of the
hole-particle symmetry, the phase diagrams are shown in
Fig.~\ref{ani-epspdf} -- for fermions, and in Fig.~\ref{ani-epspdb}
-- for bosons.

\begin{figure}[ph]
\centering \includegraphics[width=0.9\textwidth]{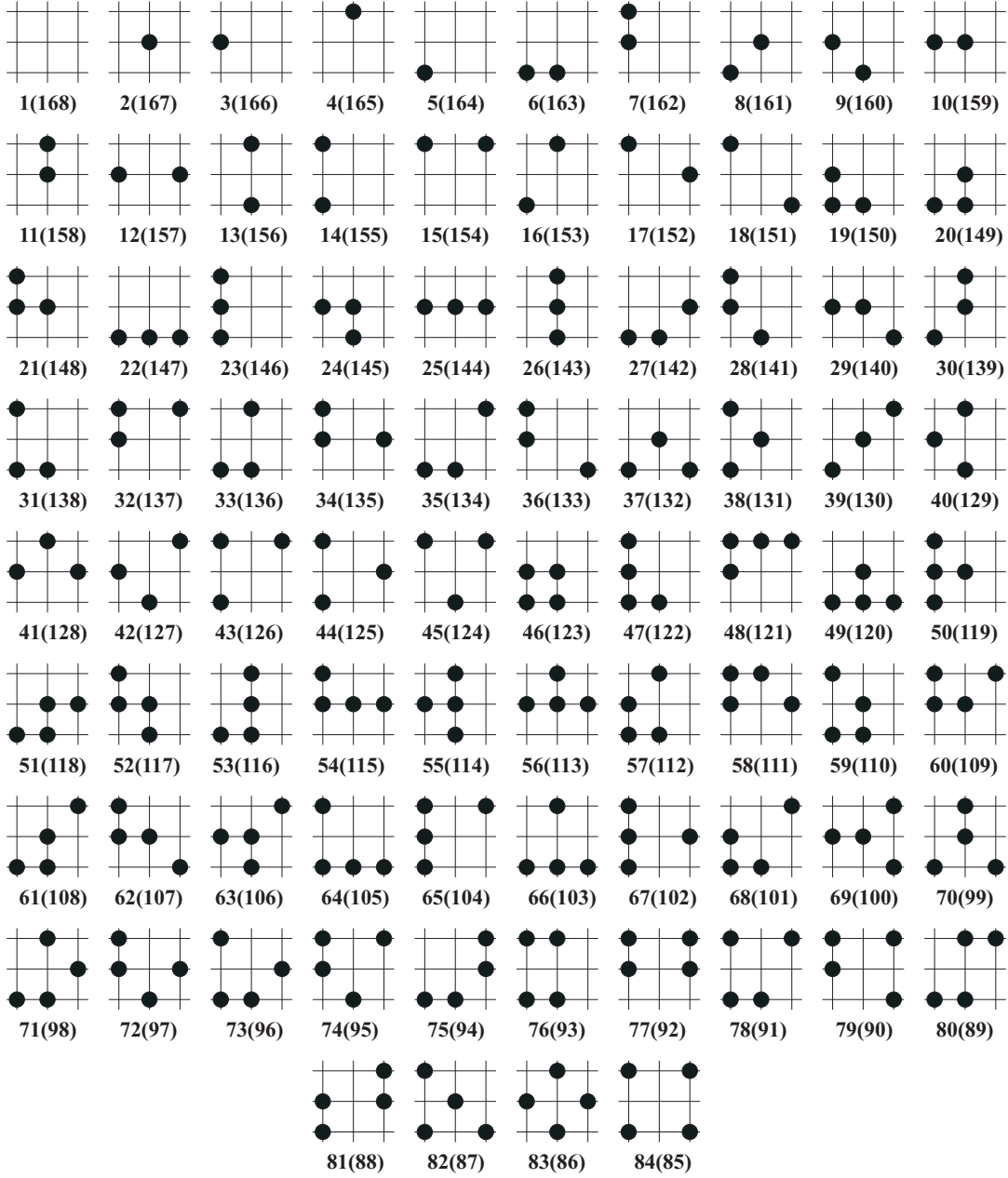}
\caption{\scriptsize{All the $T$-plaquette configurations in the
case of anisotropic interactions of n.n. hopping, up to rotations by
$\pi$ and reflections in lattice lines parallel to the axes. The
configurations that can be obtained from the displayed ones by the
hole-particle transformation are not shown, only the numbers
assigned to them are given in the brackets.}} \label{bc168}
\end{figure}

The phases that appear in these phase diagrams can conveniently be
described in terms of the phases found in ground-state phase
diagrams of the isotropic systems, with Hamiltonian $H$ given by
(\ref{ourhamilt}) (see Fig.~\ref{conf1}). The phases of anisotropic
systems, considered here, either coincide with or are simple
modifications of the isotropic phases. In fact, in Fig.~\ref{conf1}
only representative configurations of the phases of isotropic
systems are displayed. The remaining configurations can be obtained
by means of the spatial symmetry operations of isotropic
Hamiltonian, like translations and rotations by $\pi /2$. The
numbers in curly brackets, placed by the symbols of phases, stand
for the numbers of the $T$-plaquette configurations (according to
Fig.~\ref{bc168}) that are obtained by restricting the
configurations of a phase to a $T$-plaquette. By the same symbols as
the phases we denote also their domains.

Among the phases of isotropic systems, we can distinguish a class of
{\em dimeric phases}, ${\mathcal{S}}_{d1}$, \ldots,
${\mathcal{S}}_{d4}$, and a class of {\em axial-stripe phases},
${\mathcal{S}}_{v1}$, ${\mathcal{S}}_{v2}$, and
${\mathcal{S}}_{v3}$. The configurations of dimeric phases consist
of isolated pairs of n.n. occupied sites (in the sequel called {\em
the dimers}). In the configurations of axial-stripe phases, the ions
fill completely some, parallel to one of the axes, lattice lines, so
that a periodic pattern of {\em stripes} is formed. Out of the
dimeric or axial-stripe configurations of Fig.~\ref{conf1}, only
those with dimers or stripes, respectively, oriented vertically
appear in the phase diagrams of anisotropic systems. Such a
restricted phases are marked in anisotropic phase diagrams by the
additional superscript, $v$.

\begin{figure}[ht]
\begin{center}
\centering \includegraphics[width=0.5\textwidth]{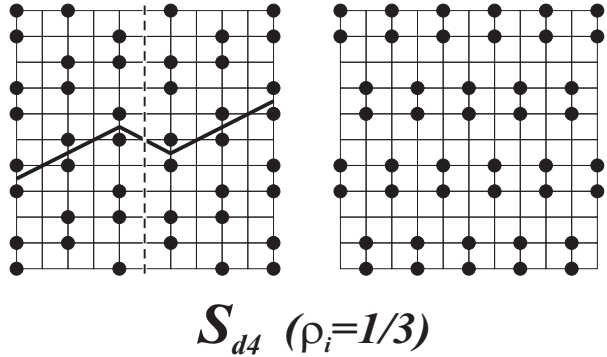}
\caption{\scriptsize{The representative configurations of the set
${\mathcal{S}}^{v}_{d4}$ whose degeneracy grows like $\exp{(const
\sqrt{\Lambda})}$. The left configuration is an example of
configurations with defect lines. Here, there is one defect line
(the dashed line, the continuous line is a guide for the eye): the
vertical lattice line separating a periodic configuration of
vertical dimers from its vertical translate by one lattice constant.
The right configuration is a periodic configuration of dimers. For
more comments see text. }} \label{conf2}
\end{center}
\end{figure}

In comparison with the isotropic phase, a new domain
${\mathcal{S}}_{d4}$ appears in the phase diagrams shown in
Fig.~\ref{ani-epspdf}. Similarly to ${\mathcal{S}}_{d2}$, its
degeneracy grows indefinitely with the size of the lattice. This
phase consists of three classes of periodic configurations of
dimers. In the class ${\mathcal{S}}_{d4a}$, the elementary cell can
be chosen as a parallelogram whose two sides of length $3$ are
parallel to dimers (which are vertical or horizontal). If the dimers
are oriented vertically, then the other two sides have the slope
$1/2$ and the length $\sqrt{5}$. By reflecting an elementary cell of
${\mathcal{S}}_{d4a}$ in a lattice line passing through its side
that is parallel to dimers, we obtain an elementary cell of the
class ${\mathcal{S}}_{d4b}$. In the third class,
${\mathcal{S}}_{d4c}$, an elementary cell can be chosen as a rhomb
formed by the centers of dimers, with the sides of length
$\sqrt{10}$. Two configurations, one from ${\mathcal{S}}_{d4a}$ and
one from ${\mathcal{S}}_{d4b}$, having the same kind of dimers
(vertical or horizontal), can be merged together along a ``defect
line'' parallel to dimers (dashed line in Fig.~\ref{conf2}) without
increasing the energy. In this way, a numerous family of
configurations can be constructed, with the number of configurations
growing like $\exp{(const \sqrt{\Lambda})}$.

It follows from the polyhedral shape of the phase domains, that the
set of values of $\beta_2$ is partitioned into open intervals, where
the boundary lines of phase domains do not change their direction,
only their distance to the origin varies in an affine way. The
boundary points of these open intervals are special values of the
anisotropy for the phase diagrams, and in the sequel we call them
{\em the critical values}. As a critical value of anisotropy is
approached, some boundary lines merge into a line or a point, which
results in disappearance of some phase domains. And vice versa, some
points and lines break off, creating new phase domains.

In particular, in the hole-particle symmetric case it can be
inferred from (Fig.~\ref{ani-sppd}) that at least up to $\beta_2=10$
-- for hopping fermions, and at least up to $\beta_2=5$ -- for
hardcore bosons, there is only one critical value of $\beta_2$. For
fermions, it amounts to $\beta_2 =7$, where the phase
${\mathcal{S}}_{dd}$ disappears, while for hardcore bosons it is
$\beta_2 =2$, where the phase ${\mathcal{S}}_{pcb}$ disappears.

Off the hole-particle symmetry, for bosons, there are no critical
values of $\beta_{2}$, at least up to $\beta_{2}=3$. On the other
hand, for fermions and for $\beta_{2}\leq 6$ there are as many as
three critical values of $\beta_{2}$. In increasing order, the first
is $\beta_{2}=1$, where the  phases ${\mathcal{S}}^{v}_{v1}$,
${\mathcal{S}}^{v}_{d1}$, and ${\mathcal{S}}^{v}_{d2}$ appear. The
second is $\beta_{2}=2$, where the phases ${\mathcal{S}}^{v}_{d3}$,
${\mathcal{S}}^{v}_{d4}$, and ${\mathcal{S}}^{v}_{v3}$ appear. And
the last one is $\beta_{2}=3$, where ${\mathcal{S}}_{dd}$ is
replaced by ${\mathcal{S}}^{v}_{v2}$.

If there are no more critical values of the anisotropy parameter
$\beta_2$, the shape of the fourth-order phase diagrams for any
value $\beta_2$ larger than the greatest considered in
Figs.~\ref{ani-sppd},~\ref{ani-epspdf},~\ref{ani-epspdb} remains the
same. On increasing the anisotropy parameter, the diagrams undergo
only some translations. To verify whether for the values of
$\beta_2$ larger than those considered in this section new critical
values do appear, we proceed to investigating stronger, i.e.
intermediate, deviations from the isotropic case.

\subsection{The intermediate deviation from the isotropic case}
\begin{figure}[th]
\includegraphics[width=0.42\textwidth]{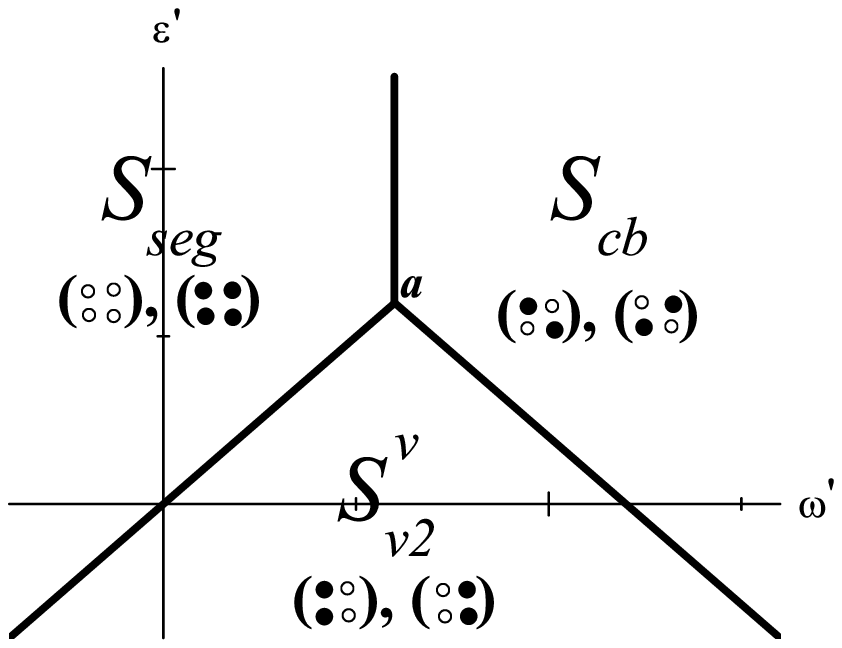}
\hfill
\includegraphics[width=0.42\textwidth]{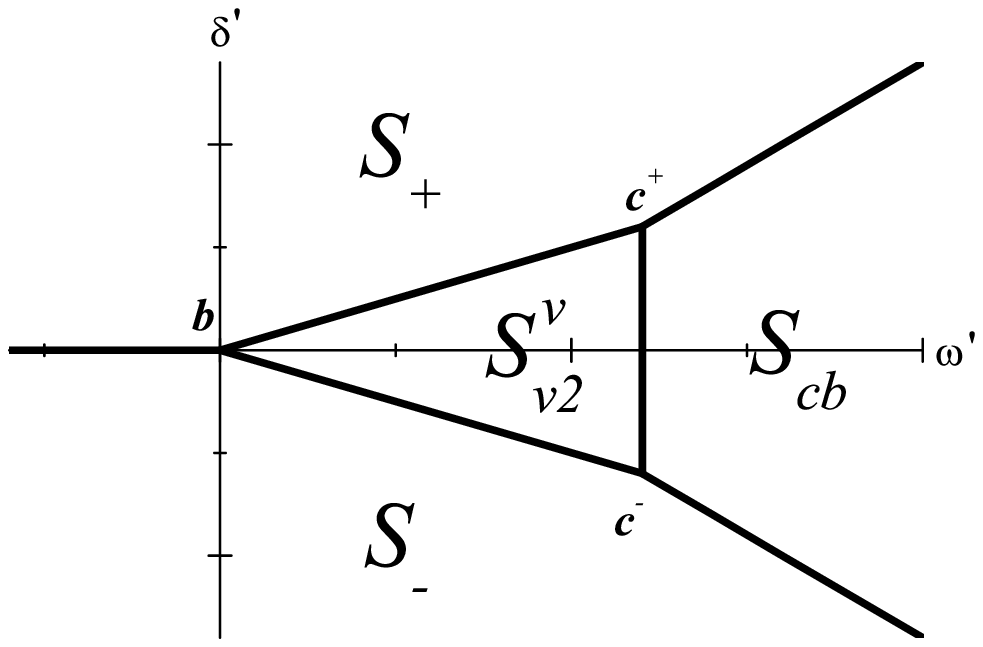}
\\
\parbox[t]{0.47\textwidth}{\caption{{\scriptsize{
The case of an intermediate anisotropy ($0<a<2$) and of
hole-particle symmetric systems ($\mu =0$). The phase diagram
(common for hopping fermion and hardcore boson systems) of
$H_{P_{2}}^{(2+a)}$. The coordinates of point ${\bf a}$ are
$\omega^{\prime}=\beta_{a}$, $\varepsilon^{\prime}=\beta_{a}$. The
boundary lines of ${\mathcal{S}}^{v}_{v2}$, from left to right, are:
$\varepsilon^{\prime}=\omega^{\prime}$ and $\varepsilon^{\prime}
=-\omega^{\prime}+2\beta_{a}$. The boundary between
${\mathcal{S}}_{seg}$ and ${\mathcal{S}}_{cb}$ is
$\omega^{\prime}=\beta_{a}$.}}} \label{ani-pd3-2}} \hfill
\parbox[t]{0.47\textwidth}{\caption{{\scriptsize{
The case of an intermediate anisotropy ($0<a<2$), off the
hole-particle symmetry, with $\varepsilon^{\prime}=0$. The phase
diagram (common for hopping fermion and hardcore boson systems) of
$H_{P_{2}}^{(2+a)}$. In the
$(\omega^{\prime},\delta^{\prime})$-plane, ${\bf b}=(0,0)$, ${\bf
c^{+}}=(2\beta_{a},\beta_{a})$, ${\bf
c^{-}}=(2\beta_{a},-\beta_{a})$. The boundary lines of
${\mathcal{S}}_{-}$, from left to right, are: $\delta^{\prime}=0$,
$\delta^{\prime}=-\frac{\omega^{\prime}}{2}$ and
$\delta^{\prime}=-\omega^{\prime}+\beta_{a}$. The boundary lines of
${\mathcal{S}}_{+}$ are obtained by changing $\delta^{\prime}
\rightarrow -\delta^{\prime}$. }}} \label{ani-pd3-1}}
\end{figure}

Now, the fourth-order effective Hamiltonian for fermions assumes the
form:
\begin{eqnarray}
\left( E^{f}_{S} \right)^{(4)} &=& \left.\left( E^{f}_{S}
\right)^{(4)}\right|_{\gamma=1} - \beta_{2} \frac{t^{2+a}}{4}
\sum\limits_{\langle x,y \rangle_{1,v}} s_{x}s_{y},
\end{eqnarray}
and for hardcore bosons an analogous formula holds true. Obviously,
the phase diagrams in the zeroth and second orders remain the same
as in the isotropic case, described above. Therefore, we proceed to
constructing the phase diagram in next order, which is ($2+a$)-order
with $0<a<2$, and the corresponding effective Hamiltonian reads:
\begin{equation}
E^{(2+a)}_{S} = E^{(2)}_{S} - \beta_{2} \frac{t^{2+a}}{4}
\sum\limits_{\langle x,y \rangle_{1,v}} s_{x}s_{y},
\end{equation}
where $ E^{(2)}_{S}$ stands for the, common for fermions and bosons,
second-order effective Hamiltonian. For the reasons given in the
previous subsection, we consider a neighborhood of the point
$W=-2t^{2}$, $\mu =0$, $\tilde{\varepsilon}=0$, where the energies
of all the configurations are equal. In this neighborhood it is
convenient to introduce new variables, $\delta^{\prime}$,
$\varepsilon^{\prime}$, and $\omega^{\prime}$,
\begin{eqnarray}
\mu=t^{2+a} \delta^{\prime} , \qquad \tilde{\varepsilon}=t^{2+a}
\varepsilon^{\prime} , \qquad W=-2t^{2}+t^{2+a} \omega^{\prime} ,
\end{eqnarray}
and rewrite the expansion up to the order $2+a$:
\begin{eqnarray}
E^{(2+a)}_{S}  & = & \frac{t^{2+a}}{2} \left\{ - \delta^{\prime}
\sum\limits_{x} \left( s_{x} + 1 \right) + \frac{\omega^{\prime}}{4}
\sum\limits_{\langle x,y \rangle_{1}} s_{x}s_{y} -
\frac{\varepsilon^{\prime}}{8} \sum\limits_{\langle x,y \rangle_{2}}
s_{x}s_{y}- \frac{\beta_{a}}{2} \sum\limits_{\langle x,y
\rangle_{1,v}} s_{x}s_{y} \right\} \nonumber \\
& = & \frac{t^{2+a}}{2} \sum\limits_{P_{2}} H_{P_{2}}^{(2+a)},
\end{eqnarray}
where
\begin{eqnarray}
H_{P_{2}}^{(2+a)} = - \frac{\delta^{\prime}}{4}
\sideset{}{'}\sum\limits_{x} \left( s_{x} + 1 \right) +
\frac{\omega^{\prime}}{8} \sideset{}{'}\sum\limits_{\langle x,y
\rangle_{1}} s_{x}s_{y} - \frac{\varepsilon^{\prime}}{8}
\sideset{}{'}\sum\limits_{\langle x,y \rangle_{2}} s_{x}s_{y} -
\frac{\beta_{a}}{4} \sideset{}{'}\sum\limits_{\langle x,y
\rangle_{1,v}} s_{x}s_{y}.
\end{eqnarray}
The summations in the primed sums run over a plaquette $P_{2}$. The
plaquette potentials $H_{P_{2}}^{(2+a)}$ have to be minimized over
all the plaquette configurations. As in the previous case, we are
interested in phase diagrams of hole-particle symmetric systems
($\delta^{\prime} =0$) or unsymmetrical systems with
$\varepsilon^{\prime}=0$. It turns out that for such energy
parameters and plaquette configurations the potentials
$H_{P_{2}}^{(2+a)}$ are m-potentials. The resulting phase diagrams
are shown in Fig.~\ref{ani-pd3-2} and Fig.~\ref{ani-pd3-1}.
We note that each of the points ${\bf a}$:
$\omega^{\prime}=\beta_{a}$, $\varepsilon^{\prime}=\beta_{a}$, ${\bf
b}$: $\omega^{\prime}=0$, $\delta^{\prime}=0$, ${\bf c^{-}}$:
$\omega^{\prime}=2\beta_{a}$, $\delta^{\prime}=-\beta_{a}$, and
${\bf c^{+}}$: $\omega^{\prime}=2\beta_{a}$,
$\delta^{\prime}=\beta_{a}$, is the coexistence point of three
periodic phases. That is, the only plaquette configurations
minimizing the potential $H_{P_{2}}^{(2+a)}$ at such a point are
those obtained by restricting the configurations of coexisting
phases to a plaquette (these plaquette configurations are shown in
Fig.~\ref{ani-pd3-2})

Now, following our recursive procedure of constructing phase
diagrams to some order, we are ready to investigate the effect of
fourth-order interactions. As in earlier steps, it is enough to
consider neighborhoods of the coexistence points ${\bf a}$, ${\bf
b}$, and ${\bf c^{-}}$ of Figs.~\ref{ani-pd3-2},~\ref{ani-pd3-1}
(the diagram in a neighborhood of ${\bf c^{+}}$ can be obtained from
that around ${\bf c^{-}}$ by a symmetry operation).

\begin{figure}[t]
\includegraphics[width=0.40\textwidth]{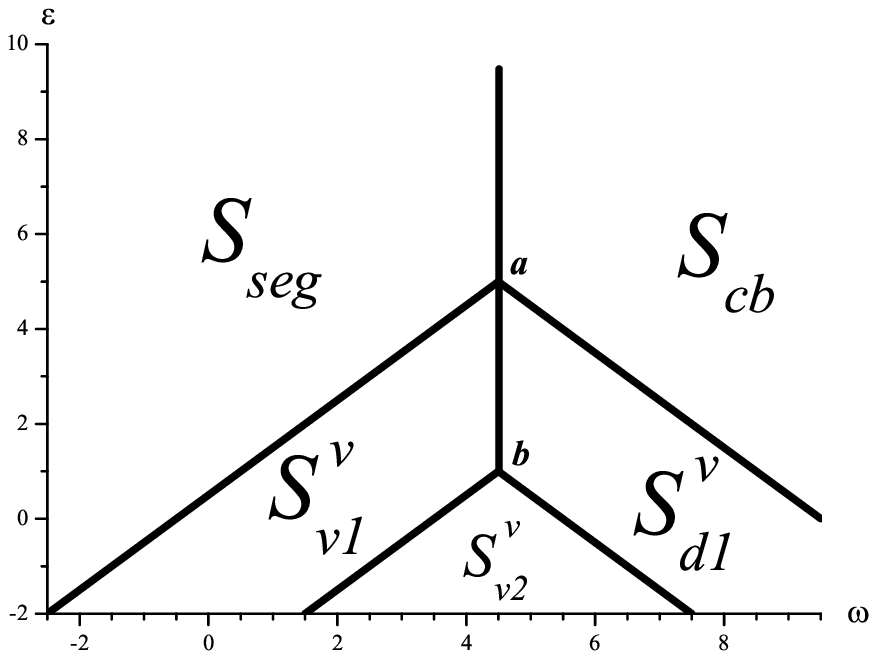}
\hfill
\includegraphics[width=0.40\textwidth]{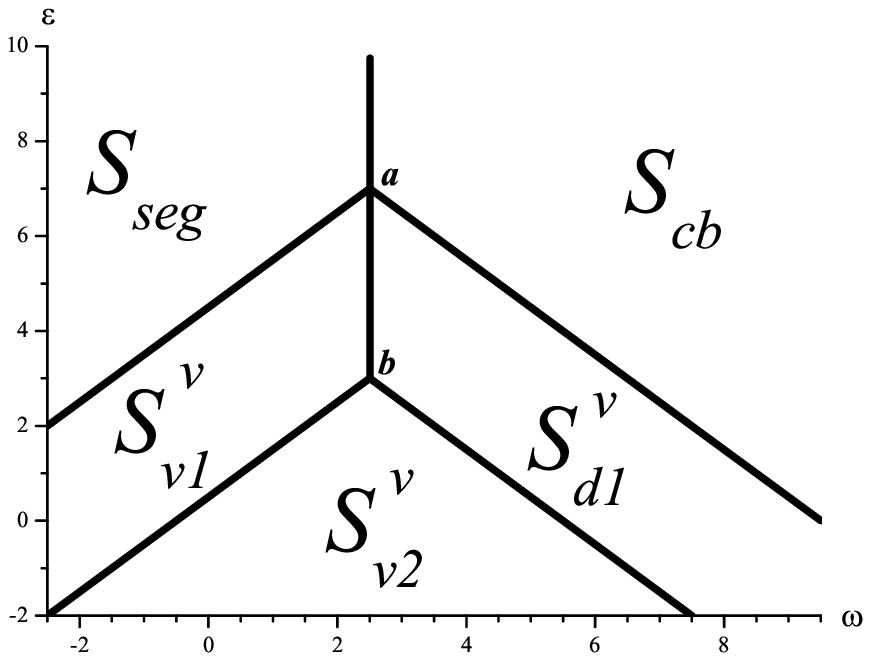}
\\
\parbox[t]{0.46\textwidth}{\caption{{\scriptsize{
The case of an intermediate anisotropy ($0<a<2$) and of
hole-particle symmetric systems ($\mu =0$). The phase diagram of
$\left( E_{S}^{f} \right)^{(4)}$ in a neighborhood of point ${\bf
a}$ ($\omega^{\prime}=\beta_{a}$, $\varepsilon^{\prime}=\beta_{a}$)
(see formula (\ref{ani-Ef4a})). The equations defining the boundary
lines of the phase domains are given in Tab.~\ref{D-tb5} of Appendix
D. }}} \label{ani-p1fmnd0}} \hfill
\parbox[t]{0.46\textwidth}{\caption{{\scriptsize{
The case of an intermediate anisotropy ($0<a<2$) and of
hole-particle symmetric systems ($\mu =0$). The phase diagram of
$\left( E_{S}^{b} \right)^{(4)}$ in a neighborhood of point ${\bf
a}$ ($\omega^{\prime}=\beta_{a}$, $\varepsilon^{\prime}=\beta_{a}$)
(see formula (\ref{ani-Ef4a})). The equations defining the boundary
lines of the phase domains are given in Tab.~\ref{D-tb5} of Appendix
D. }}} \label{ani-p1hcbd0}}
\end{figure}

\begin{figure}[ph]
\centering \includegraphics[width=0.9\textwidth]{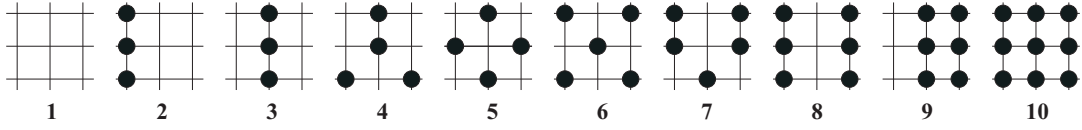}
\caption{\scriptsize{All the admissible $T$-plaquette
configurations, up to rotations by $\pi$, that are used in
constructing the fourth-order phase diagram in a neighborhood of
point {\bf a} of Fig.~\ref{ani-pd3-2}, i.e. the elements of
${\mathcal{S}}^{{\bf a}}_{T}$. }} \label{bc10}
\end{figure}

\paragraph{A. A neighborhood of ${\bf a}$}$\\$
Here $\delta^{\prime}=0$, hence the systems are hole-particle
invariant. A convenient change of variables is:
\begin{eqnarray}
\omega^{\prime} = \beta_{a} + t^{2-a}\omega ,  \qquad
\varepsilon^{\prime}=\beta_{a}+ t^{2-a}\varepsilon.
\end{eqnarray}
In these variables, the fourth-order effective Hamiltonian reads:
\begin{eqnarray}
\left( E_{S}^{f} \right)^{(4)} =  \frac{t^{2+a}}{2}
\sum\limits_{P_{2}}
\left. H^{(2+a)}_{P_{2}}\right|_{\substack{\delta^{\prime}=0\\
\varepsilon^{\prime}=\omega^{\prime}=\beta_{a}}}+\frac{t^{4}}{2}\sum\limits_{T}\left.\left(
H^{f}_{T} \right)^{(4)}\right|_{\gamma=1}, \label{ani-Ef4a}
\end{eqnarray}
with a similar formula for hard-core bosons, where
\begin{eqnarray}
\left. H^{(2+a)}_{P_{2}}\right|_{\substack{\delta^{\prime}=0\\
\varepsilon^{\prime}=\omega^{\prime}=\beta_{a}}} =
\frac{\beta_{a}}{8} \left(\sideset{}{'}\sum\limits_{\langle x,y
\rangle_{1,h}} s_{x}s_{y} - \sideset{}{'}\sum\limits_{\langle x,y
\rangle_{1,v}} s_{x}s_{y}- \sideset{}{'}\sum\limits_{\langle x,y
\rangle_{2}} s_{x}s_{y}\right).
\end{eqnarray}
In the previous order, it has been established that the only
plaquette configurations
minimizing $\left. H^{(2+a)}_{P_{2}}\right|_{\substack{\delta^{\prime}=0\\
\varepsilon^{\prime}=\omega^{\prime}=\beta_{a}}}$ are the ones
obtained by restricting to a plaquette $P_2$ the periodic
configurations ${\mathcal{S}}_{+}$, ${\mathcal{S}}_{-}$,
${\mathcal{S}}_{cb}$, and ${\mathcal{S}}^{v}_{v2}$. Let us denote
this set of plaquette configurations by ${\mathcal{S}}^{{\bf
a}}_{P_{2}}$. Consequently, the minimization of the fourth-order
potentials $\left.\left( H^{f}_{T} \right)^{(4)}\right|_{\gamma=1}$,
$\left.\left( H^{b}_{T} \right)^{(4)}\right|_{\gamma=1}$, should be
carried out only over the set ${\mathcal{S}}^{{\bf a}}_{T}$ of those
$T$-plaquette configurations whose restriction to a plaquette
$P_{2}$ belongs to ${\mathcal{S}}^{{\bf a}}_{P_{2}}$. The
configurations of the set ${\mathcal{S}}^{{\bf a}}_{T}$ are
displayed in Fig.~\ref{bc10}. It appears that on the set
${\mathcal{S}}^{{\bf a}}_{T}$, the potentials $\left.\left(
H^{f}_{T} \right)^{(4)}\right|_{\gamma=1}$, $\left.\left( H^{b}_{T}
\right)^{(4)}\right|_{\gamma=1}$, are $m$-potentials. The obtained
phase diagrams, independent of the anisotropy parameter $\beta_{a}$,
are shown in Fig.~\ref{ani-p1fmnd0} and Fig.~\ref{ani-p1hcbd0}.

\begin{figure}[p]
\includegraphics[width=0.40\textwidth]{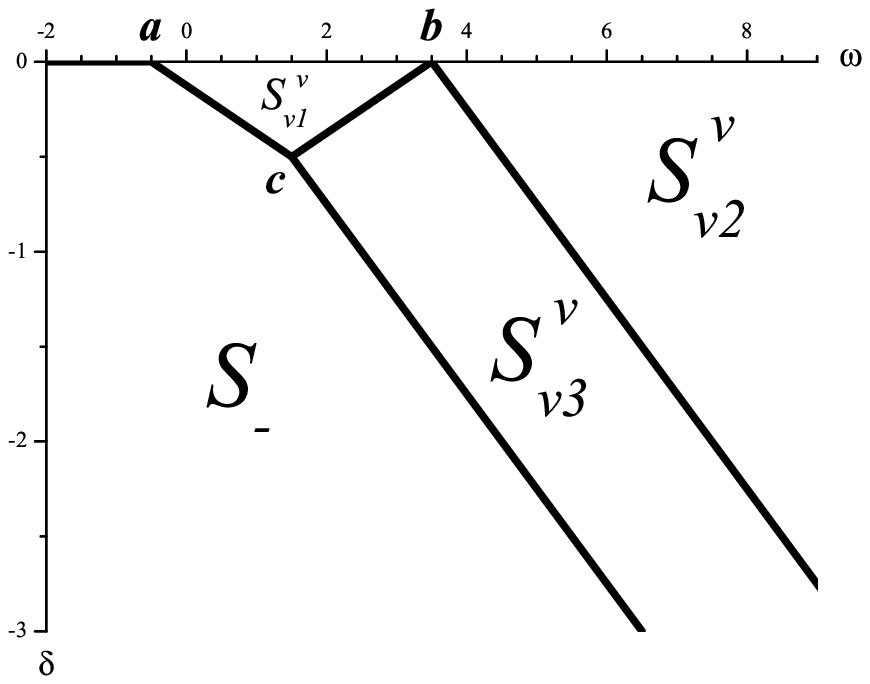}
\hfill
\includegraphics[width=0.40\textwidth]{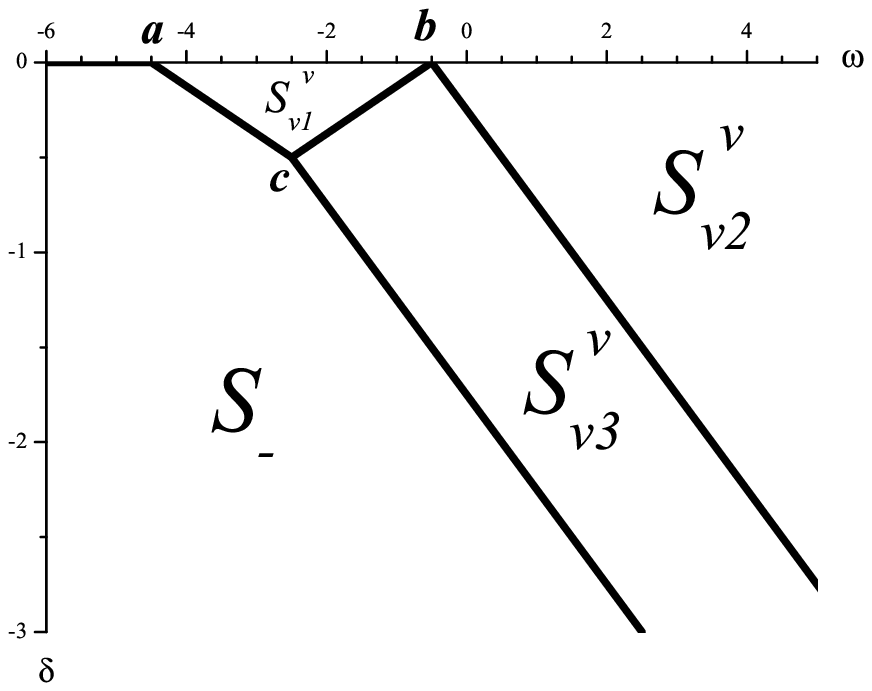}
\\
\parbox[t]{0.46\textwidth}{\caption{{\scriptsize{
The case of an intermediate anisotropy ($0<a<2$), off the
hole-particle symmetry, with $\varepsilon^{\prime}=0$. The phase
diagram of $\left( E_{S}^{f} \right)^{(4)}$ in a neighborhood of
point ${\bf b}$ ($\omega^{\prime}=0$, $\delta^{\prime}=0$), see
formula (\ref{ani-Ef4b}). The equations defining the boundary lines
of the phase domains are given in Tab.~\ref{D-tb6} of Appendix D,
while the corresponding zero-potential coefficients $\{ \alpha_{i}
\}$ in Tab.~\ref{E-tb25} of Appendix E. }}} \label{ani-p1fmne0}}
\hfill
\parbox[t]{0.46\textwidth}{\caption{{\scriptsize{
The case of an intermediate anisotropy ($0<a<2$), off the
hole-particle symmetry, with $\varepsilon^{\prime}=0$. The phase
diagram of $\left( E_{S}^{b} \right)^{(4)}$ in a neighborhood of
point ${\bf b}$ ($\omega^{\prime}=0$, $\delta^{\prime}=0)$), see
formula (\ref{ani-Ef4b}). The equations defining the boundary lines
of the phase domains are given in Tab.~\ref{D-tb6} of Appendix D,
while the corresponding zero-potential coefficients $\{ \alpha_{i}
\}$ in Tab.~\ref{E-tb26} of Appendix E. }}} \label{ani-p1hcbe0}}
\end{figure}

\begin{figure}[ph]
\centering \includegraphics[width=0.55\textwidth]{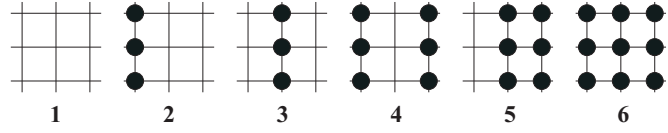}
\caption{\scriptsize{All the admissible $T$-plaquette configurations
that are used in constructing the fourth-order phase diagram, up to
rotations by $\pi$, in a neighborhood  of point {\bf b} of
Fig.~\ref{ani-pd3-1}, i.e. the elements of ${\mathcal{S}}^{{\bf
b}}_{T}$ (the vertical n.n. pairs that are occupied by one ion are
forbidden). }} \label{bc6}
\end{figure}

\paragraph{B. A neighborhood of ${\bf b}$}$\\$
Here the system is not hole-particle symmetric and
$\varepsilon^{\prime}=0$. A convenient change of variables is:
\begin{eqnarray}
\omega^{\prime} = t^{2-a}\omega ,  \qquad \delta^{\prime} = t^{2-a}
\delta .
\end{eqnarray}
Then, the fourth-order effective Hamiltonian takes the form,
\begin{eqnarray}
\left( E_{S}^{f} \right)^{(4)} = \frac{t^{2+a}}{2}
\sum\limits_{P_{2}}
\left. H^{(2+a)}_{P_{2}}\right|_{\substack{\delta^{\prime}=0\\
\varepsilon^{\prime}=\omega^{\prime}=0}}
+\frac{t^{4}}{2}\sum\limits_{T}\left.\left( H^{f}_{T}
\right)^{(4)}\right|_{\gamma=1}, \label{ani-Ef4b}
\end{eqnarray}
where
\begin{equation}
\left. H^{(2+a)}_{P_{2}}\right|_{\substack{\delta^{\prime}=0\\
\varepsilon^{\prime}=\omega^{\prime}=0}}= -  \frac{\beta_{a}}{4}
\sideset{}{'}\sum\limits_{\langle x,y \rangle_{1,v}} s_{x}s_{y}
\end{equation}
with a  similar formula for hard-core bosons.
The minimum of $\left. H^{(2+a)}_{P_{2}}\right|_{\substack{\delta^{\prime}=0\\
\varepsilon^{\prime}=\omega^{\prime}=0}}$ is attained at the
configurations belonging to ${\mathcal{S}}^{{\bf b}}_{P_{2}}$, i.e.
the plaquette configurations obtained by restricting the periodic
configurations ${\mathcal{S}}_{+}$, ${\mathcal{S}}_{-}$, and
${\mathcal{S}}^{v}_{v2}$ to a plaquette $P_2$. Let
${\mathcal{S}}^{{\bf b}}_{T}$ be the corresponding set of
$T$-plaquette configurations (there are no vertical pairs of n.n.
sites occupied by one ion). This set is shown in Fig.~\ref{bc6}.
Here the potentials $\left.\left( H^{f}_{T}
\right)^{(4)}\right|_{\gamma=1}$ are not the $m$-potentials. The
obtained phase diagrams  are shown in
Figs.~\ref{ani-p1fmne0},~\ref{ani-p1hcbe0}.

\begin{figure}[hp]
\includegraphics[width=0.45\textwidth]{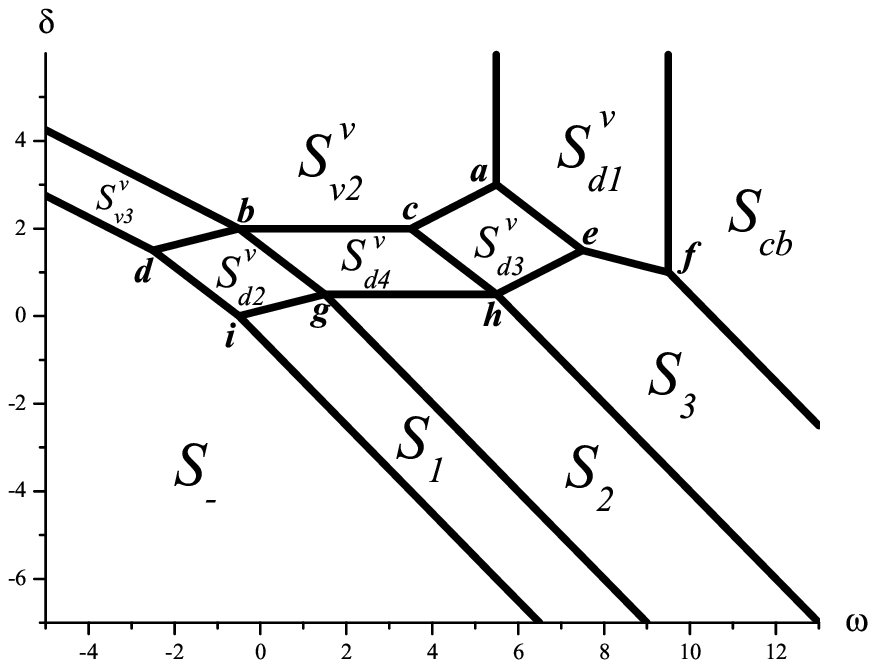}
\hfill
\includegraphics[width=0.45\textwidth]{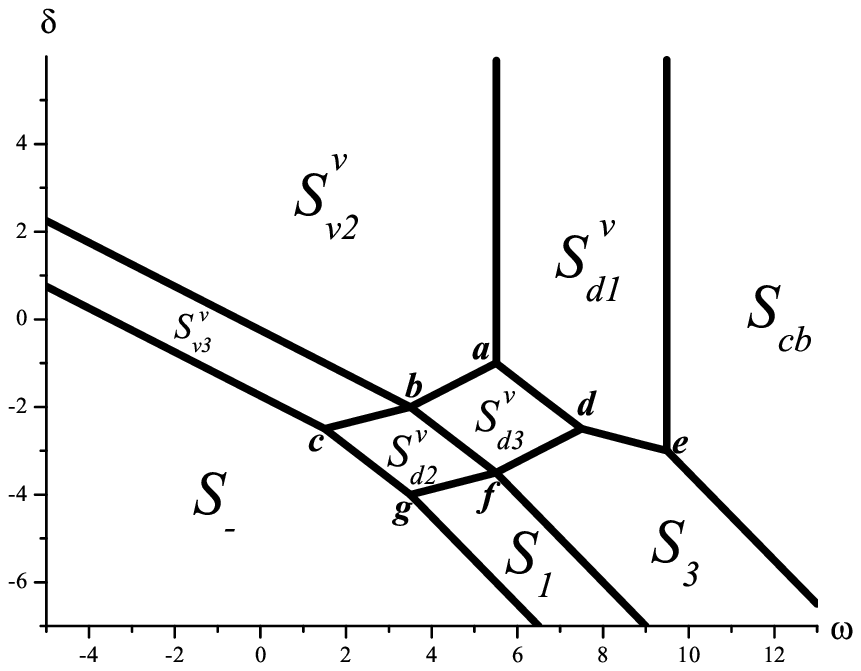}
\\
\parbox[t]{0.46\textwidth}{\caption{{\scriptsize{
The case of an intermediate anisotropy ($0<a<2$), off the
hole-particle symmetry, with $\varepsilon^{\prime}=0$. The phase
diagram of $\left( E_{S}^{f} \right)^{(4)}$ in a neighborhood of
point ${\bf c^{-}}$ ($\omega^{\prime}=2\beta_{a}$,
$\delta^{\prime}=-\beta_{a}$), see formula (\ref{ani-Ef4c}). The
equations defining the boundary lines of the phase domains are given
in Tab.~\ref{D-tb7} of Appendix D, while the corresponding
zero-potential coefficients $\{ \alpha_{i} \}$ in Tab.~\ref{E-tb27}
of Appendix E. }}} \label{ani-p2fmne0}} \hfill
\parbox[t]{0.46\textwidth}{\caption{{\scriptsize{
The case of an intermediate anisotropy ($0<a<2$), off the
hole-particle symmetry, with $\varepsilon^{\prime}=0$. The phase
diagram of $\left( E_{S}^{b} \right)^{(4)}$ in a neighborhood of
point ${\bf c^{-}}$ ($\omega^{\prime}=2\beta_{a}$,
$\delta^{\prime}=-\beta_{a}$), see formula (\ref{ani-Ef4c}). The
equations defining the boundary lines of the phase domains are given
in Tab.~\ref{D-tb7} of Appendix D, while the corresponding
zero-potential coefficients $\{ \alpha_{i} \}$ in Tab.~\ref{E-tb28}
of Appendix E. }}} \label{ani-p2hcbe0}}
\end{figure}

\begin{figure}[hpt]
\centering \includegraphics[width=0.9\textwidth]{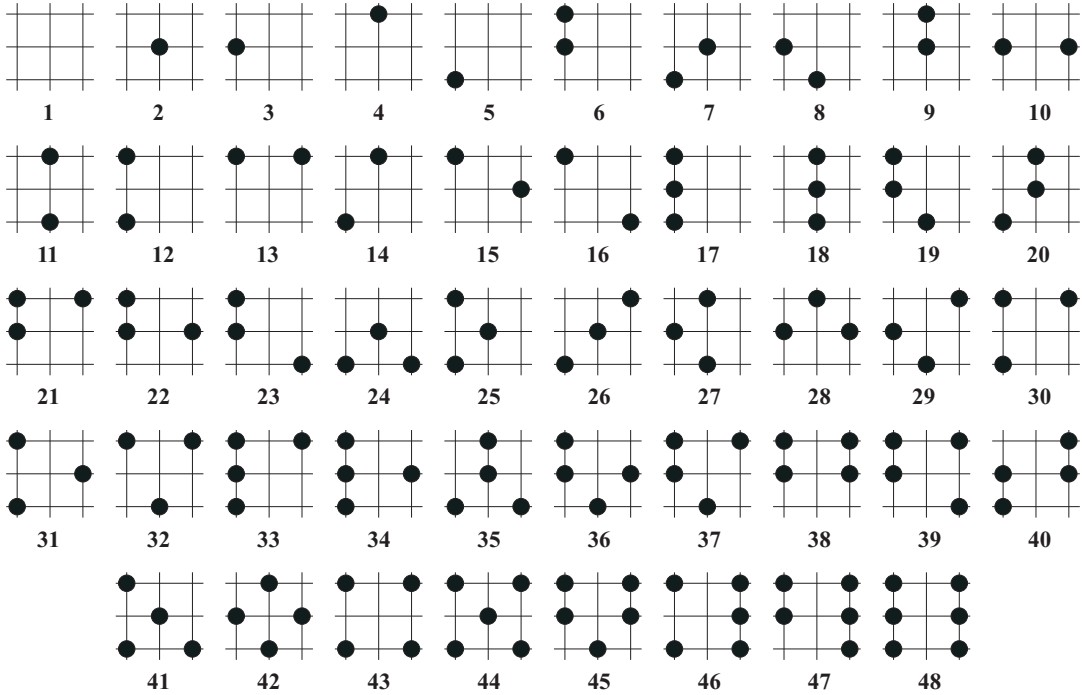}
\caption{\scriptsize{All the admissible $T$-plaquette
configurations, up to rotations by $\pi$ and reflections in lattice
lines parallel to the axes, that are used in constructing the
fourth-order phase diagram in a neighborhood  of point ${\bf c}^{-}$
of Fig.~\ref{ani-pd3-1}, i.e. the elements of ${\mathcal{S}}^{{\bf
c}^{-}}_{T}$ (the horizontal n.n. pairs that are occupied by two
ions are forbidden). }} \label{bc48}
\end{figure}

\paragraph{C. A neighborhood of ${\bf c}^{-}$}$\\$
Here the system is not hole-particle symmetric and
$\varepsilon^{\prime}=0$. A convenient change of variables is:
\begin{eqnarray}
\omega^{\prime} = 2\beta_{a}+t^{2-a}\omega ,  \qquad \delta^{\prime}
= - \beta_{a} + t^{2-a} \delta .
\end{eqnarray}
The fourth-order effective Hamiltonian reads:
\begin{eqnarray}
\left( E_{S}^{f} \right)^{(4)} = \frac{t^{2+a}}{2}
\sum\limits_{P_{2}}
\left. H^{(2+a)}_{P_{2}}\right|_{\substack{\varepsilon^{\prime}=0\\
\delta^{\prime}=-\beta_a, \omega^{\prime}=2\beta_a}}
+\frac{t^{4}}{2}\sum\limits_{T} \left.\left( H^{f}_{T}
\right)^{(4)}\right|_{\gamma=1} \label{ani-Ef4c}
\end{eqnarray}
with a similar formula for hard-core bosons, where
\begin{equation}
\left. H^{(2+a)}_{P_{2}}\right|_{\substack{\varepsilon^{\prime}=0\\
\delta^{\prime}=-\beta_a, \omega^{\prime}=2\beta_a}}=
\frac{\beta_{a}}{4} \sideset{}{'}\sum\limits_{\langle x,y
\rangle_{1,h}} \left( s_{x}+ s_{y} + s_{x}s_{y}+ 1 \right)
\end{equation}
The potentials $\left. H^{(2+a)}_{P_{2}}\right|_{\substack{\varepsilon^{\prime}=0\\
\delta^{\prime}=-\beta_a, \omega^{\prime}=2\beta_a}}$ are minimized
by restrictions to a plaquette $P_{2}$ of periodic configurations
${\mathcal{S}}_{-}$, ${\mathcal{S}}_{cb}$, and
${\mathcal{S}}^{v}_{v2}$, that constitute the set
${\mathcal{S}}^{c^{-}}_{P_{2}}$. The corresponding set
${\mathcal{S}}^{c^{-}}_{T}$ of $T$-plaquette configurations consists
of configurations where no horizontal pair of n.n. sites is occupied
by two ions (see Fig.~\ref{bc48}). Here the potentials $\left.\left(
H^{f}_{T} \right)^{(4)}\right|_{\gamma=1}$ are not the
$m$-potentials. The corresponding phase diagrams are shown in
Fig.~\ref{ani-p2fmne0} and Fig.~\ref{ani-p2hcbe0}.

\subsection{Discussion of the phase diagrams and conclusions}
In previous subsection, we have obtained phase diagrams according to
the truncated effective Hamiltonians of fourth order, with an
intermediate anisotropy of hopping. These diagrams have been
constructed to see what happens to the fourth-order phase diagrams
with the weakest anisotropy if the anisotropy parameter $\beta_2$
grows beyond the values taken into account in subsection 4.2. Let us
consider firstly the diagrams in a neighborhood of point ${\bf a}$.
Apparently, up to a rescaling the phase diagram in
Fig.~\ref{ani-p1fmnd0} looks the same as the part in the dotted
circle of the fermionic phase diagram in Fig.~\ref{ani-sppd}.
Similarly, up to a rescaling the phase diagram in
Fig.~\ref{ani-p1hcbd0} looks the same as the the part in the dotted
circle of the bosonic phase diagram in Fig.~\ref{ani-sppd}. The same
remarks apply to the phase diagrams in neighborhoods of points ${\bf
b}$ and ${\bf c^{-}}$. The phase diagram in Fig.~\ref{ani-p1fmne0}
reproduces the part in the upper dotted ellipse in
Fig.~\ref{ani-epspdf}, while the diagram in Fig.~\ref{ani-p1hcbe0}
-- the part in the upper dotted ellipse in Fig.~\ref{ani-epspdb}.
Then, the phase diagram in Fig.~\ref{ani-p2fmne0} reproduces the
part in the lower dotted ellipse in Fig.~\ref{ani-epspdf}, while the
diagram in Fig.~\ref{ani-p2hcbe0} -- the part in the lower dotted
ellipse in Fig.~\ref{ani-epspdb}. Therefore, we conclude that in the
case of the weakest anisotropy in the fourth-order effective
Hamiltonians, we have determined all the critical values of the
anisotropy parameter $\beta_2$. For the both kinds of hopping
particles, if $\beta_2$ exceeds the greatest critical value, the
obtained phase diagram undergoes only a translation (varying with
$\beta_2$).

Now, the basic question to be answered is concerned with the
relation between these fourth-order phase diagrams and the phase
diagrams of quantum systems described by Hamiltonians $H$, given by
(\ref{ourhamilt}). By adapting the arguments presented in Refs.
\cite{Kennedy1,GMMU}, we can demonstrate, see for instance Ref.
\cite{DJ1}, that if the remainders, $\left(R_{S}^{f}\right)^{(4)}$
and $\left(R_{S}^{b}\right)^{(4)}$, are taken into account, then
there is a sufficiently small $t_{0}$ such that for $t<t_{0}$ the
phase diagrams of quantum systems look the same as the phase
diagrams according  to the effective Hamiltonians truncated at the
fourth order, except some narrow regions, of width $O(t^2)$ (at the
diagrams displayed above), located along the phase-domains
boundaries, and except the domains ${\mathcal{S}}^{v}_{d2}$ and
${\mathcal{S}}^{v}_{d4}$. For $t<t_{0}$ and for each domain
${\mathcal{S}}_{D}$, which is different from
${\mathcal{S}}^{v}_{d2}$ and ${\mathcal{S}}^{v}_{d4}$, there is a
nonempty two-dimensional open domain ${\mathcal{S}}_{D}^{\infty}$
that is contained in the domain ${\mathcal{S}}_{D}$ and such that in
${\mathcal{S}}_{D}^{\infty}$ the set of ground-state configurations
coincides with ${\mathcal{S}}_{D}$. Moreover, in comparison with the
critical values of the anisotropy parameter $\beta_2$, determined
according to the fourth-order effective Hamiltonians, the
corresponding critical values of the quantum systems described by
Hamiltonians $H$ differ by $O(t^2)$, i.e. for a quantum system
$\gamma = 1 - \beta_2 t^2 + O(t^4)$.

Remarkably, in the fourth-order the hole-particle symmetric phase
diagrams of fermions and of hardcore bosons are geometrically
similar. That is, a phase diagram of hard core bosons, with any
$\beta_2 \geq 0$, can be obtained from a phase diagram of fermions,
with $\beta_2 \geq 5$, by the translation whose vector reads:
$\omega =-7$, $\varepsilon =-3$, and $\beta_{2} =-5$. The existence
of this translation vector is related to the fact that for both
kinds of systems there is one critical value of the anisotropy
parameter. Additionally, for fermions with $\beta_2 < 7$, it is
necessary to replace the phase ${\mathcal{S}}_{dd}$ in the central
domain by ${\mathcal{S}}_{pcb}$. Off the hole-particle symmetry, the
relation between the bosonic and fermionic phase diagrams is not
that close. For fermions, there are three critical values of
$\beta_2$, while for hardcore bosons there is no critical values.
Thus, the system with hopping hardcore bosons is less sensitive to
the anisotropy of hopping, than the system with hopping fermions.
Nevertheless, if $\varepsilon =0$, then the phase diagrams of both
kinds of systems are topologically similar, except that in the
bosonic phase diagrams the phases ${\mathcal{S}}_{d3}^{v}$ and
${\mathcal{S}}_{2}$ are missing. However, we know from
Ref.~\cite{DJ1} that this deficiency can be removed by switching on
the n.n.n. interactions with negative $\varepsilon$.

In the fourth-order effective Hamiltonians (\ref{4ordsmallani}), the
weakest anisotropy of n.n. hopping assumes the form of a
fourth-order attractive n.n. interaction in vertical direction (i.e.
the direction of a weaker hopping). This interaction favors n.n
pairs of occupied or empty sites that are oriented vertically. As a
result, the dimeric and axial-stripe phases oriented vertically are
stabilized for any value of the anisotropy parameter $\beta_2$,
while ${\mathcal{S}}_{pcb}$ and ${\mathcal{S}}_{dd}$ are replaced by
${\mathcal{S}}_{v2}^{v}$ above a critical value of $\beta_2$. Note
however, that at any higher order $2k$, $k=3,4 \ldots$ the weakest
anisotropy of n.n. hopping will cause the same effects, in the
effective Hamiltonians as well as in the corresponding phase
diagrams. This implies that in the quantum systems described by $H$,
arbitrary small anisotropy of n.n. hopping orients the dimeric and
axial-stripe phases in the direction of a weaker hopping.

\clearpage
\section{Influence of next-nearest-neighbor anisotropy on diagonally-striped phases}
In the previous Section the influence of n.n. anisotropy of hopping
intensity on axial striped phases, was investigated. At the regime
where stripes are stable, we have proved rigorously that for both
systems, of hopping fermions and hard-core bosons, an arbitrarily
small anisotropy of n.n. hopping orients the axial striped phases in
the direction of a weaker hopping. The analogous, arising naturally,
question is how the anisotropy of the next-nearest-neighbor (n.n.n.)
hopping influences the degeneracy of diagonal-striped phases. To
answer this question, we investigate the influence of n.n.n. hopping
on the phase $\mathcal{S}_{dd}$ (Fig.~\ref{t1-40f}), whose stability
was proved for fermions in Section 3.

In this Section\footnote{This section is based on \cite{Derzhko}.}
we introduce a next-nearest-neighbor hopping, small enough not to
destroy the striped structure and examine rigorously how the
presence of the next-nearest-neighbor hopping anisotropy reduces the
$\pi/2$-rotation degeneracy of the diagonal-striped phase. The
effect appears to be similar to that in the case of anisotropy of
the nearest-neighbor hopping: the stripes are oriented in the
direction of the weaker next-nearest-neighbor hopping.

\subsection{The effective interaction up to the fourth order}
We consider our model in the case when only next-nearest-neighbor
hopping is anisotropic, and nearest-neighbor hopping amplitude
remains the same in different directions, i.e. $t_{h}=t_{v}=t$. For
simplicity $\tilde{\varepsilon}=0$. In this case, the ground-state
energy expansion (\ref{effh}) with the expansion terms up to the
fourth order, i.e. $a+b+c\leqslant4$, reads:
\begin{align}
E_{S}\left( \mu \right) =& E_{S}^{(4)}\left( \mu \right) +
R^{(4)}, \nonumber \\
E_{S}^{(4)}\left( \mu \right) =& - \left[ \frac{\mu}{2}+
\frac{3}{4}t^{2}\left( t_{+}+t_{-} \right) \right]
\sum\limits_{x}s_{x} +
\nonumber \displaybreak[0]\\
&+ \left[ \frac{1}{4}t^{2} -\frac{9}{16}t^{4}
-\frac{1}{16}t^{2}\left( 3t_{+}^{2}+10t_{+}t_{-}+3t_{-}^{2} \right)
+\frac{W}{8} \right] \sum\limits_{\langle x,y \rangle_{1}}
s_{x}s_{y} +
\nonumber \displaybreak[0]\\
&+ \left[ \frac{1}{4}t_{+}^{2} +\frac{3}{16}t^{4}
-\frac{3}{8}t^{2}\left(2t_{+}^{2}+t_{+}t_{-}\right)
-\frac{3}{16}t_{+}^{4} -\frac{3}{8}t^{2}_{+}t^{2}_{-} \right]
\sum\limits_{\langle x,y \rangle_{2,+}} s_{x}s_{y} +
\nonumber \displaybreak[0]\\
&+ \left[ \frac{1}{4}t_{-}^{2} +\frac{3}{16}t^{4}
-\frac{3}{8}t^{2}\left(2t_{-}^{2}+t_{+}t_{-}\right)
-\frac{3}{16}t_{-}^{4} -\frac{3}{8}t^{2}_{+}t^{2}_{-} \right]
\sum\limits_{\langle x,y \rangle_{2,-}} s_{x}s_{y} +
\nonumber \displaybreak[0]\\
&+ \left[ \frac{1}{8}t^{4} -\frac{1}{8}t^{2}t_{+}t_{-}
+\frac{3}{16}t^{2}_{+}t^{2}_{-} \right] \sum\limits_{\langle x,y
\rangle_{3}} s_{x}s_{y} + \frac{3}{16}t^{2}t_{+}^{2}
\sum\limits_{\langle x,y \rangle_{4,+}} s_{x}s_{y} +
\frac{3}{16}t^{2}t_{-}^{2} \sum\limits_{\langle x,y \rangle_{4,-}}
s_{x}s_{y} +
\nonumber \displaybreak[0]\\
&+ \frac{1}{8}t_{+}^{4} \sum\limits_{\langle x,y \rangle_{5,+}}
s_{x}s_{y} + \frac{1}{8}t_{-}^{4} \sum\limits_{\langle x,y
\rangle_{5,-}} s_{x}s_{y} + \frac{3}{8}t^{2}t_{+}
\sum\limits_{P^{+}_{1}}s_{P^{+}_{1}} + \frac{3}{8}t^{2}t_{-}
\sum\limits_{P^{-}_{1}}s_{P^{-}_{1}} +
\nonumber\displaybreak[0]\\
&+ \frac{5}{16} \left[ t^{4}+ 2t^{2}t_{+}t_{-} \right]
\sum\limits_{P_{2}} s_{P_{2}} + \frac{5}{16} t^{2}_{+}t^{2}_{-}
\sum\limits_{P_{3}} s_{P_{3}} + \frac{5}{16} t^{2}t_{+}t_{-}
\sum\limits_{P_{4}} s_{P_{4}} +
\nonumber\displaybreak[0]\\
&+ \frac{5}{16} t^{2}t^{2}_{+} \sum\limits_{P^{+}_{5}} s_{P^{+}_{5}}
+ \frac{5}{16} t^{2}t^{2}_{-} \sum\limits_{P^{-}_{5}} s_{P^{-}_{5}}
. \label{anianigse}
\end{align}
The remainder $R^{(4)}$ is independent of the chemical potentials
and $W$, and collects all the terms proportional to
$t^{a}t^{b}_{+}t^{c}_{-}$, with $a+b+c=5,6,\ldots$. The above
expansion is absolutely convergent for sufficiently small $t$,
$t_{+}$ and $t_{-}$, uniformly in $\Lambda$. In the special case of
$t_{+}=t_{-}=t^{\prime}$ and $W=0$ it was obtained in
\cite{Wojtkiewicz}.

In Section 3, we have obtained the phase diagram of the isotropic
model without n.n.n. hopping, i.e. for $t_{+}=t_{-}=0$. Here, our
aim is to determine the influence of the n.n.n.-hopping anisotropy
on the diagonal-striped phase $\mathcal{S}_{dd}$ (see
Fig.~\ref{t1-40f}). For this job, the value of the n.n.n.-hopping
intensities, $t_{+}$, $t_{-}$, cannot be too large, in order to
preserve the phase diagram up to 4th order. On the other hand, the
n.n.n.-hopping intensities cannot be too small, in order to appear
in the fourth-order effective Hamiltonian. In an attempt to satisfy
the both requirements, we choose the smallest n.n.n.-hopping
intensities $t_{+}$, $t_{-}$, i.e. such that they do not appear in
the expansion terms of order smaller than four: $t_{+}=a_{+}t^{2}$
and $t_{-}=a_{-}t^{2}$. In this case, the effective Hamiltonian
assumes the form:
\begin{align}
E_{S}^{(4)}\left( \mu \right) =& - \left[ \frac{1}{2}\mu+
\frac{3}{4}t^{4}\left( a_{+}+a_{-} \right) \right]
\sum\limits_{x}s_{x} + \left[ \frac{1}{4}t^{2} -\frac{9}{16}t^{4}
+\frac{W}{8} \right] \sum\limits_{\langle x,y \rangle_{1}}
s_{x}s_{y} +
\nonumber \displaybreak[0]\\
&+ \left[ \frac{1}{4}t^{4}a_{+}^{2} +\frac{3}{16}t^{4} \right]
\sum\limits_{\langle x,y \rangle_{2,+}} s_{x}s_{y} + \left[
\frac{1}{4}t^{4}a_{-}^{2} +\frac{3}{16}t^{4} \right]
\sum\limits_{\langle x,y \rangle_{2,-}} s_{x}s_{y} +
\nonumber \displaybreak[0]\\
&+ \frac{1}{8}t^{4} \sum\limits_{\langle x,y \rangle_{3}} s_{x}s_{y}
+ \frac{3}{8}t^{4}a_{+} \sum\limits_{P^{+}_{1}}s_{P^{+}_{1}} +
\frac{3}{8}t^{4}a_{-} \sum\limits_{P^{-}_{1}}s_{P^{-}_{1}} +
\frac{5}{16} t^{4} \sum\limits_{P_{2}} s_{P_{2}}, \label{anianihaa}
\end{align}
i.e. the second requirement is satisfied. To answer the question
concerning the influence of anisotropy of n.n.n. hopping on the
degeneracy of the phase $\mathcal{S}_{dd}$, there is no need to
consider the whole phase diagram. For $t_{+}=t_{-}=0$, we fix a
point, well inside the domain of the diagonal-striped phase
$S_{dd}$, say $\mu=0$ and $W=-2t^{2}+9/2t^{4}$, i.e. $\omega=9/2$
(see Fig.~\ref{t1-40f}). Then, with the fixed point in
$(\mu,W)$-plain, we introduce a n.n.n. hopping which does not change
the ground-state configurations. Calculations show that
$\mathcal{S}_{dd}$ has the minimal energy for $a=|a_{+}|=|a_{-}|$,
where $-1/4\leqslant a\leqslant1/4$ (we suppose that the difference
between $a_{+}$ and $a_{-}$ is not large, so they are of the same
sign). Therefore, with our choice of n.n.n.-hopping intensities, the
first of the above two requirements can also be satisfied.
Eventually, we fix the values of n.n.n.-hopping intensities:
$a_{+}=1/8$, $a_{-}=\gamma a_{+}$, with $\gamma$ varying about $1$
(say, $0\leqslant\gamma\leqslant2$).

Since all the energy parameters, except the parameter $\gamma$ of
n.n.n.-hopping anisotropy, have been fixed, the effective
Hamiltonian (\ref{anianihaa}) depends only on $\gamma$. In the
following subsection, we examine how n.n.n.-hopping anisotropy,
$\gamma \neq 1$, influences the degeneracy of the diagonal-striped
phase $\mathcal{S}_{dd}$.

\subsection{Diagonally-striped phase versus n.n.n.-hopping anisotropy}
For technical reasons (see paragraph~2.4.5), it is convenient to
deal with such energies of configurations that are affine functions
of the parameters of the effective Hamiltonian. However, the
effective Hamiltonian (\ref{anianihaa}) contains the terms
proportional to $\gamma$ and $\gamma^{2}$. To get rid of
nonlinearities, we replace $\gamma$ and $\gamma^{2}$ by two
independent parameters $d_{1}$ and $d_{2}$, respectively, with
$d_1,d_2$ varying in the rectangle $0\leq d_1 \leq 2$ and $0\leq d_2
\leq 4$, in which the Hamiltonian is affine. After constructing the
phase diagram in $(d_{1},d_{2})$-plane, we restrict it to the
$d_{2}=d^{2}_{1}$ curve.

To compare the energies of configurations, we rewrite
$E_{S}^{(4)}(d_{1},d_{2})$ as the sum,
\begin{eqnarray}
E_{S}^{(4)}\left( d_{1},d_{2}
\right)=\frac{t^{4}}{2}\sum\limits_{T}H^{(4)}_{T},
\label{anianiepot}
\end{eqnarray}
over $T$-plaquettes. The potential $H^{(4)}_{T}$ is of the form:
\begin{eqnarray}
H^{(4)}_{T}&=&-\frac{3}{16}\left(d_{1}+1\right)s_{5} +
\frac{49}{512}\sideset{}{^{\prime\prime}}\sum\limits_{\langle x,y
\rangle_{2,+}}s_{x}s_{y} + \frac{1}{32}\left( \frac{1}{16}d_{2}+3
\right)\sideset{}{^{\prime\prime}}\sum\limits_{\langle x,y
\rangle_{2,-}}s_{x}s_{y}+\frac{1}{12}\sideset{}{^{\prime\prime}}\sum\limits_{\langle
x,y \rangle_{3}}s_{x}s_{y}+\nonumber\\
&&+\frac{3}{128}\sideset{}{^{\prime\prime}}\sum\limits_{P^{+}_{1}}s_{P^{+}_{1}}
+\frac{3}{128}d_{1}\sideset{}{^{\prime\prime}}\sum\limits_{P^{-}_{1}}s_{P^{-}_{1}}
+\frac{5}{32}\sideset{}{^{\prime\prime}}\sum\limits_{P_{2}}s_{P_{2}},
\label{anianiht4}
\end{eqnarray}
where $s_{5}$ is the central site of a $T$-plaquette
(Fig.~\ref{tblock}).

Unfortunately, the potential (\ref{anianiht4}) is not an
$m$-potential in the rectangle of considered values of $d_1$ and
$d_2$. Therefore, we introduce appropriate zero-potentials given in
Appendix A. In order to obtain the phase diagram, we have to compare
the energies of all the possible $T$-plaquette configurations. The
zero-potential coefficients $\alpha$ needed for this are given in
Tab.~\ref{tb4-tab1} in the Appendix. We provide their values only at
certain generating points, since we can assume that the coefficients
$\alpha$ are affine functions of parameters $(d_{1},d_{2})$.

\begin{figure}[t]
\centering \includegraphics[width=0.4\textwidth]{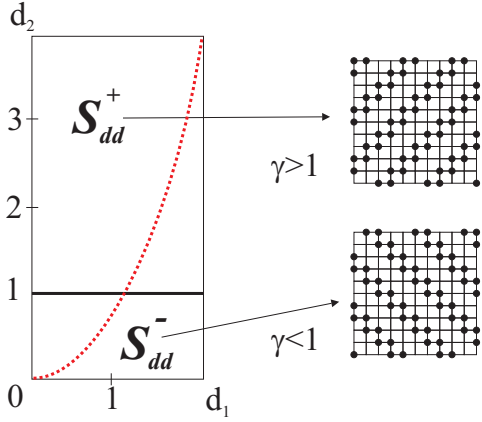}
\caption{The phase diagram of $E^{(4)}(d_{1},d_{2})$, with
$0\leqslant d_{1}\leqslant2$ and $0\leqslant d_{2}\leqslant4$. The
rectangle breaks down into two regions: above $d_{2}=1$ the
configurations $\mathcal{S}^{+}_{dd}$ are the ground-state
configurations, while $\mathcal{S}^{+}_{dd}$ are replaced by
$\mathcal{S}^{-}_{dd}$ below $d_{2}=1$. The dashed line represents
the condition $d_{2}=d_{1}^{2}$. The representative configurations
(up to translations) of $\mathcal{S}^{+}_{dd}$,
$\mathcal{S}^{-}_{dd}$ are shown on the right.} \label{anianipdd1d2}
\end{figure}

The phase diagram of $E^{(4)}$ is shown in Fig.~\ref{anianipdd1d2}.
The rectangle of considered points $(d_{1},d_{2})$ breaks down into
two domains. In the lower one, where $\gamma<1$ and $t_{-}<t_{+}$,
it is the phase $\mathcal{S}^{-}_{dd}$, with stripes being parallel
to the direction of $t_{-}$-hopping, that is stable. The analogous
situation is in the upper domain, where $\gamma>1$ and
$t_{-}>t_{+}$: the stable phase, $\mathcal{S}^{+}_{dd}$, consists of
stripes oriented along $t_{+}$-hopping. At $\gamma =1$, we have the
isotropic phase $\mathcal{S}_{dd}$ whose configurations consist of
$\mathcal{S}^{+}_{dd}$ and $\mathcal{S}^{-}_{dd}$.

So we see that, at least for the truncated effective Hamiltonian,
switching on of a n.n.n.-hopping anisotropy reduces the rotational
degeneracy of diagonal-striped phases: they become oriented in the
direction of the weaker hopping.

This result is similar to that described in Section 4, where the
influence of n.n.-hopping anisotropy on axial-striped phases was
investigated. In that case, not only for a truncated effective
Hamiltonian but also for the corresponding quantum one, it was
proved that for any nonzero value of n.n.-hopping anisotropy the
rotational degeneracy of axial-striped phases is reduced by making
them oriented along the direction of the weaker n.n. hopping. Now in
turn, the natural question is whether the conclusions we arrived at,
concerning n.n.n.-hopping anisotropy, hold true for the quantum
model, described by Hamiltonian (\ref{ourhamilt}). Applying the
arguments presented in \cite{GMMU,Kennedy1}, it could be
demonstrated, that the stable phases of the obtained above phase
diagram remain stable for the model (\ref{ourhamilt}), but in some
smaller domains. That is, if the remainder $R^{(4)}$ is taken into
account, then there exist such a small $t_{0}$, that for $t<t_{0}$
the phase diagram looks the same for the quantum model, excepting of
some narrow regions (of width $O(t)$ in the scale of the
fourth-order phase diagram shown in Fig.~\ref{anianipdd1d2}),
located along the phase-boundary lines. In our case that means that
the breaking of the rotational symmetry occurs for $\gamma=1+O(t)$,
when the n.n.n.-hopping intensities are $O(t^{2})$. Unfortunately,
we cannot claim that any non-zero n.n.n.-hopping anisotropy reduces
the rotational degeneracy of the quantum model, as it was the case
for n.n. hopping (see Section 4). Here it seems, at least for small
n.n.n.-hopping intensities, that there is certain critical value of
$|\gamma - 1|$, above which the degeneracy of phase
$\mathcal{S}_{dd}$ is reduced.

\clearpage
\section{Conclusions}
In Sections 3 and 4, we have succeeded in constructing ground-state
phase diagrams of some classical (i.e. diagonal in some basis)
Hamiltonians, in the whole range of parameters (coupling constants,
chemical potentials). These Hamiltonians have been derived as
effective Hamiltonians in certain orders of quantum Hamiltonians
(termed the extended Falicov--Kimball Hamiltonians), in the
strong-coupling regime. In other words, we have proved that the
considered classical Hamiltonians can be expressed by
$m$-potentials, for arbitrary value of parameters. Such Hamiltonians
are sometimes referred to as frustrationless. It should be
emphasized however, that, a priori, nothing guarantees that the
method we used here will ``produce'' any phase diagram: one can fail
to construct $m$-potentials or they might not exist.

Those ground-state phase diagrams of effective Hamiltonians
constitute the basic input of the quantum extension of the
Pirogov--Sinai theory of phase diagrams, developed in \cite{DFF2}.
This theory enables us to claim that not only the periodic ground
states of effective classical Hamiltonians are also ground states of
the corresponding quantum Hamiltonians, but that, at sufficiently
low temperatures, to each periodic ground-state phase there
corresponds a thermodynamic phase characterized by the same kind of
long-range order. The above mentioned properties of periodic
ground-states of effective Hamiltonians are referred to as stability
with respect to higher-order effective interactions and to small
thermal fluctuations.

Several basic problems concerning striped phases in a system of
strongly correlated electrons were considered. First of all, we
addressed the problem of the stability of charge-stripe phases
versus the phase separated state --- the segregated phase, in
strongly-interacting systems of fermions or hardcore bosons. There
are numerous works devoted to this problem, where it is studied, by
approximate methods, in the framework of models relevant for
experiments. Unfortunately, due to tiny energy differences involved,
it is difficult to settle this problem by means of approximate
methods, which bias the calculated energies with hardly controllable
errors of various nature. We have studied simple models, that by
many physicists can be considered less realistic, but which, in
return, are amenable to a rigorous analysis. Our models are
described by extended Falicov--Kimball Hamiltonians, with hopping
particles being spinless fermions or hardcore bosons. The
ground-state phase diagrams of these models have been constructed
rigorously in the regime of strong coupling and half-filling. In the
both cases, of fermions and hardcore bosons, we have found
transitions from a crystalline chessboard phase to the segregated
phase via striped phases.

After the stability of striped phases had been proven, both system
(with the hopping particles being either spinless fermions or
hardcore bosons) have been studied in order to show rigorously the
influence of nearest-neighbor and next-nearest-neighbor hopping
anisotropy on phase diagrams. In the case of nearest-neighbor
hopping only, two main conclusions have been drawn. Firstly,
arbitrary small anisotropy of nearest-neighbor hopping orients the
dimeric and axial-stripe phases in the direction of a weaker
hopping. Secondly, even a weak anisotropy of hopping reveals a
tendency of fermionic phase diagrams to become similar to the
bosonic ones.

In the case of next-nearest-neighbor hopping we have shown that a
weak anisotropy of the next-nearest-neighbor hopping reduces the
degeneracy of a diagonal-striped phase, it orients the stripes in
the direction of the weaker next-nearest-neighbor hopping.

\clearpage
\section*{Acknowledgments}
\addcontentsline{toc}{section}{Acknowledgments}

The author is deeply indebted to Prof. Janusz J\c{e}drzejewski for
his unstinting support over years.


The author thanks {\L}ukasz Andrzejewski and Pawe{\l} Musia{\l} for
their comments on the manuscript.

The author is grateful to the University of Wroc{\l}aw for Scientific
Research Grant 2479/W/IFT, and to the Institute of Theoretical
Physics for financial support. The Max Born Scholarship is
gratefully acknowledged.

\clearpage
\addcontentsline{toc}{section}{References}

\clearpage
\section*{Appendix A}
\addcontentsline{toc}{section}{Appendix A}


Here we present $T$-block zero-potentials in the isotropic case
(considered in Section 3), which are invariant with respect to
$\pi/2$-rotations and reflections.
\begin{align*}
k^{(1)}_{T}(s)=&s_{1}+ s_{3}+ s_{7}+ s_{9} - 4s_{5}, \displaybreak[0] \\
k^{(2)}_{T}(s)=&s_{2}+ s_{4}+ s_{6}+ s_{8} - 4s_{5}; \displaybreak[0] \\
k^{(3)}_{T}(s)=&s_{1}s_{2} +s_{2}s_{3} +s_{3}s_{6}
+s_{6}s_{9} +s_{8}s_{9} +s_{7}s_{8} +s_{4}s_{7} +s_{1}s_{4} \\
& -2s_{2}s_{5} -2s_{5}s_{6} -2s_{5}s_{8} -2s_{4}s_{5}, \displaybreak[0] \\
k^{(4)}_{T}(s)=&s_{1}s_{5} +s_{3}s_{5} +s_{5}s_{9}
+s_{5}s_{7} -s_{2}s_{4} -s_{4}s_{8} -s_{8}s_{6} -s_{2}s_{6}, \displaybreak[0] \\
k^{(5)}_{T}(s)=&s_{1}s_{3} +s_{3}s_{9} +s_{7}s_{9} +s_{1}s_{7}
-2s_{4}s_{6} -2s_{2}s_{8}; \displaybreak[0] \\
k^{(6)}_{T}(s)=&s_{1}s_{2}s_{3} +s_{3}s_{6}s_{9} +s_{7}s_{8}s_{9}
+s_{1}s_{4}s_{7} - 2s_{4}s_{5}s_{6} -2s_{2}s_{5}s_{8}, \displaybreak[0] \\
k^{(7)}_{T}(s)=&s_{1}s_{2}s_{5} +s_{4}s_{5}s_{7} +s_{5}s_{8}s_{9}
+s_{3}s_{5}s_{6} +s_{2}s_{3}s_{5}
+s_{1}s_{4}s_{5} +s_{5}s_{7}s_{8} +s_{5}s_{6}s_{9} \\
&-2s_{1}s_{2}s_{4} -2s_{4}s_{7}s_{8} -2s_{6}s_{8}s_{9}
-2s_{2}s_{3}s_{6}, \displaybreak[0] \\
k^{(8)}_{T}(s)=&s_{1}s_{2}s_{5} +s_{4}s_{5}s_{7} +s_{5}s_{8}s_{9}
+s_{3}s_{5}s_{6} +s_{2}s_{3}s_{5}
+s_{1}s_{4}s_{5} +s_{5}s_{7}s_{8} +s_{5}s_{6}s_{9} \\
&-2s_{2}s_{4}s_{5} -2s_{4}s_{5}s_{8} -2s_{5}s_{6}s_{8}
-2s_{2}s_{5}s_{6}, \displaybreak[0] \\
k^{(9)}_{T}(s)=&s_{1}s_{2}s_{6} +s_{2}s_{4}s_{7} +s_{4}s_{8}s_{9}
+s_{3}s_{6}s_{8} +s_{6}s_{7}s_{8}
+s_{2}s_{6}s_{9} +s_{2}s_{3}s_{4} +s_{1}s_{4}s_{8}\\
&-s_{3}s_{4}s_{5} -s_{1}s_{5}s_{8} -s_{5}s_{6}s_{7} -s_{2}s_{5}s_{9}
-s_{4}s_{5}s_{9} -s_{3}s_{5}s_{8} -s_{1}s_{5}s_{6} -s_{2}s_{5}s_{7},
\displaybreak[0] \\
k^{(10)}_{T}(s)=&s_{1}s_{2}s_{7} +s_{4}s_{7}s_{9} +s_{3}s_{8}s_{9}
+s_{1}s_{3}s_{6} +s_{2}s_{3}s_{9}
+s_{1}s_{3}s_{4} +s_{1}s_{7}s_{8} +s_{6}s_{7}s_{9}\\
&-s_{1}s_{2}s_{8} -s_{4}s_{6}s_{7} -s_{2}s_{8}s_{9} -s_{3}s_{4}s_{6}
-s_{2}s_{3}s_{8} -s_{1}s_{4}s_{6}
-s_{2}s_{7}s_{8} -s_{4}s_{6}s_{9}, \displaybreak[0] \\
k^{(11)}_{T}(s)=&s_{1}s_{3}s_{5} +s_{1}s_{5}s_{7} +s_{5}s_{7}s_{9}
+s_{3}s_{5}s_{9} -s_{2}s_{4}s_{8} -s_{4}s_{6}s_{8} -s_{2}s_{6}s_{8}
-s_{2}s_{4}s_{6}; \displaybreak[0] \\
k^{(12)}_{T}(s)=&s_{1}s_{2}s_{4}s_{7} +s_{4}s_{7}s_{8}s_{9}
+s_{3}s_{6}s_{8}s_{9} +s_{1}s_{2}s_{3}s_{6}
+s_{2}s_{3}s_{6}s_{9} +s_{1}s_{2}s_{3}s_{4} \\
&+s_{1}s_{4}s_{7}s_{8} +s_{6}s_{7}s_{8}s_{9} -s_{1}s_{2}s_{5}s_{8}
-s_{4}s_{5}s_{6}s_{7} -s_{2}s_{5}s_{8}s_{9}
-s_{3}s_{4}s_{5}s_{6} \\
&-s_{2}s_{3}s_{5}s_{8} -s_{1}s_{4}s_{5}s_{6}
-s_{2}s_{5}s_{7}s_{8} -s_{4}s_{5}s_{6}s_{9}, \displaybreak[0] \\
k^{(13)}_{T}(s)=&s_{1}s_{2}s_{3}s_{5} +s_{1}s_{4}s_{5}s_{7}
+s_{5}s_{7}s_{8}s_{9} +s_{3}s_{5}s_{6}s_{9}
-s_{2}s_{4}s_{5}s_{8} -s_{4}s_{5}s_{6}s_{8} \\
&-s_{2}s_{5}s_{6}s_{8} -s_{2}s_{4}s_{5}s_{6}, \displaybreak[0] \\
k^{(14)}_{T}(s)=&s_{1}s_{2}s_{4}s_{8} +s_{4}s_{6}s_{7}s_{8}
+s_{2}s_{6}s_{8}s_{9} +s_{2}s_{3}s_{4}s_{6}
+s_{2}s_{3}s_{6}s_{8} +s_{1}s_{2}s_{4}s_{6} \\
&+s_{2}s_{4}s_{7}s_{8} +s_{4}s_{6}s_{8}s_{9} -s_{1}s_{2}s_{5}s_{7}
-s_{4}s_{5}s_{7}s_{9} -s_{3}s_{5}s_{8}s_{9}
-s_{1}s_{3}s_{5}s_{6} \\
&-s_{2}s_{3}s_{5}s_{9} -s_{1}s_{3}s_{4}s_{5}
-s_{1}s_{5}s_{7}s_{8} -s_{5}s_{6}s_{7}s_{9}; \displaybreak[0] \\
k^{(15)}_{T}(s)=&s_{4}s_{6}s_{7}s_{8}s_{9}
+s_{2}s_{3}s_{6}s_{8}s_{9} +s_{1}s_{2}s_{3}s_{4}s_{6}
+s_{1}s_{2}s_{4}s_{7}s_{8} -s_{4}s_{5}s_{6}s_{7}s_{9} \\
&-s_{2}s_{3}s_{5}s_{8}s_{9} -s_{1}s_{3}s_{4}s_{5}s_{6}
-s_{1}s_{2}s_{5}s_{7}s_{8}, \displaybreak[0] \\
k^{(16)}_{T}(s)=&s_{3}s_{5}s_{6}s_{8}s_{9}
+s_{1}s_{4}s_{5}s_{7}s_{8} +s_{1}s_{2}s_{3}s_{5}s_{6}
+s_{1}s_{2}s_{4}s_{5}s_{7} +s_{4}s_{5}s_{7}s_{8}s_{9} \\
&+s_{5}s_{6}s_{7}s_{8}s_{9} +s_{2}s_{3}s_{5}s_{6}s_{9}
+s_{1}s_{2}s_{3}s_{4}s_{5} -s_{4}s_{5}s_{6}s_{8}s_{9}
-s_{2}s_{3}s_{5}s_{6}s_{8} \\
&-s_{1}s_{2}s_{4}s_{5}s_{6} -s_{2}s_{4}s_{5}s_{7}s_{8}
-s_{2}s_{3}s_{4}s_{5}s_{6} -s_{1}s_{2}s_{4}s_{5}s_{8}
-s_{4}s_{5}s_{6}s_{7}s_{8} \\
&-s_{2}s_{5}s_{6}s_{8}s_{9}.
\end{align*}

In the case of n.n. hopping anisotropy (considered in Section 4),
the zero-potentials invariant with respect to $\pi$-rotations and
reflections are of the form:
\begin{align*}
k_{{\rm{T}}}^{(1)} = & s_{1} +s_{3} +s_{7} +s_{9} -4s_{5}, \displaybreak[0] \\
k_{{\rm{T}}}^{(2)} = & s_{2} +s_{8} -2s_{5}, \displaybreak[0] \\
k_{{\rm{T}}}^{(3)} = & s_{4} +s_{6} -2s_{5}; \displaybreak[0] \\
k_{{\rm{T}}}^{(4)} = & s_{1}s_{2} +s_{2}s_{3} +s_{7}s_{8}
+s_{8}s_{9} -2s_{4}s_{5}-2s_{5}s_{6},\displaybreak[0] \\
k_{{\rm{T}}}^{(5)} = & s_{1}s_{4} +s_{3}s_{6} +s_{4}s_{7}
+s_{6}s_{9} -2s_{2}s_{5}-2s_{5}s_{8},\displaybreak[0] \\
k_{{\rm{T}}}^{(6)} = & s_{1}s_{5} +s_{3}s_{5} +s_{5}s_{9}
+s_{5}s_{7} -s_{2}s_{4} -s_{4}s_{8} -s_{8}s_{6} -s_{2}s_{6},\displaybreak[0] \\
k_{{\rm{T}}}^{(7)} = & s_{1}s_{3} +s_{7}s_{9}-2s_{4}s_{6},\displaybreak[0] \\
k_{{\rm{T}}}^{(8)} = & s_{1}s_{7} +s_{3}s_{9}-2s_{2}s_{8}; \displaybreak[0] \\
k_{{\rm{T}}}^{(9)} = & s_{1}s_{2}s_{4} +s_{6}s_{8}s_{9}
+s_{2}s_{3}s_{6} +s_{4}s_{7}s_{8}
- s_{2}s_{4}s_{5} -s_{5}s_{6}s_{8} - s_{2}s_{5}s_{6} -s_{4}s_{5}s_{8},\displaybreak[0] \\
k_{{\rm{T}}}^{(10)} = & s_{1}s_{2}s_{5} +s_{5}s_{8}s_{9}
+s_{2}s_{3}s_{5} +s_{5}s_{7}s_{8}
- s_{2}s_{4}s_{5} -s_{5}s_{6}s_{8} - s_{2}s_{5}s_{6} -s_{4}s_{5}s_{8},\displaybreak[0] \\
k_{{\rm{T}}}^{(11)} = & s_{4}s_{5}s_{7} +s_{3}s_{5}s_{6}
+s_{5}s_{6}s_{9} +s_{1}s_{4}s_{5}
- s_{2}s_{4}s_{5} -s_{5}s_{6}s_{8} - s_{2}s_{5}s_{6} -s_{4}s_{5}s_{8},\displaybreak[0] \\
k_{{\rm{T}}}^{(12)} = & s_{1}s_{2}s_{3} +s_{7}s_{8}s_{9} -2s_{4}s_{5}s_{6},\displaybreak[0] \\
k_{{\rm{T}}}^{(13)} = & s_{1}s_{4}s_{7} +s_{3}s_{6}s_{9} -2s_{2}s_{5}s_{8},\displaybreak[0] \\
k_{{\rm{T}}}^{(14)} = & s_{1}s_{2}s_{6} +s_{4}s_{8}s_{9}
+s_{2}s_{3}s_{4}
+s_{6}s_{7}s_{8} -s_{3}s_{4}s_{5} -s_{5}s_{6}s_{7} -s_{1}s_{5}s_{6} -s_{4}s_{5}s_{9},\displaybreak[0] \\
k_{{\rm{T}}}^{(15)} = & s_{2}s_{4}s_{7} +s_{3}s_{6}s_{8}
+s_{2}s_{6}s_{9}
+s_{1}s_{4}s_{8} -s_{1}s_{5}s_{8} -s_{2}s_{5}s_{9} -s_{3}s_{5}s_{8} -s_{2}s_{5}s_{7},\displaybreak[0] \\
k_{{\rm{T}}}^{(16)} = & s_{1}s_{2}s_{7} +s_{3}s_{8}s_{9}
+s_{2}s_{3}s_{9}
+s_{1}s_{7}s_{8} -s_{1}s_{2}s_{8} -s_{2}s_{8}s_{9} -s_{2}s_{3}s_{8} -s_{2}s_{7}s_{8},\displaybreak[0] \\
k_{{\rm{T}}}^{(17)} = & s_{4}s_{7}s_{9} +s_{1}s_{3}s_{6}
+s_{6}s_{7}s_{9}
+s_{1}s_{3}s_{4} -s_{4}s_{6}s_{7} -s_{3}s_{4}s_{6} -s_{4}s_{6}s_{9} -s_{1}s_{4}s_{6},\displaybreak[0] \\
k_{{\rm{T}}}^{(18)} = & s_{1}s_{3}s_{5}+s_{5}s_{7}s_{9}-s_{4}s_{6}s_{8}-s_{2}s_{4}s_{6},\displaybreak[0] \\
k_{{\rm{T}}}^{(19)} = &
s_{1}s_{5}s_{7}+s_{3}s_{5}s_{9}-s_{2}s_{4}s_{8}-s_{2}s_{6}s_{8};\displaybreak[0] \\
k_{{\rm{T}}}^{(20)} = & s_{1}s_{2}s_{4}s_{7} +s_{3}s_{6}s_{8}s_{9}
+s_{2}s_{3}s_{6}s_{9} +s_{1}s_{4}s_{7}s_{8} -s_{1}s_{2}s_{5}s_{8}
-s_{2}s_{5}s_{8}s_{9} -s_{2}s_{3}s_{5}s_{8}
-s_{2}s_{5}s_{7}s_{8},\displaybreak[0] \\
k_{{\rm{T}}}^{(21)} = & s_{4}s_{7}s_{8}s_{9} +s_{1}s_{2}s_{3}s_{6}
+s_{6}s_{7}s_{8}s_{9} +s_{1}s_{2}s_{3}s_{4} -s_{4}s_{5}s_{6}s_{7}
-s_{3}s_{4}s_{5}s_{6} -s_{4}s_{5}s_{6}s_{9}
-s_{1}s_{4}s_{5}s_{6},\displaybreak[0] \\
k_{{\rm{T}}}^{(22)} = & s_{1}s_{2}s_{3}s_{5} +s_{5}s_{7}s_{8}s_{9}
-s_{4}s_{5}s_{6}s_{8} -s_{2}s_{4}s_{5}s_{6},\displaybreak[0] \\
k_{{\rm{T}}}^{(23)} = & s_{1}s_{4}s_{5}s_{7} +s_{3}s_{5}s_{6}s_{9}
-s_{2}s_{4}s_{5}s_{8} -s_{2}s_{5}s_{6}s_{8},\displaybreak[0] \\
k_{{\rm{T}}}^{(24)} = & s_{1}s_{2}s_{4}s_{8} +s_{2}s_{6}s_{8}s_{9}
+s_{2}s_{3}s_{6}s_{8} +s_{2}s_{4}s_{7}s_{8} -s_{1}s_{2}s_{5}s_{7}
-s_{3}s_{5}s_{8}s_{9} -s_{2}s_{3}s_{5}s_{9}
-s_{1}s_{5}s_{7}s_{8},\displaybreak[0] \\
k_{{\rm{T}}}^{(25)} = & s_{4}s_{6}s_{7}s_{8} +s_{2}s_{3}s_{4}s_{6}
+s_{4}s_{6}s_{8}s_{9} +s_{1}s_{2}s_{4}s_{6} -s_{4}s_{5}s_{7}s_{9}
-s_{1}s_{3}s_{5}s_{6} -s_{5}s_{6}s_{7}s_{9} -s_{1}s_{3}s_{4}s_{5};\displaybreak[0] \\
k_{{\rm{T}}}^{(26)} = & s_{2}s_{3}s_{5}s_{8}s_{9}
+s_{1}s_{2}s_{5}s_{7}s_{8}
-s_{2}s_{3}s_{6}s_{8}s_{9} -s_{1}s_{2}s_{4}s_{7}s_{8},\displaybreak[0] \\
k_{{\rm{T}}}^{(27)} = & s_{4}s_{5}s_{6}s_{7}s_{9}
+s_{1}s_{3}s_{4}s_{5}s_{6}
-s_{4}s_{6}s_{7}s_{8}s_{9} -s_{1}s_{2}s_{3}s_{4}s_{6},\displaybreak[0] \\
k_{{\rm{T}}}^{(28)} = & s_{2}s_{3}s_{5}s_{6}s_{8}
+s_{2}s_{4}s_{5}s_{7}s_{8}
+s_{1}s_{2}s_{4}s_{5}s_{8} +s_{2}s_{5}s_{6}s_{8}s_{9}\\
& -s_{3}s_{5}s_{6}s_{8}s_{9} -s_{1}s_{2}s_{4}s_{5}s_{7}
-s_{1}s_{4}s_{5}s_{7}s_{8} -s_{2}s_{3}s_{5}s_{6}s_{9},\displaybreak[0] \\
k_{{\rm{T}}}^{(29)} = & s_{4}s_{5}s_{6}s_{8}s_{9}
+s_{1}s_{2}s_{4}s_{5}s_{6}
+s_{4}s_{5}s_{6}s_{7}s_{8} +s_{2}s_{3}s_{4}s_{5}s_{6}\\
&-s_{1}s_{2}s_{3}s_{5}s_{6} -s_{4}s_{5}s_{7}s_{8}s_{9}
-s_{1}s_{2}s_{3}s_{4}s_{5} -s_{5}s_{6}s_{7}s_{8}s_{9}.
\end{align*}
In Section 4, it is sufficient to consider only first nine
zero-potentials, i.e. we put $\alpha_{i}=0$ for $i\geqslant10$.

In the case of n.n. and n.n.n. hopping anisotropy (considered in
Section 5), the zero-potentials invariant with respect to
$\pi$-rotations are of the form:

Here the zero-potentials, in the case considered in Section 5, are
presented:
\begin{align*}
k_{{\rm{T}}}^{(1)} =&s_{1} +s_{9} -2s_{5}, \displaybreak[0] \\
k_{{\rm{T}}}^{(2)} =&s_{2} +s_{8} -2s_{5}, \displaybreak[0] \\
k_{{\rm{T}}}^{(3)} =&s_{3} +s_{7} -2s_{5}, \displaybreak[0] \\
k_{{\rm{T}}}^{(4)} =&s_{4} +s_{6} -2s_{5}; \displaybreak[0] \\
k_{{\rm{T}}}^{(5)} =&s_{1}s_{2} +s_{8}s_{9} -s_{4}s_{5} -s_{5}s_{6}, \displaybreak[0] \\
k_{{\rm{T}}}^{(6)} =&s_{2}s_{3} +s_{7}s_{8} -s_{4}s_{5} -s_{5}s_{6}, \displaybreak[0] \\
k_{{\rm{T}}}^{(7)} =&s_{1}s_{4} +s_{6}s_{9} -s_{2}s_{5} -s_{5}s_{8}, \displaybreak[0] \\
k_{{\rm{T}}}^{(8)} =&s_{3}s_{6} +s_{4}s_{7} -s_{2}s_{5} -s_{5}s_{8}, \displaybreak[0] \\
k_{{\rm{T}}}^{(9)} =&s_{2}s_{4} +s_{6}s_{8} -s_{3}s_{5} -s_{5}s_{7}, \displaybreak[0] \\
k_{{\rm{T}}}^{(10)}=&s_{2}s_{6} +s_{4}s_{8} -s_{1}s_{5} -s_{5}s_{9}, \displaybreak[0] \\
k_{{\rm{T}}}^{(11)}=&s_{1}s_{3} +s_{7}s_{9}-2s_{4}s_{6}, \displaybreak[0] \\
k_{{\rm{T}}}^{(12)}=&s_{1}s_{7} +s_{3}s_{9}-2s_{2}s_{8}; \displaybreak[0] \\
k_{{\rm{T}}}^{(13)}=&s_{1}s_{2}s_{5}+s_{5}s_{8}s_{9}-s_{2}s_{3}s_{6}-s_{4}s_{7}s_{8}, \displaybreak[0] \\
k_{{\rm{T}}}^{(14)}=&s_{2}s_{5}s_{6}+s_{4}s_{5}s_{8}-s_{2}s_{3}s_{6}-s_{4}s_{7}s_{8}, \displaybreak[0] \\
k_{{\rm{T}}}^{(15)}=&s_{1}s_{4}s_{5}+s_{5}s_{6}s_{9}-s_{2}s_{3}s_{6}-s_{4}s_{7}s_{8}, \displaybreak[0] \\
k_{{\rm{T}}}^{(16)}=&s_{2}s_{3}s_{5}+s_{5}s_{7}s_{8}-s_{1}s_{2}s_{4}-s_{6}s_{8}s_{9}, \displaybreak[0] \\
k_{{\rm{T}}}^{(17)}=&s_{3}s_{5}s_{6}+s_{4}s_{5}s_{7}-s_{1}s_{2}s_{4}-s_{6}s_{8}s_{9}, \displaybreak[0] \\
k_{{\rm{T}}}^{(18)}=&s_{2}s_{4}s_{5}+s_{5}s_{6}s_{8}-s_{1}s_{2}s_{4}-s_{6}s_{8}s_{9}, \displaybreak[0] \\
k_{{\rm{T}}}^{(19)}=&s_{1}s_{5}s_{6}+s_{4}s_{5}s_{9}-s_{1}s_{2}s_{6}-s_{4}s_{8}s_{9}, \displaybreak[0] \\
k_{{\rm{T}}}^{(20)}=&s_{2}s_{3}s_{4}+s_{6}s_{7}s_{8}-s_{3}s_{4}s_{5}-s_{5}s_{6}s_{7}, \displaybreak[0] \\
k_{{\rm{T}}}^{(21)}=&s_{2}s_{4}s_{7}+s_{3}s_{6}s_{8}-s_{2}s_{5}s_{7}-s_{3}s_{5}s_{8}, \displaybreak[0] \\
k_{{\rm{T}}}^{(22)}=&s_{1}s_{4}s_{8}+s_{2}s_{6}s_{9}-s_{2}s_{5}s_{9}-s_{1}s_{5}s_{8}, \displaybreak[0] \\
k_{{\rm{T}}}^{(23)}=&s_{1}s_{2}s_{3}+s_{7}s_{8}s_{9}-2s_{4}s_{5}s_{6}, \displaybreak[0] \\
k_{{\rm{T}}}^{(24)}=&s_{1}s_{4}s_{7}+s_{3}s_{6}s_{9}-2s_{2}s_{5}s_{8}, \displaybreak[0] \\
k_{{\rm{T}}}^{(25)}=&s_{1}s_{2}s_{7}+s_{3}s_{8}s_{9}-s_{2}s_{3}s_{8}-s_{2}s_{7}s_{8}, \displaybreak[0] \\
k_{{\rm{T}}}^{(26)}=&s_{1}s_{3}s_{6}+s_{4}s_{7}s_{9}-s_{1}s_{4}s_{6}-s_{4}s_{6}s_{9}, \displaybreak[0] \\
k_{{\rm{T}}}^{(27)}=&s_{3}s_{4}s_{6}+s_{4}s_{6}s_{7}-s_{1}s_{3}s_{4}-s_{6}s_{7}s_{9}, \displaybreak[0] \\
k_{{\rm{T}}}^{(28)}=&s_{1}s_{2}s_{8}+s_{2}s_{8}s_{9}-s_{1}s_{7}s_{8}-s_{2}s_{3}s_{9}, \displaybreak[0] \\
k_{{\rm{T}}}^{(29)}=&s_{1}s_{3}s_{5}+s_{5}s_{7}s_{9}-s_{2}s_{4}s_{6}-s_{4}s_{6}s_{8}, \displaybreak[0] \\
k_{{\rm{T}}}^{(30)}=&s_{1}s_{5}s_{7}+s_{3}s_{5}s_{9}-s_{2}s_{4}s_{8}-s_{2}s_{6}s_{8}; \displaybreak[0] \\
k_{{\rm{T}}}^{(31)}=&s_{1}s_{2}s_{4}s_{5}+s_{5}s_{6}s_{8}s_{9}-s_{2}s_{3}s_{5}s_{6}-s_{4}s_{5}s_{7}s_{8}, \displaybreak[0] \\
k_{{\rm{T}}}^{(32)}=&s_{1}s_{2}s_{4}s_{7}+s_{3}s_{6}s_{8}s_{9}-s_{2}s_{3}s_{5}s_{8}-s_{2}s_{5}s_{7}s_{8}, \displaybreak[0] \\
k_{{\rm{T}}}^{(33)}=&s_{1}s_{4}s_{7}s_{8}+s_{2}s_{3}s_{6}s_{9}-s_{1}s_{2}s_{5}s_{8}-s_{2}s_{5}s_{8}s_{9}, \displaybreak[0] \\
k_{{\rm{T}}}^{(34)}=&s_{1}s_{2}s_{3}s_{6}+s_{4}s_{7}s_{8}s_{9}-s_{1}s_{4}s_{5}s_{6}-s_{4}s_{5}s_{6}s_{9}, \displaybreak[0] \\
k_{{\rm{T}}}^{(35)}=&s_{1}s_{2}s_{3}s_{4}+s_{6}s_{7}s_{8}s_{9}-s_{3}s_{4}s_{5}s_{6}-s_{4}s_{5}s_{6}s_{7}, \displaybreak[0] \\
k_{{\rm{T}}}^{(36)}=&s_{1}s_{2}s_{3}s_{5}+s_{5}s_{7}s_{8}s_{9}-s_{2}s_{4}s_{5}s_{6}-s_{4}s_{5}s_{6}s_{8}, \displaybreak[0] \\
k_{{\rm{T}}}^{(37)}=&s_{1}s_{4}s_{5}s_{7}+s_{3}s_{5}s_{6}s_{9}-s_{2}s_{4}s_{5}s_{8}-s_{2}s_{5}s_{6}s_{8}, \displaybreak[0] \\
k_{{\rm{T}}}^{(38)}=&s_{1}s_{2}s_{4}s_{8}+s_{2}s_{6}s_{8}s_{9}-s_{1}s_{5}s_{7}s_{8}-s_{2}s_{3}s_{5}s_{9}, \displaybreak[0] \\
k_{{\rm{T}}}^{(39)}=&s_{2}s_{3}s_{6}s_{8}+s_{2}s_{4}s_{7}s_{8}-s_{1}s_{2}s_{5}s_{7}-s_{3}s_{5}s_{8}s_{9}, \displaybreak[0] \\
k_{{\rm{T}}}^{(40)}=&s_{2}s_{3}s_{4}s_{6}+s_{4}s_{6}s_{7}s_{8}-s_{1}s_{3}s_{4}s_{5}-s_{5}s_{6}s_{7}s_{9}, \displaybreak[0] \\
k_{{\rm{T}}}^{(41)}=&s_{1}s_{2}s_{4}s_{6}+s_{4}s_{6}s_{8}s_{9}-s_{1}s_{3}s_{5}s_{6}-s_{4}s_{5}s_{7}s_{9}; \displaybreak[0] \\
k_{{\rm{T}}}^{(42)}=&s_{1}s_{2}s_{3}s_{4}s_{6}+s_{4}s_{6}s_{7}s_{8}s_{9}-s_{1}s_{3}s_{4}s_{5}s_{6}-s_{4}s_{5}s_{6}s_{7}s_{9}, \displaybreak[0] \\
k_{{\rm{T}}}^{(43)}=&s_{1}s_{2}s_{5}s_{7}s_{8}+s_{2}s_{3}s_{5}s_{8}s_{9}-s_{1}s_{2}s_{4}s_{7}s_{8}-s_{2}s_{3}s_{6}s_{8}s_{9}, \displaybreak[0] \\
k_{{\rm{T}}}^{(44)}=&s_{1}s_{2}s_{4}s_{5}s_{7}+s_{3}s_{5}s_{6}s_{8}s_{9}-s_{2}s_{3}s_{5}s_{6}s_{8}-s_{2}s_{4}s_{5}s_{7}s_{8}, \displaybreak[0] \\
k_{{\rm{T}}}^{(45)}=&s_{1}s_{2}s_{4}s_{5}s_{8}+s_{2}s_{5}s_{6}s_{8}s_{9}-s_{1}s_{4}s_{5}s_{7}s_{8}-s_{2}s_{3}s_{5}s_{6}s_{9}, \displaybreak[0] \\
k_{{\rm{T}}}^{(46)}=&s_{1}s_{2}s_{3}s_{5}s_{6}+s_{4}s_{5}s_{7}s_{8}s_{9}-s_{1}s_{2}s_{4}s_{5}s_{6}-s_{4}s_{5}s_{6}s_{8}s_{9}, \displaybreak[0] \\
k_{{\rm{T}}}^{(47)}=&s_{1}s_{2}s_{3}s_{4}s_{5}+s_{5}s_{6}s_{7}s_{8}s_{9}-s_{2}s_{3}s_{4}s_{5}s_{6}-s_{4}s_{5}s_{6}s_{7}s_{8}.
\end{align*}
In Section 5, it is sufficient to consider only first nine
zero-potentials, i.e. we put $\alpha_{i}=0$ for $i\geqslant10$.

\clearpage
\section*{Appendix B}
\addcontentsline{toc}{section}{Appendix B}


\begin{figure}[th]
\centering \includegraphics[width=0.9\textwidth]{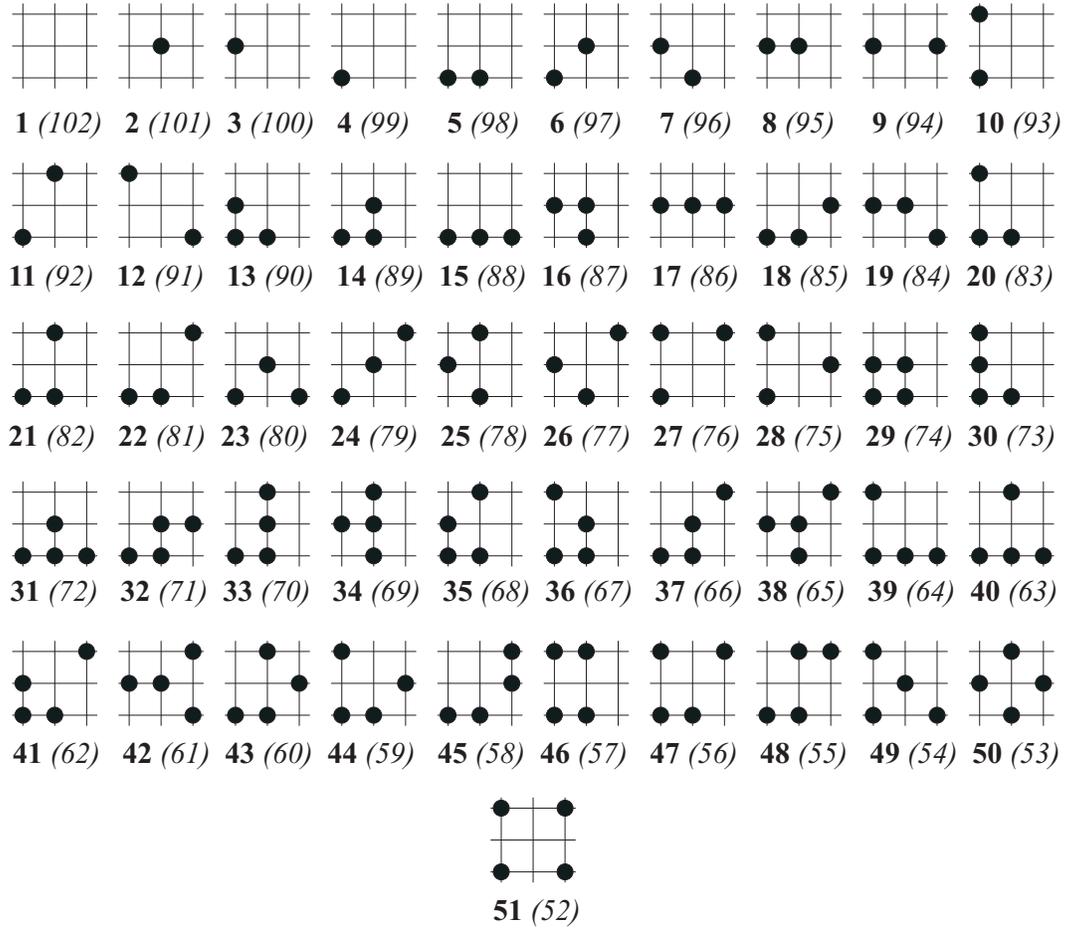}
\caption{{\scriptsize{All the $T$-plaquette configurations, over
which the potentials (\ref{H4fmn-1}) and (\ref{H4hcb-1}) are
minimized, up to the symmetries of $H$ (in the isotropic case of
n.n. hopping only). The configurations that can be obtained from the
displayed ones by the hole-particle transformation are not shown,
only the numbers assigned to them are given in brackets.}}}
\label{blconf102}
\end{figure}
\clearpage

\clearpage

\clearpage
\section*{Appendix C}
\addcontentsline{toc}{section}{Appendix C}


In this section, at each point $p=(\omega,\varepsilon)$ we give the
sets $\{\alpha_{i}(p)\}$ of the zero-potential coefficients
$\alpha_{i}$ that we used to construct the four phase diagrams,
presented in the previous section.  Specifically, for each of the
four phase diagrams we define a partition, called
$\alpha$-partition, of the $(\omega,\varepsilon)$-plane (see
Figs.~\ref{part1-1}--\ref{part1-4} displayed below and
Tables~\ref{tb1-1}--\ref{tb1-4} in Appendix E) into sets that
differ, in general, from the domains ${\mathcal{S}}_{D}$. To each
set of such a partition we assign a set $\{(a_{i},b_{i},c_{i})\}$ of
triplets of rational numbers, such that the coefficients
$\alpha_{i}(p)$ (with $p$ in the considered set) are affine
functions of the form $a_{i}\omega + b_{i}\varepsilon + c_{i}$. We
have not been able to find one set $\{(a_{i},b_{i},c_{i})\}$, that
turns the fourth order potentials into $m$-potentials in the whole
$(\omega,\varepsilon)$-plane. We have not also succeeded in
assigning to each domain ${\mathcal{S}}_{D}$ an exactly one set
$\{(a_{i},b_{i},c_{i})\}$. The same remarks apply to the
$(\omega,\delta)$-plane.
\begin{figure}[hbp]
\centering\includegraphics[height=0.3\textheight]{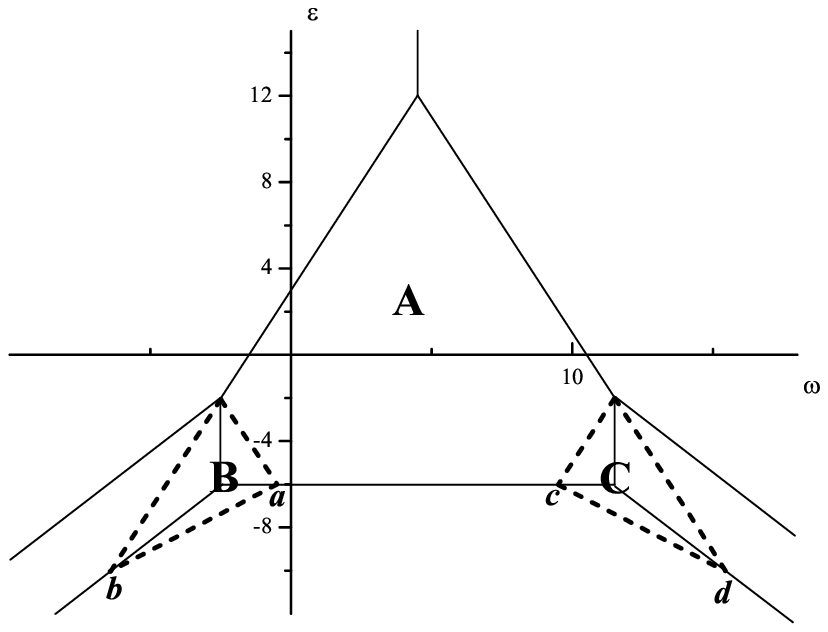}
\parbox[t]{\textwidth}
{\caption{{\small{The sets {\bf{A}}, {\bf{B}}, {\bf{C}}, whose
boundaries are marked with dashed lines, used in Table~\ref{tb1-1}
to define the $\alpha$-partition of the fourth order phase diagram
in the case of fermions and for $\delta=0$.The dashed-line segments
are determined by their intersection points: ${\bf a}=(-1/2,-6)$,
${\bf b}=(-13/2,-10)$, ${\bf c}=(19/2,-6)$, ${\bf d}=(31/2,-10)$.
The coordinates of the remaining points can be read from
Fig.~\ref{t1-40f}.}}} \label{part1-1}} \vfill
\centering\includegraphics[height=0.3\textheight]{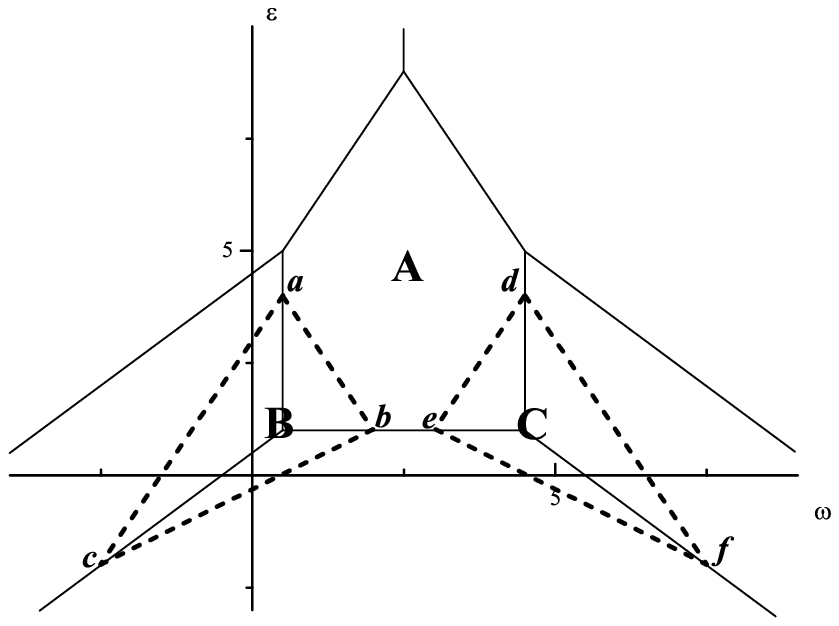}
\parbox[t]{\textwidth}
{\caption{{\small{The sets {\bf{A}}, {\bf{B}}, {\bf{C}}, whose
boundaries are marked with dashed lines, used in Table~\ref{tb1-2}
to define the $\alpha$-partition of the fourth order phase diagram
in the case of hardcore bosons and for $\delta=0$. The dashed-line
segments are determined by their intersection points: ${\bf
a}=(1/2,4)$, ${\bf b}=(2,1)$, ${\bf c}=(-5/2,-2)$, ${\bf
d}=(9/2,4)$, ${\bf e}= (3,1)$, ${\bf f} = (15/2,-2)$.}}}
\label{part1-2}}
\end{figure}

\begin{figure}[ht]
\centering\includegraphics[height=0.3\textheight]{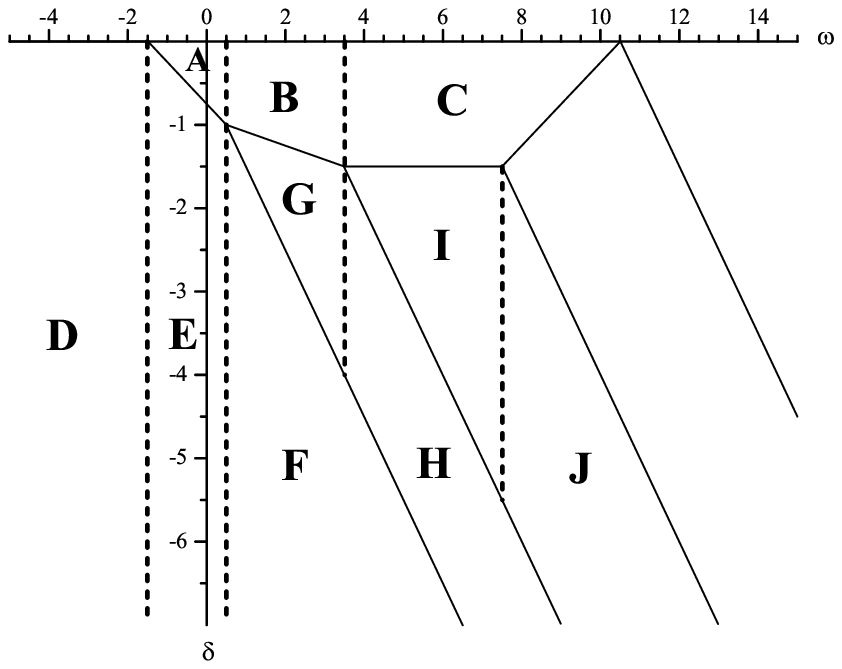}
\parbox[t]{\textwidth}
{\caption{{\small{The sets {\bf A,B},\ldots, whose boundaries are
marked with dashed and continuous lines (the continuous lines are
the boundaries of the phase diagram), used in Table~\ref{tb1-3} to
define the $\alpha$-partition of the fourth order phase diagram in
the case of fermions and for $\varepsilon=0$.}}} \label{part1-3}}
\vfill \centering\includegraphics[height=0.3\textheight]{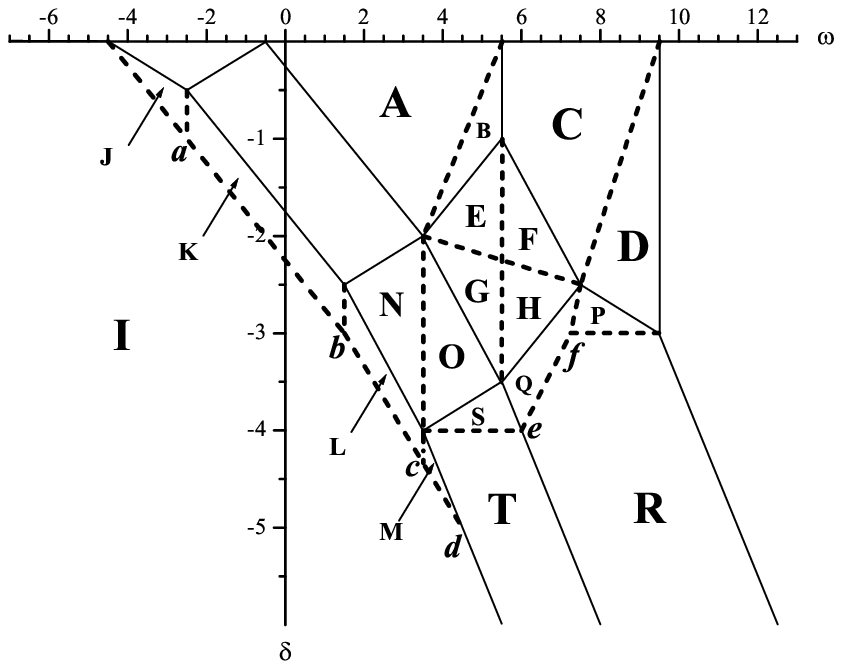}
\parbox[t]{\textwidth}
{\caption{{\small{The sets {\bf A,B},\ldots, whose boundaries are
marked with dashed and continuous lines (the continuous lines are
the boundaries of the phase diagram), used in Table~\ref{tb1-4} to
define the $\alpha$-partition of the fourth order phase diagram in
the case of hardcore bosons and for $\varepsilon=0$. The dashed-line
segments are determined by their intersection points: ${\bf a} =
(-5/2,-1)$, ${\bf b} = (3/2,-3)$, ${\bf c} = (7/2,-13/3)$, ${\bf d}
= (9/2,-5)$, ${\bf e} = (6,-4)$, ${\bf f} = (29/4,-3)$.}}}
\label{part1-4}}
\end{figure}

\clearpage
\section*{Appendix D}
\addcontentsline{toc}{section}{Appendix D}

Here we provide equations of line boundaries between the phase
domains, and coordinates of the crossing points of the line
boundaries that appear in the phase diagrams presented in Section 4.
The symbols like ${\mathcal{S}}_{1} | {\mathcal{S}}_{2}$ stand for
the line boundary between the phases ${\mathcal{S}}_{1}$ and
${\mathcal{S}}_{2}$, etc.

\renewcommand\baselinestretch{1,5}\small\normalsize
\begin{table}[h]
\begin{center}
\small \caption{Domain boundaries of the phase diagrams for
fermions, shown in Fig.~\ref{ani-sppd}.} \label{D-tb1}
\begin{tabular}{|l|c|c|}
\hline Lines & $0<\beta_{2}\leqslant7$ & $7\leqslant
\beta_{2}\leqslant10$
\\
\hline ${\mathcal{S}}_{seg} | {\mathcal{S}}_{cb}$ &
\multicolumn{2}{|c|}{$\omega=\beta_{2}+\frac{9}{2}$}
\\
\hline ${\mathcal{S}}_{seg} | {\mathcal{S}}_{dd}$ &
$\varepsilon=2\omega-2\beta_{2}+3$ & ---
\\
\hline ${\mathcal{S}}_{dd} | {\mathcal{S}}_{cb}$ &
$\varepsilon=-2\omega+2\beta_{2}+21$ & ---
\\
\hline ${\mathcal{S}}_{seg} | {\mathcal{S}}^{v}_{v1}$ &
\multicolumn{2}{|c|}{$\varepsilon=\omega+\frac{1}{2}$}
\\
\hline ${\mathcal{S}}^{v}_{d1} | {\mathcal{S}}_{cb}$ &
\multicolumn{2}{|c|}{$\varepsilon=-\omega+2\beta_{2}+\frac{19}{2}$}
\\
\hline ${\mathcal{S}}^{v}_{v1} | {\mathcal{S}}_{dd}$ &
$\omega=2\beta_{2}-\frac{5}{2}$ & ---
\\
\hline ${\mathcal{S}}_{dd} | {\mathcal{S}}^{v}_{d1}$ &
$\omega=\frac{23}{2}$ & ---
\\
\hline ${\mathcal{S}}^{v}_{v2} | {\mathcal{S}}_{dd}$ &
$\varepsilon=2\beta_{2}-6$ & ---
\\
\hline ${\mathcal{S}}^{v}_{v1} | {\mathcal{S}}^{v}_{v2}$ &
\multicolumn{2}{|c|}{$\varepsilon=\omega-\frac{7}{2}$}
\\
\hline ${\mathcal{S}}^{v}_{v2} | {\mathcal{S}}^{v}_{d1}$ &
\multicolumn{2}{|c|}{$\varepsilon=-\omega+2\beta_{2}+\frac{11}{2}$}
\\
\hline ${\mathcal{S}}^{v}_{v1} | {\mathcal{S}}^{v}_{d1}$ & --- &
$\omega=\beta_{2}+\frac{9}{2}$
\\
\hline
\end{tabular}
\\
\vspace{10mm}
\begin{tabular}{|l|c|c|}
\hline Points & $0<\beta_{2}\leqslant7$ & $7 \leqslant
\beta_{2}\leqslant10$
\\
\hline ${\bf a}$ & $\left( \beta_{2}+\frac{9}{2},12 \right)$ &
$\left( \beta_{2}+\frac{9}{2},\beta_{2}+5 \right)$
\\
\hline ${\bf b}$ & $\left( 2\beta_{2}-\frac{5}{2},2\beta_{2}-2
\right)$ & $\left( \beta_{2}+\frac{9}{2},\beta_{2}+1 \right)$
\\
\hline ${\bf c}$ & $\left( \frac{23}{2},2\beta_{2}-2 \right)$ & ---
\\
\hline ${\bf d}$ & $\left( 2\beta_{2}-\frac{5}{2},2\beta_{2}-6
\right)$ & ---
\\
\hline ${\bf e}$ & $\left( \frac{23}{2},2\beta_{2}-6 \right)$ & ---
\\
\hline
\end{tabular}
\end{center}
\end{table}

\begin{table}[ph]
\begin{center}
\small \caption{Domain boundaries of the phase diagrams for hardcore
bosons, shown in  Fig.~\ref{ani-sppd}.} \label{D-tb2}
\begin{tabular}{|l|c|c|}
\hline Lines & $0<\beta_{2}\leqslant2$ & $2<\beta_{2}\leqslant5$
\\
\hline ${\mathcal{S}}_{seg} | {\mathcal{S}}_{cb}$ &
\multicolumn{2}{|c|}{$\omega=\beta_{2}+\frac{5}{2}$}
\\
\hline ${\mathcal{S}}_{seg} | {\mathcal{S}}_{pcb}$ &
$\varepsilon=2\omega-2\beta_{2}+4$ & ---
\\
\hline ${\mathcal{S}}_{pcb} | {\mathcal{S}}_{cb}$ &
$\varepsilon=-2\omega+2\beta_{2}+14$ & ---
\\
\hline ${\mathcal{S}}_{seg} | {\mathcal{S}}^{v}_{v1}$ &
\multicolumn{2}{|c|}{$\varepsilon=\omega+\frac{9}{2}$}
\\
\hline ${\mathcal{S}}^{v}_{d1} | {\mathcal{S}}_{cb}$ &
\multicolumn{2}{|c|}{$\varepsilon=-\omega+2\beta_{2}+\frac{19}{2}$}
\\
\hline ${\mathcal{S}}^{v}_{v1} | {\mathcal{S}}_{pcb}$ &
$\omega=2\beta_{2}+\frac{1}{2}$ & ---
\\
\hline ${\mathcal{S}}_{pcb} | {\mathcal{S}}^{v}_{d1}$ &
$\omega=\frac{9}{2}$ & ---
\\
\hline ${\mathcal{S}}^{v}_{v2} | {\mathcal{S}}_{pcb}$ &
$\varepsilon=2\beta_{2}+1$ & ---
\\
\hline ${\mathcal{S}}^{v}_{v1} | {\mathcal{S}}^{v}_{v2}$ &
\multicolumn{2}{|c|}{$\varepsilon=\omega+\frac{1}{2}$}
\\
\hline ${\mathcal{S}}^{v}_{v2} | {\mathcal{S}}^{v}_{d1}$ &
\multicolumn{2}{|c|}{$\varepsilon=-\omega+2\beta_{2}+\frac{11}{2}$}
\\
\hline ${\mathcal{S}}^{v}_{v1} | {\mathcal{S}}^{v}_{d1}$ & --- &
$\omega=\beta_{2}+\frac{5}{2}$
\\
\hline
\end{tabular}
\\
\vspace{10mm}
\begin{tabular}{|l|c|c|}
\hline Points & $0<\beta_{2}\leqslant2$ & $2<\beta_{2}\leqslant5$
\\
\hline ${\bf a}$ & $\left( \beta_{2}+\frac{5}{2},9 \right)$ &
$\left( \beta_{2}+\frac{5}{2},\beta_{2}+7 \right)$
\\
\hline ${\bf b}$ & $\left( 2\beta_{2}+\frac{1}{2},2\beta_{2}+5
\right)$ & $\left( \beta_{2}+\frac{5}{2},\beta_{2}+3 \right)$
\\
\hline ${\bf c}$ & $\left( \frac{9}{2},2\beta_{2}+5 \right)$ & ---
\\
\hline ${\bf d}$ & $\left( 2\beta_{2}+\frac{1}{2},2\beta_{2}+1
\right)$ & ---
\\
\hline ${\bf e}$ & $\left( \frac{9}{2},2\beta_{2}+1 \right)$ & ---
\\
\hline
\end{tabular}
\end{center}
\end{table}

\begin{table}[h]
\begin{center}
\scriptsize \caption{Domain boundaries of the phase diagrams shown
in  Fig.~\ref{ani-epspdf}.} \label{D-tb3}
\begin{tabular}{|l|c|c|c|c|}
\hline Lines & $0<\beta_{2}\leqslant1$ & $1<\beta_{2}\leqslant2$ &
$2<\beta_{2}\leqslant3$ & $3<\beta_{2}\leqslant6$
\\
\hline ${\mathcal{S}}_{+} | {\mathcal{S}}_{-}$ &
\multicolumn{4}{|c|}{$\delta=0$}
\\
\hline ${\mathcal{S}}_{dd} | {\mathcal{S}}_{-}$ &
\multicolumn{2}{|c|}{$\delta=-\frac{\omega}{2}+\frac{\beta_{2}}{2}-\frac{3}{4}$}
& \multicolumn{2}{|c|}{---}
\\
\hline ${\mathcal{S}}_{dd} | {\mathcal{S}}_{1}$ &
\multicolumn{2}{|c|}{
$\delta=-\frac{\omega}{6}+\frac{\beta_{2}}{6}-\frac{11}{12}$} &
\multicolumn{2}{|c|}{---}
\\
\hline ${\mathcal{S}}_{dd} | {\mathcal{S}}_{2}$ &
\multicolumn{2}{|c|}{$\delta=-\frac{3}{2}$} &
\multicolumn{2}{|c|}{---}
\\
\hline ${\mathcal{S}}_{dd} | {\mathcal{S}}_{3}$ &
\multicolumn{2}{|c|}{$\delta=\frac{\omega}{2}-\frac{\beta_{2}}{2}-\frac{21}{4}$}
& \multicolumn{2}{|c|}{---}
\\
\hline ${\mathcal{S}}_{-} | {\mathcal{S}}_{1}$ &
\multicolumn{4}{|c|}{$\delta=-\omega+\beta_{2}-\frac{1}{2}$}
\\
\hline ${\mathcal{S}}_{1} | {\mathcal{S}}_{2}$ &
\multicolumn{4}{|c|}{$\delta=-\omega+\beta_{2}+2$}
\\
\hline
\end{tabular}
\end{center}
\end{table}

\begin{table}[h]
\begin{center}
\begin{tabular}{|l|c|c|c|c|}
\hline Points & $0<\beta_{2}\leqslant1$ & $1<\beta_{2}\leqslant2$ &
$2<\beta_{2}\leqslant3$ & $3<\beta_{2}\leqslant6$
\\
\hline {\bf a} & $\left( \beta_{2}-\frac{3}{2},0 \right)$ &
\multicolumn{3}{|c|}{$\left( -\frac{1}{2},0 \right)$}
\\
\hline {\bf b} & $\left( \beta_{2}+\frac{1}{2},-1 \right)$ & $\left(
2\beta_{2}-\frac{5}{2},-\frac{\beta_{2}}{2}+\frac{1}{2} \right)$ &
$\left( 2\beta_{2}-\frac{5}{2},\frac{\beta_{2}}{2}-\frac{3}{2}
\right)$ & $\left( \frac{7}{2},0 \right)$
\\
\hline {\bf c} & $\left( \beta_{2}+\frac{7}{2},-\frac{3}{2} \right)$
& $\left( \frac{23}{2},-\frac{\beta_{2}}{2}+\frac{1}{2} \right)$ &
$\left( \frac{23}{2},\frac{\beta_{2}}{2}-\frac{3}{2} \right)$ &
$\left( \frac{3}{2},-\frac{1}{2} \right)$
\\
\hline {\bf d} & $\left( \beta_{2}+\frac{15}{2},-\frac{3}{2}
\right)$ & $\left( 2\beta_{2}+\frac{19}{2},-\beta_{2}+1 \right)$ &
$\left( \frac{3}{2},-\frac{1}{2} \right)$ & $\left(
2\beta_{2}+\frac{11}{2},-\beta_{2}+3 \right)$
\\
\hline {\bf e} & $\left( \beta_{2}+\frac{21}{2},0 \right)$ & $\left(
\frac{3}{2},\frac{\beta_{2}}{2}-\frac{3}{2} \right)$ & $\left(
4\beta_{2}-\frac{13}{2},-\frac{\beta_{2}}{2}+\frac{1}{2} \right)$ &
$\left( 2\beta_{2}-\frac{1}{2},-\beta_{2}+2 \right)$
\\
\hline {\bf f} & --- & $\left(
4\beta_{2}-\frac{5}{2},-\frac{\beta_{2}}{2}-\frac{1}{2} \right)$ &
$\left( 2\beta_{2}-\frac{5}{2},-\beta_{2}+\frac{3}{2} \right)$ &
$\left( 2\beta_{2}+\frac{7}{2},-\beta_{2}+2 \right)$
\\
\hline {\bf g} & --- & $\left( 2\beta_{2}-\frac{1}{2},-\beta_{2}
\right)$ & $\left( 2\beta_{2}+\frac{15}{2},-\beta_{2}+\frac{3}{2}
\right)$ & $\left( 2\beta_{2}-\frac{5}{2},-\beta_{2}+\frac{3}{2}
\right)$
\\
\hline
\end{tabular}
\end{center}
\end{table}
\clearpage

\begin{table}[tph]
\begin{center}
\small \caption{Domain boundaries of the phase diagrams shown in
Fig.~\ref{ani-epspdb}.} \label{D-tb4}
\begin{tabular}{|l|c|}
\hline Lines & $0<\beta_{2}\leqslant3$
\\
\hline ${\mathcal{S}}_{-} | {\mathcal{S}}_{+}$ & $\delta=0$
\\
\hline ${\mathcal{S}}_{-} | {\mathcal{S}}^{v}_{v1}$ &
$\delta=-\frac{\omega}{4}-\frac{9}{8}$
\\
\hline ${\mathcal{S}}^{v}_{v1}|{\mathcal{S}}^{v}_{v3}$ &
$\delta=\frac{\omega}{4}+\frac{1}{8}$
\\
\hline ${\mathcal{S}}^{v}_{v3}|{\mathcal{S}}^{v}_{v2}$ &
$\delta=-\frac{\omega}{2}-\frac{1}{4}$
\\
\hline ${\mathcal{S}}_{-}|{\mathcal{S}}^{v}_{v3}$ &
$\delta=-\frac{\omega}{2}-\frac{7}{4}$
\\
\hline ${\mathcal{S}}^{v}_{v3}|{\mathcal{S}}^{v}_{d2}$ &
$\delta=\frac{\omega}{4}-\frac{3\beta_{2}}{2}-\frac{23}{8}$
\\
\hline ${\mathcal{S}}^{v}_{d2}|{\mathcal{S}}^{v}_{d3}$ &
$\delta=-\frac{3\omega}{4}+\frac{\beta_{2}}{2}+\frac{5}{8}$
\\
\hline ${\mathcal{S}}_{-}|{\mathcal{S}}^{v}_{d2}$ &
$\delta=-\frac{3\omega}{4}+\frac{\beta_{2}}{2}-\frac{11}{8}$
\\
\hline ${\mathcal{S}}^{v}_{d2}|{\mathcal{S}}_{1}$ &
$\delta=\frac{\omega}{4}-\frac{3\beta_{2}}{2}-\frac{39}{8}$
\\
\hline ${\mathcal{S}}^{v}_{v2}|{\mathcal{S}}^{v}_{d3}$ &
$\delta=\frac{\omega}{2}-2\beta_{2}-\frac{15}{4}$
\\
\hline ${\mathcal{S}}^{v}_{d3}|{\mathcal{S}}_{3}$ &
$\delta=\frac{\omega}{2}-2\beta_{2}-\frac{25}{4}$
\\
\hline ${\mathcal{S}}^{v}_{d3}|{\mathcal{S}}^{v}_{d1}$ &
$\delta=-\frac{3\omega}{4}+\frac{\beta_{2}}{2}+\frac{25}{8}$
\\
\hline ${\mathcal{S}}^{v}_{d1}|{\mathcal{S}}_{3}$ &
$\delta=-\frac{\omega}{4}-\frac{\beta_{2}}{2}-\frac{5}{8}$
\\
\hline ${\mathcal{S}}^{v}_{v2}|{\mathcal{S}}^{v}_{d1}$ &
$\omega=2\beta_{2}+\frac{11}{2}$
\\
\hline ${\mathcal{S}}^{v}_{d1}|{\mathcal{S}}_{cb}$ &
$\omega=2\beta_{2}+\frac{19}{2}$
\\
\hline ${\mathcal{S}}_{-}|{\mathcal{S}}_{1}$ &
$\delta=-\omega+\beta_{2}-\frac{1}{2}$
\\
\hline ${\mathcal{S}}_{1}|{\mathcal{S}}_{3}$ &
$\delta=-\omega+\beta_{2}+2$
\\
\hline ${\mathcal{S}}_{3}|{\mathcal{S}}_{cb}$ &
$\delta=-\omega+\beta_{2}+\frac{13}{2}$
\\
\hline
\end{tabular}
\hspace{10mm}
\begin{tabular}{|l|c|}
\hline Points & $0<\beta_{2}\leqslant3$
\\
\hline ${\bf a}$ & $\left( -\frac{9}{2},0 \right)$
\\
\hline ${\bf b}$ & $\left( -\frac{1}{2},0 \right)$
\\
\hline ${\bf c}$ & $\left( -\frac{5}{2}, -\frac{1}{2} \right)$
\\
\hline ${\bf d}$ & $\left( 2\beta_{2}+\frac{11}{2}, -\beta_{2}-1
\right)$
\\
\hline ${\bf e}$ & $\left( 2\beta_{2}+\frac{7}{2}, -\beta_{2}-2
\right)$
\\
\hline ${\bf f}$ & $\left( 2\beta_{2}+\frac{3}{2},
-\beta_{2}-\frac{5}{2} \right)$
\\
\hline ${\bf g}$ & $\left( 2\beta_{2}+\frac{15}{2},
-\beta_{2}-\frac{5}{2} \right)$
\\
\hline ${\bf h}$ & $\left( 2\beta_{2}+\frac{19}{2}, -\beta_{2}-3
\right)$
\\
\hline ${\bf i}$ & $\left( 2\beta_{2}+\frac{11}{2}, -\beta_{2}-7/2
\right)$
\\
\hline ${\bf j}$ & $\left( 2\beta_{2}+\frac{7}{2}, -\beta_{2}-4
\right)$
\\
\hline
\end{tabular}
\end{center}
\end{table}

\begin{table}[ph]
\begin{center}
\small \caption{Domain boundaries of the phase diagrams shown in
Fig.~\ref{ani-p1fmnd0} and Fig.~\ref{ani-p1hcbd0}.} \label{D-tb5}
\begin{tabular}{|l|c|c|}
\hline Lines & fermions & bosons
\\
\hline ${\mathcal{S}}_{seg} | {\mathcal{S}}_{cb}$ &
$\omega=\frac{9}{2}$ & $\omega=\frac{5}{2}$
\\
\hline ${\mathcal{S}}^{v}_{v1} | {\mathcal{S}}^{v}_{d1}$ &
$\omega=\frac{9}{2}$ & $\omega=\frac{5}{2}$
\\
\hline ${\mathcal{S}}^{v}_{v1}|{\mathcal{S}}_{seg}$ &
$\varepsilon=\omega+\frac{1}{2}$ & $\varepsilon=\omega+\frac{9}{2}$
\\
\hline ${\mathcal{S}}^{v}_{d1}|{\mathcal{S}}_{cb}$ &
\multicolumn{2}{|c|}{$\varepsilon=-\omega+\frac{19}{2}$}
\\
\hline ${\mathcal{S}}^{v}_{v2}|{\mathcal{S}}^{v}_{v1}$ &
$\varepsilon=\omega-\frac{7}{2}$ & $\varepsilon=\omega+\frac{1}{2}$
\\
\hline ${\mathcal{S}}^{v}_{v2}|{\mathcal{S}}^{v}_{d1}$ &
\multicolumn{2}{|c|}{$\varepsilon=-\omega+\frac{11}{2}$}
\\
\hline
\end{tabular}
\hspace{10mm}
\begin{tabular}{|l|c|c|}
\hline Points & fermions & bosons
\\
\hline ${\bf a}$ & $\left( \frac{9}{2},5 \right)$ & $\left(
\frac{5}{2},7 \right)$
\\
\hline ${\bf b}$ & $\left( \frac{9}{2},1 \right)$ & $\left(
\frac{5}{2},3 \right)$
\\
\hline
\end{tabular}
\end{center}
\end{table}

\begin{table}[pb]
\begin{center}
\small \caption{Domain boundaries of the phase diagrams shown in
Fig.~\ref{ani-p1fmne0} and Fig.~\ref{ani-p1hcbe0}.} \label{D-tb6}
\begin{tabular}{|l|c|c|}
\hline Lines & fermions & bosons
\\
\hline ${\mathcal{S}}_{-} | {\mathcal{S}}_{+}$ &
\multicolumn{2}{|c|}{$\delta=0$}
\\
\hline ${\mathcal{S}}^{v}_{v1} | {\mathcal{S}}_{-}$ &
$\delta=-\frac{\omega}{4}-\frac{1}{8}$ &
$\delta=-\frac{\omega}{4}-\frac{9}{8}$
\\
\hline ${\mathcal{S}}^{v}_{v1}|{\mathcal{S}}^{v}_{v3}$ &
$\delta=\frac{\omega}{4}-\frac{7}{8}$ &
$\delta=\frac{\omega}{4}+\frac{1}{8}$
\\
\hline ${\mathcal{S}}^{v}_{v3}|{\mathcal{S}}_{-}$ &
$\delta=-\frac{\omega}{2}+\frac{1}{4}$ &
$\delta=-\frac{\omega}{2}-\frac{7}{4}$
\\
\hline ${\mathcal{S}}^{v}_{v3}|{\mathcal{S}}^{v}_{v2}$ &
$\delta=-\frac{\omega}{2}+\frac{7}{4}$ &
$\delta=-\frac{\omega}{2}-\frac{1}{4}$
\\
\hline
\end{tabular}
\hspace{10mm}
\begin{tabular}{|l|c|c|}
\hline Points & fermions & bosons
\\
\hline ${\bf a}$ & $\left( -\frac{1}{2},0 \right)$ & $\left(
-\frac{9}{2},0 \right)$
\\
\hline ${\bf b}$ & $\left( \frac{7}{2},0 \right)$ & $\left(
-\frac{1}{2},0 \right)$
\\
\hline ${\bf c}$ & $\left( \frac{3}{2},-\frac{1}{2} \right)$ &
$\left( -\frac{5}{2},-\frac{1}{2} \right)$
\\
\hline
\end{tabular}
\end{center}
\end{table}

\begin{table}[ph]
\begin{center}
\small \caption{Domain boundaries of the phase diagrams shown in
Fig.~\ref{ani-p2fmne0} and Fig.~\ref{ani-p2hcbe0}.} \label{D-tb7}
\begin{tabular}{|l|c|c|}
\hline Lines & fermions & bosons
\\
\hline ${\mathcal{S}}^{v}_{v3} | {\mathcal{S}}_{-}$ &
$\delta=-\frac{\omega}{2}+\frac{1}{4}$ &
$\delta=-\frac{\omega}{2}-\frac{7}{4}$
\\
\hline ${\mathcal{S}}^{v}_{v3} | {\mathcal{S}}^{v}_{v2}$ &
$\delta=-\frac{\omega}{2}+\frac{7}{4}$ &
$\delta=-\frac{\omega}{2}-\frac{1}{4}$
\\
\hline ${\mathcal{S}}^{v}_{v3}|{\mathcal{S}}^{v}_{d2}$ &
$\delta=\frac{\omega}{4}+\frac{17}{8}$ &
$\delta=\frac{\omega}{4}-\frac{23}{8}$
\\
\hline ${\mathcal{S}}^{v}_{d2}|{\mathcal{S}}_{-}$ &
$\delta=-\frac{3\omega}{4}-\frac{3}{8}$ &
$\delta=-\frac{3\omega}{4}-\frac{11}{8}$
\\
\hline ${\mathcal{S}}^{v}_{d2}|{\mathcal{S}}^{v}_{d4}$ &
$\delta=-\frac{3\omega}{4}+\frac{13}{8}$ & ---
\\
\hline ${\mathcal{S}}^{v}_{d2}|{\mathcal{S}}^{v}_{d3}$ & --- &
$\delta=-\frac{3\omega}{4}+\frac{5}{8}$
\\
\hline ${\mathcal{S}}^{v}_{d2}|{\mathcal{S}}_{1}$ &
$\delta=\frac{\omega}{4}+\frac{1}{8}$ &
$\delta=\frac{\omega}{4}-\frac{39}{8}$
\\
\hline
\end{tabular}
\hspace{10mm}
\begin{tabular}{|l|c|c|}
\hline Points & fermions & bosons
\\
\hline ${\bf a}$ & $\left( \frac{11}{2}, 3 \right)$ & $\left(
\frac{11}{2}, -1 \right)$
\\
\hline ${\bf b}$ & $\left( -\frac{1}{2}, 2 \right)$ & $\left(
\frac{7}{2}, -2 \right)$
\\
\hline ${\bf c}$ & $\left( \frac{7}{2}, 2 \right)$ & $\left(
\frac{3}{2}, -\frac{5}{2} \right)$
\\
\hline ${\bf d}$ & $\left( -\frac{5}{2}, \frac{3}{2} \right)$ &
$\left( \frac{15}{2}, -\frac{5}{2} \right)$
\\
\hline ${\bf e}$ & $\left( \frac{15}{2}, \frac{3}{2} \right)$ &
$\left( \frac{19}{2}, -3 \right)$
\\
\hline ${\bf f}$ & $\left( \frac{19}{2}, 1 \right)$ & $\left(
\frac{11}{2}, -\frac{7}{2} \right)$
\\
\hline ${\bf g}$ & $\left( \frac{3}{2}, \frac{1}{2} \right)$ &
$\left( \frac{7}{2}, -4 \right)$
\\
\hline ${\bf h}$ & $\left( \frac{11}{2}, \frac{1}{2} \right)$ & ---
\\
\hline ${\bf i}$ & $\left( -\frac{1}{2}, 0 \right)$ & ---
\\
\hline
\end{tabular}
\end{center}
\end{table}
\renewcommand\baselinestretch{1}\small\normalsize

\clearpage
\section*{Appendix E}
\addcontentsline{toc}{section}{Appendix E}


\begin{table}[hbp]
\scriptsize
\begin{center}
\caption{The case of isotropic n.n. hopping only. The set of
zero-potential coefficients $\{\alpha_{i}\}$ in the case of fermions
and for $\delta=0$. In the first column the sets of the
$\alpha$-partition are specified. For more comments see the text in
Appendix C.} \label{tb1-1}
\renewcommand\baselinestretch{1,5}\small\normalsize
\begin{tabular}{|l|c|c|c|c|c|}
\hline & $\alpha_{1}$ & $\alpha_{2}$ & $\alpha_{3}$ & $\alpha_{4}$ &
$\alpha_{5}$
\\
\hline {\bf{A}} & $0$ & $0$ & $-\frac{\omega}{96}+\frac{3}{64}$ &
$0$ & $-\frac{1}{48}$
\\
\hline ${\mathcal{S}}_{dd}\cap${\bf{B}} & $0$ & $0$ & $-\frac{5
\omega}{192} -\frac{\varepsilon}{64} -\frac{3}{128}$ &
$\frac{\omega}{32} +\frac{\varepsilon}{32} +\frac{9}{64}$ &
$\frac{\omega}{64} +\frac{\varepsilon}{64} +\frac{19}{384}$
\\
\hline ${\mathcal{S}}_{v1}\cap${\bf{B}} & $0$ & $0$ &
$\frac{\omega}{192} -\frac{\varepsilon}{64} +\frac{7}{128}$ &
$-\frac{\omega}{32} +\frac{\varepsilon}{32} -\frac{1}{64}$ &
$-\frac{\omega}{64} +\frac{\varepsilon}{64} -\frac{11}{384}$
\\
\hline ${\mathcal{S}}_{v2}\cap${\bf{B}} & $0$ & $0$ & $-\frac{5
\omega}{192} +\frac{\varepsilon}{64} +\frac{21}{128}$ &
$\frac{\omega}{32} -\frac{\varepsilon}{32} -\frac{15}{64}$ &
$\frac{\omega}{64} -\frac{\varepsilon}{64} -\frac{53}{384}$
\\
\hline ${\mathcal{S}}_{dd}\cap${\bf{C}} & $0$ & $0$ & $-\frac{5
\omega}{192} +\frac{\varepsilon}{64} +\frac{33}{128}$ &
$-\frac{\omega}{32} +\frac{\varepsilon}{32} +\frac{27}{64}$ &
$-\frac{\omega}{64} +\frac{\varepsilon}{64} +\frac{73}{384}$
\\
\hline ${\mathcal{S}}_{v2}\cap${\bf{C}} & $0$ & $0$ & $-\frac{5
\omega}{192} -\frac{\varepsilon}{64} +\frac{9}{128}$ &
$-\frac{\omega}{32} -\frac{\varepsilon}{32} +\frac{3}{64}$ &
$-\frac{\omega}{64} -\frac{\varepsilon}{64} +\frac{1}{384}$
\\
\hline ${\mathcal{S}}_{d1}\cap${\bf{C}} & $0$ & $0$ &
$\frac{\omega}{192} +\frac{\varepsilon}{64} -\frac{13}{128}$ &
$\frac{\omega}{32} +\frac{\varepsilon}{32} -\frac{19}{64}$ &
$\frac{\omega}{64} +\frac{\varepsilon}{64} -\frac{65}{384}$
\\
\hline
\end{tabular}
\end{center}
\end{table}

\begin{table}[ht]
\scriptsize
\begin{center}
\caption{The case of isotropic n.n. hopping only. The set of
zero-potential coefficients $\{\alpha_{i}\}$ in the case of hardcore
bosons and for $\delta=0$. In the first column the sets of the
$\alpha$-partition are specified. For more comments see the text in
Appendix C.} \label{tb1-2}
\renewcommand\baselinestretch{1,5}\small\normalsize
\begin{tabular}{|l|c|c|c|c|c|}
\hline & $\alpha_{1}$ & $\alpha_{2}$ & $\alpha_{3}$ & $\alpha_{4}$ &
$\alpha_{5}$
\\
\hline {\bf{A}} & $0$ & $0$ & $-\frac{\omega}{96}+\frac{5}{192}$ &
$0$ & $-\frac{1}{48}$
\\
\hline ${\mathcal{S}}_{pcb}\cap${\bf{B}} & $0$ & $0$ &
$-\frac{\omega}{24} -\frac{\varepsilon}{64} +\frac{23}{192}$ &
$\frac{\omega}{16} +\frac{7 \varepsilon}{192} -\frac{37}{192}$ &
$\frac{\omega}{32} +\frac{5 \varepsilon}{256} -\frac{91}{768}$
\\
\hline ${\mathcal{S}}_{v1}\cap${\bf{B}} & $0$ & $0$ &
$\frac{\omega}{192} -\frac{\varepsilon}{64} +\frac{37}{384}$ &
$-\frac{7 \omega}{192} +\frac{7 \varepsilon}{192} -\frac{55}{384}$ &
$-\frac{5 \omega}{256} +\frac{5 \varepsilon}{256} -\frac{143}{1536}$
\\
\hline ${\mathcal{S}}_{v2}\cap${\bf{B}} & $0$ & $0$ &
$-\frac{\omega}{24} +\frac{\varepsilon}{32} +\frac{7}{96}$ &
$\frac{\omega}{16} -\frac{\varepsilon}{16} -\frac{3}{32}$ &
$\frac{\omega}{32} -\frac{\varepsilon}{32} -\frac{13}{192}$
\\
\hline ${\mathcal{S}}_{pcb}\cap${\bf{C}} & $0$ & $0$ &
$-\frac{\omega}{24} +\frac{\varepsilon}{64} +\frac{17}{192}$ &
$-\frac{\omega}{16} +\frac{7 \varepsilon}{192} +\frac{23}{192}$ &
$-\frac{\omega}{32} +\frac{5 \varepsilon}{256} +\frac{29}{768}$
\\
\hline ${\mathcal{S}}_{v2}\cap${\bf{C}} & $0$ & $0$ &
$-\frac{\omega}{24} -\frac{\varepsilon}{32} +\frac{13}{96}$ &
$-\frac{\omega}{16} -\frac{\varepsilon}{16} +\frac{7}{32}$ &
$-\frac{\omega}{32} -\frac{\varepsilon}{32} +\frac{17}{192}$
\\
\hline ${\mathcal{S}}_{d1}\cap${\bf{C}} & $0$ & $0$ &
$\frac{\omega}{192} +\frac{\varepsilon}{64} -\frac{47}{384}$ &
$\frac{7 \omega}{192} +\frac{7 \varepsilon}{192} -\frac{125}{384}$ &
$\frac{5 \omega}{256} +\frac{5 \varepsilon}{256} -\frac{293}{1536}$
\\
\hline
\end{tabular}
\end{center}
\end{table}

\begin{table}[t]
\scriptsize
\begin{center}
\caption{The case of isotropic n.n. hopping only. The set of
zero-potential coefficients $\{\alpha_{i}\}$ in the case of fermions
and for $\varepsilon=0$. In the first column the sets of the
$\alpha$-partition are specified. For more comments see the text in
Appendix C.} \label{tb1-3}
\renewcommand\baselinestretch{1,5}\small\scriptsize
\begin{tabular}{|l|c|c|c|c|c|}
\hline & $\alpha_{1}$ & $\alpha_{2}$ & $\alpha_{3}$ & $\alpha_{4}$ &
$\alpha_{5}$
\\
\hline ${\mathcal{S}}_{-}\cap${\bf{D}} & $-\frac{201 \delta}{3200}$
& $-\frac{\delta}{8}$ & $-\frac{\omega}{96} +\frac{\delta}{6400}
+\frac{3}{64}$ & $0$ & $-\frac{1}{48}$
\\
\hline ${\mathcal{S}}_{-}\cap${\bf{E}} & $\frac{\omega}{32}
-\frac{\delta}{16} +\frac{3}{64}$ & $-\frac{\delta}{8}$ &
$-\frac{\omega}{96} +\frac{3}{64}$ & $0$ & $-\frac{1}{48}$
\\
\hline ${\mathcal{S}}_{-}\cap${\bf{F}} & $\frac{\omega}{48}
-\frac{\delta}{16} +\frac{5}{96}$ & $-\frac{\delta}{8}$ &
$\frac{1}{24}$ & $0$ & $-\frac{1}{48}$
\\
\hline ${\mathcal{S}}_{1}\cap${\bf{G}} & $\frac{11
\omega}{960}-\frac{9 \delta}{160}+ \frac{121}{1920}$ &
$-\frac{\delta}{8}$ & $-\frac{7 \omega}{5760}+ \frac{\delta}{320}+
\frac{403}{11520}$ &
$-\frac{\omega}{180}-\frac{\delta}{80}-\frac{11}{360}$ &
$-\frac{\omega}{144}+\frac{1}{288}$
\\
\hline ${\mathcal{S}}_{1}\cap${\bf{H}} & $-\frac{3
\omega}{160}-\frac{13 \delta}{160}+ \frac{11}{160}$ &
$\frac{\omega}{40}-\frac{\delta}{10}+\frac{1}{80}$ & $-\frac{13
\omega}{960}-\frac{\delta}{320}+\frac{17}{320}$ &
$-\frac{\omega}{40}-\frac{\delta}{40}-\frac{1}{80}$ &
$-\frac{1}{48}$
\\
\hline ${\mathcal{S}}_{2}\cap${\bf{I}} & $-\frac{9
\delta}{128}+\frac{21}{256}$ & $-\frac{5 \delta}{64}+\frac{9}{128}$
& $-\frac{\omega}{96}+\frac{\delta}{128}+\frac{19}{256}$ & $-\frac{3
\delta}{128}-\frac{17}{256}$ & $\frac{\omega}{512}-\frac{13}{3072}$
\\
\hline ${\mathcal{S}}_{2}\cap${\bf{J}} & $\frac{\omega}{128}
-\frac{7 \delta}{128}+\frac{3}{64}$ & $-\frac{\omega}{64}-\frac{9
\delta}{64}+\frac{3}{32}$ & $-\frac{\omega}{96}+\frac{1}{16}$ &
$\frac{\omega}{128}+\frac{\delta}{128}-\frac{5}{64}$ &
$\frac{\omega}{128}+\frac{\delta}{128}-\frac{7}{192}$
\\
\hline ${\mathcal{S}}_{3}$ & $-\frac{7 \omega}{384} -\frac{17
\delta}{192}+ \frac{49}{256}$ & $-\frac{\delta}{8}$ & $-\frac{29
\omega}{2304}- \frac{7 \delta}{1152}+\frac{107}{1536}$ &
$\frac{\omega}{144}+ \frac{\delta}{144}-\frac{7}{96}$ &
$-\frac{\omega}{144}-\frac{\delta}{144}+\frac{5}{96}$
\\
\hline ${\mathcal{S}}_{cb}$ & $-\frac{359 \delta}{5728}$ &
$-\frac{\delta}{8}$ & $-\frac{173 \omega}{17184}+ \frac{3
\delta}{11456}+ \frac{495}{11456}$ & $0$ & $-\frac{1}{48}$
\\
\hline ${\mathcal{S}}_{dd}\cap${\bf{A}} & $-\frac{\delta}{8}$ &
$-\frac{\delta}{8}$ & $-\frac{\omega}{96}+ \frac{\delta}{96}+
\frac{3}{64}$ & $\frac{\delta}{48}$ &
$-\frac{\delta}{48}-\frac{1}{48}$
\\
\hline ${\mathcal{S}}_{dd}\cap${\bf{B}} & $-\frac{\delta}{8}$ &
$-\frac{\delta}{8}$ & $\frac{\delta}{96}+ \frac{1}{24}$ &
$\frac{\delta}{48}$ & $-\frac{\delta}{48}-\frac{1}{48}$
\\
\hline ${\mathcal{S}}_{dd}\cap${\bf{C}} & $-\frac{\delta}{8}$ &
$-\frac{\delta}{8}$ & $- \frac{\omega}{96}- \frac{\delta}{96}+
\frac{3}{64}$ & $\frac{\delta}{48}$ &
$-\frac{\delta}{48}-\frac{1}{48}$
\\
\hline
\end{tabular}
\end{center}
\end{table}
\begin{table}[t]
\scriptsize
\begin{center}
\caption{The case of isotropic n.n. hopping only. The set of
zero-potential coefficients $\{\alpha_{i}\}$ in the case of hardcore
bosons and for $\varepsilon=0$. In the first column the sets of the
$\alpha$-partition are specified. The cases, where the set
${\mathcal{S}}_{D|T}$ is a proper subset of ${\mathcal{S}}_{TD}(p)$
are marked by the asterisk. For more comments see the text in
Appendix C.} \label{tb1-4}
\renewcommand\baselinestretch{1,5}\small\scriptsize
\begin{tabular}{|l|c|c|c|c|c|}
\hline & $\alpha_{1}$ & $\alpha_{2}$ & $\alpha_{3}$ & $\alpha_{4}$ &
$\alpha_{5}$
\\
\hline ${\mathcal{S}}_{-}\cap${\bf{I}} & $-\frac{\delta}{16}$ &
$-\frac{\delta}{8}$ & $-\frac{\omega}{96}-\frac{1}{192}$ & $0$ &
$-\frac{1}{48}$
\\
\hline ${\mathcal{S}}_{-}\cap${\bf{J}} & $\frac{\omega}{32}
-\frac{\delta}{16} +\frac{9}{64}$ & $-\frac{\delta}{8}$ & $\frac{7
\omega}{192} +\frac{3 \delta}{32} +\frac{91}{384}$ & $-\frac{3
\omega}{64} -\frac{3 \delta}{32} -\frac{27}{128}$ & $-\frac{1}{48}$
\\
\hline ${\mathcal{S}}_{-}\cap${\bf{K}} & $-\frac{3 \omega}{128}
-\frac{\delta}{16} -\frac{7}{256}$ & $\frac{3 \omega}{64}
-\frac{\delta}{8} +\frac{23}{128}$ & $\frac{\omega}{48}
+\frac{\delta}{16} +\frac{19}{96}$ & $-\frac{\omega}{32} -\frac{3
\delta}{64} -\frac{21}{128}$ & $-\frac{15 \omega}{512} -\frac{7
\delta}{256} -\frac{331}{3072}$
\\
\hline ${\mathcal{S}}_{-}\cap${\bf{L}} & $\frac{\omega}{16}
-\frac{\delta}{16} -\frac{5}{32}$ & $-\frac{\omega}{8}
-\frac{\delta}{8} +\frac{7}{16}$ & $\frac{11 \omega}{384} +\frac{3
\delta}{32} +\frac{203}{768}$ & $\frac{21 \omega}{128} +\frac{5
\delta}{32} +\frac{13}{256}$ & $-\frac{41 \omega}{512} -\frac{19
\delta}{128} -\frac{1027}{3072}$
\\
\hline ${\mathcal{S}}_{-}\cap${\bf{M}} & $-\frac{\delta}{16}
+\frac{1}{16}$ & $-\frac{\omega}{32} -\frac{5 \delta}{32}
-\frac{1}{64}$ & $-\frac{23 \omega}{768} -\frac{5 \delta}{256}
+\frac{25}{1536}$ & $\frac{7 \omega}{128} +\frac{7 \delta}{128}
+\frac{7}{256}$ & $-\frac{\omega}{128} -\frac{\delta}{128}
-\frac{19}{768}$
\\
\hline ${\mathcal{S}}_{1}\cap${\bf{S}} & $-\frac{9 \omega}{160}
-\frac{11 \delta}{40} -\frac{189}{320}$ & $\frac{\omega}{20}
+\frac{\delta}{20} +\frac{21}{40}$ & $-\frac{5 \omega}{192}
-\frac{\delta}{16} -\frac{65}{384}$ & $-\frac{3 \omega}{80}
-\frac{\delta}{10} -\frac{43}{160}$ & $-\frac{\omega}{80} -\frac{3
\delta}{40} -\frac{133}{480}$
\\
\hline ${\mathcal{S}}_{1}\cap${\bf{T}} & $-\frac{\omega}{40}
-\frac{7 \delta}{80} +\frac{1}{20}$ & $\frac{\omega}{40}
-\frac{\delta}{10} +\frac{1}{80}$ & $-\frac{\omega}{60}
-\frac{\delta}{160} +\frac{11}{480}$ & $-\frac{\omega}{40}
-\frac{\delta}{40} -\frac{1}{80}$ & $-\frac{1}{48}$
\\
\hline ${\mathcal{S}}_{3}\cap${\bf{P}} & $-\frac{\omega}{192}
-\frac{\delta}{12} -\frac{5}{384}$ & $-\frac{\delta}{8}$ &
$-\frac{\omega}{288} -\frac{5 \delta}{144} -\frac{83}{576}$ &
$\frac{\omega}{192} -\frac{\delta}{24} -\frac{67}{384}$ &
$-\frac{\omega}{288} -\frac{\delta}{72} -\frac{17}{576}$
\\
\hline ${\mathcal{S}}_{3}\cap${\bf{Q}} & $\frac{25 \omega}{192}
-\frac{\delta}{6} -\frac{475}{384}$ & $-\frac{\omega}{8}
-\frac{\delta}{16} +\frac{35}{32}$ & $\frac{3 \omega}{128} -\frac{7
\delta}{192} -\frac{269}{768}$ & $\frac{\omega}{16}
-\frac{\delta}{32} -\frac{37}{64}$ & $\frac{5 \omega}{96}
-\frac{\delta}{24} -\frac{33}{64}$
\\
\hline ${\mathcal{S}}_{3}\cap${\bf{R}} & $-\frac{\delta}{16}$ &
$-\frac{\omega}{72} -\frac{5 \delta}{36} +\frac{13}{144}$ &
$-\frac{\omega}{144} +\frac{\delta}{288} +\frac{1}{288}$ &
$\frac{\omega}{72} +\frac{\delta}{72} -\frac{13}{144}$ &
$-\frac{1}{48}$
\\
\hline ${\mathcal{S}}_{cb}$ & $-\frac{\delta}{16}$ &
$-\frac{\delta}{8}$ & $-\frac{\omega}{96} +\frac{5}{192}$ & $0$ &
$-\frac{1}{48}$
\\
\hline ${\mathcal{S}}_{v1}$ & $-\frac{3 \delta}{16}$ &
$-\frac{\delta}{8}$ & $\frac{\omega}{192} +\frac{37}{384}$ &
$-\frac{\omega}{32} -\frac{\delta}{16} -\frac{9}{64}$ &
$-\frac{\omega}{64} -\frac{\delta}{8} -\frac{35}{384}$
\\
\hline ${\mathcal{S}}_{v2}\cap${\bf{A}} & $-\frac{\delta}{32}$ &
$-\frac{3 \delta}{16}$ & $-\frac{\omega}{32} -\frac{\delta}{96}
+\frac{5}{64}$ & $-\frac{1}{8}$ & $-\frac{1}{12}$
\\
\hline ${\mathcal{S}}_{v2}\cap${\bf{B}} $^{*}$ & $ \frac{\omega}{32}
-\frac{\delta}{16} -\frac{11}{64}$ & $-\frac{\omega}{16}
-\frac{\delta}{8} +\frac{11}{32}$ & $-\frac{\omega}{24}
+\frac{13}{96}$ & $-\frac{1}{8}$ & $-\frac{1}{12}$
\\
\hline ${\mathcal{S}}_{v3}$ & $-\frac{\omega}{192}
-\frac{\delta}{24} -\frac{1}{384}$ & $-\frac{\omega}{32}
-\frac{\delta}{4} -\frac{1}{64}$ & $-\frac{7 \omega}{192}
-\frac{\delta}{48} +\frac{29}{384}$ & $-\frac{\omega}{96}
-\frac{\delta}{48} -\frac{25}{192}$ & $-\frac{1}{12}$
\\
\hline ${\mathcal{S}}_{d1}\cap${\bf{C}} & $-\frac{\delta}{16}$ &
$-\frac{\delta}{8}$ & $\frac{\omega}{192} -\frac{47}{384}$ &
$\frac{3 \omega}{64} -\frac{49}{128}$ & $\frac{\omega}{32}
-\frac{49}{192}$
\\
\hline ${\mathcal{S}}_{d1}\cap${\bf{D}} & $-\frac{\delta}{16}$ &
$-\frac{\delta}{8}$ & $\frac{\omega}{192} -\frac{47}{384}$ &
$\frac{\omega}{64} -\frac{19}{128}$ & $-\frac{1}{48}$
\\
\hline ${\mathcal{S}}_{d2}\cap${\bf{N}} & $\frac{\omega}{64}
-\frac{\delta}{8} -\frac{31}{128}$ & $-\frac{5 \omega}{64}
-\frac{\delta}{16} +\frac{67}{128}$ & $-\frac{\omega}{24}
+\frac{13}{96}$ & $-\frac{\delta}{16} -\frac{1}{4}$ &
$\frac{\omega}{128} -\frac{\delta}{32} -\frac{133}{768}$
\\
\hline ${\mathcal{S}}_{d2}\cap${\bf{O}} & $-\frac{3 \omega}{32}
-\frac{\delta}{8} +\frac{9}{64}$ & $\frac{5 \omega}{64}
-\frac{\delta}{16} -\frac{3}{128}$ & $-\frac{\omega}{24}
+\frac{13}{96}$ & $-\frac{3 \omega}{64} -\frac{\delta}{16}
-\frac{11}{128}$ & $-\frac{3 \omega}{128} -\frac{\delta}{32}
-\frac{49}{768}$
\\
\hline ${\mathcal{S}}_{d3}\cap${\bf{E}} $^{*}$ & $\frac{\omega}{32}
-\frac{\delta}{16} -\frac{11}{64}$ & $-\frac{\omega}{20} -\frac{3
\delta}{20} +\frac{1}{4}$ & $-\frac{\omega}{60} -\frac{\delta}{20}
-\frac{5}{96}$ & $\frac{\omega}{32} -\frac{\delta}{16}
-\frac{23}{64}$ & $\frac{3 \omega}{160} -\frac{3 \delta}{80}
-\frac{43}{192}$
\\
\hline ${\mathcal{S}}_{d3}\cap${\bf{F}} $^{*}$ &
$-\frac{\delta}{16}$ & $-\frac{3 \omega}{160} -\frac{3 \delta}{20}
+\frac{5}{64}$ & $-\frac{31 \omega}{960} -\frac{\delta}{20}
+\frac{13}{384}$ & $-\frac{\delta}{16} -\frac{3}{16}$ &
$\frac{\omega}{320} -\frac{3 \delta}{80} -\frac{53}{384}$

\\
\hline ${\mathcal{S}}_{d3}\cap${\bf{G}} $^{*}$ & $\frac{3
\omega}{64} +\frac{\delta}{16} +\frac{3}{128}$ & $-\frac{\omega}{16}
-\frac{\delta}{4} +\frac{3}{32}$ & $-\frac{\omega}{240}
+\frac{\delta}{20} +\frac{5}{48}$ & $\frac{3 \omega}{64}
+\frac{\delta}{16} -\frac{21}{128}$ & $\frac{9 \omega}{320} +\frac{3
\delta}{80} -\frac{41}{384}$
\\
\hline ${\mathcal{S}}_{d3}\cap${\bf{H}} $^{*}$ & $\frac{\omega}{64}
+\frac{\delta}{16} +\frac{25}{128}$ & $-\frac{\omega}{32}
-\frac{\delta}{4} -\frac{5}{64}$ & $-\frac{19 \omega}{960}
+\frac{\delta}{20} +\frac{73}{384}$ & $\frac{\omega}{64}
+\frac{\delta}{16} +\frac{1}{128}$ & $\frac{\omega}{80} +\frac{3
\delta}{80} -\frac{1}{48}$
\\
\hline
\end{tabular}
\end{center}
\end{table}
\clearpage

\renewcommand\baselinestretch{1}\small\normalsize
Below, in a series of tables, we provide the coefficients
$\{\alpha_i \}$, $i=1, \ldots, 9$, of the zero-potentials for the
phase diagrams presented in Section 4. The coefficients that are
missing in a table are equal to zero. As in Appendix D, the symbol
${\mathcal{S}}_{1} | {\mathcal{S}}_{2}$ denotes the boundary between
the phases ${\mathcal{S}}_{1}$ and ${\mathcal{S}}_{2}$, etc.
\begin{table}[h]
\begin{center}
\small \caption{Zero-potentials coefficients for the phase diagram
of fermions, shown in Fig.~\ref{ani-sppd}, for $\beta_{2}=0$.}
\label{E-tb8}
\renewcommand\baselinestretch{1,4}\small\normalsize
\begin{tabular}{|l|c|c|c|c|c|}
\hline $\{ \alpha_{i} \}$ & ${\mathcal{S}}_{seg} |
{\mathcal{S}}_{cb}$ & ${\mathcal{S}}_{seg} | {\mathcal{S}}_{v1}$ &
${\mathcal{S}}_{cb} | {\mathcal{S}}_{d1}$ & ${\mathcal{S}}_{v2} |
{\mathcal{S}}_{v1}$ & ${\mathcal{S}}_{v2} | {\mathcal{S}}_{d1}$
\\
\hline $\alpha_{1}$ & \multicolumn{5}{|c|}{$0$}
\\
\hline $\alpha_{2}$ & \multicolumn{5}{|c|}{$0$}
\\
\hline $\alpha_{3}$ & \multicolumn{5}{|c|}{$0$}
\\
\hline $\alpha_{4}$ &
\multicolumn{3}{|c|}{$-\frac{\omega}{96}+\frac{3}{64}$} &
$-\frac{\omega}{96}+\frac{7}{64}$ &
$-\frac{\omega}{96}-\frac{1}{64}$
\\
\hline $\alpha_{5}$ &
\multicolumn{3}{|c|}{$-\frac{\omega}{96}+\frac{3}{64}$} &
$-\frac{\omega}{96}+\frac{7}{64}$ &
$-\frac{\omega}{96}-\frac{1}{64}$
\\
\hline $\alpha_{6}$ & \multicolumn{3}{|c|}{$0$} &
\multicolumn{2}{|c|}{$-\frac{1}{8}$}
\\
\hline $\alpha_{7}$ & \multicolumn{3}{|c|}{$-\frac{1}{48}$} &
\multicolumn{2}{|c|}{$-\frac{1}{12}$}
\\
\hline $\alpha_{8}$ & \multicolumn{3}{|c|}{$-\frac{1}{48}$} &
\multicolumn{2}{|c|}{$-\frac{1}{12}$}
\\
\hline
\end{tabular}
\end{center}
\end{table}
\begin{table}[h]
\begin{center}
\small \caption{Zero-potentials coefficients for the phase diagram
of fermions, shown in Fig.~\ref{ani-sppd}, for $\beta_{2}=7/2$.}
\label{E-tb9}
\renewcommand\baselinestretch{1,4}\small\normalsize
\begin{tabular}{|l|c|c|c|c|c|}
\hline $\{ \alpha_{i} \}$ & ${\mathcal{S}}_{seg} |
{\mathcal{S}}_{cb}$ & ${\mathcal{S}}_{seg} | {\mathcal{S}}^{v}_{v1}$
& ${\mathcal{S}}_{cb} | {\mathcal{S}}^{v}_{d1}$ &
${\mathcal{S}}^{v}_{v2} | {\mathcal{S}}^{v}_{v1}$ &
${\mathcal{S}}^{v}_{v2} | {\mathcal{S}}^{v}_{d1}$
\\
\hline $\alpha_{1}$ & \multicolumn{5}{|c|}{$0$}
\\
\hline $\alpha_{2}$ & \multicolumn{5}{|c|}{$0$}
\\
\hline $\alpha_{3}$ & \multicolumn{5}{|c|}{$0$}
\\
\hline $\alpha_{4}$ &
\multicolumn{5}{|c|}{$-\frac{\omega}{96}+\frac{3}{64}$}
\\
\hline $\alpha_{5}$ &
\multicolumn{5}{|c|}{$-\frac{\omega}{96}+\frac{23}{192}$}
\\
\hline $\alpha_{6}$ & \multicolumn{5}{|c|}{$0$}
\\
\hline $\alpha_{7}$ & \multicolumn{5}{|c|}{$-\frac{1}{48}$}
\\
\hline $\alpha_{8}$ & \multicolumn{5}{|c|}{$-\frac{1}{48}$}
\\
\hline
\end{tabular}
\end{center}
\end{table}

\clearpage

\begin{table}[ph]
\begin{center}
\small \caption{Zero-potentials coefficients for the phase diagram
of fermions, shown in Fig.~\ref{ani-sppd}, for $\beta_{2}=7$.}
\label{E-tb10}
\renewcommand\baselinestretch{1,4}\small\normalsize
\begin{tabular}{|l|c|c|c|c|c|}
\hline $\{ \alpha_{i} \}$ & ${\mathcal{S}}_{seg} |
{\mathcal{S}}_{cb}$ & ${\mathcal{S}}_{seg} | {\mathcal{S}}^{v}_{v1}$
& ${\mathcal{S}}_{cb} | {\mathcal{S}}^{v}_{d1}$ &
${\mathcal{S}}^{v}_{v2} | {\mathcal{S}}^{v}_{v1}$ &
${\mathcal{S}}^{v}_{v2} | {\mathcal{S}}^{v}_{d1}$
\\
\hline $\alpha_{1}$ & \multicolumn{5}{|c|}{$0$}
\\
\hline $\alpha_{2}$ & \multicolumn{5}{|c|}{$0$}
\\
\hline $\alpha_{3}$ & \multicolumn{5}{|c|}{$0$}
\\
\hline $\alpha_{4}$ &
\multicolumn{3}{|c|}{$-\frac{\omega}{96}-\frac{1}{64}$} &
\multicolumn{2}{|c|}{$-\frac{\omega}{96}+\frac{3}{64}$}
\\
\hline $\alpha_{5}$ &
\multicolumn{3}{|c|}{$-\frac{\omega}{96}+\frac{49}{192}$} &
\multicolumn{2}{|c|}{$-\frac{\omega}{96}+\frac{37}{192}$}
\\
\hline $\alpha_{6}$ & \multicolumn{3}{|c|}{$\frac{1}{8}$} &
\multicolumn{2}{|c|}{$0$}
\\
\hline $\alpha_{7}$ & \multicolumn{3}{|c|}{$-\frac{1}{12}$} &
\multicolumn{2}{|c|}{$-\frac{1}{48}$}
\\
\hline $\alpha_{8}$ & \multicolumn{3}{|c|}{$-\frac{1}{12}$} &
\multicolumn{2}{|c|}{$-\frac{1}{48}$}
\\
\hline
\end{tabular}
\end{center}
\end{table}

\begin{table}[ph]
\begin{center}
\small \caption{Zero-potentials coefficients for the phase diagram
of fermions, shown in Fig.~\ref{ani-sppd}, for $\beta_{2}=10$.}
\label{E-tb11}
\renewcommand\baselinestretch{1,4}\small\normalsize
\begin{tabular}{|l|c|c|c|c|c|}
\hline $\{ \alpha_{i} \}$ & ${\mathcal{S}}_{seg} |
{\mathcal{S}}_{cb}$ & ${\mathcal{S}}_{seg} | {\mathcal{S}}^{v}_{v1}$
& ${\mathcal{S}}_{cb} | {\mathcal{S}}^{v}_{d1}$ &
${\mathcal{S}}^{v}_{v2} | {\mathcal{S}}^{v}_{v1}$ &
${\mathcal{S}}^{v}_{v2} | {\mathcal{S}}^{v}_{d1}$
\\
\hline $\alpha_{1}$ & \multicolumn{5}{|c|}{$0$}
\\
\hline $\alpha_{2}$ & \multicolumn{5}{|c|}{$0$}
\\
\hline $\alpha_{3}$ & \multicolumn{5}{|c|}{$0$}
\\
\hline $\alpha_{4}$ & \multicolumn{3}{|c|}{$-\frac{1}{6}$} &
\multicolumn{2}{|c|}{$-\frac{\omega}{96}+\frac{3}{64}$}
\\
\hline $\alpha_{5}$ & \multicolumn{3}{|c|}{$\frac{1}{6}$} &
\multicolumn{2}{|c|}{$-\frac{\omega}{96}+\frac{49}{192}$}
\\
\hline $\alpha_{6}$ & \multicolumn{3}{|c|}{$\frac{1}{8}$} &
\multicolumn{2}{|c|}{$0$}
\\
\hline $\alpha_{7}$ & \multicolumn{3}{|c|}{$-\frac{1}{12}$} &
\multicolumn{2}{|c|}{$-\frac{1}{48}$}
\\
\hline $\alpha_{8}$ & \multicolumn{3}{|c|}{$-\frac{1}{12}$} &
\multicolumn{2}{|c|}{$-\frac{1}{48}$}
\\
\hline
\end{tabular}
\end{center}
\end{table}

\begin{table}[ph]
\begin{center}
\small \caption{Zero-potentials coefficients for the phase diagram
of hardcore bosons, shown in Fig.~\ref{ani-sppd}, for
$\beta_{2}=0$.} \label{E-tb12}
\renewcommand\baselinestretch{1,5}\small\normalsize
\begin{tabular}{|l|c|c|c|c|c|}
\hline $\{ \alpha_{i} \}$ & ${\mathcal{S}}_{seg} |
{\mathcal{S}}_{cb}$ & ${\mathcal{S}}_{seg} | {\mathcal{S}}_{v1}$ &
${\mathcal{S}}_{cb} | {\mathcal{S}}_{d1}$ & ${\mathcal{S}}_{v2} |
{\mathcal{S}}_{v1}$ & ${\mathcal{S}}_{v2} | {\mathcal{S}}_{d1}$
\\
\hline $\alpha_{1}$ & \multicolumn{5}{|c|}{$0$}
\\
\hline $\alpha_{2}$ & \multicolumn{5}{|c|}{$0$}
\\
\hline $\alpha_{3}$ & \multicolumn{5}{|c|}{$0$}
\\
\hline $\alpha_{4}$ &
\multicolumn{3}{|c|}{$-\frac{\omega}{96}+\frac{5}{192}$} &
$-\frac{\omega}{96}+\frac{17}{192}$ &
$-\frac{\omega}{96}-\frac{7}{192}$
\\
\hline $\alpha_{5}$ &
\multicolumn{3}{|c|}{$-\frac{\omega}{96}+\frac{5}{192}$} &
$-\frac{\omega}{96}+\frac{17}{192}$ &
$-\frac{\omega}{96}-\frac{7}{192}$
\\
\hline $\alpha_{6}$ & \multicolumn{3}{|c|}{$0$} &
\multicolumn{2}{|c|}{$-\frac{1}{8}$}
\\
\hline $\alpha_{7}$ & \multicolumn{3}{|c|}{$-\frac{1}{48}$} &
\multicolumn{2}{|c|}{$-\frac{1}{12}$}
\\
\hline $\alpha_{8}$ & \multicolumn{3}{|c|}{$-\frac{1}{48}$} &
\multicolumn{2}{|c|}{$-\frac{1}{12}$}
\\
\hline
\end{tabular}
\end{center}
\end{table}

\begin{table}[ph]
\begin{center}
\small \caption{Zero-potentials coefficients for the phase diagram
of hardcore bosons, shown in Fig.~\ref{ani-sppd}, for
$\beta_{2}=1$.} \label{E-tb13}
\renewcommand\baselinestretch{1,5}\small\normalsize
\begin{tabular}{|l|c|c|c|c|c|}
\hline $\{ \alpha_{i} \}$ & ${\mathcal{S}}_{seg} |
{\mathcal{S}}_{cb}$ & ${\mathcal{S}}_{seg} | {\mathcal{S}}^{v}_{v1}$
& ${\mathcal{S}}_{cb} | {\mathcal{S}}^{v}_{d1}$ &
${\mathcal{S}}^{v}_{v2} | {\mathcal{S}}^{v}_{v1}$ &
${\mathcal{S}}^{v}_{v2} | {\mathcal{S}}^{v}_{d1}$
\\
\hline $\alpha_{1}$ & \multicolumn{5}{|c|}{$0$}
\\
\hline $\alpha_{2}$ & \multicolumn{5}{|c|}{$0$}
\\
\hline $\alpha_{3}$ & \multicolumn{5}{|c|}{$0$}
\\
\hline $\alpha_{4}$ & $-\frac{1}{24}$ &
\multicolumn{4}{|c|}{$-\frac{\omega}{96}+\frac{5}{192}$}
\\
\hline $\alpha_{5}$ & $\frac{1}{24}$ &
\multicolumn{4}{|c|}{$-\frac{\omega}{96}+\frac{3}{64}$}
\\
\hline $\alpha_{6}$ & \multicolumn{5}{|c|}{$0$}
\\
\hline $\alpha_{7}$ & \multicolumn{5}{|c|}{$-\frac{1}{48}$}
\\
\hline $\alpha_{8}$ & \multicolumn{5}{|c|}{$-\frac{1}{48}$}
\\
\hline
\end{tabular}
\end{center}
\end{table}
\clearpage

\begin{table}[ph]
\begin{center}
\small \caption{Zero-potentials coefficients for the phase diagram
of hardcore bosons, shown in Fig.~\ref{ani-sppd}, for
$\beta_{2}=2$.} \label{E-tb14}
\renewcommand\baselinestretch{1,4}\small\normalsize
\begin{tabular}{|l|c|c|c|c|c|}
\hline $\{ \alpha_{i} \}$ & ${\mathcal{S}}_{seg} |
{\mathcal{S}}_{cb}$ & ${\mathcal{S}}_{seg} | {\mathcal{S}}^{v}_{v1}$
& ${\mathcal{S}}_{cb} | {\mathcal{S}}^{v}_{d1}$ &
${\mathcal{S}}^{v}_{v2} | {\mathcal{S}}^{v}_{v1}$ &
${\mathcal{S}}^{v}_{v2} | {\mathcal{S}}^{v}_{d1}$
\\
\hline $\alpha_{1}$ & \multicolumn{5}{|c|}{$0$}
\\
\hline $\alpha_{2}$ & \multicolumn{5}{|c|}{$0$}
\\
\hline $\alpha_{3}$ & \multicolumn{5}{|c|}{$0$}
\\
\hline $\alpha_{4}$ & $-\frac{1}{12}$ &
\multicolumn{2}{|c|}{$-\frac{\omega}{96}-\frac{1}{192}$} &
\multicolumn{2}{|c|}{$-\frac{\omega}{96}+\frac{1}{24}$}
\\
\hline $\alpha_{5}$ & $\frac{1}{12}$ &
\multicolumn{4}{|c|}{$-\frac{\omega}{96}+\frac{19}{192}$}
\\
\hline $\alpha_{6}$ & $\frac{1}{8}$ &
\multicolumn{2}{|c|}{$\frac{1}{32}$} &
\multicolumn{2}{|c|}{$-\frac{1}{32}$}
\\
\hline $\alpha_{7}$ & $-\frac{1}{12}$ &
\multicolumn{2}{|c|}{$-\frac{1}{48}$} &
\multicolumn{2}{|c|}{$-\frac{1}{192}$}
\\
\hline $\alpha_{8}$ & $-\frac{1}{12}$ &
\multicolumn{2}{|c|}{$-\frac{1}{48}$} &
\multicolumn{2}{|c|}{$-\frac{5}{96}$}
\\
\hline
\end{tabular}
\end{center}
\end{table}

\begin{table}[ph]
\begin{center}
\small \caption{Zero-potentials coefficients for the phase diagram
of hardcore bosons, shown in Fig.~\ref{ani-sppd}, for
$\beta_{2}=5$.} \label{E-tb15}
\renewcommand\baselinestretch{1,4}\small\normalsize
\begin{tabular}{|l|c|c|c|c|c|}
\hline $\{ \alpha_{i} \}$ & ${\mathcal{S}}_{seg} |
{\mathcal{S}}_{cb}$ & ${\mathcal{S}}_{seg} | {\mathcal{S}}^{v}_{v1}$
& ${\mathcal{S}}_{cb} | {\mathcal{S}}^{v}_{d1}$ &
${\mathcal{S}}^{v}_{v2} | {\mathcal{S}}^{v}_{v1}$ &
${\mathcal{S}}^{v}_{v2} | {\mathcal{S}}^{v}_{d1}$
\\
\hline $\alpha_{1}$ & \multicolumn{5}{|c|}{$0$}
\\
\hline $\alpha_{2}$ & \multicolumn{5}{|c|}{$0$}
\\
\hline $\alpha_{3}$ & \multicolumn{5}{|c|}{$0$}
\\
\hline $\alpha_{4}$ & \multicolumn{3}{|c|}{$-\frac{11}{96}$} &
\multicolumn{2}{|c|}{$-\frac{\omega}{96}+\frac{1}{24}$}
\\
\hline $\alpha_{5}$ & \multicolumn{3}{|c|}{$\frac{11}{96}$} &
\multicolumn{2}{|c|}{$-\frac{\omega}{96}+\frac{19}{192}$}
\\
\hline $\alpha_{6}$ & \multicolumn{3}{|c|}{$\frac{1}{8}$} &
\multicolumn{2}{|c|}{$-\frac{1}{32}$}
\\
\hline $\alpha_{7}$ & \multicolumn{3}{|c|}{$-\frac{1}{12}$} &
\multicolumn{2}{|c|}{$-\frac{1}{192}$}
\\
\hline $\alpha_{8}$ & \multicolumn{3}{|c|}{$-\frac{1}{12}$} &
\multicolumn{2}{|c|}{$-\frac{5}{96}$}
\\
\hline
\end{tabular}
\end{center}
\end{table}

\begin{table}[ph]
\begin{center}
\small \caption{Zero-potentials coefficients for the phase diagram
shown in Fig.~\ref{ani-epspdf}, for $\beta_{2}=0$.} \label{E-tb16}
\renewcommand\baselinestretch{1,5}\small\normalsize
\begin{tabular}{|l|c|c|c|c|c|c|}
\hline $\{ \alpha_{i} \}$ & ${\mathcal{S}}_{+} | {\mathcal{S}}_{-}$
& ${\mathcal{S}}_{-} | {\mathcal{S}}_{1}$ & ${\mathcal{S}}_{1} |
{\mathcal{S}}_{2}$ & ${\mathcal{S}}_{2} | {\mathcal{S}}_{3}$ &
${\mathcal{S}}_{3} | {\mathcal{S}}_{cb}$ & $\delta=0$,
$\omega\geqslant\frac{21}{2}$
\\
\hline $\alpha_{1}$ & $0$ & $\frac{\omega}{16}+\frac{3}{32}$ &
$\frac{\omega}{16}-\frac{3}{32}$ & $\frac{\omega}{16}-\frac{9}{32}$
& $\frac{\omega}{16}-\frac{21}{32}$ & $0$
\\
\hline $\alpha_{2}$ & $0$ & $\frac{\omega}{8}+\frac{1}{16}$ &
$\frac{\omega}{8}-\frac{3}{16}$ & $\frac{\omega}{8}-\frac{3}{4}$ &
$\frac{\omega}{8}-\frac{21}{16}$ & $0$
\\
\hline $\alpha_{3}$ & $0$ & $\frac{\omega}{8}+\frac{1}{16}$ &
$\frac{\omega}{8}-\frac{3}{16}$ & $\frac{\omega}{8}-\frac{3}{4}$ &
$\frac{\omega}{8}-\frac{21}{16}$ & $0$
\\
\hline $\alpha_{4}$ &
\multicolumn{3}{|c|}{$-\frac{\omega}{96}+\frac{3}{64}$} &
$-\frac{\omega}{96}+\frac{1}{16}$ &
\multicolumn{2}{|c|}{$-\frac{\omega}{96}+\frac{3}{64}$}
\\
\hline $\alpha_{5}$ &
\multicolumn{3}{|c|}{$-\frac{\omega}{96}+\frac{3}{64}$} &
$-\frac{\omega}{96}+\frac{1}{16}$ &
\multicolumn{2}{|c|}{$-\frac{\omega}{96}+\frac{3}{64}$}
\\
\hline $\alpha_{6}$ & \multicolumn{2}{|c|}{$0$} & $-\frac{1}{16}$ &
$-\frac{1}{32}$ & \multicolumn{2}{|c|}{$0$}
\\
\hline $\alpha_{7}$ & \multicolumn{3}{|c|}{$-\frac{1}{48}$} &
$\frac{1}{96}$ & \multicolumn{2}{|c|}{$-\frac{1}{48}$}
\\
\hline $\alpha_{8}$ & \multicolumn{3}{|c|}{$-\frac{1}{48}$} &
$\frac{1}{96}$ & \multicolumn{2}{|c|}{$-\frac{1}{48}$}
\\
\hline
\end{tabular}
\end{center}
\end{table}

\begin{table}[h]
\begin{center}
\small \caption{Zero-potentials coefficients for the phase diagram
shown in Fig.~\ref{ani-epspdf}, for $\beta_{2}=1$.} \label{E-tb17}
\renewcommand\baselinestretch{1,5}\small\normalsize
\begin{tabular}{|l|c|c|c|c|c|c|}
\hline $\{ \alpha_{i} \}$ & ${\mathcal{S}}_{+} | {\mathcal{S}}_{-}$
& ${\mathcal{S}}_{-} | {\mathcal{S}}_{1}$ & ${\mathcal{S}}_{1} |
{\mathcal{S}}_{2}$ & ${\mathcal{S}}_{2} | {\mathcal{S}}_{3}$ &
${\mathcal{S}}_{3} | {\mathcal{S}}_{cb}$ & $\delta=0$,
$\omega\geqslant\frac{23}{2}$
\\
\hline $\alpha_{1}$ & $0$ & $\frac{\omega}{16}+\frac{1}{32}$ &
$\frac{\omega}{16}-\frac{3}{32}$ & $\frac{\omega}{16}-\frac{11}{32}$
& $\frac{\omega}{16}-\frac{23}{32}$ & $0$
\\
\hline $\alpha_{2}$ & $0$ & $\frac{\omega}{8}-\frac{1}{16}$ &
$\frac{\omega}{8}-\frac{11}{32}$ & $\frac{\omega}{8}-\frac{41}{48}$
& $\frac{\omega}{8}-\frac{23}{16}$ & $0$
\\
\hline $\alpha_{3}$ & $0$ & $\frac{\omega}{8}-\frac{1}{16}$ &
$\frac{\omega}{8}-\frac{13}{32}$ & $\frac{\omega}{8}-\frac{43}{48}$
& $\frac{\omega}{8}-\frac{23}{16}$ & $0$
\\
\hline $\alpha_{4}$ &
\multicolumn{2}{|c|}{$-\frac{\omega}{96}+\frac{3}{64}$} &
$-\frac{\omega}{96}+\frac{9}{128}$ &
$-\frac{\omega}{96}+\frac{13}{192}$ &
\multicolumn{2}{|c|}{$-\frac{\omega}{96}+\frac{3}{64}$}
\\
\hline $\alpha_{5}$ &
\multicolumn{2}{|c|}{$-\frac{\omega}{96}+\frac{13}{192}$} &
$-\frac{\omega}{96}+\frac{29}{384}$ &
$-\frac{\omega}{96}+\frac{5}{64}$ &
\multicolumn{2}{|c|}{$-\frac{\omega}{96}+\frac{13}{192}$}
\\
\hline $\alpha_{6}$ & \multicolumn{2}{|c|}{$0$} &
\multicolumn{2}{|c|}{$-\frac{1}{32}$} & \multicolumn{2}{|c|}{$0$}
\\
\hline $\alpha_{7}$ & \multicolumn{2}{|c|}{$-\frac{1}{48}$} &
$\frac{1}{96}$ & $-\frac{1}{192}$ &
\multicolumn{2}{|c|}{$-\frac{1}{48}$}
\\
\hline $\alpha_{8}$ & \multicolumn{2}{|c|}{$-\frac{1}{48}$} &
$\frac{1}{96}$ & $\frac{5}{192}$ &
\multicolumn{2}{|c|}{$-\frac{1}{48}$}
\\
\hline
\end{tabular}
\end{center}
\end{table}

\begin{table}[h]
\begin{center}
\small \caption{Zero-potentials coefficients for the phase diagram
shown in Fig.~\ref{ani-epspdf}, for $\beta_{2}=3/2$.} \label{E-tb18}
\renewcommand\baselinestretch{1,5}\small\footnotesize
\begin{tabular}{|l|c|c|c|c|c|c|c|c|c|}
\hline lines \& points & $\alpha_{1}$ & $\alpha_{2}$ & $\alpha_{3}$
& $\alpha_{4}$ & $\alpha_{5}$ & $\alpha_{6}$ & $\alpha_{7}$ &
$\alpha_{8}$ & $\alpha_{9}$
\\
\hline ${\mathcal{S}}_{+} | {\mathcal{S}}_{-}$ & $0$ & $0$ & $0$ &
$-\frac{\omega}{96}+\frac{3}{64}$ &
$-\frac{\omega}{96}+\frac{5}{64}$ & $\frac{1}{64}$ &
$-\frac{7}{192}$ & $-\frac{1}{48}$ & 0
\\
\hline ${\mathcal{S}}_{-} | {\mathcal{S}}_{1}$ & $\frac{\omega}{16}$
& $\frac{\omega}{8}-\frac{1}{8}$ & $\frac{\omega}{8}-\frac{3}{32}$ &
$-\frac{\omega}{96}+\frac{15}{256}$ &
$-\frac{\omega}{96}+\frac{5}{64}$ & $-\frac{1}{256}$ &
$-\frac{17}{384}$ & $-\frac{1}{192}$ & 0
\\
\hline ${\mathcal{S}}_{1} | {\mathcal{S}}_{2}$ &
$\frac{\omega}{16}-\frac{49}{256}$ &
$\frac{\omega}{8}-\frac{31}{64}$ & $\frac{\omega}{8}-\frac{13}{32}$
& $-\frac{\omega}{96}+\frac{7}{128}$ &
$-\frac{\omega}{96}+\frac{27}{256}$ & $-\frac{7}{256}$ &
$-\frac{23}{384}$ & $-\frac{23}{384}$ & $\frac{19}{256}$
\\
\hline ${\mathcal{S}}_{2} | {\mathcal{S}}_{3}$ &
$\frac{\omega}{16}-\frac{105}{256}$ &
$\frac{\omega}{8}-\frac{115}{128}$ &
$\frac{\omega}{8}-\frac{29}{32}$ &
$-\frac{\omega}{96}+\frac{7}{128}$ &
$-\frac{\omega}{96}+\frac{43}{512}$ & $-\frac{25}{512}$ &
$-\frac{157}{3072}$ &  $\frac{113}{3072}$ & $0$
\\
\hline ${\mathcal{S}}_{3} | {\mathcal{S}}_{cb}$ &
$\frac{\omega}{16}-\frac{3}{4}$ & $\frac{\omega}{8}-\frac{3}{2}$ &
$\frac{\omega}{8}-\frac{3}{2}$ &  $-\frac{\omega}{96}+\frac{3}{64}$
& $-\frac{\omega}{96}+\frac{5}{64}$ & $0$ & $-\frac{1}{48}$ &
$-\frac{1}{48}$ & $0$
\\
\hline $\delta=0$, $\omega\geqslant\frac{25}{2}$ & $0$ & $0$ & $0$ &
$-\frac{\omega}{96}+\frac{3}{64}$ &
$-\frac{\omega}{96}+\frac{5}{64}$ & $\frac{1}{64}$ & $-\frac{1}{48}$
& $-\frac{7}{192}$ & $0$
\\
\hline $\left( \frac{1}{2},0 \right)$ & $0$ & $0$ & $0$ &
$\frac{1}{24}$ & $\frac{7}{96}$ & $0$ & $-\frac{1}{48}$ &
$-\frac{1}{48}$ & $0$
\\
\hline
\end{tabular}
\end{center}
\end{table}

\begin{table}[h]
\begin{center}
\small \caption{Zero-potentials coefficients for the phase diagram
shown in Fig.~\ref{ani-epspdf}, for $\beta_{2}=2$.} \label{E-tb19}
\renewcommand\baselinestretch{1,5}\small\footnotesize
\begin{tabular}{|l|c|c|c|c|c|c|c|c|c|}
\hline lines \& points & $\alpha_{1}$ & $\alpha_{2}$ & $\alpha_{3}$
& $\alpha_{4}$ & $\alpha_{5}$ & $\alpha_{6}$ & $\alpha_{7}$ &
$\alpha_{8}$ & $\alpha_{9}$
\\
\hline ${\mathcal{S}}_{+} | {\mathcal{S}}_{-}$ & $0$ & $0$ & $0$ &
$-\frac{\omega}{96}+\frac{3}{64}$ &
$-\frac{\omega}{96}+\frac{17}{192}$ & $\frac{1}{32}$ &
$-\frac{5}{96}$ & $-\frac{1}{48}$ & 0
\\
\hline ${\mathcal{S}}_{-} | {\mathcal{S}}_{1}$ &
$\frac{\omega}{16}-\frac{1}{32}$ & $\frac{\omega}{8}-\frac{3}{16}$ &
$\frac{\omega}{8}-\frac{3}{16}$ & $-\frac{\omega}{96}+\frac{3}{64}$
& $-\frac{\omega}{96}+\frac{17}{192}$ & $0$ & $-\frac{1}{48}$ &
$-\frac{1}{48}$ & 0
\\
\hline ${\mathcal{S}}_{1} | {\mathcal{S}}_{2}$ &
$\frac{\omega}{16}-\frac{1}{4}$ & $\frac{\omega}{8}-\frac{9}{16}$ &
$\frac{\omega}{8}-\frac{7}{16}$ & $-\frac{\omega}{96}+\frac{3}{64}$
& $-\frac{\omega}{96}+\frac{23}{192}$ & $-\frac{1}{32}$ &
$-\frac{1}{12}$ & $-\frac{1}{12}$ & $\frac{3}{32}$
\\
\hline ${\mathcal{S}}_{2} | {\mathcal{S}}_{3}$ &
$\frac{\omega}{16}-\frac{15}{32}$ & $\frac{\omega}{8}-\frac{15}{16}$
& $\frac{\omega}{8}-\frac{15}{16}$ &
$-\frac{\omega}{96}+\frac{3}{64}$ &
$-\frac{\omega}{96}+\frac{17}{192}$ & $-\frac{1}{16}$ &
$-\frac{1}{12}$ &  $\frac{1}{24}$ & $0$
\\
\hline ${\mathcal{S}}_{3} | {\mathcal{S}}_{cb}$ &
$\frac{\omega}{16}-\frac{25}{32}$ & $\frac{\omega}{8}-\frac{25}{16}$
& $\frac{\omega}{8}-\frac{25}{16}$ &
$-\frac{\omega}{96}+\frac{3}{64}$ &
$-\frac{\omega}{96}+\frac{17}{192}$ & $0$ & $-\frac{1}{48}$ &
$-\frac{1}{48}$ & $0$
\\
\hline $\delta=0$, $\omega\geqslant\frac{27}{2}$ & $0$ & $0$ & $0$ &
$-\frac{\omega}{96}+\frac{3}{64}$ &
$-\frac{\omega}{96}+\frac{17}{192}$ & $\frac{1}{32}$ &
$-\frac{1}{48}$ & $-\frac{5}{96}$ & $0$
\\
\hline $\left( \frac{3}{2},0 \right)$ & $0$ & $0$ & $0$ &
$\frac{1}{32}$ & $\frac{7}{96}$ & $0$ & $-\frac{1}{48}$ &
$-\frac{1}{48}$ & $0$
\\
\hline
\end{tabular}
\end{center}
\end{table}

\begin{table}[h]
\begin{center}
\small \caption{Zero-potentials coefficients for the phase diagram
shown in Fig.~\ref{ani-epspdf}, for $\beta_{2}=5/2$.} \label{E-tb20}
\renewcommand\baselinestretch{1,5}\small\footnotesize
\begin{tabular}{|l|c|c|c|c|c|c|c|c|c|}
\hline lines \& points & $\alpha_{1}$ & $\alpha_{2}$ & $\alpha_{3}$
& $\alpha_{4}$ & $\alpha_{5}$ & $\alpha_{6}$ & $\alpha_{7}$ &
$\alpha_{8}$ & $\alpha_{9}$
\\
\hline ${\mathcal{S}}_{+} | {\mathcal{S}}_{-}$ & $0$ & $0$ & $0$ &
$-\frac{\omega}{96}+\frac{3}{64}$ &
$-\frac{\omega}{96}+\frac{19}{192}$ & $\frac{1}{32}$ &
$-\frac{5}{96}$ & $-\frac{1}{48}$ & 0
\\
\hline ${\mathcal{S}}_{-} | {\mathcal{S}}_{1}$ &
$\frac{\omega}{16}-\frac{1}{16}$ & $\frac{\omega}{8}-\frac{1}{4}$ &
$\frac{\omega}{8}-\frac{1}{4}$ & $-\frac{\omega}{96}+\frac{3}{64}$ &
$-\frac{\omega}{96}+\frac{19}{192}$ & $0$ & $-\frac{1}{48}$ &
$-\frac{1}{48}$ & 0
\\
\hline ${\mathcal{S}}_{1} | {\mathcal{S}}_{2}$ &
$\frac{\omega}{16}-\frac{9}{32}$ & $\frac{\omega}{8}-\frac{5}{8}$ &
$\frac{\omega}{8}-\frac{1}{2}$ & $-\frac{\omega}{96}+\frac{3}{64}$ &
$-\frac{\omega}{96}+\frac{25}{192}$ & $-\frac{1}{32}$ &
$-\frac{1}{12}$ & $-\frac{1}{12}$ & $\frac{3}{32}$
\\
\hline ${\mathcal{S}}_{2} | {\mathcal{S}}_{3}$ &
$\frac{\omega}{16}-\frac{33}{64}$ & $\frac{\omega}{8}-\frac{31}{32}$
& $\frac{\omega}{8}-1$ & $-\frac{\omega}{96}+\frac{3}{64}$ &
$-\frac{\omega}{96}+\frac{35}{384}$ & $-\frac{9}{128}$ &
$-\frac{1}{12}$ &  $\frac{5}{192}$ & $0$
\\
\hline ${\mathcal{S}}_{3} | {\mathcal{S}}_{cb}$ &
$\frac{\omega}{16}-\frac{13}{16}$ & $\frac{\omega}{8}-\frac{13}{8}$
& $\frac{\omega}{8}-\frac{13}{8}$ &
$-\frac{\omega}{96}+\frac{3}{64}$ &
$-\frac{\omega}{96}+\frac{19}{192}$ & $0$ & $-\frac{1}{48}$ &
$-\frac{1}{48}$ & $0$
\\
\hline $\delta=0$, $\omega\geqslant\frac{29}{2}$ & $0$ & $0$ & $0$ &
$-\frac{\omega}{96}+\frac{3}{64}$ &
$-\frac{\omega}{96}+\frac{19}{192}$ & $\frac{1}{32}$ &
$-\frac{1}{48}$ & $-\frac{5}{96}$ & $0$
\\
\hline $\left( \frac{5}{2},0 \right)$ & $0$ & $0$ & $0$ &
$\frac{1}{48}$ & $\frac{7}{96}$ & $0$ & $-\frac{1}{48}$ &
$-\frac{1}{48}$ & $0$
\\
\hline
\end{tabular}
\end{center}
\end{table}

\begin{table}[h]
\begin{center}
\small \caption{Zero-potentials coefficients for the phase diagram
shown in Fig.~\ref{ani-epspdf}, for $\beta_{2}=3$.} \label{E-tb21}
\renewcommand\baselinestretch{1,5}\small\normalsize
\begin{tabular}{|l|c|c|c|c|c|c|c|c|}
\hline lines \& points & $\alpha_{1}$ & $\alpha_{2}$ & $\alpha_{3}$
& $\alpha_{4}$ & $\alpha_{5}$ & $\alpha_{6}$ & $\alpha_{7}$ &
$\alpha_{8}$
\\
\hline ${\mathcal{S}}_{+} | {\mathcal{S}}_{-}$ & $0$ & $0$ & $0$ &
$-\frac{\omega}{96}+\frac{3}{64}$ &
$-\frac{\omega}{96}+\frac{7}{64}$ & $\frac{1}{32}$ & $-\frac{5}{96}$
& $-\frac{1}{48}$
\\
\hline ${\mathcal{S}}_{-} | {\mathcal{S}}_{1}$ &
$\frac{\omega}{16}-\frac{3}{32}$ & $\frac{\omega}{8}-\frac{5}{16}$ &
$\frac{\omega}{8}-\frac{5}{16}$ & $-\frac{\omega}{96}+\frac{3}{64}$
& $-\frac{\omega}{96}+\frac{7}{64}$ & $0$ & $-\frac{1}{48}$ &
$-\frac{1}{48}$
\\
\hline ${\mathcal{S}}_{1} | {\mathcal{S}}_{2}$ &
$\frac{\omega}{16}-\frac{11}{32}$ & $\frac{\omega}{8}-\frac{7}{16}$
& $\frac{\omega}{8}-\frac{11}{16}$ &
$-\frac{\omega}{96}+\frac{3}{64}$ &
$-\frac{\omega}{96}+\frac{3}{64}$ & $-\frac{1}{16}$ &
$-\frac{1}{12}$ & $-\frac{1}{12}$
\\
\hline ${\mathcal{S}}_{2} | {\mathcal{S}}_{3}$ &
$\frac{\omega}{16}-\frac{9}{16}$ & $\frac{\omega}{8}-1$ &
$\frac{\omega}{8}-\frac{17}{16}$ & $-\frac{\omega}{96}+\frac{3}{64}$
& $-\frac{\omega}{96}+\frac{3}{32}$ & $-\frac{5}{64}$ &
$-\frac{1}{12}$ &  $\frac{1}{96}$
\\
\hline ${\mathcal{S}}_{3} | {\mathcal{S}}_{cb}$ &
$\frac{\omega}{16}-\frac{27}{32}$ & $\frac{\omega}{8}-\frac{27}{16}$
& $\frac{\omega}{8}-\frac{27}{16}$ &
$-\frac{\omega}{96}+\frac{3}{64}$ &
$-\frac{\omega}{96}+\frac{7}{64}$ & $0$ & $-\frac{1}{48}$ &
$-\frac{1}{48}$
\\
\hline $\delta=0$, $\omega\geqslant\frac{31}{2}$ & $0$ & $0$ & $0$ &
$-\frac{\omega}{96}+\frac{3}{64}$ &
$-\frac{\omega}{96}+\frac{7}{64}$ & $\frac{1}{32}$ & $-\frac{1}{48}$
& $-\frac{5}{96}$
\\
\hline $\left( \frac{7}{2},0 \right)$ (${\bf{b}}$) & $0$ & $0$ & $0$
& $\frac{1}{96}$ & $\frac{7}{96}$ & $0$ & $-\frac{1}{48}$ &
$-\frac{1}{48}$
\\
\hline
\end{tabular}
\end{center}
\end{table}

\begin{table}[h]
\begin{center}
\small \caption{Zero-potentials coefficients for the phase diagram
shown in Fig.~\ref{ani-epspdf}, for $\beta_{2}=6$.} \label{E-tb22}
\renewcommand\baselinestretch{1,5}\small\normalsize
\begin{tabular}{|l|c|c|c|c|c|c|c|c|}
\hline lines \& points & $\alpha_{1}$ & $\alpha_{2}$ & $\alpha_{3}$
& $\alpha_{4}$ & $\alpha_{5}$ & $\alpha_{6}$ & $\alpha_{7}$ &
$\alpha_{8}$
\\
\hline ${\mathcal{S}}_{+} | {\mathcal{S}}_{-}$ & $0$ & $0$ & $0$ &
$-\frac{\omega}{96}+\frac{3}{64}$ &
$-\frac{\omega}{96}+\frac{11}{64}$ & $\frac{1}{32}$ &
$-\frac{5}{96}$ & $-\frac{1}{48}$
\\
\hline ${\mathcal{S}}_{-} | {\mathcal{S}}_{1}$ &
$\frac{\omega}{16}-\frac{9}{32}$ & $\frac{\omega}{8}-\frac{11}{16}$
& $\frac{\omega}{8}-\frac{11}{16}$ &
$-\frac{\omega}{96}+\frac{3}{64}$ &
$-\frac{\omega}{96}+\frac{11}{64}$ & $0$ & $-\frac{1}{48}$ &
$-\frac{1}{48}$
\\
\hline ${\mathcal{S}}_{1} | {\mathcal{S}}_{2}$ &
$\frac{\omega}{16}-\frac{19}{32}$ & $\frac{\omega}{8}-\frac{11}{16}$
& $\frac{\omega}{8}-\frac{15}{16}$ &
$-\frac{\omega}{96}+\frac{3}{64}$ &
$-\frac{\omega}{96}+\frac{7}{64}$ & $-\frac{1}{8}$ & $-\frac{1}{12}$
& $-\frac{1}{12}$
\\
\hline ${\mathcal{S}}_{2} | {\mathcal{S}}_{3}$ &
$\frac{\omega}{16}-\frac{25}{32}$ & $\frac{\omega}{8}-\frac{21}{16}$
& $\frac{\omega}{8}-\frac{23}{16}$ &
$-\frac{\omega}{96}+\frac{3}{64}$ &
$-\frac{\omega}{96}+\frac{9}{64}$ & $-\frac{3}{32}$ &
$-\frac{1}{12}$ &  $-\frac{1}{48}$
\\
\hline ${\mathcal{S}}_{3} | {\mathcal{S}}_{cb}$ &
$\frac{\omega}{16}-\frac{33}{32}$ & $\frac{\omega}{8}-\frac{33}{16}$
& $\frac{\omega}{8}-\frac{33}{16}$ &
$-\frac{\omega}{96}+\frac{3}{64}$ &
$-\frac{\omega}{96}+\frac{11}{64}$ & $0$ & $-\frac{1}{48}$ &
$-\frac{1}{48}$
\\
\hline $\delta=0$, $\omega\geqslant\frac{43}{2}$ & $0$ & $0$ & $0$ &
$-\frac{\omega}{96}+\frac{3}{64}$ &
$-\frac{\omega}{96}+\frac{11}{64}$ & $\frac{1}{32}$ &
$-\frac{1}{48}$ & $-\frac{5}{96}$
\\
\hline ${\bf{b}}$ & $0$ & $0$ & $0$ & $\frac{1}{96}$ &
$\frac{13}{96}$ & $-\frac{1}{32}$ & $-\frac{5}{96}$ &
$-\frac{1}{48}$
\\
\hline
\end{tabular}
\end{center}
\end{table}

\begin{table}[h]
\begin{center}
\small \caption{Zero-potentials coefficients for the phase diagram
shown in Fig.~\ref{ani-epspdb}, for $\beta_{2}=0$.} \label{E-tb23}
\renewcommand\baselinestretch{1,5}\small\normalsize
\begin{tabular}{|l|c|c|c|c|c|c|c|c|}
\hline lines \& points & $\alpha_{1}$ & $\alpha_{2}$ & $\alpha_{3}$
& $\alpha_{4}$ & $\alpha_{5}$ & $\alpha_{6}$ & $\alpha_{7}$ &
$\alpha_{8}$
\\
\hline ${\mathcal{S}}_{+} | {\mathcal{S}}_{-}$ & $0$ & $0$ & $0$ &
$-\frac{\omega}{96}+\frac{5}{192}$ &
$-\frac{\omega}{96}+\frac{5}{192}$ & $0$ & $-\frac{1}{48}$ &
$-\frac{1}{48}$
\\
\hline ${\mathcal{S}}_{-} | {\mathcal{S}}_{1}$ &
$\frac{\omega}{16}+\frac{3}{32}$ & $\frac{\omega}{8}+\frac{1}{16}$ &
$\frac{\omega}{8}+\frac{1}{16}$ & $-\frac{\omega}{96}+\frac{5}{192}$
& $-\frac{\omega}{96}+\frac{5}{192}$ & $0$ & $-\frac{1}{48}$ &
$-\frac{1}{48}$
\\
\hline ${\mathcal{S}}_{1} | {\mathcal{S}}_{3}$ &
$\frac{\omega}{16}-\frac{9}{32}$ & $\frac{\omega}{8}-\frac{1}{16}$ &
$\frac{\omega}{8}-\frac{1}{16}$ & $-\frac{\omega}{96}-\frac{7}{192}$
& $-\frac{\omega}{96}-\frac{7}{192}$ & $-\frac{1}{8}$ &
$-\frac{1}{12}$ &  $-\frac{1}{12}$
\\
\hline ${\mathcal{S}}_{3} | {\mathcal{S}}_{cb}$ &
$\frac{\omega}{16}-\frac{13}{32}$ & $\frac{\omega}{8}-\frac{13}{16}$
& $\frac{\omega}{8}-\frac{13}{16}$ &
$-\frac{\omega}{96}+\frac{5}{192}$ &
$-\frac{\omega}{96}+\frac{5}{192}$ & $0$ & $-\frac{1}{48}$ &
$-\frac{1}{48}$
\\
\hline $\delta=0$, $\omega\geqslant\frac{19}{2}$ & $0$ & $0$ & $0$ &
$-\frac{\omega}{96}+\frac{5}{192}$ &
$-\frac{\omega}{96}+\frac{5}{192}$ & $0$ & $-\frac{1}{48}$ &
$-\frac{1}{48}$
\\
\hline ${\bf{b}}$ & $0$ & $0$ & $0$ & $\frac{3}{32}$ &
$\frac{3}{32}$ & $-\frac{1}{8}$ & $-\frac{1}{12}$ & $-\frac{1}{12}$
\\
\hline $\left( \frac{11}{2},0 \right)$ & $0$ & $0$ & $0$ &
$-\frac{3}{32}$ & $-\frac{3}{32}$ & $-\frac{1}{8}$ & $-\frac{1}{12}$
& $-\frac{1}{12}$
\\
\hline
\end{tabular}
\end{center}
\end{table}

\begin{table}[h]
\begin{center}
\small \caption{Zero-potentials coefficients for the phase diagram
shown in Fig.~\ref{ani-epspdb}, for $\beta_{2}=3$.} \label{E-tb24}
\renewcommand\baselinestretch{1,5}\small\normalsize
\begin{tabular}{|l|c|c|c|c|c|c|c|c|}
\hline lines \& points & $\alpha_{1}$ & $\alpha_{2}$ & $\alpha_{3}$
& $\alpha_{4}$ & $\alpha_{5}$ & $\alpha_{6}$ & $\alpha_{7}$ &
$\alpha_{8}$
\\
\hline ${\mathcal{S}}_{+} | {\mathcal{S}}_{-}$ & $0$ & $0$ & $0$ &
$-\frac{\omega}{96}+\frac{5}{192}$ &
$-\frac{\omega}{96}+\frac{17}{192}$ & $\frac{1}{32}$ &
$-\frac{5}{96}$ & $-\frac{1}{48}$
\\
\hline ${\mathcal{S}}_{-} | {\mathcal{S}}_{1}$ &
$\frac{\omega}{16}-\frac{3}{32}$ & $\frac{\omega}{8}-\frac{5}{16}$ &
$\frac{\omega}{8}-\frac{5}{16}$ & $-\frac{\omega}{96}+\frac{5}{192}$
& $-\frac{\omega}{96}+\frac{17}{192}$ & $0$ & $-\frac{1}{48}$ &
$-\frac{1}{48}$
\\
\hline ${\mathcal{S}}_{1} | {\mathcal{S}}_{3}$ &
$\frac{\omega}{16}-\frac{15}{32}$ & $\frac{\omega}{8}-\frac{7}{16}$
& $\frac{\omega}{8}-\frac{7}{16}$ &
$-\frac{\omega}{96}-\frac{7}{192}$ &
$-\frac{\omega}{96}+\frac{5}{192}$ & $-\frac{1}{8}$ &
$-\frac{1}{12}$ &  $-\frac{1}{12}$
\\
\hline ${\mathcal{S}}_{3} | {\mathcal{S}}_{cb}$ &
$\frac{\omega}{16}-\frac{19}{32}$ & $\frac{\omega}{8}-\frac{19}{16}$
& $\frac{\omega}{8}-\frac{19}{16}$ &
$-\frac{\omega}{96}+\frac{5}{192}$ &
$-\frac{\omega}{96}+\frac{17}{192}$ & $0$ & $-\frac{1}{48}$ &
$-\frac{1}{48}$
\\
\hline $\delta=0$, $\omega\geqslant\frac{31}{2}$ & $0$ & $0$ & $0$ &
$-\frac{\omega}{96}+\frac{5}{192}$ &
$-\frac{\omega}{96}+\frac{17}{192}$ & $\frac{1}{32}$ &
$-\frac{1}{48}$ & $-\frac{5}{96}$
\\
\hline ${\bf{b}}$ & $0$ & $0$ & $0$ & $\frac{1}{32}$ &
$\frac{3}{32}$ & $-\frac{1}{32}$ & $-\frac{5}{96}$ & $-\frac{1}{48}$
\\
\hline $\left( \frac{23}{2},0 \right)$ & $0$ & $0$ & $0$ &
$-\frac{3}{32}$ & $-\frac{1}{32}$ & $-\frac{1}{32}$ &
$-\frac{1}{48}$ & $-\frac{5}{96}$
\\
\hline
\end{tabular}
\end{center}
\end{table}

\begin{table}[h]
\begin{center}
\small \caption{Zero-potentials coefficients for the phase diagram
shown in Fig.~\ref{ani-p1fmne0}.} \label{E-tb25}
\renewcommand\baselinestretch{1,5}\small\normalsize
\begin{tabular}{|l|c|}
\hline lines & $\alpha_{1}$
\\
\hline ${\mathcal{S}}_{+} | {\mathcal{S}}_{-}$ & $0$
\\
\hline ${\mathcal{S}}_{-} | {\mathcal{S}}_{v3}^{v}$ &
$\frac{\omega}{16}+\frac{1}{32}$
\\
\hline ${\mathcal{S}}_{v3}^{v} | {\mathcal{S}}_{v2}^{v}$ &
$\frac{\omega}{16}-\frac{7}{32}$
\\
\hline $\delta=0$, $\omega\geqslant\frac{7}{2}$ & $0$
\\
\hline
\end{tabular}
\end{center}
\end{table}

\begin{table}[h]
\begin{center}
\small \caption{Zero-potentials coefficients for the phase diagram
shown in  Fig.~\ref{ani-p1hcbe0}.} \label{E-tb26}
\renewcommand\baselinestretch{1,5}\small\normalsize
\begin{tabular}{|l|c|}
\hline lines & $\alpha_{1}$
\\
\hline ${\mathcal{S}}_{+} | {\mathcal{S}}_{-}$ & $0$
\\
\hline ${\mathcal{S}}_{-} | {\mathcal{S}}_{v3}^{v}$ &
$\frac{\omega}{16}+\frac{9}{32}$
\\
\hline ${\mathcal{S}}_{v3}^{v} | {\mathcal{S}}_{v2}^{v}$ &
$\frac{\omega}{16}+\frac{1}{32}$
\\
\hline $\delta=0$, $\omega\geqslant-\frac{1}{2}$ & $0$
\\
\hline
\end{tabular}
\end{center}
\end{table}

\begin{table}[h]
\begin{center}
\small \caption{Zero-potentials coefficients for the phase diagram
shown in Fig.~\ref{ani-p2fmne0}.} \label{E-tb27}
\renewcommand\baselinestretch{1,5}\small\normalsize
\begin{tabular}{|l|c|c|c|c|c|c|c|c|}
\hline lines \& points & $\alpha_{1}$ & $\alpha_{2}$ & $\alpha_{3}$
& $\alpha_{4}$ & $\alpha_{5}$ & $\alpha_{6}$ & $\alpha_{7}$ &
$\alpha_{8}$
\\
\hline ${\mathcal{S}}_{-} | {\mathcal{S}}_{v3}^{v}$ & $0$ & $0$ &
$\frac{\omega}{8}+\frac{1}{16}$ & $-\frac{\omega}{24}+\frac{1}{16}$
& $-\frac{\omega}{96}+\frac{3}{64}$ & $-\frac{1}{32}$ &
$-\frac{1}{48}$ & $-\frac{1}{48}$
\\
\hline ${\mathcal{S}}_{-} | {\mathcal{S}}_{1}$ & $0$ &
$-\frac{1}{8}$ & $\frac{\omega}{4}+\frac{1}{4}$ &
$-\frac{7\omega}{96}-\frac{3}{64}$ &
$-\frac{\omega}{96}+\frac{3}{64}$ & $0$ & $-\frac{1}{48}$ &
$-\frac{1}{48}$
\\
\hline ${\mathcal{S}}_{1} | {\mathcal{S}}_{2}$ &
$\frac{7\omega}{96}-\frac{17}{64}$ &
$\frac{7\omega}{48}-\frac{1}{32}$ &
$\frac{5\omega}{48}-\frac{3}{32}$ & $0$ &
$-\frac{\omega}{96}-\frac{1}{64}$ & $-\frac{1}{8}$ & $-\frac{1}{12}$
& $-\frac{1}{12}$
\\
\hline ${\mathcal{S}}_{2} | {\mathcal{S}}_{3}$ &
$\frac{7\omega}{96}-\frac{29}{64}$ &
$\frac{7\omega}{48}-\frac{21}{32}$ &
$\frac{5\omega}{48}-\frac{19}{32}$ & $0$ &
$-\frac{\omega}{96}+\frac{1}{64}$ & $-\frac{3}{32}$ &
$-\frac{1}{12}$ & $-\frac{1}{48}$
\\
\hline ${\mathcal{S}}_{3} | {\mathcal{S}}_{cb}$ & $0$ & $0$ &
$\frac{\omega}{4}-\frac{21}{8}$ &
$-\frac{7\omega}{96}+\frac{45}{64}$ &
$-\frac{\omega}{96}+\frac{3}{64}$ & $0$ & $-\frac{1}{48}$ &
$-\frac{1}{48}$
\\
\hline ${\mathcal{S}}_{d1}^{v} | {\mathcal{S}}_{cb}$ & $0$ & $0$ &
$-\frac{\delta}{4}$ & $\frac{\delta}{16}-\frac{5}{96}$ &
$-\frac{5}{96}$ & $0$ & $-\frac{1}{48}$ & $-\frac{1}{48}$
\\
\hline ${\mathcal{S}}_{v2}^{v} | {\mathcal{S}}_{d1}^{v}$ & $0$ & $0$
& $-\frac{\delta}{4}$ & $\frac{\delta}{16}-\frac{1}{96}$ &
$-\frac{5}{192}$ & $-\frac{1}{64}$ & $-\frac{1}{48}$ &
$-\frac{1}{48}$
\\
\hline
\end{tabular}
\end{center}
\end{table}

\begin{table}[h]
\begin{center}
\small \caption{Zero-potentials coefficients for the phase diagram
shown in Fig.~\ref{ani-p2hcbe0}.} \label{E-tb28}
\renewcommand\baselinestretch{1,5}\small\normalsize
\begin{tabular}{|l|c|c|c|c|c|c|c|c|}
\hline lines \& points & $\alpha_{1}$ & $\alpha_{2}$ & $\alpha_{3}$
& $\alpha_{4}$ & $\alpha_{5}$ & $\alpha_{6}$ & $\alpha_{7}$ &
$\alpha_{8}$
\\
\hline ${\mathcal{S}}_{-} | {\mathcal{S}}_{v3}^{v}$ & $0$ &
$-\frac{1}{4}$ & $\frac{\omega}{8}+\frac{9}{16}$ &
$-\frac{\omega}{24}-\frac{7}{48}$ &
$-\frac{\omega}{96}+\frac{5}{192}$ & $\frac{1}{32}$ &
$-\frac{1}{48}$ & $-\frac{1}{48}$
\\
\hline ${\mathcal{S}}_{-} | {\mathcal{S}}_{1}$ & $0$ &
$-\frac{1}{8}$ & $\frac{\omega}{4}+\frac{1}{4}$ &
$-\frac{7\omega}{96}-\frac{13}{192}$ &
$-\frac{\omega}{96}+\frac{5}{192}$ & $0$ & $-\frac{1}{48}$ &
$-\frac{1}{48}$
\\
\hline ${\mathcal{S}}_{1} | {\mathcal{S}}_{3}$ & $-\frac{1}{4}$ &
$0$ & $\frac{\omega}{4}-\frac{1}{8}$ &
$-\frac{7\omega}{96}-\frac{1}{192}$ &
$-\frac{\omega}{96}-\frac{7}{192}$ & $-\frac{1}{8}$ &
$-\frac{1}{12}$ & $-\frac{1}{12}$
\\
\hline ${\mathcal{S}}_{3} | {\mathcal{S}}_{cb}$ & $0$ & $0$ &
$\frac{\omega}{4}-\frac{13}{8}$ &
$-\frac{7\omega}{96}+\frac{83}{192}$ &
$-\frac{\omega}{96}+\frac{5}{192}$ & $0$ & $-\frac{1}{48}$ &
$-\frac{1}{48}$
\\
\hline ${\mathcal{S}}_{d1}^{v} | {\mathcal{S}}_{cb}$ & $0$ & $0$ &
$-\frac{\delta}{4}$ & $\frac{\delta}{16}-\frac{7}{96}$ &
$-\frac{7}{96}$ & $0$ & $-\frac{1}{48}$ & $-\frac{1}{48}$
\\
\hline ${\mathcal{S}}_{v2}^{v} | {\mathcal{S}}_{d1}^{v}$ & $0$ & $0$
& $-\frac{\delta}{4}$ & $\frac{\delta}{16}-\frac{1}{32}$ &
$-\frac{3}{64}$ & $-\frac{1}{64}$ & $-\frac{1}{48}$ &
$-\frac{1}{48}$
\\
\hline ${\mathcal{S}}_{v3}^{v} | {\mathcal{S}}_{v2}^{v}$ &
$-\frac{1}{8}$ & $0$ & $\frac{\omega}{8}+\frac{5}{16}$ &
$-\frac{\omega}{24}-\frac{5}{96}$ &
$-\frac{\omega}{96}+\frac{5}{192}$ & $-\frac{1}{8}$ &
$-\frac{1}{12}$ & $-\frac{1}{12}$
\\
\hline {\bf{d}} & $0$ & $0$ & $\frac{5}{8}$ & $-\frac{23}{96}$ &
$-\frac{13}{192}$ & $-\frac{1}{64}$ & $-\frac{1}{48}$ &
$-\frac{1}{48}$
\\
\hline
\end{tabular}
\end{center}
\end{table}

\clearpage

Here we present the zero-potential coefficients $\alpha_{i}$ in the
generating points of phase diagram shown in Fig.~\ref{anianipdd1d2}.

\begin{table}[h]
\begin{center}
\small \caption{Zero-potentials coefficients for the phase diagram
shown in Fig.~\ref{anianipdd1d2}.} \label{tb4-tab1}
\renewcommand\baselinestretch{1,5}\small\normalsize
\begin{tabular}{|l|c|c|c|c|c|c|}
\hline Points $(d_{1},d_{2})$ & $(0,0)$ & $(2,0)$ & $(0,1)$ &
$(2,1)$ & $(0,4)$ & $(2,4)$
\\
\hline $\alpha_{1}$ & $-\frac{83}{1536}$ & $-\frac{259}{3072}$ &
$-\frac{79}{1536}$ & $-\frac{253}{3072}$ & $-\frac{67}{1536}$ &
$-\frac{217}{3072}$
\\
\hline $\alpha_{2}$ & $-\frac{25}{1536}$ & $-\frac{245}{3072}$ &
$-\frac{29}{1536}$ & $-\frac{251}{3072}$ & $-\frac{41}{1536}$ &
$-\frac{287}{3072}$
\\
\hline $\alpha_{3}$ & $-\frac{47}{1536}$ & $-\frac{331}{3072}$ &
$-\frac{43}{1536}$ & $-\frac{325}{3072}$ & $-\frac{31}{1536}$ &
$-\frac{289}{3072}$
\\
\hline $\alpha_{4}$ & $-\frac{25}{1536}$ & $-\frac{245}{3072}$ &
$-\frac{29}{1536}$ & $-\frac{251}{3072}$ & $-\frac{41}{1536}$ &
$-\frac{287}{3072}$
\\
\hline $\alpha_{5}$ & $\frac{53}{3072}$ & $\frac{137}{8192}$ &
$\frac{19}{1024}$ & $\frac{139}{8192}$ & $\frac{3}{128}$ &
$\frac{91}{8192}$
\\
\hline $\alpha_{6}$ & $-\frac{53}{3072}$ & $-\frac{335}{24576}$ &
$-\frac{19}{1024}$ & $-\frac{365}{24576}$ & $-\frac{3}{128}$ &
$-\frac{91}{8192}$
\\
\hline $\alpha_{7}$ & $\frac{53}{3072}$ & $\frac{401}{8192}$ &
$\frac{19}{1024}$ & $\frac{403}{8192}$ & $\frac{3}{128}$ &
$\frac{1117}{24576}$
\\
\hline $\alpha_{8}$ & $-\frac{53}{3072}$ & $-\frac{571}{24576}$ &
$-\frac{19}{1024}$ & $-\frac{589}{24576}$ & $-\frac{3}{128}$ &
$-\frac{139}{6144}$
\\
\hline $\alpha_{9}$ & $-\frac{19}{3072}$ & $\frac{101}{2048}$ &
$-\frac{19}{3072}$ & $\frac{13}{256}$ & $-\frac{19}{3072}$ &
$\frac{1451}{24576}$
\\
\hline
\end{tabular}
\end{center}
\end{table}
\renewcommand\baselinestretch{1}\small\normalsize


\begin{thebibliography}{99}
\bibitem{TBSL1}
J.M. Tranquada, D.J. Buttrey, V. Sachan, and J.E. Lorenzo,
{\em{Simultaneous Ordering of Holes and Spins in
$La_{2}NiO_{4.125}$}}, Phys. Rev. Lett. {\bf 73}, 1003 (1994)

\bibitem{TSANU1}
J.M. Tranquada, B.J. Sternlieb, J.D. Axe, Y. Nakamura, and S.
Uchida, {\em{Evidence for stripe correlations of spins and holes in
copper oxide superconductors}}, Nature (London) {\bf 375}, 561
(1995)

\bibitem{KNSZGC1}
K. Kern, H. Niehus, A. Schatz, P. Zeppenfeld, J. Goerge, and G.
Comsa, {\em{Long-Range Spatial Self-Organization in the
Adsorbate-Induced Restructuring of Surfaces: Cu$\{ 110 \} - (2
\times 1 )$O}}, Phys. Rev. Lett. {\bf 67}, 855 (1991)

\bibitem{SW1}
M. Seul and R. Wolfe,
{\em{Evolution of disorder in magnetic stripe domains. I. Transverse instabilities
and disclination unbinding in lamellar patterns}},
Phys. Rev. A {\bf 46}, 7519 (1992);
M. Seul and R. Wolfe,
{\em{volution of disorder in two-dimensional stripe patterns: ``smectic'' instabilities
and disclination unbinding}},
Phys. Rev. Lett. {\bf 68}, 2460 (1992);
R. Allenspach, M. Stampanoni, and A. Bischof,
{\em{Magnetic domains in thin epitaxial $Co/Au(111)$ films}},
Phys. Rev. Lett. {\bf 65}, 3344 (1990);
R. Allenspach and A. Bischof,
{\em{Magnetization direction switching in $Fe/Cu(100)$ epitaxial films:
temperature and thickness dependence}},
Phys. Rev. Lett. {\bf 69}, 3385 (1992)

\bibitem{MP1}
G. Malescio and G. Pellicane, {\em{Stripe phases from isotropic
repulsive interactions}}, Nature Materials {\bf 2}, 97 (2003)

\bibitem{Sasaki1}
K. Sasaki, {\em{Lattice gas model for striped structures of adatom
rows on surfaces}}, Surf. Sci. {\bf 318}, L1230 (1994)

\bibitem{BMWB1}
I. Booth, A. B. MacIsaac, J. P. Whitehead, K. De'Bell,
{\em{Domain Structures in Ultrathin Magnetic Films}},
Phys. Rev. Lett. {\bf 75}, 950 (1995);
J. Arlett, J. P. Whitehead, A. B. MacIsaac, K. De'Bell,
{\em{Phase diagram for the striped phase in the two-dimensional dipolar
Ising model}},
Phys. Rev. B {\bf 54}, 3394 (1996);
A. D. Stoycheva and S. J. Singer,
{\em{Stripe melting in a two-dimensional system with competing interactions}},
Phys. Rev. Lett. {\bf 84}, 4657 (2000)

\bibitem{SDSS1}
A.W. Sandvik, S. Daul, R.R.P. Singh, and D.J. Scalapino,
{\em{Striped Phase in a Quantum $XY$ Model with Ring Exchange}},
Phys. Rev. Lett. {\bf 89}, 247201 (2002)

\bibitem{Spivak1}
B. Spivak,
{\em{Phase separation in the two-dimensional electron liquid
in MOSFET's}},
Phys. Rev. B {\bf 67}, 125205 (2003);
B. Spivak and S. A. Kivelson,
{\em{Phases intermediate between a two-dimensional electron liquid
and  Wigner crystal}},
Phys. Rev. B {\bf 70}, 155114 (2004)

\bibitem{PR1}
D. Poilblanc and T.M. Rice, {\em{Charged solitons in the
Hartree-Fock approximation to the large-$U$ Hubbard model}}, Phys.
Rev. B {\bf 39}, 9749 (1989)

\bibitem{ZG1}
J. Zaanen and O. Gunnarsson, {\em{Charged magnetic domain lines and
the magnetism of high-$T_{c}$ oxides}}, Phys. Rev. B {\bf 40}, 7391
(1989)

\bibitem{Machida1}
K. Machida, {\em{Magnetism in $La_2CuO_4$ based compounds}}, Physica
C {\bf 158}, 192 (1989)

\bibitem{KMNF1}
M. Kato, K. Machida, H. Nakanishi, and M. Fujita, {\em{Soliton
lattice modulation of incommensurate spin density wave in two
dimensional Hubbard model --- a mean field study}}, J. Phys. Soc.
Jpn., {\bf 59}, 1047 (1990)

\bibitem{Oles1}
A.M. Ole{\'{s}}, {\em{Stripe phases in high-temperature
superconductors}}, Acta Physica Polonica B {\bf 31}, 2963 (2000)

\bibitem{WS1}
S.R. White and D.J. Scalapino, {\em{Density Matrix Renormalization
Group Study of the Striped Phase in the 2D $t$--$J$ Model}}, Phys.
Rev. Lett. {\bf 80}, 1272 (1998)

\bibitem{WS2}
S.R. White and D.J. Scalapino, {\em{Energetics of Domain Walls in
the 2D $t$--$J$ Model}}, Phys. Rev. Lett. {\bf 81}, 3227 (1998)

\bibitem{TV1}
J. M. Tipper and K. J. E. Vos,
{\em{Formation of stripes and incommensurate peaks in the orthorombic phase
of underdoped $La_{2-x}Sr_{x}CuO_{4}$}},
Phys. Rev. B {\bf{67}}, 144511 (2003)

\bibitem{EKT1}
V.J. Emery, S.A. Kivelson, and J.M. Tranquada, {\em{Stripe phases in
high-temperature superconductors}}, Proc. Natl. Acad. Sci. USA {\bf
96}, 8814 (1999)

\bibitem{ZH1}
N.G. Zhang and C.L. Henley, {\em{Stripes and holes in a
two-dimensional model of spinless fermions or hardcore bosons}},
Phys. Rev. B {\bf{68}}, 014506 (2003)

\bibitem{LFB1}
R. Lema{\'{n}}ski, J.K. Freericks, and G. Banach, {\em{Stripe Phases
in the Two-Dimensional Falicov--Kimball Model}}, Phys. Rev. Lett.
{\bf 89}, 196403 (2002)

\bibitem{LEFK1}
U. L\"{o}w, V. J. Emery, K. Fabricius, and S. A. Kivelson,
{\em{Study of an Ising model with competing long- and short-range interactions}},
Phys. Rev. Lett. {\bf{72}}, 1918 (1994)

\bibitem{VS1}
D. Valdez-Balderas and D. Stroud,
{\em{Superconductivity versus phase separation, stripes, and
checkerboard ordering: a two-dimensional Monte Carlo study}},
Phys. Rev. B {\bf{72}}, 214501 (2005)

\bibitem{BYM1}
C. Buhler, S. Yunoki, and A. Moreo, {\em{Magnetic Domains and
Stripes in a Spin-Fermion Model for Cuprates}}, Phys. Rev. Lett.
{\bf 84}, 2690 (2000)

\bibitem{LFB2}
R. Lema{\'{n}}ski, J.K. Freericks, and G. Banach, {\em{Charge
Stripes due to Electron Correlations in the Two-Dimensional Spinless
Falicov--Kimball Model}}, J. Stat. Phys. {\bf 116}, 699 (2004)

\bibitem{ZH2}
C.L. Henley and N.G. Zhang, {\em{Spinless fermions and charged
stripes at the strong-coupling limit}}, Phys. Rev. B {\bf{63}},
233107 (2001)

\bibitem{FLU1}
J.K.Freericks, E.H. Lieb, and D. Ueltschi, {\em{Phase separation due
to quantum mechanical correlations}}, Phys. Rev. Lett. {\bf 88},
106401 (2002).

\bibitem{RNO1}
M. Raczkowski, B. Normand, and A. M. Ole{\'{s}}, {\em{Vertical and
diagonal stripes in the extended Hubbard model}}, Phys. Stat. Sol.
(b) {\bf{236}}, 376 (2003)

\bibitem{DJ2}
V. Derzhko, J. J{\c{e}}drzejewski, {\em{Formation of charge-stripe
phases in a system of spinless fermions or hardcore bosons}},
Physica A {\bf{349}}, 511 (2005).

\bibitem{DJ3}
V. Derzhko, J. J{\c{e}}drzejewski, {\em{Charge-stripe phases versus
a weak anisotropy of nearest-neighbor hopping}},
arXiv:cond-mat/0509698.

\bibitem{Derzhko}
V. Derzhko, {\em{Influence of anisotropic next-nearest-neighbor
hopping on diagonal charge-striped phases}}, arXiv:cond-mat/0511557.

\bibitem{Hubbard1}
J. Hubbard, Proc. Roy. Soc. London A {\bf{276}}, 238 (1963);
{\bf{277}}, 237 (1964); {\bf{281}}, 401 (1964).

\bibitem{Gutzwiller1}
M. C. Gutzwiller, {\em{Correlation of electrons in a narrow $s$
band}}, Phys. Rev. {\bf{137}}, A1726 (1965).

\bibitem{FK1}
L.M. Falicov and J.C. Kimball, {\em {Simple model for
semiconductor-metal transitions: $SmB_{6}$ and transition-metal
oxides}}, Phys. Rev. Lett. {\bf 22}, 997 (1969).

\bibitem{KL1}
T. Kennedy and E. H. Lieb, {\em{An itinerant electron model with
crystalline or magnetic long range order}}, Physica A {\bf{138}},
320 (1986).

\bibitem{Lieb1}
E. H. Lieb, {\em{A model for crystallization: a variation on the
Hubbard model}}, Physica A {\bf{140}}, 240 (1986)

\bibitem{FZ1}
J. K. Freericks, V. Zlati{\'{c}}, {\em{Exact dynamical mean-field
theory of the Falicov-Kimball model}}, Rev. Mod. Phys. {\bf{75}},
1333 (2003)

\bibitem{CF1}
H. {\v{C}}en{\v{c}}arikov{\'{a}}, P. Farka{\v{s}}ovsk{\'{y}},
{\em{The influence of correlated hopping on the ground-state
properties of the two-dimensional Falicov--Kimball model}}, Phys.
Stat. Sol. (b) {\bf{242}}, 2061 (2005)

\bibitem{FCT1}
P. Farka{\v{s}}ovsk{\'{y}}, H. {\v{C}}en{\v{c}}arikov{\'{a}}, N.
Toma\v{s}ovi\v{c}ov\'{a}, {\em{Ground-states of the
three-dimensional Falicov--Kimball model}}, Eur. J. Phys. B
{\bf{45}}, 479 (2005)

\bibitem{Farkasovsky1}
P. Farka{\v{s}}ovsk{\'{y}}, {\em{Ground-state properties of the
Falicov-Kimball model in one and two dimensions}}, Eur. J. Phys. B
{\bf{20}}, 209 (2001)

\bibitem{GMMU}
C. Gruber, N. Macris, A. Messager, D. Ueltschi, {\em{Ground states
and flux configurations of the two-dimensional Falicov--Kimball
model}}, J. Stat. Phys. {\bf 86}, 57 (1997).

\bibitem{GL1}
Z. Gajek, R. Lema{\'{n}}ski, {\em{Correlated hopping in the 1D
Falicov--Kimball model}}, Acta Phys. Polon. B {\bf{32}}, 3473 (2001)

\bibitem{WL1}
J. Wojtkiewicz, R. Lema{\'{n}}ski, {\em{Ground states of the
Falicov--Kimball model with correlated hopping}}, Phys. Rev. B
{\bf{64}}, 233103 (2001)

\bibitem{WL2}
J. Wojtkiewicz, R. Lema{\'{n}}ski, {\em{2D Falicov--Kimball model
with correlated hopping in the large $U$ limit}}, Acta Phys. Polon.
B {\bf{32}}, 3467 (2001)

\bibitem{BFH1}
U. Brandt, A. Fledderjohann, and G. H{\"{u}}lsenbeck, {\em{New
phases in a spin-1/2 Falicov–-Kimball model}}, Z. Phys. B 81, 409
(1990).

\bibitem{BF1}
U. Brandt, A. Fledderjohann, {\em{Existence of a phase transition in
a spin-1/2 Falicov–-Kimball model}}, Z. Phys. B 87 (1992) 111.

\bibitem{LW1}
R. Lema{\'{n}}ski, J. Wojtkiewicz, {\em{Ground states of the
spin-1/2 Falicov--Kimball model}}, Phys. Stat. Sol. (b) {\bf{236}},
408 (2003)

\bibitem{Lemanski1}
R. Lema{\'{n}}ski, {\em{Model of charge and magnetic order formation
in itinerant electron system}}, Phys. Rev. B {\bf{71}}, 035107
(2005)

\bibitem{JL1}
J. J{\c{e}}drzejewski and R. Lema{\'{n}}ski, {\em{Falicov--Kimball
models of collective phenomena in solids (a concise guide)}}, Acta
Phys. Pol. B {\bf{32}}, 3243 (2001).

\bibitem{MM1}
A. Messager and S. Miracle-Sol{\'{e}}, {\em{Low temperature states
in the Falicov--Kimball model}}, Rev. Math. Phys. {\bf 8}, 271
(1996).

\bibitem{GJL}
C. Gruber, J. J\c{e}drzejewski and P. Lemberger, {\em{Ground States
of the Spinless Falicov--Kimball Model. II}}, J. Stat. Phys. {\bf
66}, 913 (1992).

\bibitem{GM1}
C. Gruber and N. Macris, {\em{The Falicov--Kimball model: a review
of exact results and extensions}}, Helv. Phys. Acta {\bf{69}}, 850
(1996).

\bibitem{Gruber1}
C. Gruber, {\em{Falicov--Kimball models: a partial review of the
ground states problem}}, arXiv:cond-mat/9811299

\bibitem{GU1}
C. Gruber and D. Ueltschi, {\em{The Falicov--Kimball model}},
arXiv:math-ph/0502041 (2005).

\bibitem{DFF1}
N. Datta, R. Fern{\'{a}}ndez, J. Fr{\"{o}}hlich,
{\em{Low-temperature Phase Diagrams of quantum lattice systems. I.
Stability for quantum pertrubations of classical systems with
finitely-many ground states}}, J. Stat. Phys. {\bf{84}}, 455 (1996).

\bibitem{DFFR1}
N. Datta, R. Fern{\'{a}}ndez, J. Fr{\"{o}}hlich, L. Rey-Bellet,
{\em{Low-temperature Phase Diagrams of quantum lattice systems. II.
Convergent pertrubation expansions and stability in systems with
infenite degeneracy}}, Helv. Phys. Acta. {\bf{69}}, 752 (1996).

\bibitem{FR1}
J. Fr{\"{o}}hlich, L. Rey-Bellet, {\em{Low-temperature Phase
Diagrams of quantum lattice systems. III. Examples}}, Helv. Phys.
Acta. {\bf{69}}, 821 (1996).

\bibitem{DFF2}
N. Datta, R. Fern{\'{a}}ndez, J. Fr{\"{o}}hlich, {\em{Effective
Hamiltonians and phase diagrams for tight-binding models}}, J. Stat.
Phys. {\bf{96}}, 545 (1999).

\bibitem{Slawny}
J. Slawny, {\em{Low-temperature properties of classical lattice
systems: phase transitions and phase diagrams}}, in: C. Domb, J.
Lebowitz (Eds.), Phase Transitions and Critical Phenomena, vol. 11,
Academic Press, London, New York, 1985.

\bibitem{Kennedy1}
Tom Kennedy, {\em{Some rigorous results on the ground states of the
Falicov--Kimball model}}, Rev. Math. Phys. {\bf 6}, 901 (1994).

\bibitem{FLU2}
J.K.Freericks, E.H. Lieb, and D. Ueltschi, {\em{Segregation in the
Falicov--Kimball model}}, Commun. Math. Phys. {\bf 227}, 243 (2002).

\bibitem{DJ1}
V. Derzhko and J. J{\c{e}}drzejewski, {\em{From phase separation to
long-range order in a system of interacting electrons}}, Physica A
{\bf 328}, 449 (2003)

\bibitem{Kennedy2}
T. Kennedy, {\em{Phase separation in the neutral Falicov--Kimball
model}}, J. Stat. Phys. {\bf 91}, 829 (1998)

\bibitem{SDHC1}
C. Morais Smith, Yu.A. Dimashko, N. Hasselmann, and A.O. Caldeira,
{\em{Dynamics of stripes in doped antiferromagnets}}, Phys. Rev. B
{\bf 58}, 453 (1998)

\bibitem{Wojtkiewicz}
J. Wojtkiewicz, {\em{Phase diagram of the two-dimensional
$t$-$t^{\prime}$ Falicov-Kimball model}}, arXiv:cond-mat/0310043.





\end{thebibliography}
\end{document}